\documentclass[3p]{elsarticle}

\usepackage{amsmath, amssymb, epsfig, bm, graphicx}
\usepackage[breaklinks]{hyperref}

\journal{Physics Reports}

\begin{document}

\begin{frontmatter}

  \title{Coherent and dissipative dynamics at quantum phase transitions}

  \author{Davide Rossini}
  \ead{davide.rossini@unipi.it}

  \author{Ettore Vicari}
  \ead{ettore.vicari@unipi.it}

  \address{Dipartimento di Fisica dell’Universit\`a di Pisa and INFN,
    Largo Pontecorvo 3, I-56127 Pisa, Italy}

  \begin{abstract}
    The many-body physics at quantum phase transitions shows a subtle
    interplay between quantum and thermal fluctuations, emerging in
    the low-temperature limit. In this review, we first give a
    pedagogical introduction to the equilibrium behavior of systems in
    that context, whose scaling framework is essentially developed by
    exploiting the quantum-to-classical mapping and the
    renormalization-group theory of critical phenomena at continuous
    phase transitions. Then we specialize to protocols entailing the
    out-of-equilibrium quantum dynamics, such as instantaneous
    quenches and slow passages across quantum transitions. These are
    mostly discussed within dynamic scaling frameworks, obtained by
    appropriately extending the equilibrium scaling laws.  We review
    phenomena at first-order quantum transitions as well, whose
    peculiar scaling behaviors are characterized by an extreme
    sensitivity to the boundary conditions, giving rise to
    exponentials or power laws for the same bulk system. In the last
    part, we cover aspects related to the effects of dissipative
    interactions with an environment, through suitable generalizations
    of the dynamic scaling at quantum transitions. The presentation is
    limited to issues related to, and controlled by, the quantum
    transition developed by closed many-body systems, treating the
    dissipation as a perturbation of the critical regimes, as for the
    temperature at the zero-temperature quantum transition. We focus
    on the physical conditions giving rise to a nontrivial interplay
    between critical modes and various dissipative mechanisms,
    generally realized when the involved mechanism excites only the
    low-energy modes of the quantum transitions.
  \end{abstract}


\end{frontmatter}

\tableofcontents

\vspace*{5mm}

\hrulefill

\section{Plan of the review}
\label{introplan}

\subsection{Introduction}
\label{intro}

The quantum evolution of many-body systems has been considered a
challenging problem for long time. The recent experimental progress
in the realization, control, and readout of the coherent dynamics of
(quasi) isolated, strongly correlated, quantum systems has made this
issue particularly relevant for experiments and realizations of
physical devices for quantum computing.

Quantum phase transitions (or, more compactly, quantum transitions)
separating different phases of closed systems are striking signatures
of many-body collective behaviors (see \cite{Sachdev-book} for an
introduction to this issue).  They are essentially related to the
properties of the low-energy spectrum of the system, and in particular
the ground state.  They give rise to notable long-range quantum
correlations and scaling behaviors similar to those observed at
classical phase transitions. These emerging critical scenarios entail
corresponding out-of-equilibrium phenomena around the quantum
transition.  Indeed the universal features of the quantum transitions
could also be probed by out-of-equilibrium dynamic protocols, for
example analyzing the effects of changes of the Hamiltonian parameters
across them, which may be instantaneous or extremely slow. The
out-of-equilibrium dynamics at quantum transitions can be also
addressed within dynamic scaling frameworks, which allow us to
identify the critical regimes controlled by the global and universal
properties of the quantum transition, such as the nature of the order
parameter and the associated symmetry-breaking pattern.

Phenomena related to quantum transitions are important to understand
the physics behind several low-energy situations, which include
fermionic and bosonic gases, atomic systems in optical lattices,
high-Tc superconductivity, quantum-Hall systems, low lying magnetic
and spin fluctuations of some insulators and crystals, such as heavy
fermion compounds, etc. (see, e.g., Refs.~\cite{Sachdev-book, SGCS-97}
and references therein).

We review issues related to the equilibrium and out-of-equilibrium
dynamics of many-body systems at quantum transitions. They are mostly
discussed within dynamic scaling frameworks, which have been
essentially developed by exploiting the quantum-to-classical mapping
and the renormalization-group theory of critical phenomena at
continuous phase transitions. In this context, first-order quantum
transitions are addressed as well, with emphasis on the dynamic
scaling behaviors emerging in finite-size systems: these appear even
more complex than those at continuous quantum transitions, due to
their extreme sensitivity to the boundary conditions, which may give
rise to exponentials or power laws for the same bulk system.

We also cover aspects related to the effects of dissipative
interactions with an environment, which are unavoidable in actual
experiments. The analysis of dissipative perturbations is presented
within appropriate extensions of the dynamic scaling framework at
quantum transitions, allowing us to identify a low-dissipation regime
where the dissipative perturbations can be incorporated into an
extended dynamic scaling theory controlled by the universality class
of the quantum transition of the isolated system.  We limit our
presentation to issues related to, and controlled by, the quantum
transition developed by closed many-body systems, treating dissipative
mechanisms as perturbations, like the temperature at the
zero-temperature quantum transition.  In other words, we will not
treat dynamic critical phenomena arising from the dissipative
mechanism itself.  The outlined dynamic scaling framework, even the
one extended to allow for dissipative mechanisms, applies when the
involved mechanism excites only the low-energy critical modes of the
quantum transitions: hereafter we will essentially focus on the
corresponding perturbed critical regimes.

To be more precise, we consider dynamical phenomena emerging in the critical
regimes of quantum-many body systems, i.e., when the Hamiltonian driving the
unitary dynamics is close to a quantum transition and only low-energy
critical modes are effectively excited.  Within this regime,
dynamic scaling frameworks, similar to those developed for equilibrium
quantum criticality, provide valuable insight for out-of-equilibrium
quantum evolutions as well, such as when quenching the Hamiltonian
parameters and/or in the presence of dissipative perturbations.
A full understanding of the nonequilibrium quantum
dynamics in more general conditions calls for further tools and
concepts, which are currently object of intensive investigation.

\subsection{Plan}
\label{plan}

This review focuses on the quantum dynamics of many-body systems close
to continuous or first-order quantum transitions (CQTs and FOQTs,
respectively), separating their zero-temperature quantum phases.  The
review can be divided into three parts: In
sections~\ref{cfoqts}-\ref{quinfoatqts}, we discuss the equilibrium
features of quantum transitions (QTs).
Sections~\ref{dynqts}-\ref{centralspin} deal with the
out-of-equilibrium unitary dynamics, controlled by the global and
universal properties of the QTs.  In the last
sections~\ref{sec:MEq}-\ref{measQT}, we extend the analysis to quantum
systems subject to dissipative interactions with an environment,
discussing the effects of their perturbations when the many-body
systems are close to a QT.  The detailed plan of the review
follows.%~\footnote{Most of the acronyms that will be used throughout
%  the whole review are defined in this plan.}

\begin{itemize}
  
\item[$\bullet$] In Sec.~\ref{cfoqts}, we begin with a compact
  introduction to QTs, distinguishing between CQTs and FOQTs.  We
  outline their main features, such as the typical universal power
  laws characterizing continuous transitions, and the infinite-volume
  discontinuities of the first-order transitions.  We discuss the
  Landau-Ginzburg-Wilson (LGW) framework to study critical phenomena,
  and some examples of QTs which may depart from this scenario, such
  as the so-called topological transitions.

\item[$\bullet$] In Sec.~\ref{models} we present a number of
  prototypical models undergoing CQTs and FOQTs, useful to place the
  quantum phenomena discussed in this review into concrete grounds. We
  introduce quantum Ising-like or XY systems, discussing their quantum
  phase diagram and the nature of their transitions; in the case of
  one-dimensional (1$d$) models, we comment on their relation with the
  Kitaev fermionic wire. We also present other physically interesting
  systems, such as Bose-Hubbard (BH) models, with their bosonic
  condensation phenomena and Mott phases, as well as quantum rotor and
  Heisenberg spin models.

\item[$\bullet$] In Sec.~\ref{escalingcqt} we outline the scaling
  theory describing the universal behaviors of systems at CQTs, in
  equilibrium conditions around the quantum critical point (QCP). For
  this purpose, the quantum-to-classical mapping and the
  renormalization-group (RG) theory of critical phenomena are employed
  as guiding ideas.  We present a detailed analysis of the RG scaling
  ansatz in the thermodynamic limit and in the finite-size scaling
  (FSS) limit. We also extend the theory to the case the system is
  subject to an external inhomogeneous potential, such as for trapped
  particle gases in cold-atom experiments.

\item[$\bullet$] In Sec.~\ref{escalingfoqt} we outline the appropriate
  scaling theory for FOQTs in equilibrium conditions. In particular,
  we discuss how systems at FOQTs develop a peculiar FSS,
  characterized by an extreme sensitivity to the boundary conditions
  (BC), which may give rise to exponential or power-law behaviors with
  respect to the system size. We provide a unified view of the FSS at
  QTs, including both CQTs and FOQTs, for which the main difference
  between CQTs and FOQTs is essentially related to their sensitivity
  to the BC.  The latter feature of FOQTs lies at the basis of the
  existence of CQTs induced by localized defects, whose tuning can
  change the bulk phase, unlike the standard scenarios at CQTs.
  
\item[$\bullet$] In Sec.~\ref{qinfosec} we present an overview of some
  concepts founded on the recently developing quantum information
  science, which have been proven useful to spotlight the presence of
  singularities at QTs in many-body systems, such as the ground-state
  fidelity, the Loschmidt amplitude, the quantum Fisher information,
  various indicators of quantum correlations (as entanglement or
  quantum discord), etc.

\item[$\bullet$] In Sec.~\ref{quinfoatqts} we discuss the behaviors of
  the quantum-information related quantities introduced in
  Sec.~\ref{qinfosec} at CQTs and FOQTs, showing that they present
  peculiar scaling behaviors, in particular focusing on the
  ground-state fidelity and the bipartite entanglement.

\item[$\bullet$] Sec.~\ref{dynqts} is the first of the sections
  dedicated to the out-of-equilibrium quantum dynamics at QTs.  We
  outline the dynamic scaling theory describing the out-of-equilibrium
  dynamics arising from soft instantaneous quantum quenches at CQTs,
  when only critical low-energy modes get excited by the process. For
  this purpose, we assume a general hypothesis of homogeneous scaling
  laws, which are then specialized to the thermodynamic limit and the
  FSS limit.  Beside the standard quantum correlations of local
  operators, we discuss the behavior of more complex quantities such as
  the Loschmidt echo, the work fluctuations, the bipartite entanglement.
  We finally shed light on the possible signatures of QTs in hard quantum
  quenches, where excitations with higher energy are also involved in
  the process.

\item[$\bullet$] Sec.~\ref{KZdynamics} deals with another class of
  dynamic protocols, where the Hamiltonian parameters are slowly
  changed across CQTs, analogous to those considered to generate the
  so-called Kibble-Zurek (KZ) mechanism, related to the abundance of
  defects after crossing a CQT.  Starting again from a set of dynamic
  homogeneous scaling laws, we outline the corresponding dynamic KZ
  scaling theory in the thermodynamic limit and in the FSS limit, controlled
  by the universality class of the underlying CQT. This describes the
  growth of an out-of-equilibrium dynamics even in the limit of very slow
  changes of the Hamiltonian parameters, because large-scale modes are
  unable to equilibrate as the system changes phase. We also overview
  the analytical and numerical evidence of the KZ scaling theory,
  reported in the literature.

\item[$\bullet$] In Sec.~\ref{foqtdynamics} we extend the
  out-of-equilibrium dynamic scaling theory to FOQTs: analogously as
  at equilibrium, the dynamic behavior across a FOQT is dramatically
  sensitive to the BC, giving rise to nonequilibrium evolutions with
  exponential or power-law time scales (unlike CQTs, where the time
  scaling is generally independent of the type of boundaries). We
  address both soft quantum quenches and slow KZ protocols. We also
  discuss the conditions under which the system at the FOQT
  effectively behaves rigidly as a few-level quantum system.

\item[$\bullet$] In Sec.~\ref{centralspin} we begin considering
  decoherence phenomena, which generally arise when a given quantum
  system interacts with an environmental many-body system.  We start
  from the simplest central spin model, describing a single qubit
  interacting with a many-body system. We discuss how the decoherence
  dynamics develops when the many-body system is close to a QT,
  emphasizing the qualitative changes with respect to systems in
  normal conditions.  Again, the analysis is essentially developed
  within a dynamic scaling framework.

\item[$\bullet$] Sec.~\ref{sec:MEq} is the first of a series of
  sections discussing the behavior of quantum systems in the
  presence of dissipative interactions with an environment. We
  introduce the so-called Lindblad framework allowing for the 
  dissipation through a system-bath coupling scheme that respects
  some assumptions (typically based on the weak coupling
  approximation). These lead to a well behaved Markovian master
  equation for the density matrix operator of the system.
  
\item[$\bullet$] In Sec.~\ref{dissQT} we address the effects of
  dissipative interactions on the out-of-equilibrium dynamics of
  systems at QTs, both CQTs and FOQTs.  We discuss how the dynamic
  scaling theory of closed systems can be extended to take into
  account the perturbations arising from the dissipative
  interactions. We concentrate on the effects emerging at quantum
  quenches, identifying a low-dissipation regime where a dissipative
  dynamic scaling behavior emerges as well.

\item[$\bullet$] In Sec.~\ref{KZdiss} we extend the discussion to slow
  KZ-like protocols. In particular, we describe the conditions under
  which a dynamic KZ scaling behavior can be still observed, even in
  the presence of dissipation, showing that this requires a particular
  low-dissipation limit.

\item[$\bullet$] Sec.~\ref{measQT} deals with the effects of local
  measurements on systems at QTs. This issue is discussed within a
  dynamic scaling framework, which shows that a peculiar dynamic
  scaling behavior emerges even in the presence of such type of
  decoherence perturbations, that is controlled by the universality
  class of the CQT.

\item[$\bullet$] The concluding Sec.~\ref{applications} contains a
  brief summary of this review, with possible motivations that could
  stimulate new experimental investigations in the field of quantum
  many-body physics.  A brief list of the envisioned platforms where
  such studies are possible is finally presented.
\end{itemize}

We shall emphasize that the topics covered in this review deal with
phenomena that genuinely arise from the existence of a QT in closed
many-body systems.  On top of that, we consider general mechanisms,
such as dissipative interactions or local measurements, that may
destroy the critical features of the QTs, and discuss the regimes when
they act as a perturbation that can only effectively excite the
low-energy critical modes of the underlying quantum transition
bringing them out of equilibrium.
As a consequence, some interesting and flourishing issues in the realm
of quantum many-body physics will not be touched.  We warmly direct
the interested reader to the good and recent review papers cited
below.

\begin{table}[!b]
  \small
  \begin{center}
    \begin{tabular}{ll|ll}
      \hline\hline
      ANNNI & Anisotropic next-to-nearest neighbor Ising &
      LDA   & Local density approximation \\
      ABC   & Antiperiodic boundary conditions &
      LOCC  & Local operations and classical communication \\
      BKT   & Berezinskii-Kosterlitz-Thouless &
      MC    & Monte Carlo  \\ 
      BEC   & Bose-Einstein condensation &
      MI    & Mott insulator  \\
      BH    & Bose-Hubbard &
      NPRG  & Nonperturbative renormalization-group  \\
      BC    & Boundary conditions &
      1$d$  & One-dimensional  \\
      CFT   & Conformal field theory &
      OBC   & Open boundary conditions  \\
      CQT   & Continuous quantum transition &
      OFBC  & Opposite and fixed boundary conditions  \\
      DMRG  & Density-matrix renormalization group &
      PBC   & Periodic boundary conditions  \\
      ETH   & Eigenstate thermalization hypothesis  &
      QCP   & Quantum critical point  \\
      EPR   & Einstein-Podolsky-Rosen &
      QED   & Quantum electrodynamics  \\      
      EFBC  & Equal and fixed boundary conditions  &
      QT    & Quantum transition  \\
      FSS   & Finite-size scaling &
      QLRO  & Quasi-long-range order  \\
      FOQT  & First-order quantum transition  &
      RG    & Renormalization-group \\
      FBC   & Fixed boundary conditions  &
      SFT   & Statistical field theory  \\
      GKLS  & Gorini-Kossakowski-Lindblad-Sudarshan  &
      3$d$  & Three-dimensional  \\
      HI    & Haldane insulator  &
      TSS   & Trap-size scaling \\
      HT    & High-temperature  &
      2$d$  & Two-dimensional  \\
      KZ    & Kibble-Zurek &
      VBS   & Valence-bond-solid  \\
      LGW   & Landau-Ginzburg-Wilson && \\
      \hline\hline
    \end{tabular}
    \caption{List of abbreviations used in this review.}
    \label{Table:abbreviations}
  \end{center}
\end{table}

We are not going to discuss QTs in the presence of disorder, which
have been already addressed in Refs.~\cite{Sachdev-book, Vojta-19}.
Other important developments, concerning the quantum dynamics of
many-body systems, are not directly related to the presence of an
equilibrium QT, and thus will not be covered hereafter.  Among them,
we mention issues related to the thermalization of closed
systems~\cite{Dziarmaga-10, PSSV-11, EFG-15, GE-16, DKPR-16, MIKU-18},
whether it is
eventually realized or some properties of the system prevent it, as in
the case of many-body localization~\cite{NH-15, AV-15, VM-16, AP-17,
  AL-18, AABS-19} or integrability of the system~\cite{EF-16, VR-16,
  IMPZ-16, LGS-16}.  We also do not address dynamic transitions that
may be observed after hard quenches~\cite{Zvyagin-16, Heyl-18}, and
phenomena arising from periodically-driven systems~\cite{GH-98,
  BDP-15, Holthaus-15} such as the time crystals~\cite{SZ-17, EMNY-20}.
Moreover, it is not our purpose to deal with dynamic QTs intimately
driven by the presence of dissipation~\cite{CC-13, NA-16, Hartmann-16}
or of local measurements (see, e.g., the pioneering works in
Refs.~\cite{LCF-18, LCF-19, CNPS-19, SRN-19}).

In Table~\ref{Table:abbreviations} we list all the abbreviations
used throughout this review.

\section{Quantum transitions}
\label{cfoqts}
\subsection{General features of continuous and first-order quantum transitions}
\label{qtsintro}

Zero-temperature QTs are phenomena of great interest in modern
physics, both theoretically and experimentally.  In the context of
many-body systems, they are associated with an infinite-volume
nonanalyticity of the ground state of their Hamiltonian with respect
to one of its parameters, coupling constants, etc.  At QTs, such
systems usually undergo a qualitative change in the nature of the
low-energy and large-distance correlations. Therefore, different
quantum phases emerge, characterized by distinctive quantum
properties.  Below we address the most important features of QTs,
which turn out to be useful in the remainder of this review. The
interested reader can find an excellent and exhaustive introduction to
this issue in Sachdev's book~\cite{Sachdev-book}, and also in
Refs.~\cite{SGCS-97, Vojta-03, BKV-05, SK-11, Dutta-etal-book}.

Analogously to thermal (finite-temperature) transitions, it is
generally possible to distinguish the nonanalytic behaviors at QTs
between FOQTs and CQTs. Specifically, QTs are of the first order when the
ground-state properties in the infinite-volume (thermodynamic) limit
are discontinuous across the transition point.  On the other hand,
they are continuous when the ground-state features change continuously
at the transition point, quantum correlation functions develop
a divergent length scale, and the energy spectrum is gapless
in the infinite-volume limit.

At low temperature and close to a CQT, the interplay between quantum
and thermal fluctuations give rise to peculiar scaling behaviors,
which can be described by appropriately extending the
RG theory of critical phenomena (see, e.g.,
Refs.~\cite{WK-74, Fisher-74, Ma-book, BGZ-76, Wegner-76, PP-book,
  Wilson-83,ID-book1, ZinnJustin-book, Cardy-book, Fisher-98, PV-02,
  AM-book}), through the addition of a further imaginary-time
direction associated with the temperature~\cite{Sachdev-book, SGCS-97}.
Like thermal transitions, in many cases CQTs are associated with the
spontaneous breaking of a symmetry, thus related to condensation
phenomena.

Many-body systems at continuous classical and quantum phase
transitions present notable universal critical behaviors, which are
largely independent of the local details of the system. According to
the RG theory of critical phenomena, their universal features are
essentially determined by global properties, such as the spatial
dimensionality, the nature of the order parameter, the symmetry, and
the symmetry-breaking pattern.  Therefore, the universal asymptotic
critical behaviors are shared by a large (universality) class of
models, which may include very different systems, such as particle
gases and spin networks.

Systems at QTs develop equilibrium and dynamic scaling behaviors, in
the thermodynamic and also FSS limits.  Such scaling behaviors at CQTs
are generally characterized by power laws involving universal critical
exponents, such as the divergence of the correlation length, $\xi \sim
|g-g_c|^{-\nu}$, and the corresponding suppression of the energy
difference (gap) of the lowest levels, $\Delta\sim \xi^{-z}$, when a
given Hamiltonian parameter $g$ approaches the critical-point value
$g_c$.  However, there are also notable CQTs characterized by
exponential laws as, for example, the Berezinskii-Kosterlitz-Thouless
(BKT) transition~\cite{B-71, B-72, KT-73, Kosterlitz-74}.  Systems at
FOQTs show equilibrium and dynamic peculiar scaling behaviors, as
well.  However, unlike at CQTs whose critical laws are independent of
the BC, systems at FOQTs may develop power or exponential FSS laws
depending on the BC~\cite{CNPV-14, CPV-15-def, CPV-15-bf}.  For
example one generally observes an exponentially suppressed gap when the
BC do not favor any of the two phases separated by the transition, while
power-law FSS behaviors may arise from other BC, in particular those
giving rise to domain walls.  Within FSS frameworks, the sensitivity
to the BC may be considered as the main difference between CQTs and
FOQTs.

\subsection{The Landau-Ginzburg-Wilson approach to continuous
phase transitions}
\label{LGWappro}

The Landau paradigm~\cite{Landau-37a, Landau-37b} provides an
effective framework to describe phase transitions. The main idea is to
identify the symmetry breaking related to an ordered phase and the
corresponding order parameter, which is nonzero within the ordered
phase and vanishes at the critical point and in the disordered
phase. This theory can be used both for classical and for quantum
phase transitions, and has been widely applied to predict the phase
diagram of several systems, such as phases of water and magnetic
systems.  Within the Landau framework, the main ideas to describe the
critical behavior at a continuous phase transition are:

\begin{itemize}
\item[$\bullet$] Existence of an {\em order-parameter field} which
  effectively describes the critical modes, whose condensation
  determines the symmetry-breaking pattern.

\item[$\bullet$] {\em Scaling hypothesis}: singularities arise from the
  long-range correlations of the order-parameter field, which develop a
  diverging length scale.

\item[$\bullet$] {\em Universality}: the critical behavior is
  essentially determined by a few global properties, such as the space
  dimensionality, the nature and the symmetry of the order parameter,
  the symmetry breaking, the range of the effective interactions.
\end{itemize}

The RG theory of critical phenomena~\cite{Wilson-71a, Wilson-71b,
  WK-74, Fisher-74} provides a general framework where these features
naturally arise. It considers a RG flow in a Hamiltonian space.  The
critical behavior is associated with a fixed point of the RG flow,
where only a few perturbations are relevant.  The corresponding
positive eigenvalues of the linearized theory around the fixed point
are related to the critical exponents $\nu$, $\eta$, etc.

In that framework, a
quantitative description of many continuous phase transitions
can be obtained by considering an effective LGW
$\Phi^4$ field theory, constructed using the order-parameter field
$\Phi({\bm x})$ and containing up to fourth-order powers of the field
components. Several continuous phase transitions are associated with
LGW theories realizing the same symmetry-breaking pattern.  The
simplest example is the O($N$)-symmetric $\Phi^4$ theory, defined by
the Lagrangian density
\begin{equation}
  {\cal L}_{{\rm O}(N)} = {1\over 2} \partial_\mu {\bm \Phi}({\bm x})\cdot
\partial_\mu {\bm \Phi}({\bm x}) + {1\over 2} r \, {\bm \Phi}({\bm
  x})\cdot {\bm \Phi}({\bm x}) + {1\over 4!} u \big[ {\bm \Phi}({\bm
    x})\cdot {\bm \Phi}({\bm x})\big]^2 + {\bm h}\cdot {\bm \Phi}({\bm
  x}) \,,
\label{HON}
\end{equation}
where ${\bm \Phi}$ is a $N$-component real field. They represent the
so-called $N$-vector universality class.  These $\Phi^4$ theories
describe phase transitions characterized by the symmetry breaking
O($N$)$\rightarrow$O($N-1$).  We mention the Ising universality class
for $N=1$ (which is relevant for the liquid-vapor transition in simple
fluids, for the Curie transition in uniaxial magnetic systems, etc.),
the $XY$ universality class for $N=2$ (which describes the superfluid
transition in $^4$He, the formation of Bose-Einstein condensates in
interacting bosonic gases, transitions in magnets with easy-plane
anisotropy and in superconductors), the Heisenberg universality class
for $N=3$ (describing the Curie transition in isotropic magnets), the
hadronic finite-temperature transition with two light quarks in the
chiral limit for $N=4$.  Moreover, the limit $N\rightarrow 0$
describes the behavior of dilute homopolymers in a good solvent, in
the limit of large polymerization.~\footnote{See, e.g.,
  Refs.~\cite{ZinnJustin-book, PV-02} for reviews of applications of
  the O($N$)-symmetric $\Phi^4$ field theories.}  Beside the
transitions described by O($N$) models, there are also other
physically interesting transitions described by more general LGW
$\Phi^4$ field theories, characterized by complex symmetries and
symmetry-breaking patterns, arising from more involved quartic terms
(see, e.g., Refs.~\cite{ZinnJustin-book, PV-02, VZ-06, Vicari-07}).
For example, this approach has been applied to investigate the
critical behavior of magnets with anisotropy~\cite{Aharony-76,
  CPV-00}, disordered systems~\cite{Harris-74, PV-00}, frustrated
systems~\cite{Kawamura-98, PRV-01, CPS-02, DMT-04, CPPV-04}, spin and
density wave models~\cite{ZDS-02, Sachdev-03, DPV-04, DPV-06, PSV-08,
  Kim-etal-08}, competing orderings giving rise to multicritical
behaviors~\cite{FN-74, NKF-74, KNF-76, CPV-03}, and also the
finite-temperature chiral transition in hadronic matter~\cite{PW-84,
  BPV-03, PV-13-qcd}.

In the field-theoretical LGW approach, the RG flow is determined by a
set of equations for the correlation functions of the order
parameter.  The so-called {\em quantum-to-classical mapping} (see,
e.g., Ref.~\cite{Sachdev-book}) allows us to extend the classical
applications to QTs, so that $d$-dimensional QTs are described
by $(d+1)$-dimensional statistical (quantum) field theories.

\subsection{Power laws approaching the critical point at
  continuous transitions}
\label{cqte}

To fix the ideas, consider a prototypical $d$-dimensional many-body
system at a CQT, characterized by two relevant parameters $r$ and $h$,
which can be defined in such a way that they vanish at the critical
point. The odd parameter $h$ is generally associated with the order
parameter driving the symmetry breaking, as in the
Lagrangian~\eqref{HON}.  The zero-temperature QCP is thus located at
$r = h = 0$ and, of course, $T=0$.  Also assume the presence of a
parity-like ${\mathbb Z}_2$-symmetry, as it occurs, e.g., in QTs
belonging to the Ising or O($N$) vector universality classes, which
separate a paramagnetic phase with $r > 0$ from a ferromagnetic phase
with $r < 0$.  The parameter $r \equiv g-g_c$ is thus associated with
a RG perturbation that is invariant under the symmetry, while $h$ is
associated with the leading odd perturbation.  When approaching the
critical point, the length scale $\xi$ of the critical modes diverges
as
\begin{subequations}
\begin{eqnarray}
  \xi \, \sim & \!\!\! |r|^{-\nu} \quad\:
  & {\rm for} \quad h=0\,, \; T=0 \,, \label{xidiv1}\\
  \xi \, \sim & \!\!\! T^{-1/z}   \quad
  & {\rm for} \quad r=0\,, \; h=0 \,. \label{xidiv2}
\end{eqnarray}
\end{subequations}
These power laws are characterized by universal critical exponents,
namely, the correlation-length exponent $\nu$ and the dynamic exponent
$z$ associated with the time and the temperature, respectively.
Moreover, the low-energy scales vanish. In particular, the
ground-state gap $\Delta$ gets suppressed as
\begin{equation}
  \Delta \, \sim \, \xi^{-z} \, \sim \,
  |r|^{z\nu}\qquad  {\rm for}\quad h=0\,,\; T=0\,.
  \label{gapsuppr}
\end{equation}
The RG dimensions of the perturbations associated with $r$ and $h$ are
related to the critical exponents $\nu$ and $\eta$,
as~\cite{Sachdev-book, ZinnJustin-book, PV-02, CPV-14}
\begin{equation}
  y_r = 1 / \nu \,, \qquad y_h = (d+z+2-\eta)/2 \, .
  \label{ymuh}
\end{equation}
The critical exponent $\eta$ is traditionally introduced to
characterize the space dependence of the critical two-point function
of the order-parameter operator associated with $h$,
\begin{equation}
  G(\bm{x}_1 - \bm{x}_2)\big|_{r=h=T=0}  \sim
  |\bm{x_1} - \bm{x_2}|^{-\zeta}\,,
  \qquad \zeta = 2(d+z - y_h) =  d + z - 2 + \eta\,.   
      \label{gx1x2crit}
\end{equation}

The universal critical exponents $z$, $\nu$, and $\eta$ are shared by
the given universality class of the CQTs, essentially dependent on
some global properties such as the spatial dimensions and the
symmetry-breaking pattern. They are generally associated with quantum
(statistical) field theories in $d+1$ dimensions~\cite{Sachdev-book},
such as those in Eq.~\eqref{HON}.  Using scaling and hyperscaling
arguments (see, e.g., Refs.~\cite{Ma-book, ZinnJustin-book, PV-02,
  Sachdev-book}), the exponents associated with the critical power laws
of other observables, such as those of the magnetization and the
critical equation of state, can be derived in terms of the independent
exponents $z$, $\nu$, and $\eta$.  The corrections to the asymptotic
power laws are usually controlled by irrelevant perturbations, whose
leading one determines their asymptotic suppression as
$\xi^{-\omega}$; the universal exponent $\omega$ is related to its RG
dimension~\cite{ZinnJustin-book, PV-02}.  We will return to this issue
in Sec.~\ref{escalingcqt}.

\subsection{Topological transitions}
\label{toptra}

Within the LGW framework, the standard examples of CQTs involve a
gapped disordered phase separated from a broken phase with a Landau
order parameter. In this case, the critical phenomena may be described
within the framework of a LGW theory sharing the same symmetry and
symmetry-breaking pattern.  However, there are also examples of CQTs
that lie beyond the Landau paradigm. For example, one or both phases
may not have a Landau order, while they may have {\em topological
order}. In such case, a description based on an order-parameter field,
such as the LGW approach, would fail to capture the universal features
of the critical behavior.  Therefore, the critical phenomenon cannot
be reproduced by the standard LGW paradigm (see,
e.g., Refs.~\cite{Xu-12, Sachdev-19, BKV-02, BLS-20}).

Unconventional scenarios, departing from the conventional Landau
paradigm, generally involve topological phases (see, e.g.,
Refs.~\cite{HK-10, QZ-11} for reviews) and emerging gauge fields (see,
e.g., Ref.~\cite{Sachdev-19}).~\footnote{Quantum gauge theories
provide effective descriptions of deconfined quantum phases of matter
which exhibit fractionalization of low-energy excitations, topological
order, and long-range entanglement~\cite{Wen-book, Fradkin-book,
  ZCZW-book}.}  Namely, one may have unconventional QTs~\cite{Xu-12,
  Sachdev-19} between different topological phases~\cite{GJWK-16}, in
particular from topologically trivial and nontrivial phases, between
topological order and spin-ordered phases, between topological order
and valence-bond-solid (VBS) phases~\cite{RS-90}, and direct
transitions between differently ordered phases, such as the
N\'eel-to-VBS transition of 2$d$ quantum magnets~\cite{SBSVF-04,
  SVBSF-04}.  Within this type of QTs, there are also transitions in
the context of integer and fractional quantum Hall
effect~\cite{Huckestein-95, HHSV-17, Giesbers-etal-09, OGE-12}.
The Wegner ${\mathbb Z}_2$ lattice gauge theory in three
dimensions~\cite{Wegner-71} provides a paradigmatic statistical model
undergoing a topological transition without a local symmetry-breaking
order parameter~\cite{Sachdev-19}.

These unconventional phase transitions share the main properties of
CQTs driven by a local order parameter, such as the divergence of the
length scale $\xi$ of the corresponding critical modes, and the
suppression of the gap. Critical length scales $\xi$ can be defined
through extended objects, such as the area law of Wilson loops in
gauge theories, or arise from surface states.  One may again
introduce a universal critical exponent $\nu$ associated with the
divergence $\xi\sim |g-g_c|^{-\nu}$ of such length scale at the
transition point $g_c$ separating the different phases.  The dynamic
critical exponent $z$ is still defined from the power law describing
the suppression of the gap.  On the other hand, due to the lack of an
effectively local order-parameter field, the critical exponents
associated with the order-parameter field may not be appropriate, such
as $\eta$, associated with the large-distance decay of the correlation
function of the order parameter, and $\beta$, associated with
its power-law suppression at the critical point.

\section{Some paradigmatic models}
\label{models}

In this section we introduce a number of prototypical models
undergoing CQTs and FOQTs, which will turn useful to place the
quantum phenomena discussed in this review into concrete grounds.

\subsection{Quantum Ising-like models}
\label{isingmodels}

As a first paradigmatic example, we define the $d$-dimensional
spin-1/2 quantum Ising model in a transverse field, through the
following Hamiltonian on a $L^d$ cubic-like lattice:
\begin{equation}
  \hat H_{\rm Is}(g,h) = - J \sum_{\langle {\bm x},{\bm y}\rangle}
  \hat \sigma^{(1)}_{\bm x} \hat \sigma^{(1)}_{\bm y} - g \sum_{\bm x}
  \hat \sigma^{(3)}_{\bm x} - h \sum_{\bm x} \hat \sigma^{(1)}_{\bm x}\,,
  \label{hisdef}
\end{equation}
where $\hat \sigma^{(k)}$ are the usual spin-1/2 Pauli matrices
($k=1,2,3$), the first sum is over all bonds connecting
nearest-neighbor sites $\langle {\bm x},{\bm y}\rangle$, while the
other two sums are over all sites.  The parameters $g$ and $h$
represent external homogeneous transverse and longitudinal fields,
respectively.  Without loss of generality, we can assume $J=1$, $g>0$,
and a lattice spacing $a=1$.  At $h=0$, the model undergoes a CQT at a
critical point $g=g_c$, separating a disordered phase ($g>g_c$) from
an ordered one ($g<g_c$), the order parameter being the longitudinal
magnetization $\big\langle \hat \sigma_{\bm x}^{(1)} \big\rangle$.
For example, in $1d$ the critical point is located at $g_c=1$.  The
parameters $r\equiv g-g_c$ and $h$ are associated with the (even and
odd, respectively) relevant perturbations driving the critical
behavior at the CQT.  For $g<g_c$, the longitudinal field $h$ drives
FOQTs.

It is worth mentioning that quantum Ising-like models in a transverse
field are also important from a phenomenological point of view, since
they describe several physical quantum many-body systems.  A
discussion of the experimental realizations and applications can be
found in Ref.~\cite{Dutta-etal-book} and references therein.

Let us also introduce the related quantum XY extension, whose
Hamiltonian for a $1d$ chain is given by
\begin{equation}
  \hat H_{\rm XY}(\gamma,g) = - J \sum_{x}
  \left( {1+\gamma\over 2} \, \hat \sigma^{(1)}_x \hat \sigma^{(1)}_{x+1}
  + {1-\gamma\over 2} \, \hat \sigma^{(2)}_x \hat \sigma^{(2)}_{x+1} \right)
  - g \sum_{x} \hat \sigma^{(3)}_x \,.
  \label{XYchain}
\end{equation}
Once again, we set $J=1$ and assume $g>0$. For $\gamma=1$ we recover
the quantum Ising chain in a transverse field ($h=0$), while for
$\gamma=0$ we obtain the so-called XX chain. For $\gamma>0$ the model
undergoes a quantum Ising transition at $g=g_c=1$ (independently of
$\gamma$) and $h=0$, separating a quantum paramagnetic phase ($g>g_c$)
from a quantum ferromagnetic phase ($g<g_c$). Again, for any
$\gamma>0$, the presence of an additional longitudinal field $h$
drives FOQTs, for $g<g_c$.

The quantum XY Hamiltonian~\eqref{XYchain} can be mapped into a
quadratic model of spinless fermions through a Jordan-Wigner
transformation~\cite{LSM-61, Katsura-62}, obtaining the so-called
Kitaev quantum wire defined by~\cite{Kitaev-01}
\begin{equation}
  \hat H_{\rm K}(\gamma,\mu) = - J \sum_{x} \big( \hat c_x^\dagger \hat
  c_{x+1} + \gamma \, \hat c_x^\dagger \hat c_{x+1}^\dagger + {\rm h.c.}
  \big) - \mu \sum_{x} \hat n_x \,,
  \label{kitaev2}
\end{equation}
where $\hat c_x^{(\dagger)}$ is the fermionic annihilation (creation)
operator on site $x$ of the chain, $\hat n_x\equiv \hat c_x^\dagger
\hat c_x$ is the corresponding number operator, and $\mu=-2g$.  The
Hamiltonian~\eqref{kitaev2} can be straightforwardly diagonalized into
\begin{equation}
  \hat H = \sum_k {E}(k) \left( \hat a^\dagger_k \hat a_k 
- \tfrac12 \right),
  \label{H-harmonic}
\end{equation}
where $\hat a^{(\dagger)}_k$ are new fermionic annihilation (creation)
operators, which are obtained through a suitable linear transformation
of the $\hat c_x^{(\dagger)}$ operators, and
\begin{equation}
  E(k) = 2 \left[ g^2 + \gamma^2 - 2 g \cos k
    + (1 - \gamma^2) \cos^2 k \right]^{1/2}.
  \label{Ek-XY}
\end{equation}
The set of $k$ values which must be summed over and the allowed
states depend on the BC~\cite{LSM-61, Katsura-62, Pfeuty-70, BG-85}.

As we shall see later, the study of finite-size systems at QTs is
important essentially for two reasons: (i)~they are phenomenologically
relevant, because actual experiments are often performed on relatively
small systems; (ii)~numerical studies on finite-size systems are
usually most effective to determine the relevant quantities at QTs,
through extrapolations guided by the appropriate ansatze that are
invoked by the FSS theory~\cite{FB-72, Barber-83, Privman-book,
  Binder-87, PHA-91, Cardy-review, PV-02, CPV-14, CNPV-14}.  Various
BC are usually considered:
\begin{itemize}
\item Periodic boundary conditions (PBC), for which
$\hat \sigma^{(k)}_{\bm x} = \hat \sigma^{(k)}_{{\bm x}+L\vec{\mu}}$
  ($\vec{\mu}$ indicates a generic lattice direction).
\item
  Antiperiodic boundary conditions (ABC), for which
  $\hat \sigma^{(k)}_{\bm x} = - \hat \sigma^{(k)}_{{\bm x}+L\vec{\mu}}$.
\item
  Open boundary conditions (OBC).
\item
  Fixed boundary conditions (FBC), where the states corresponding to
  the lattice boundaries are fixed, for example $x=0$ and
  $x=L+1$ for $1d$ systems, typically choosing one of the eigenstates
  of the longitudinal spin operator $\hat \sigma^{(1)}$ in Ising-like
  systems. In particular, we may have equal FBC (EFBC) when the states
  at the boundaries are eigenstates of $\hat \sigma^{(1)}$ with the
  same eigenvalue, or opposite FBC (OFBC) when the eigenstates at the
  boundaries have opposite eigenvalues.

\end{itemize}
Note that PBC and ABC give rise to systems without boundaries,
where translation invariance is satisfied. On the other hand, in OBC
and FBC, translation invariance is violated by the boundaries.

\subsubsection{The phase diagram}
\label{phdiaisi}

\begin{figure}
  \begin{center}
    \includegraphics[width=0.47\columnwidth]{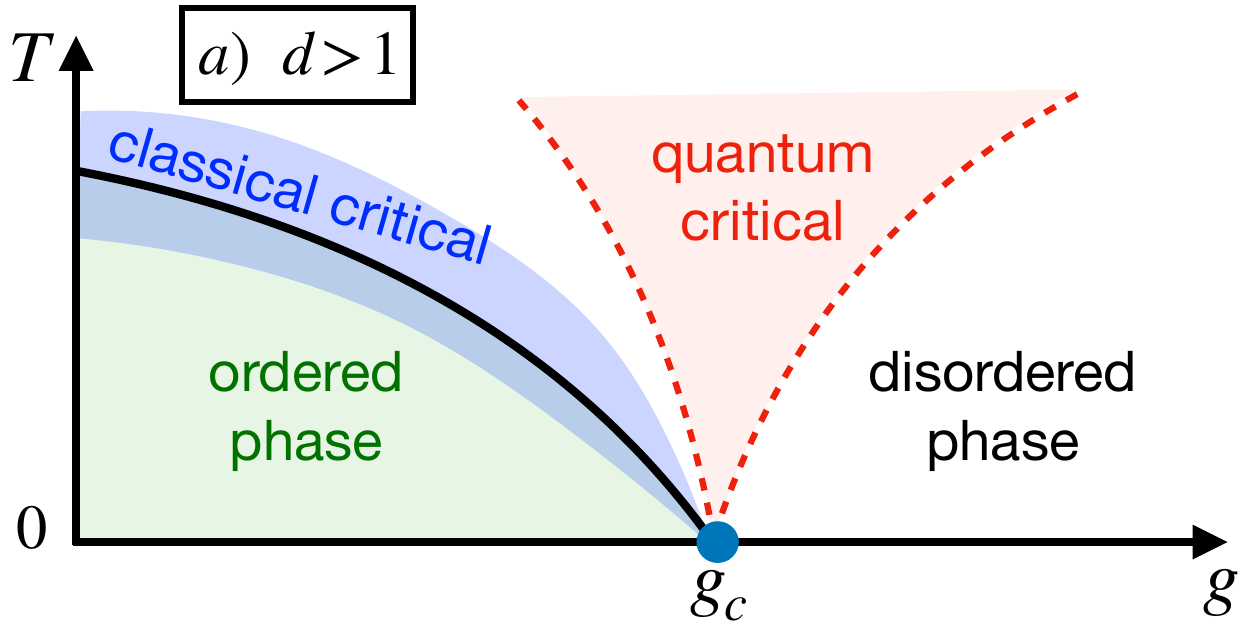}
    \hspace*{6mm}
    \includegraphics[width=0.47\columnwidth]{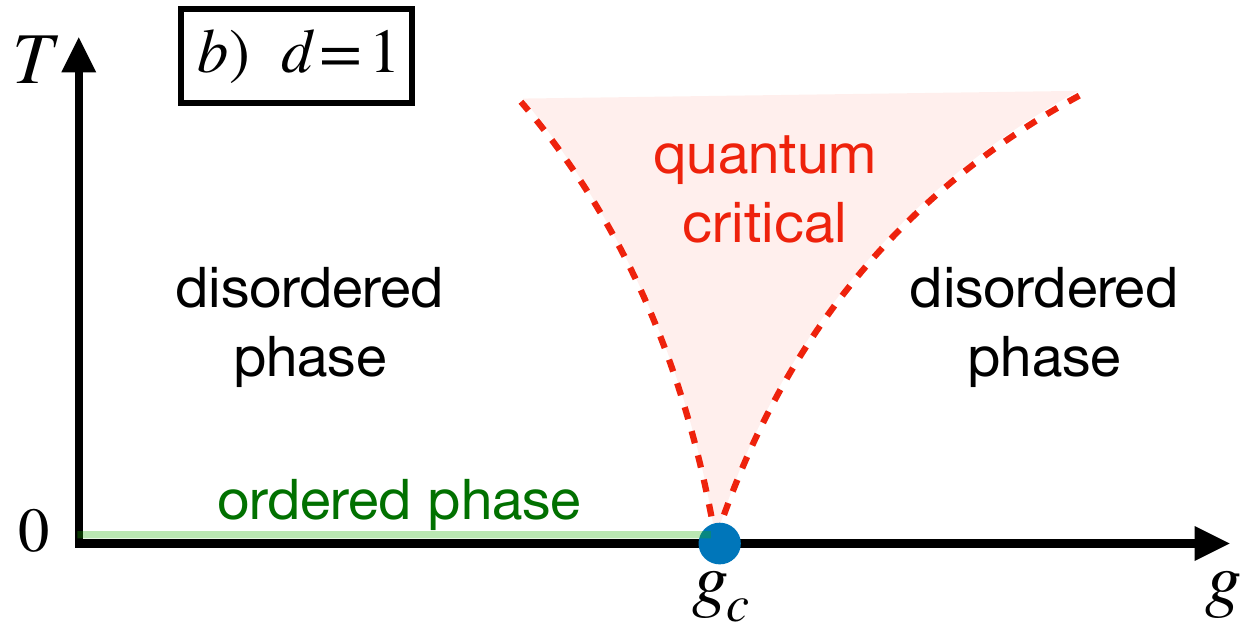}
    \caption{Sketch of the phase diagram, in the $T$-$g$ plane, for
      quantum Ising systems~\eqref{hisdef}.  Panel $a)$ is for systems
      in $d>1$ dimensions: at finite temperature, a line of classical
      transitions separates an ordered phase  from a disordered phase,
      terminating at the QCP (for $T=0$, $g=g_c$).  The theory of
      phase transitions in classical systems driven by thermal
      fluctuations can be applied within the shaded blue region across
      such transition line. Conversely, in the red region
      emerging from the QCP, quantum and thermal fluctuations are
      comparable and thus it is not possible to invoke any
      semiclassical description of the system behavior.  Panel $b)$
      shows the $1d$ case: the classical transition line disappears
      and the ordered phase survives only at zero temperature and for
      $g<g_c$ (with $g_c=1$).}
    \label{phadiaisid23}
  \end{center}
\end{figure}

The equilibrium thermodynamic behavior of the above quantum Ising
systems is described by the ca\-no\-ni\-cal partition function
\begin{equation}
  Z = {\rm Tr} \, \big[ e^{-\beta \, \hat H_{\rm Is}(g,h)} \big] \,,
  \qquad \mbox{where } \quad \beta \equiv 1/T \,.~\footnote{Throughout
    the whole review, we will assume $\hslash = k_B = 1$.} 
  \label{partfuncisi}
\end{equation}
Their phase diagram depends on the temperature $T$ and the
transverse-field parameter $g$. A sketch for $d$-dimensional quantum
Ising systems (with $d>1$) at $h=0$ is shown in
Fig.~\ref{phadiaisid23}, panel $a)$.  The QCP is located at a given
$g=g_c$ and $T=0$, separating quantum disordered and ordered
phases. The corresponding quantum critical region is characterized by
a nontrivial interplay between quantum and thermal fluctuations. At
finite temperature and for $d>1$ dimensions, the disordered and
ordered phases are separated by {\em classical} phase transitions
driven by thermal fluctuations only, belonging to the $d$-dimensional
Ising universality class, whose main features have been extensively
investigated in the literature (see, e.g., Ref.~\cite{PV-02}). The
longitudinal external field $h$ drives classical first-order
transitions within the whole finite-temperature ordered phase, and
FOQTs within the quantum zero-temperature ordered phase.

In contrast, $1d$ systems such as the Ising and the XY chain, do not
develop a finite-temperature transition, being always disordered for
any temperature $T>0$ and value of $g$ [see Fig.~\ref{phadiaisid23}, panel
  $b)$].  This is essentially related to the fact that $1d$ Ising-like
systems described by the classical Gibbs ensembles do not show
finite-temperature ferromagnetic transitions.  However, they present a
zero-temperature QCP separating quantum disordered and ordered phases.

\subsubsection{Continuous quantum transitions}
\label{cqtisi}

The critical behavior at the CQT belongs to the $(d+1)$-dimensional
Ising universality class characterized by a global $\mathbb{Z}_2$
symmetry, associated with the $(d+1)$-dimensional quantum field theory
(QFT) defined by the Lagrangian density~\eqref{HON} with a
single-component real field~\cite{Sachdev-book, Dutta-etal-book,
  ZinnJustin-book, PV-02}. While the critical exponents $\nu$ and
$\eta$ depend on $d$, the dynamic exponent $z$ is equal to
\begin{equation}
  z = 1
  \label{isiz}
\end{equation}
in any dimension.  In particular, the critical gap $\Delta(L)$ of the
1$d$ Ising chain~\eqref{hisdef} at the CQT point $g_c=1$ and $h=0$
behaves as~\cite{CPV-14, CPV-15-bf, BG-85}
\begin{equation}
  \Delta(L)_{\rm OBC} = {\pi \over L} + O(L^{-2})\,, \qquad 
  \Delta(L)_{\rm PBC} = {\pi \over 2L} + O(L^{-2})\,, \qquad 
  \Delta(L)_{\rm ABC} = {3\pi \over 2L} + O(L^{-2})\,,
  \label{gapiscqt}
\end{equation}
respectively for OBC, PBC, and ABC.

The CQT of quantum $1d$ systems belongs to the two-dimensional
($2d$) Ising universality class, hence its critical behavior is
associated with a $2d$ conformal field theory (CFT) with central
charge $c=1/2$~\cite{ID-book2, CFT-book}.  The critical exponents
assume the values
\begin{equation}
  \nu=1\,, \quad \eta=1/4\,, \qquad {\rm for} \;\; d=1 \,.
  \label{2disingexp}
\end{equation}
The structure of the scaling corrections within the $2d$ Ising
universality class has been also thoroughly discussed (see, e.g.,
Refs.~\cite{Henkel-87, Reinicke-87a, Reinicke-87b, CCCPV-00, CHPV-02,
  CPV-14}).  The leading corrections due to irrelevant operators are
generally characterized by the exponent $\omega=2$ for unitary Ising-like
theories~\cite{CCCPV-00, CHPV-02}, such as those emerging at the Ising
CQTs of the XY chain. Due to the relatively large value of $\omega$,
corrections from other sources may dominate for some observables, such
as those arising from analytical backgrounds and/or the presence of
boundaries~\cite{CHPV-02, CPV-14} (see also below). A detailed analysis
of the scaling corrections at the CQT of $1d$ quantum Ising-like
systems is reported in Ref.~\cite{CPV-14}.

For $2d$ quantum Ising models, the critical exponents are those of the
$3d$ Ising universality class, which are not known
exactly, but there are very accurate estimates by various approaches,
ranging from lattice techniques to statistical field theory (SFT)
computations. A selection of the most recent and accurate estimates of
the (classical) $3d$ Ising critical exponents is contained in
Table~\ref{isingres3d}, where we report the correlation-length
exponent $\nu$, the exponent $\eta$ related to the space-dependence of
the critical two-point function, and the scaling-correction
exponent $\omega$ related to the leading irrelevant perturbation at the
corresponding fixed point of the RG flow.
Finally, for $3d$ quantum systems, the critical exponents assume
 mean-field values, i.e., $\nu=1/2$ and $\eta=0$, apart from
logarithms (see, e.g., Refs.~\cite{ZinnJustin-book, ID-book1}).

\begin{table}
  \begin{center}
    \begin{tabular}{lllllllc}
      \hline\hline
      \multicolumn{2}{c}{$3d$ Ising}& 
      \multicolumn{1}{c}{$\nu$}& 
      \multicolumn{1}{c}{$\eta$}& 
      \multicolumn{1}{c}{$\omega$}&
      \multicolumn{1}{c}{Ref.$_{\rm year}$}
      \\\hline
      Lattice  
      & HT exp & 0.63012(16) & 0.0364(2) & 0.82(4) & \cite{CPRV-02}$_{2002}$ \\
      & MC & 0.63020(12) & 0.0372(10) & 0.82(3)
      &\cite{DB-03}$_{2003}$ \\
      & MC & 0.63002(10) & 0.03627(10) & 0.832(6)
      &\cite{Hasenbusch-10}$_{2010}$ \\
      \hline
      SFT & 6-loop 3$d$ expansion & 0.6304(13) & 0.0335(25) & 0.799(11)
      & \cite{GZ-98}$_{1998}$ \\
      & 6-loop $\epsilon$ expansion & 0.6292(5) & 0.0362(6) & 0.820(7)
      & \cite{KP-17}$_{2017}$ \\
      & NPRG & 0.63012(16) & 0.0361(11) & 0.832(14) & 
      \cite{DBTW-20}$_{2020}$ \\
      & CFT bootstrap & 0.629971(4) & 0.036298(2) & 0.8297(2)
      & \cite{KPSV-16}$_{2016}$ \\
      \hline\hline
    \end{tabular}
    \caption{Some estimates of the critical exponents of the $3d$
      Ising universality class, by lattice techniques [such as
        high-temperature (HT) expansions~\cite{CPRV-02} and Monte
        Carlo (MC) simulations~\cite{Hasenbusch-10,
          DB-03,Hasenbusch-01, Hasenbusch-99}] and statistical field
      theory (SFT) approaches [such as resummations of high-order SFT
        perturbative expansions (using 6-loop calculations within
        fixed-dimension~\cite{GZ-98, AS-95, LZ-77a, LZ-77b, BNGM-76,
          Parisi-80} and $\epsilon$-expansion~\cite{KP-17, CGLT-83,
          WF-72, tHV-72} RG schemes), nonperturbative RG (NPRG)
        approaches~\cite{DBTW-20}, and conformal field theory (CFT)
        bootstrap~\cite{KPSV-16, PRV-19-rev}].  A more complete list
      of theoretical and experimental results can be found in
      Ref.~\cite{PV-02} (at least up to 2002), however those reported
      here are the most accurate within the various approaches.  We
      recall that the critical exponents controlling the asymptotic
      exponents of other observables, such as the magnetization, can
      be generally obtained from $\nu$ and $\eta$ by using scaling and
      hyperscaling relations (see, e.g., Refs.~\cite{ZinnJustin-book,
        AM-book}). We note that there is an overall agreement among
      the results obtained by the different approaches. Moreover,
      there is also a good agreement with experiments in various
      physical systems at continuous transitions such as those in
      liquid-vapor systems, binary systems, uniaxial magnetic systems,
      Coulombic systems, etc. (see, e.g., Ref.~\cite{PV-02} for a list
      of experimental results).}
    \label{isingres3d}
  \end{center}
\end{table}

Experimental evidence of the CQT emerging in quantum Ising-like
systems has been obtained through nuclear magnetic resonance in
various contexts, in particular see Refs.~\cite{Zhang-etal-09,
  Coldea-etal-10, Morris-etal-14, Kinross-etal-14, Liang-etal-15,
  BNB-19}.

\subsubsection{First-order quantum transitions}
\label{foqtisi}

Within the quantum Ising model of Eq.~\eqref{hisdef}, the longitudinal
field $h$ drives FOQTs along the line $g<g_c$.  As already mentioned,
many-body systems at FOQTs turn out to be extremely sensitive to the type
of BC, whether they favor one of the phases or they are neutral,
giving rise to exponential or power-law behaviors (we will return to
this point later). The FOQTs of systems with BC that do not favor any
of the two magnetized phases, such as PBC and OBC, are characterized
by the level crossing of the two lowest states $| \!  \uparrow
\rangle$ and $| \! \downarrow \rangle$ for $h=0$, such that
\begin{equation}
  \langle \uparrow \! |\hat \sigma_{\bm x}^{(1)} | \! \uparrow \rangle = m_0\,,
  \qquad  \langle \downarrow \! | \hat \sigma_{\bm
    x}^{(1)} | \! \downarrow \rangle = -m_0\,,
  \label{sigmasingexp}
\end{equation}
with $m_0>0$ and independently of ${\bm x}$~\footnote{For example, in
$1d$ systems one has~\cite{Pfeuty-70}: $m_0 = (1 -
g^2)^{1/8}$. \label{foot1}}.  The degeneracy of these states is lifted
by the longitudinal field $h$.  Therefore $h = 0$ is a FOQT point,
where the (average) longitudinal magnetization $M$,
\begin{equation}
  M \equiv L^{-d} \sum_{\bm x} m_{\bm x}\,,\qquad m_{\bm x} \equiv
  \big\langle \hat \sigma_{\bm x}^{(1)} \big\rangle\,,
  \label{longmagn}
\end{equation}
becomes discontinuous in the infinite-volume limit.  The transition
separates two different phases characterized by opposite values of the
magnetization $m_0$, i.e.,
\begin{equation}
  \lim_{h \to 0^\pm} \lim_{L\to\infty} M = \pm m_0\,.
  \label{m0def}
\end{equation}
In a finite system of size $L$, the two lowest states are superpositions
of magnetized states $| + \rangle$ and $| - \rangle$ such that
$\langle \pm | \hat \sigma_{\bm x}^{(1)} | \pm \rangle = \pm \,m_0$,
for all sites ${\bm x}$.  Due to tunneling effects, the
energy difference (gap) $\Delta$ of the lowest states at $h=0$
vanishes exponentially as $L$ increases~\cite{PF-83, CNPV-14},
\begin{equation}
  \Delta(L,h=0) \equiv E_1(L,h=0)-E_0(L,h=0) \sim e^{-c L^d}\,,
  \label{del}
\end{equation}
apart from powers of $L$, where $c>0$ depends on $g$.
In particular, for a
$1d$ quantum Ising chain with $g<1$, it is exponentially suppressed
as~\cite{Pfeuty-70, CJ-87}
\begin{subequations}
  \begin{eqnarray}
    \Delta(L,h=0) \, = & \!\!\!\! 2 \, (1-g^2) \; g^L \, [1+ O(g^{2L})]
    & \quad \mbox{ for} \quad \mbox{OBC} \,, \label{deltaobc} \\
    \Delta(L,h=0) \, \approx & \!\!\!\! 2 \,
    \sqrt{(1-g^2)/(\pi L)} \; g^L \hspace*{0.6cm}
    & \quad \mbox{ for} \quad \mbox{PBC} \,.
    \label{deltapbc}
  \end{eqnarray}
\end{subequations}
The differences
\begin{equation}
  \Delta_n(L,h) \equiv E_n(L,h)-E_0(L,h)\;, \qquad n>1\,,
  \label{deltan}
\end{equation}
for the higher excited states are finite for $L\to \infty$.~\footnote{For
  the sake of compactness in the notations, we will avoid indicating
  the explicit dependence of energy levels $E_n$ and gaps $\Delta_n$ on
  the parameter associated with the perturbation invariant
  under the symmetry (for example, $g$ in Ising-like models and $\mu$
  in the Kitaev wire), keeping only their dependence on $L$ and on
  the symmetry-breaking perturbation $h$.}

The above picture based on the quasi-level crossing of the lowest
states, giving rise to exponentially suppressed gaps, strongly depends
on the choice of the BC.  Indeed other scenarios emerge with BC
forcing domain walls in the system, such as ABC and OFBC.  In those
two cases, the lowest-energy states are associated with domain walls
(kinks), i.e., with nearest-neighbor pairs of antiparallel spins,
which can be considered as one-particle states with $O(L^{-1})$
momenta. Hence, there is an infinite number of excitations with a gap
of order $L^{-2}$.  In particular, for $1d$ systems
\begin{equation}
  \Delta(L,h=0)
  = c \,{g\over 1-g} \, {\pi^2\over L^2} + O(L^{-3}) \,,
  \label{deltaabcfobc}
\end{equation}
with $c=1$ for ABC~\cite{CJ-87} and $c=3$ for
OFBC~\cite{CNPV-14, CPV-15-bf}.

\subsubsection{Relation between the quantum Ising chain and the Kitaev wire}
\label{sec:Kitaev}

At the beginning of this section, we stated that the XY chain of
Eq.~\eqref{XYchain} can be exactly mapped into a fermionic quantum
wire, cf.~Eq.~\eqref{kitaev2}, through a Jordan-Wigner transformation
which maps the spin-1/2 operators into spinless fermions.  However,
although the bulk behaviors of the two models in the infinite-volume
limit (and thus their phase diagram) are analogous, some features of
finite-size systems may significantly differ.
More in detail, we should stress that the BC play an important role in
this mapping. As a matter of fact, the nonlocal Jordan-Wigner
transformation of the XY chain~\eqref{XYchain} with PBC or ABC does
not map into the fermionic model~\eqref{kitaev2} with PBC or
ABC. Indeed further considerations apply~\cite{Katsura-62, Pfeuty-70},
leading to a less straightforward correspondence, which also depends
on the parity of the particle-number eigenvalue.

For example, the Kitaev quantum wire with ABC turns out to be gapped
in both of the phases separated by the QT at $\mu_c=-2$.  Indeed, the
energy difference $\Delta$ of the two lowest states is given by
\begin{equation}
  \Delta(L) = \sqrt{\bar\mu^2 + 4 (2-\bar\mu) \, [1-{\rm cos}(\pi/L)]}\,,
\end{equation}
where $\bar\mu=\mu-\mu_c$, such that
\begin{equation}
  \Delta(L) = \left\{ \begin{array}{ll} \displaystyle
    |\bar\mu| + {\pi^2 (2-\bar\mu)\over |\bar\mu| L^2} + O(L^{-4}) 
    & {\rm for} \quad |\bar\mu| > 0 \,, \vspace*{2mm} \\
    \displaystyle {2\pi \over L}
    + O(L^{-3}) & {\rm for} \quad |\bar\mu| = 0 \,.
  \end{array} \right.
\end{equation}
Therefore, the Kitaev quantum wire with ABC does not exhibit the
lowest-state degeneracy of the ordered phase of the quantum Ising
chain (namely, the exponential suppression of the gap with increasing
$L$).  The reason for such substantial difference resides in the fact
that the Hilbert space of the former is restricted with respect to
that of the latter, so that it is not possible to restore the
competition between the two vacua belonging to the
symmetric/antisymmetric sectors of the Ising model~\cite{Katsura-62,
  Kitaev-01, CPV-14}.

Note that for the simplest OBC, the XY chain can be exactly mapped
into the Kitaev model with OBC.  In this case, the degenerate lowest
magnetized states of the XY chain for $g<1$ and $h=0$ are mapped into
Majorana fermionic states localized at the boundaries~\cite{Kitaev-01,
  Alicea-12}.  In finite systems, the lowest eigenstates of the
Hamiltonian are combinations of the two magnetized states,
corresponding to superpositions of the localized Majorana states.
Indeed, in finite systems their overlap does not vanish, giving rise
to the splitting $\Delta \sim e^{-L/l_0}$.  The coherence length $l_0$
diverges when approaching the order-disorder transition $g\to 1^-$,
where $l_0^{-1}\sim |{\rm ln}\: g|$, thus diverging as
$l_0\sim|g-g_c|^{-\nu}$ with $\nu=1$ when approaching the critical
point.

\subsubsection{Quantum antiferromagnetic Ising chains}
\label{antiXY}

The quantum antiferromagnetic Ising chain, defined by the
Hamiltonian~\eqref{hisdef} in $1d$ with $J<0$, shows a phase diagram
analogous to that of ferromagnetic models, but with peculiar
finite-size effects. Assuming $J=-1$ and $h=0$, we write its
Hamiltonian as
\begin{equation}
  \hat H_{\rm AIs}(g,h) = \sum_x  \left(
  \hat \sigma^{(1)}_x \hat \sigma^{(1)}_{x+1} - g 
  \hat \sigma^{(3)}_x \right)\,.
  \label{antihisdef}
\end{equation}
The antiferromagnetic chain~\eqref{antihisdef} can be easily mapped
into the ferromagnetic one~\eqref{hisdef} with $J=1$,
by an appropriate transformation of the spin operators, 
\begin{equation}
  \hat\sigma^{(1,2)}_x \to (-1)^x \hat\sigma^{(1,2)}_x \,, \qquad
  \hat\sigma^{(3)}_x \to \hat\sigma^{(3)}_x \,,
  \label{antimap}
\end{equation}
which preserves the commutation rules.  Correspondingly, we also have
that the staggered magnetization $M_{\rm st} \equiv L^{-1} \sum_x
(-1)^x \big\langle \hat \sigma_x^{(1)} \big \rangle$ maps into the
magnetization defined in Eq.~\eqref{longmagn}.  Therefore the bulk
properties, and the nature of the QTs, must be the same: we recover a
CQT at $g_c=1$, and FOQTs for $g<g_c$ driven by a staggered external
field $h_x=(-1)^x h$.

For finite-size systems of size $L$, the model~\eqref{antihisdef} with
OBC maps into the ferromagnetic model with OBC as well, thus they
exactly share the same spectrum. However, an anomalous behavior
emerges along the FOQT line when choosing PBC.  Indeed, for even size
$L$ the mapping~\eqref{antimap} brings to a ferromagnetic model with
PBC, thus having an exponentially suppressed energy difference of the
lowest states [cf., Eq.~\eqref{deltapbc}] along the FOQT line $g<g_c$
and $h=0$. On the other hand, for odd $L$, the antiferromagnetic model
maps into the ferromagnetic one with ABC, whose lowest states are
characterized by the presence of kinks, and the gap is only power-law
suppressed as in Eq.~\eqref{deltaabcfobc}. This does not allow us to
define a FSS limit at the FOQT of antiferromagnetic models with PBC,
unless we distinguish even and odd sizes (see
Sec.~\ref{escalingfoqt}).~\footnote{Note that also at the CQT one
  observes anomalous behaviors due to the different amplitudes of the
  $1/L$ behavior of the gap, cf. Eq.~\eqref{gapiscqt}.}  The
frustration induced by PBC in antiferromagnetic spin chains with an
odd number of sites has been also discussed in Refs.~\cite{LMSS-12,
  DLC-16, MGKF-20, TMFG-21}.

\subsection{Bose-Hubbard models}
\label{bhmodel}

Another physically relevant system is the Bose-Hubbard (BH)
model~\cite{FWGF-89}, which provides a realistic description of a gas
of bosonic atoms loaded into an optical lattice~\cite{JBCGZ-98}. Its
Hamiltonian reads:
\begin{equation}
  \hat H_{\rm BH}(U,\mu) = - t \sum_{\langle {\bm x},{\bm y}\rangle} 
  \big( \hat b_{\bm x}^\dagger \hat b^{\phantom\dagger}_{\bm y}
  + \hat b_{\bm y}^\dagger \hat b^{\phantom\dagger}_{\bm x} \big)
  +{U \over 2} \sum_{\bm x} \hat n_{\bm x} (\hat n_{\bm x} - 1) 
  -\mu \sum_{\bm x} \hat n_{\bm x} \,,
  \label{bhm}
\end{equation}
where $\hat b_{\bm x}^{(\dagger)}$ annihilates (creates) a boson on
site ${\bm x}$ of a $L^d$ cubic-like lattice,
$\hat n_{\bm x} \equiv \hat b_{\bm x}^\dagger \hat b^{\phantom\dagger}_{\bm x}$
is the particle-density operator, the first sum runs over
nearest-neighbor bonds ${\langle {\bm x},{\bm y}\rangle}$,
while the two other sums run over all sites.
Moreover the parameter $t$ denotes the hopping strength, $U$ the
interaction strength, and $\mu$ the onsite chemical potential.
In this model the total number of bosons is conserved, indeed the
particle-number operator $\hat{N} \equiv \sum_{\bm x} \hat n_{\bm x}$
commutes with the Hamiltonian $\hat H_{\rm BH}$.
In the following, we set $t=1$.

\begin{figure}
  \begin{center}
    \includegraphics[width=0.5\columnwidth]{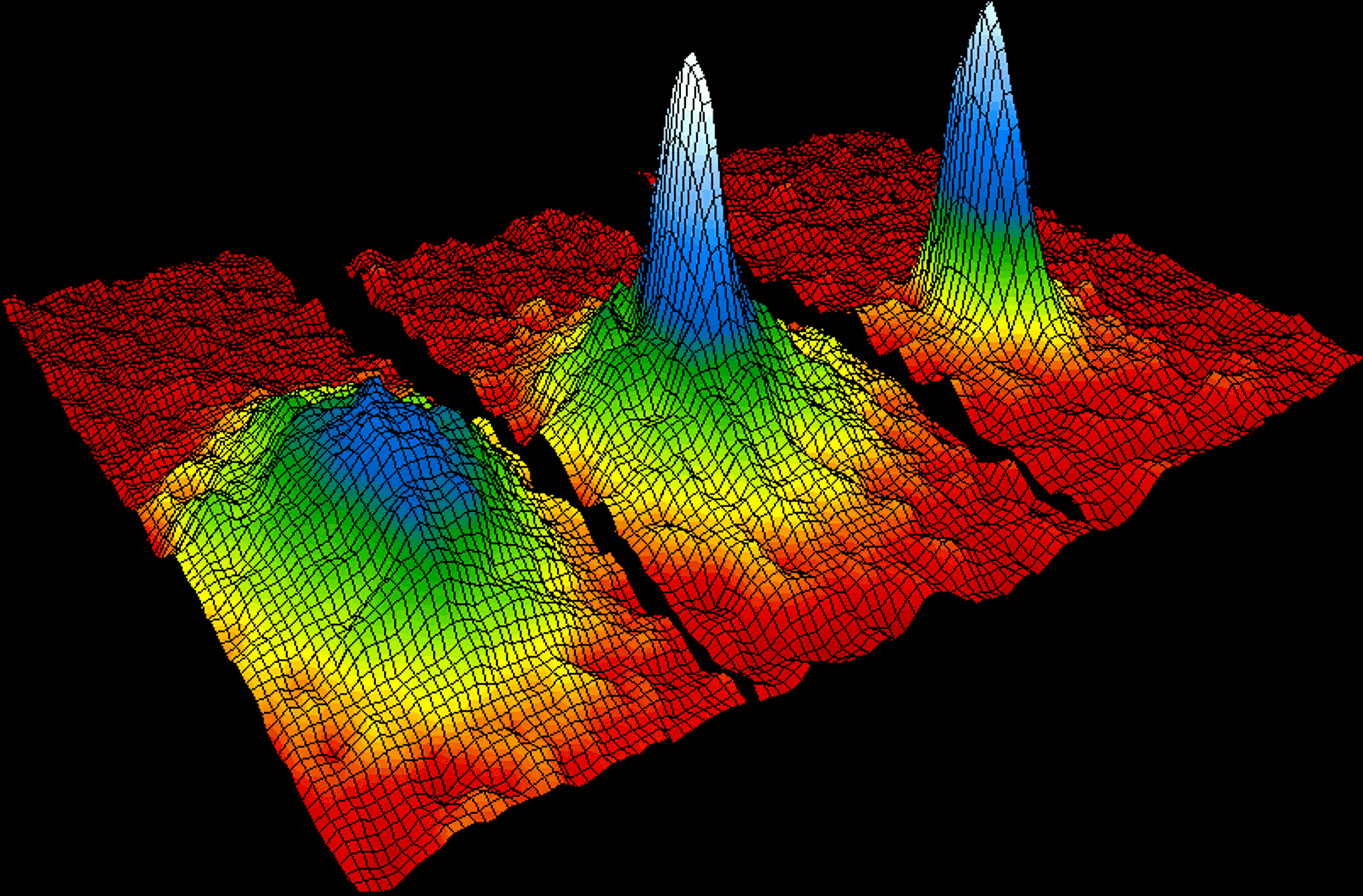}
    \caption{Three dimensional successive snapshots showing the
      velocity distribution data for a gas of rubidium atoms, which is
      progressively cooled from $200$ nK (left) down to $20$ nK
      (right). The coherent BEC emerges as a blue peak representing a
      group of atoms with the same velocity, surrounded by a field of
      noncondensed atoms with random velocities.  Image from the
      National Institute of Standards and Technology (public domain).}
    \label{fig:BEC}
  \end{center}
\end{figure}

In the infinitely repulsive (hard-core) $U \to +\infty$ limit, the
particle number can only take the values $n_{\bm x}=0,\,1$.  In such
case, the BH Hamiltonian can be exactly mapped into the XX
model~\cite{Sachdev-book}
\begin{equation}
  \hat H_{\rm XX}(\mu) = - 2 \sum_{\langle {\bm x},{\bm y}\rangle}
  \left( \hat S^{(1)}_{\bm x} \hat S^{(1)}_{\bm y} + \hat S^{(2)}_{\bm
    x} \hat S^{(2)}_{\bm y}\right) + \mu \sum_{\bm x} \Big( \hat
  S^{(3)}_{\bm x} - \tfrac12 \Big) \,,
  \label{XXH}
\end{equation}
where the spin operators $\hat S_{\bm x}^{(k)} \equiv \hat \sigma_{\bm
  x}^{(k)}/2$ are related to the bosonic ones by: $\hat \sigma_{\bm
  x}^{(1)} = \hat b_{\bm x}^\dagger + \hat b^{\phantom\dagger}_{\bm
  x}$, $\, \hat \sigma^{(2)}_{\bm x} = i(\hat b_{\bm x}^\dagger - \hat
b^{\phantom\dagger}_{\bm x})$, and $\hat \sigma^{(3)}_{\bm x} = 1-2
\hat b_{\bm x}^\dagger \hat b^{\phantom\dagger}_{\bm x}$.~\footnote{The
Hamiltonian~\eqref{XXH} is the generalization of Eq.~\eqref{XYchain}
for $\gamma=0$ to the $d$-dimensional case.  In $1d$ the two coincide
and can be mapped, through a Jordan-Wigner transformation, into
a model of free spinless fermions which coincides with
Eq.~\eqref{kitaev2} with $\gamma=0$, as discussed before.}

\subsubsection{The phase diagram}
\label{phdiabh}

The equilibrium thermodynamic behavior of the BH models is described by
the partition function
\begin{equation}
  Z = {\rm Tr} \, \big[ e^{-\beta H_{\rm BH}(U,\,\mu)} \big] \,,
  \qquad \quad \beta \equiv 1/T\,.
  \label{partfuncbhm}
\end{equation}
They show various phases depending on the temperature $T$, the chemical
potential $\mu$, and in particular on the spatial dimension $d$.

The low-temperature behavior of $3d$ bosonic gases is characterized by
the Bose-Einstein condensation (BEC) phenomenon, below a finite
temperature $T_c$. The BEC phase transition at $T_c$ separates the
high-temperature normal phase and the low-temperature superfluid BEC
phase. This is characterized by the accumulation of a macroscopic
number of atoms in a single quantum state, giving rise to a
phase-coherent condensate, as shown in Fig.~\ref{fig:BEC}.
The phase coherence properties of the BEC phase have been observed in
a number of spectacular experiments with ultracold gases
(see, e.g., Refs.~\cite{CW-02,
  Ketterle-02, Andrews-etal-97, Stenger-etal-99, Hagley-etal-99,
  BHE-00, Dettmer-etal-01, Hellweg-etal-02, Hellweg-etal-03,
  Ritter-etal-07, BDZ-08}).  Several theoretical and experimental
studies have also investigated the critical properties at the BEC
transition, when the condensate forms (see, e.g., Refs.~\cite{DGPS-99,
  ZSSK-06b, CPS-07, Donner-etal-07, DZZH-07, BB-09, CV-09, ZKKT-09,
  Trotzky-etal-10, HZ-10, PPS-10, NNCS-10, ZKKT-10, QSS-10, FCMCW-11,
  HM-11, Pollet-12, CR-12, CTV-13, CN-14, Corman-etal-14, NGSH-15,
  Chomaz-etal-15, CNPV-15-bec, CDMV-16, CNPV-16, DV-17, BN-17,
  DWGGP-17}).

\begin{figure}
  \begin{center}
    \includegraphics[width=1.0\columnwidth]{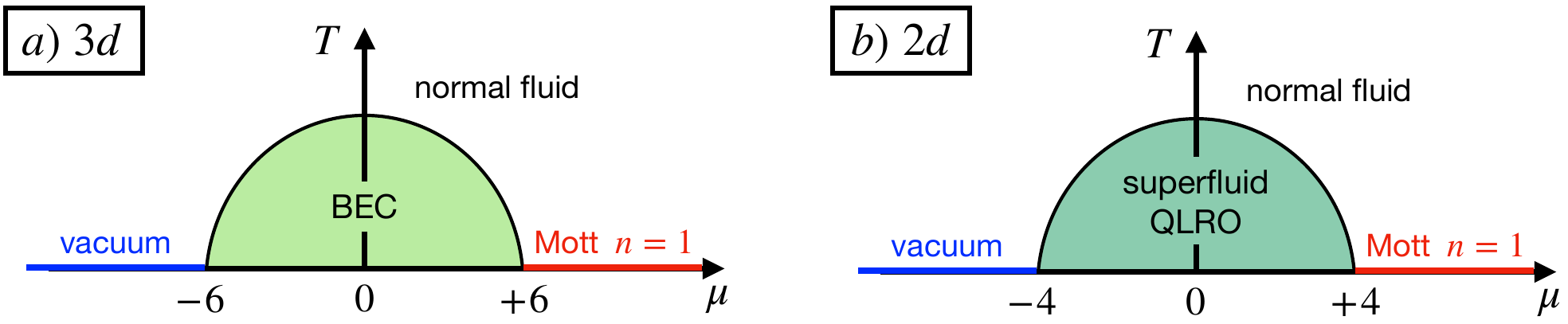}
    \caption{Sketch of the phase diagram, in the $T$-$\mu$ plane, for
      the BH model in the hard-core $U\to\infty$ limit~\eqref{XXH}.
      We adopt units of the hopping parameter $t$.  Panel $a$) shows
      the $3d$ case: The BEC phase is restricted to a finite region
      for $|\mu| \leq 6$.  It is bounded by a BEC transition line
      $T_c(\mu)$, which satisfies $T_c(\mu)=T_c(-\mu)$ due to a
      particle-hole symmetry.  Its maximum occurs at $\mu=0$,
      where~\cite{CN-14, CNPV-15-bec} $T_c(\mu=0)= 2.01599(5)$; we also
      know that~\cite{CTV-13} $T_c(\mu=\pm 4)=1.4820(2)$.  At $T=0$ two
      further quantum phases exist, i.e., the vacuum ($\mu<-6$) and
      the incompressible $n=1$ Mott insulator phase ($\mu>6$).  Panel
      $b$) shows the $2d$ case: The normal and superfluid QLRO phases
      are separated by a finite-temperature BKT transition line, which
      satisfies $T_{\rm BKT}(\mu)=T_{\rm BKT}(-\mu)$ due to a
      particle-hole symmetry.  Its maximum occurs at $\mu=0$,
      where~\cite{CNPV-13} $T_{\rm BKT}(\mu=0)= 0.6877(2)$.  The
      superfluid QLRO phase is restricted to a finite region for
      $|\mu| \leq 4$.}
    \label{phadiaBHd23}
  \end{center}
\end{figure}

The phase diagram of a $3d$ BH model, and its critical behavior, have
been deeply investigated (see, e.g., Refs.~\cite{FWGF-89, CPS-07,
  SPP-07, CR-12, CTV-13, CN-14, CNPV-15-bec}).  The $T$-$\mu$ phase
diagram presents a finite-temperature BEC transition line, as shown in
Fig.~\ref{phadiaBHd23}, panel $a$), for the hard-core $U\to\infty$
limit, where the occupation site number is limited to the cases
$n_{\bm x}=0,\,1$.  The condensate wave function provides the complex
order parameter of the BEC transition, whose critical behavior belongs
to the $3d$ U(1)-symmetric $XY$ universality class~\footnote{Note that
  this does not refer to the quantum XY model as that in
  Eq.~\eqref{XYchain}, but to the classical $N=2$ vector model with
  global symmetry O(2), or equivalently U(1), whose corresponding
  field theory is reported in Eq.~\eqref{HON}.}.  This implies that
the length scale $\xi$ of the critical modes diverges at $T_c$ as
\begin{equation}
  \xi\sim (T-T_c)^{-\nu},\qquad \nu\approx 0.6717 \,.
  \label{xi3d}
\end{equation}
A selection of the most recent and accurate estimates of the $3d$ $XY$
critical exponents is reported in Table~\ref{tablexy}.  The power
law~\eqref{xi3d} has been accurately verified by numerical studies at
the BEC phase transition (see, e.g., Refs.~\cite{CR-12, CTV-13, CN-14,
  CNPV-15-bec}).  The BEC phase extends below the BEC transition line.
In particular, in the hard-core limit $U\to\infty$ and for $\mu=0$
(corresponding to half filling), the BEC transition occurs at
$T_c=2.01599(5)$~\cite{CN-14, CNPV-15-bec}.

\begin{table}
  \begin{center}
    \begin{tabular}{lllllc}
      \hline\hline
      \multicolumn{2}{c}{$3d$ $XY$} &
      \multicolumn{1}{c}{$\nu$} & 
      \multicolumn{1}{c}{$\eta$} & 
      \multicolumn{1}{c}{$\omega$}&
      \multicolumn{1}{c}{Ref.$_{\rm year}$} 
      \\\hline
      Lattice & HT+MC & 0.6717(1) & 0.0381(2) & 0.785(20) &
      \cite{CHPV-06}$_{2006}$ \\
      & MC & 0.6717(3) & & & \cite{BMPS-06}$_{2006}$ \\
      & MC & 0.67169(7) & 0.03810(8) & 0.789(4) &
      \cite{Hasenbusch-19}$_{2019}$ \\
      \hline
      SFT & 6-loop 3$d$ expansion& 0.6703(15) & 0.035(3) & 0.789(11)
      & \cite{GZ-98}$_{1998}$ \\
      & 6-loop $\epsilon$ expansion & 0.6690(10) & 0.0380(6) & 0.804(3)
      & \cite{KP-17}$_{2017}$ \\
      & NPRG & 0.6716(6) & 0.0380(13) & 0.791(8) & 
      \cite{DBTW-20}$_{2020}$ \\
      & CFT bootstrap &
      0.67175(10) & 0.038176(44) & 0.794(8) & \cite{Chester-etal-20}$_{2020}$ \\
      \hline
      Experiment & $^4$He &
      0.6709(1) &  &  & \cite{Lipa-etal-96, Lipa-etal-00,Lipa-etal-03}$_{1996}$\\
      \hline\hline
    \end{tabular}
    \caption{Some estimates of the universal critical exponents for
      the $3d$ $XY$ universality class.  We report the
      correlation-length exponent $\nu$, the order-parameter exponent
      $\eta$, and the exponent $\omega$ associated with the leading
      scaling corrections.  They were obtained from the analysis of HT
      expansions supplemented by MC simulations~\cite{CHPV-06}, MC
      simulations~\cite{BMPS-06, Hasenbusch-19}, QFT approaches based
      on the resummation of high-order series~\cite{GZ-98,KP-17},
      nonperturbative RG (NPRG) approaches~\cite{DBTW-20}, the
      conformal-bootstrap approach~\cite{Chester-etal-20}, and
      experiments of the $^4$He superfluid transition in microgravity
      environment~\cite{Lipa-etal-03, Lipa-etal-96} (whose estimate of
      $\nu$ is obtained from the measurement of the specific-heat
      exponent $\alpha=-0.0127(3)$ and the hyperscaling relation
      $d\nu=2-\alpha$).  See Ref.~\cite{PV-02} for a more complete
      list of theoretical and experimental results.  As already noted
      in the literature, the experimental estimate of $\nu$, reported
      in the table, is not consistent with the most accurate
      theoretical estimates, likely because its error is somehow too
      optimistic. In this respect, one should take into account the
      difficulty of extracting the specific-heat exponent $\alpha$
      from the temperature dependence of the specific heat close to
      $T_c$, due to the fact that it is negative ($\alpha \approx -
      0.0150$). Thus, it is not associated with the leading behavior
      of the specific heat, which is dominated by a nonuniversal
      constant term~\cite{Lipa-etal-96, PV-02, CHPV-06}, i.e.,
      $C\approx a + b |T-T_c|^{-\alpha} + \ldots$, and for small
      $|\alpha|$ the first two leading terms are hardly
      distinguishable.}
    \label{tablexy}
  \end{center}
\end{table}

On the other hand, $2d$ bosonic gases do not display BEC phases,
because a nonvanishing order parameter cannot appear in $2d$ (or
quasi-$2d$) systems with a global U(1) symmetry~\cite{MW-66, H-67}.
However, $2d$ (or quasi-$2d$) systems with a global U(1) symmetry may
undergo a finite-temperature transition described by the BKT
theory~\cite{KT-73, B-72, Kosterlitz-74, JKKN-77, Balog-01, PV-13}.
The BKT transition separates a high-temperature normal phase and a
low-temperature phase characterized by quasi-long-range order (QLRO),
where correlations decay algebraically at large distances, without the
emergence of a nonvanishing order parameter~\cite{MW-66, H-67}.  When
approaching the BKT transition point $T_{\rm BKT}$ from the
high-temperature normal phase, these systems develop an exponentially
divergent correlation length
\begin{equation}
  \xi \sim \exp\left( {c/\sqrt{\tau}}\right),\qquad 
  \tau\equiv T/T_{\rm BKT}-1 \,,
  \label{xi2d}
\end{equation}
where $c$ is a nonuniversal constant.  Consistently with the above
picture, the $2d$ BH system undergoes a BKT transition.
Figure~\ref{phadiaBHd23}, panel $b$), shows a sketch of its phase
diagram in the hard-core $U\to\infty$ limit.  The finite-temperature
BKT transition of BH models has been numerically investigated by
several studies (see, e.g., Refs.~\cite{DM-90, Ding-92, HK-97,
  CSPS-08, CR-12, CNPV-13, DV-17}).  In particular, $T_{\rm
  BKT}=0.6877(2)$ for $U\to\infty$ and for $\mu=0$~\cite{CNPV-13}.
Below the critical temperature $T_{\rm BKT}$, $2d$ BH systems show a
QLRO phase, where the phase-coherence correlations decay
algebraically.  Experimental evidence of BKT transitions has been
also reported for quasi-$2d$ trapped atomic gases~\cite{HKCBD-06,
  KHD-07, HKCRD-08, CRRHP-09, HZGC-11, Plisson-etal-11,
  Desbuquois-etal-12}.

\subsubsection{Zero-temperature quantum transitions}
\label{qtbhmodel}

\begin{figure}
  \begin{center}
    \includegraphics[width=0.5\columnwidth]{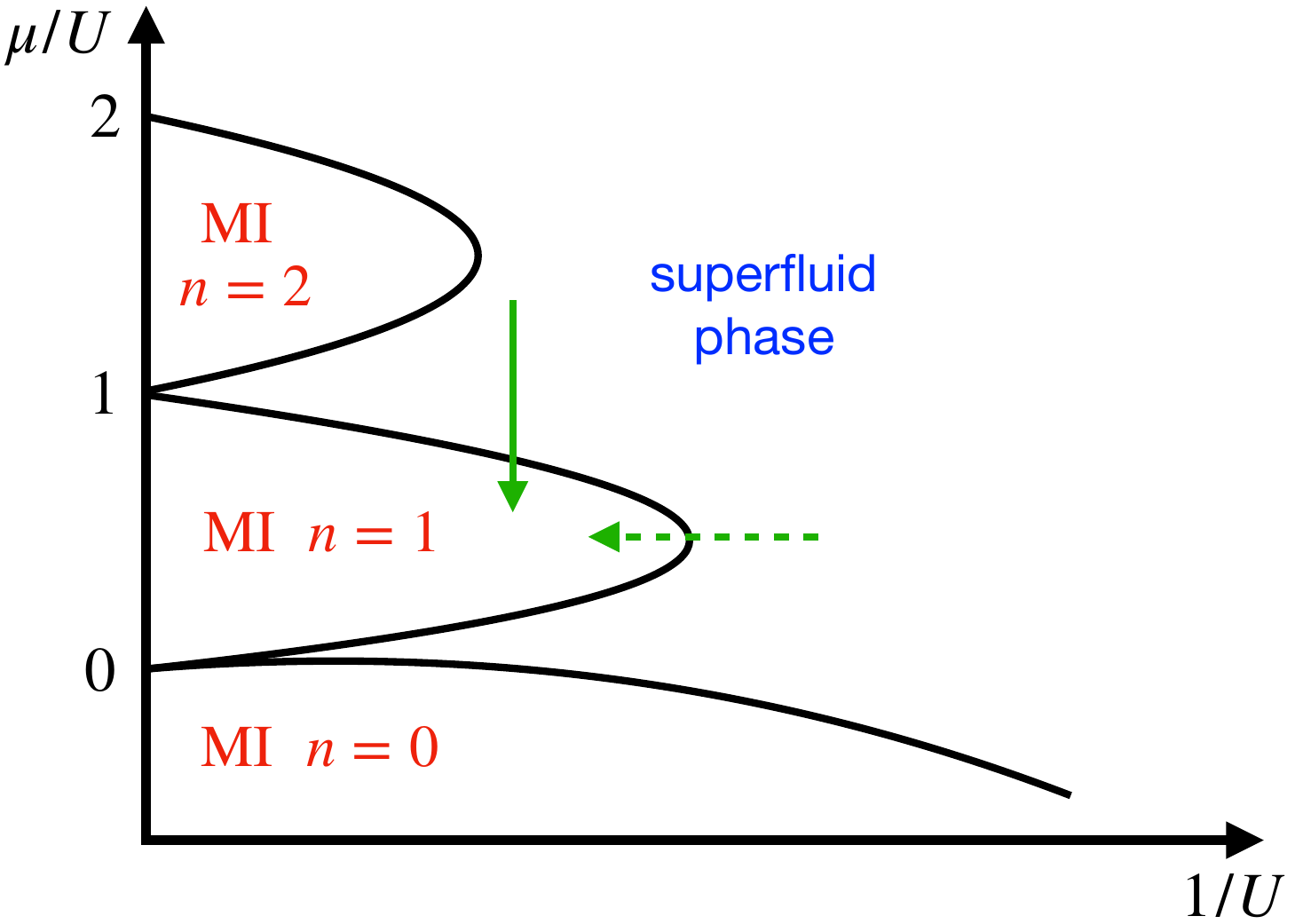}
    \caption{Sketch of the zero-temperature mean-field phase diagram,
      in the $\mu$-$(1/U)$ plane, of the BH model. We fix $t=1$.  The
      lobes, surrounded by the superfluid region, correspond to MI
      islands with integer filling factor $n$. The green lines denote
      two types of parameters variations driving CQTs from a
      superfluid to a MI phase.  The CQT along the vertical line,
      corresponding to varying the chemical potential, belongs to the
      nonrelativistic U(1)-symmetric bosonic QFT universality class.
      The CQT along the dashed line, at fixed integer density, belongs
      to the relativistic U(1)-symmetric bosonic QFT universality
      class.}
    \label{fig:BH_mf}
  \end{center}
\end{figure}

The zero-temperature phase diagram can be investigated using
mean-field approaches (see, e.g., Refs.~\cite{FWGF-89, Sachdev-book}).
In the absence of the hopping term, a uniform chemical potential fixes
the occupation site number to its integer part plus one. Once $\mu$ is
fixed, the BH Hamiltonian~\eqref{bhm} is characterized by the
competition between the onsite repulsion $U$ and the nearest-neighbor
hopping $t$.  For $U \gg t$, strong local interactions force bosons to
be localized into a gapped Mott insulator (MI) phase. In the opposite
$U \ll t$ limit, bosons are delocalized in a gapless superfluid phase.
The onset of these two phases has been observed in a clearcut way,
through experiments with ultracold atoms trapped in optical
lattices~\cite{GMEHB-02, Bakr-etal-10}.  A direct transition between
the two phases occurs at a given critical value of the ratio $t/U$,
which depends on the chemical potential, in such a way that a lobe
structure arises~\cite{FWGF-89} in the $\mu$-$(1/U)$ plane, as
sketched in Fig.~\ref{fig:BH_mf}: the higher the Mott
particle number, the smaller the lobe in the phase diagram.
The system dimensionality can affect the form and
the size of the lobes, but not this general structure~\footnote{For
  example, in $1d$ the lobes end up with a tip, as emerging from
  density-matrix renormalization group (DMRG)
  calculations~\cite{KM-98}. The lower phase boundary is bending down,
  such that the MI phase is reentrant as a function of $U^{-1}$.}.

Focusing on the hard-core limit, or equivalently on the XX
model~\eqref{XXH}, one can observe the emergence of three phases
associated with the ground-state properties~\cite{Sachdev-book}: the
vacuum ($\mu <-2d$), the superfluid ($|\mu| < 2d$), and the $n=1$ MI
phase ($\mu > 2d$).  The vacuum-to-superfluid transition at
$\mu_{\textrm{v-s}}=-2d$~\footnote{The limit
  $\mu\to\mu_{\textrm{v-s}}^+$ corresponds to the low-density regime
  $N/V\to 0$ where $N$ is the particle number, from which one can
  derive the scaling properties of the model at fixed particle number
  $N$~\cite{Sachdev-book, CV-10-offeq, ACV-14, Vicari-19}.} and the
$n=1$ superfluid-to-MI transition at $\mu_{\textrm{s-MI}}=2d$, when driven
by the chemical potential (as in the vertical green line in
Fig.~\ref{fig:BH_mf}), belong to the universality class associated
with a {\em nonrelativistic} U(1)-symmetric bosonic field
theory~\cite{FWGF-89, Sachdev-book}. Its partition function is given
by
\begin{equation}
  Z = \int [{\rm D}\phi] \, \exp \! \bigg[ - \int_0^{1/T} {\rm d}t \:
    {\rm d}^dx \,\Big( \phi^* \partial_t \phi + \tfrac{1}{ 2 m}
    |\nabla \phi|^2 + r |\phi|^2  + u |\phi|^4  \Big) \bigg] \,,
  \label{lb}
\end{equation}
where $\phi$ is a complex field and $r \sim \mu-\mu_c$.
The upper critical dimension of this
bosonic field theory is $d=2$. Thus its critical behavior is
mean-field for $d>2$. For $d=2$ the field theory is essentially free
(apart from logarithmic corrections), thus the dynamic critical
exponent is $z=2$, while the RG dimension of the relevant parameter
$r=\mu-\mu_c$ is $y_r = 2$.  In $d=1$ the theory turns out to be
equivalent to a quadratic field theory of nonrelativistic spinless
fermions~\cite{Sachdev-book}, from which one infers the RG exponents
$z=2$ and $y_r=2$.

The special transitions at fixed integer density (i.e., fixed $\mu$)
(as in the horizontal dashed green line in Fig.~\ref{fig:BH_mf})
belong to a different universality class, described by a {\em
  relativistic} U(1)-symmetric bosonic field theory~\cite{FWGF-89},
given by the $(d+1)$-dimensional O(2)-symmetric $\Phi^4$
Lagrangian~\eqref{HON} with $N=2$. Therefore, the dynamic exponent is
$z=1$, and $y_r = 1/\nu$ where $\nu$ is the correlation length
exponent of the $(d+1)$-$XY$ universality class. Thus $\nu=1/2$ for
$d=3$ (i.e., mean-field behavior apart from logarithms), $\nu\approx
0.6717$ for $d=2$ (see Table~\ref{tablexy}), and an exponential
behavior formally corresponding to $\nu=\infty$ for the BKT
transition~\cite{KT-73, PV-13} at $d=1$.

\subsubsection{Bose-Hubbard model with extended interactions}

The physics of the BH model can be considerably enriched by extending
the range of interactions beyond the onsite limit.  In particular, one
can construct the following Hamiltonian for the extended BH model:
\begin{equation}
  \hat H_{\rm EBH}(V,U,\mu) = \hat H_{\rm BH}(U,\mu)
  + V \sum_{ {\bm x} \neq {\bm y}} \frac{1}{|\bm x - \bm y|^3} \,
  \hat n_{\bm x} \hat n_{\bm y}\,,
  \label{ebhm}
\end{equation}
where $V$ denotes the nonlocal two-body interaction strength.
From a physical point of view, this model faithfully describes dipolar
bosons confined in optical lattices, in which repulsive interactions
have a long-range character and are typically decaying with the
distance as $r^{-3}$~\cite{LMSLP-09, Baier-etal-16}.

The above model has been studied in some detail in the $1d$ case,
showing that it may stabilize an insulating phase, named the bosonic
Haldane insulator (HI) phase.  This phase is of topological kind,
since it
breaks a hidden $\mathbb{Z}_2$ symmetry related to a string order
parameter.  The latter is characterized by a non-trivial ordering of
the fluctuations that appear in alternating order separated by strings
of equally populated sites of arbitrary length, being described by the
correlator~\cite{DBA-06, BDGA-08}
\begin{equation}
  R_{\rm string}(r) = \big\langle \delta \hat n_x \,
  e^{i \pi \sum_{k=x}^y \delta \hat n_k} \, \delta \hat n_y \big\rangle \,,
\end{equation}
where $\delta \hat n_x = \hat n_x - \bar n$ denotes the boson
number fluctuations from the average filling $\bar n$.

When varying $U$ and $V$, the phase diagram of the extended BH model
supports a wealth of different phases (see Ref.~\cite{Dutta-etal-15}
for a review): at commensurate fillings, the presence of interactions
between distant sites may lead to a density-modulated insulating phase
with staggered order, also named density wave. In $1d$, the
topological HI phase emerges in between the MI and the density wave,
as verified through numerical density-matrix renormalization group
(DMRG) calculations~\cite{DBA-06, BDGA-08, RF-12}.  For incommensurate
fillings, other peculiar features appear, such as the supersolid phase
and phase-separation regions.  It is worth pointing out that the MI
and HI phases can be adiabatically connected by opening a gap at the
critical point separating them, through an additional Hamiltonian term
which breaks the inversion symmetry of the system (e.g., via a
correlated tight-binding hopping).  By suitably tuning the various
Hamiltonian parameters, it is thus possible to adiabatically encircle
the MI-HI critical point and therefore to enable quantized transport
through adiabatic pumping~\cite{BLA-11, RGGF-13}.

\subsection{Quantum rotor and Heisenberg spin models}
\label{quantiferr}

Another paradigmatic model is the so-called quantum rotor model, for
which the basic orientation operator is a $N$-component unit vector
$\hat{\bm n}_{\bm x}$, such that $\hat{\bm n}_{\bm x}\cdot \hat{\bm
  n}_{\bm x}=1$, with a corresponding momentum $\hat{\bm p}_{\bm x}$,
such that $\big[ \hat{n}^a_{\bm x},\hat{p}^b_{\bm y} \big] =
i\delta_{{\bm x}{\bm y}}\delta^{ab}$.  Introducing the angular
momentum operator $\hat{L}_{\bm x}^{ab}\equiv \hat{n}^a_{\bm x}
\, \hat{p}^b_{\bm x} - \hat{n}^b_{\bm x} \, \hat{p}^a_{\bm x}$, and in
particular $\hat{L}_{\bm x}^c = {1\over 2} \epsilon^{abc}
\hat{L}^{ab}_{\bm x}$ for $N=3$, we can write the corresponding
$d$-dimensional Hamiltonian as~\cite{Sachdev-book}
\begin{equation}
  \hat H_{\rm rot} = - J \sum_{\langle {\bm x},{\bm y}\rangle}
  \hat{\bm n}_{\bm x}\cdot \hat{\bm n}_{\bm y} + 
  g \sum_{\bm x}\hat{\bm L}_{\bm x}^2 \,,
  \label{rotorham}
\end{equation}
where the first sum is over all bonds connecting nearest-neighbor
sites $\langle {\bm x},{\bm y}\rangle$ of a $L^d$ cubic-like lattice,
while the other sum is over all sites.  We fix $J=1$. For $d\ge 2$ the
system shows a quantum paramagnet phase for large values of $g$, and a
magnetized phase for small values of $g$, similarly to quantum Ising
systems~\eqref{hisdef}.  A CQT separates the two phases at a finite
coupling value $g_c$, which is associated with the symmetry-breaking
pattern O($N$) $\to$ O($N-1$).  For $d=1$ and $N\ge 3$, the rotor
model only features a quantum paramagnetic phase, i.e., $g_c=0$
formally.

The quantum criticality of these models is thoroughly discussed in
Ref.~\cite{Sachdev-book}. The corresponding QFT is provided by
a $(d+1)$-dimensional $\Phi^4$ field theory~\eqref{HON} with $N=3$. The
cases for $N=3$ are also referred to as Heisenberg spin models.  Some
accurate results for the universal critical exponents of the $3d$
Heisenberg universality class, providing the asymptotic behavior for
$2d$ QTs, are reported in Table~\ref{tableo3}.  Results for $N>3$
universality classes can be found in Refs.~\cite{PV-02, HV-11}
and large-$N$ computations in Ref.~\cite{MZ-03}.  Phase
transitions breaking the O($N$) symmetry are not expected in 1$d$
quantum Heisenberg spin models, since the corresponding 2$d$ classical
models do not undergo phase transitions, indeed they only show
disordered phases with an asymptotic critical behavior in the
zero-temperature limit, characterized by an exponential increase of
the correlation length, as $\xi\sim e^{c/T}$ (see,
e.g., Refs.~\cite{ZinnJustin-book, PV-02}).

\begin{table}
  \begin{center}
    \begin{tabular}{lllllc}
      \hline\hline \multicolumn{2}{c}{$3d$ Heisenberg} &
      \multicolumn{1}{c}{$\nu$} & \multicolumn{1}{c}{$\eta$} &
      \multicolumn{1}{c}{$\omega$}& \multicolumn{1}{c}{Ref.$_{\rm year}$}

      \\\hline Lattice & HT+MC & 0.7112(5) & 0.0375(5) &
      &\cite{CHPRV-02}$_{2002}$ \\

      & HT+MC & 0.7117(5) & 0.0378(5) &
      &\cite{HV-11}$_{2011}$ \\

      & HT+MC & 0.7116(2) & 0.0378(3) &
      &\cite{Hasenbusch-20}$_{2020}$ \\

      & MC & 0.71164(10) & 0.03784(5) &
      0.759(2) & \cite{Hasenbusch-20}$_{2020}$ \\ \hline

      SFT & 6-loop 3$d$
      expansion& 0.7073(35) & 0.0355(25) & 0.782(13) & \cite{GZ-98}$_{1998}$ \\

      & 6-loop $\epsilon$ expansion & 0.7059(20) & 0.03663(12) & 0.795(7)
      & \cite{KP-17}$_{2017}$ \\

      & NPRG & 0.7114(9) & 0.0376(13) & 0.769(11) & 
      \cite{DBTW-20}$_{2020}$ \\

      & CFT bootstrap &
      0.7117(4)
      & 0.03787(13) & & \cite{Chester-etal-20-o3}$_{2020}$ \\
      \hline\hline
    \end{tabular}
    \caption{Some estimates of the universal critical exponents for
      the $3d$ Heisenberg universality class.  We report the
      correlation-length exponent $\nu$, the order-parameter exponent
      $\eta$, and the exponent $\omega$ associated with the leading
      scaling corrections. They were obtained from the analysis of
      high-order HT expansions supplemented by MC
      simulations~\cite{CHPRV-02, HV-11, Hasenbusch-20}, MC
      simulations~\cite{Hasenbusch-20} QFT approaches based on the
      resummation of high-order series~\cite{GZ-98, KP-17},
      nonperturbative RG (NPRG) approaches~\cite{DBTW-20}, the
      conformal-bootstrap approach~\cite{Chester-etal-20-o3}.  See
      Refs.~\cite{PV-02, Hasenbusch-20} for a more complete list of
      theoretical and experimental results.}
    \label{tableo3}
  \end{center}
\end{table}

The $N=3$ quantum rotors have some connections with certain dimerized
antiferromagnetic systems of Heisenberg spins $\hat{\bm S}_{\bm
  x}\equiv\big(\hat{S}_{\bm x}^{(1)},\hat{S}_{\bm
  x}^{(2)},\hat{S}_{\bm x}^{(3)}\big)$ located at each lattice site,
belonging to the spin-$S$ representation. As argued in
Ref.~\cite{Sachdev-book}, under some conditions, quantum rotor models
provide the low-energy properties of a class of quantum
antiferromagnets.  Heisenberg quantum antiferromagnets~\cite{Bethe-31,
  Anderson-52} are generally defined as sum of terms associated with
the bonds of the lattice,
\begin{equation}
  \hat{H}_{\rm He} = \sum_{\langle {\bm x}, {\bm y}\rangle} 
J_{{\bm x},{\bm y}} \,
  \hat{\bm S}_{\bm x} \cdot \hat{\bm S}_{\bm y}\,.
\label{bondxydef}
\end{equation}
Various behaviors may arise from different lattice geometries 
and corresponding bond couplings $J_{{\bm x},{\bm y}}$.

The phase diagram and critical behaviors of quantum Heisenberg
antiferromagnets with homogeneous bonds $J_{{\bm x},{\bm y}}=J>0$ have
been largely discussed in the literature.  The quantum fluctuations do
not allow long-range order in $1d$ models: they remain always gapped
for integer $S$ or may become critical for half odd-integer
S~\cite{LP-75, Haldane-83a, Haldane-83b, Nomura-89}.  On a $2d$ square
lattice, the ground state is ordered for all $S$~\cite{DLS-78, NP-86,
  AKLT-88, RY-88}.  For $T > 0$, the order is destroyed by thermal
fluctuations~\cite{MW-66}.  However, it develops an exponential
critical behavior in the $T\to 0$ limit, described by the
asymptotically free $2d$ O(3) nonlinear $\sigma$ model (see, e.g.,
Refs.~\cite{CHN-89, HN-93, CSY-94, BW-96, KT-98, BBGW-98, PV-02}).
The ground state is ordered also in $3d$ homogeneous Heisenberg
antiferromagnets. However, the ordered phase generally persists at
low temperature, up to a finite-temperature transition to a disordered
phase, belonging to the $3d$ Heisenberg universality class
characterized by the symmetry-breaking pattern O(3) $\to$ O(2) (see, e.g.,
Refs.~\cite{Sandvik-97, Sandvik-98} and references therein).

QCPs separating quantum ordered and disordered phases can be observed
when the bond couplings are not homogeneous, such as the $S=1/2$
Heisenberg antiferromagnet on an inhomogeneous square lattice with
tunable interaction between spins belonging to different
plaquettes~\cite{ATO-08}, double-layer Heisenberg
antiferromagnets~\cite{WBS-06}, etc. (see also~\cite{Sachdev-book}
for a more detailed discussion).

We also mention another interesting issue related to the phase
transitions between N\'eel and VBS phases in Heisenberg
antiferromagnets~\cite{RS-90}.  Since both phases break a global
Hamiltonian symmetry (spin rotation and lattice rotation,
respectively), and two symmetries are unrelated to each other, the
conventional LGW theory of phase transitions implies that such phases
must be separated by a FOQT.  However, arguments in favor of a
continuous phase transition have been put forward~\cite{CSY-94, MV-04,
  SBSVF-04, SVBSF-04, Sachdev-book}, based on the concept of
deconfined criticality. Indeed, the N\'eel-to-VBS transition in $2d$
antiferromagnetic SU(2) quantum systems represent paradigmatic models
for deconfined criticality, arising from the emergence of a U(1) gauge
field, see also Refs.~\cite{TIM-05, TIM-06, Sandvik-07, MK-08,
  JNCW-08, Sandvik-10, Kaul-12, KS-12, BMK-13, KMS-13, Harada-etal-13,
  Chen-etal-13, PDA-13, NCSOS-15, SGS-16, Sachdev-16, WNMXS-17}.  The
corresponding theory at the transition is supposed to be the $3d$
Abelian-Higgs (scalar electrodynamics) theory~\cite{ZinnJustin-book}
characterized by a U(1) gauge symmetry.  In particular, the relevant
model is expected to be the lattice Abelian-Higgs model with
two-component complex scalar fields and noncompact gauge fields (in
which there are no monopoles). This is of interest in several
condensed-matter physics applications, since the presence of Berry
phases in the quantum setting gives rise to the suppression of
monopoles~\footnote{These are directly related to the Berry phases in
  the quantum case~\cite{Haldane-88}.}  (see, e.g., Ref.~\cite{BMK-13}
and references therein). Note that noncompact gauge fields give rise
to important differences with respect to lattice Abelian-Higgs models
with compact gauge fields~\cite{PV-19-AH3d}.  Theoretical and
numerical investigations of classical and quantum transitions, which
are expected to be in the same universality class as those occurring
in noncompact scalar electrodynamics with two-component scalar
fields, have provided evidence of weakly first-order or continuous
transitions belonging to a new universality class (see, e.g.,
Refs.~\cite{SBSVF-04, KPST-06, Sandvik-07, MK-08, JNCW-08, MV-08,
  KMPST-08-a, KMPST-08, CAP-08, LSK-09, CGTAB-09, CA-10, BDA-10,
  Sandvik-10, HBBS-13, Bartosch-13, Harada-etal-13, Chen-etal-13,
  PDA-13, BS-13, KMS-13, NCSOS-15, NSCOS-15, SP-15, SGS-16,
  PV-20-mfcp, SN-19, SZ-20, BPV-21}).

\section{Equilibrium scaling behavior at continuous quantum transitions}
\label{escalingcqt}

In this section we report an overview of the equilibrium scaling
properties expected at generic CQTs, as inferred by the RG theory of
critical phenomena specialized to QTs, i.e., taking into account the
peculiar features that distinguish QTs from the {\em classical} transitions
driven by thermal fluctuations.  We present a detailed analysis of the
RG scaling ansatz in the thermodynamic limit and in the FSS limit.
The asymptotic quantum critical behaviors, and the scaling corrections
characterizing the approach to the leading laws, have been thoroughly
checked in various analytical and numerical studies (see in particular
Ref.~\cite{CPV-14}, containing a detailed study for the quantum XY chains).

\subsection{Quantum-to-classical mapping}
\label{qtocl}

Several fundamental ideas of the RG theory of critical phenomena find
their origins in the seminal works on classical systems by Kadanoff,
Fisher, Wilson, among the others (see, e.g., Refs.~\cite{Fisher-71,
  WK-74, Fisher-74, Ma-book, BGZ-76, Wegner-76, Aharony-76, Wilson-83,
  Abraham-86, CJ-86, Cardy-review, Privman-book, PHA-91, Fisher-98,
  ZinnJustin-book, ID-book1, ID-book2, PV-02, Cardy-book}).  Their
extension to quantum systems is based on the quantum-to-classical
mapping, which allows one to map the quantum system on a spatial
volume $V_s$ onto a classical one defined in a box of volume $V_c =
V_s \times L_T$, with $L_T=1/T$ (using the appropriate
units)~\cite{SGCS-97, Sachdev-book, CPV-14}.  In fact, under the
quantum-to-classical mapping, the inverse temperature corresponds to
the system size in an imaginary time direction.  The BC along the
imaginary time are periodic or antiperiodic, respectively for bosonic
and fermionic excitations.  Thus, the temperature scaling at a QCP in
$d$ dimensions is analogous to FSS in a corresponding $(d+1)$-dimensional
classical system. 

Before presenting the main ideas of the RG scaling theory at CQTs, we
would like to further comment on the quantum-to-classical mapping as
guiding approach.  It is important to stress that such mapping does
not generally lead to standard classical isotropic systems in thermal
equilibrium.  Indeed, while it is true that a quantum system can be
mapped onto a classical one, the corresponding classical systems are
generally anisotropic.  In some cases, when the dynamic exponent is
$z=1$ like Ising CQTs, the anisotropy is weak, as in the classical
Ising model with direction-dependent couplings. In these cases, a
straightforward rescaling of the imaginary time allows one to recover
space-time rotationally invariant (relativistic) $\Phi^4$ theories
such as those reported in Eq.~\eqref{HON}.  There are also interesting
cases in which $z\not=1$, such as the superfluid-to-vacuum and Mott
transitions of lattice particle systems described by the Hubbard and
BH models, which have $z=2$ when they are driven by the chemical potential
(see Sec.~\ref{bhmodel}). For CQTs with $z \neq 1$, the anisotropy is
strong, i.e., correlations have different exponents in the spatial and
thermal directions.  Indeed, in the case of quantum systems of size
$L$, under a RG rescaling by a factor $b$ such that $\xi\to\xi/b$ and
$L\to L/b$, the additional spatial dimension related to the
temperature must rescale differently, as $L_T \to
L_T/b^z$.~\footnote{An extreme case of quantum-to-classical mapping
occurs at FOQTs, with very anisotropic classical
counterparts~\cite{PF-83, CPV-14}, characterized by an exponentially
larger length scale along the imaginary time (see next section).}
However, scaling and FSS is also established for classical transitions
with such anisotropies~\cite{Zia-review}.~\footnote{For example,
the classical anisotropic scenarios emerge at dynamic off-equilibrium
transitions in driven diffusive systems~\cite{Zia-review, SZ-98}.}
Therefore, the classical FSS framework~\cite{Wegner-76, Diehl-86,
  Privman-book, PHA-91, SS-00, PV-02} can be generally extended to
systems at CQTs~\cite{SGCS-97, CPV-14}.

We also mention another problem of the quantum-to-classical mapping: in
some cases, the corresponding classical system has complex-valued
Boltzmann weights, which can be hardly studied in the classical
framework. The above considerations suggest that, to achieve a
satisfactory understanding of the quantum dynamics in many-body systems,
in particular when addressing issues related to the real-time
out-of-equilibrium dynamics, a discussion of the specific problems
related to the quantum nature of the phenomena is often
required~\cite{Sachdev-book}, in addition to approaches based on the
quantum-to-classical mapping.

\subsection{Scaling law of the free energy in the thermodynamic limit}
\label{freeen}

According to the RG theory of critical phenomena, the Gibbs free
energy obeys a general scaling law. Indeed, we can write it in terms
of the nonlinear scaling fields associated with the RG eigenoperators
at the fixed point of the RG flow~\cite{Wegner-76, PV-02}.  Guided by
the quantum-to-classical mapping and the RG theory of critical
phenomena, analogous scaling laws are assumed at CQTs~\cite{SGCS-97,
  CPV-14}.  Therefore, close to a CQT the Gibbs free-energy density
\begin{equation}
  F \equiv - {T\over V} \ln Z\,,\qquad Z = {\rm Tr}\, \big[ e^{-\hat H/T} \big]\,,
  \label{gibfreeen}
\end{equation}
in the thermodynamic infinite-volume limit can be written in terms of
scaling fields~\cite{Wegner-76}, such as~\cite{CPV-14}
\begin{equation}
  F(T,r,h) = F_{\rm reg}(r,h^2) + F_{\rm sing} 
\big( u_t,u_r,u_h,\{v_i\} \big) \,.
  \label{Gsing-RG-1}
\end{equation}
Here $F_{\rm reg}$ is a nonuniversal function, which is analytic at
the critical point and must be even with respect to the parameter $h$
related to the odd perturbation; it is also generally assumed not to
depend on $T$~\cite{CPV-14}.~\footnote{For classical systems, in the
  absence of boundaries (e.g., for PBC or ABC), $F_{\rm reg}$ is
  assumed to be independent of $L$, or, more plausibly, to depend on
  $L$ only through exponentially small
  corrections~\cite{Privman-book,PHA-91,SS-00}. This conjecture can be
  naturally extended to the quantum case~\cite{CPV-14}, implying that
  $F_{\rm reg}$ does not depend on the temperature, since PBC or ABC are
  always taken in the thermal direction.}  The other contribution,
$F_{\rm sing}$, bears the nonanalyticity of the critical behavior and
its universal features. The arguments of $F_{\rm sing}$ are the
so-called nonlinear scaling fields~\cite{Wegner-76}. They are analytic
nonlinear functions of the model parameters, associated with the
eigenoperators that diagonalize the RG flow close to the RG fixed
point.  The scaling fields $u_r$ and $u_h$ are the relevant fields
related to the model parameters $r$ and $h$.  The scaling field $u_t$
associated with the temperature, $u_t\sim T$, is also relevant and has
a RG dimension $y_t = z$.  Beside the relevant scaling fields, there
is also an infinite number of irrelevant scaling fields $\{v_i\}$ with
negative RG dimensions $y_i$, which are responsible for the scaling
corrections to the asymptotic critical behavior in the infinite-volume
limit.  Using the standard notation (see, e.g., Ref.~\cite{PV-02}) and
assuming that they are ordered so that $|y_1| \le |y_2| \le \ldots$,
we set
\begin{equation}
  \omega = -y_1\,.
  \label{defomega}
\end{equation}
In general, the nonlinear scaling fields depend on the control
parameters $r$, $h$, and $T$.  However $u_r$, $u_h$, and $\{v_i\}$ are
conjectured not to depend on the temperature and the system
size~\cite{CPV-14}.~\footnote{This hypothesis is quite natural for
  systems with short-range interactions.  Under a RG transformation,
  the transformed bulk couplings only depend on the local Hamiltonian,
  hence they are independent of the boundary.}  Taking into account
the assumed ${\mathbb Z}_2$ symmetry and the respectively even and odd
properties of $r$ and $h$, close to the critical point the relevant
scaling fields $u_r$ and $u_h$ can be generally expanded as
\begin{equation}
  u_r = r + c_r r^2 + O(r^3,h^2r)\,,\qquad\qquad
  u_h = h + c_h r h + O(h^3,r^2 h)\,,
  \label{urhscal}
\end{equation}
where $c_r$ and $c_h$ are nonuniversal constants.  As for the
irrelevant scaling fields, they are usually nonvanishing at the
critical point.
The thermal scaling field $u_t$ is expected to behave as~\cite{CPV-14}
\begin{equation}
  u_t= s(r,h)\,T\,,\qquad\qquad s(r,h) = s_0 + s_r r + O(r^2,h^2)\,,
  \label{utscal}
\end{equation}
where $s(r,h)$ is an appropriate nonuniversal function, while $s_0$
and $s_r$ are constant (see also Sec.~\ref{fsstra}, where this
expression of $u_t$ is argued in the FSS context).

The singular part of the free energy~\eqref{Gsing-RG-1} is expected to
satisfy the homogeneous scaling law
\begin{equation}
  F_{\rm sing} \big( u_t,u_r,u_h,\{v_i\} \big) = b^{-(d+z)} \,
  F_{\rm sing} \big( b^z u_t, b^{y_r} u_r , b^{y_h} u_h,\{b^{y_i} v_i\} \big) \,,
  \label{Fsing-scaling}
\end{equation}
where $b$ is an arbitrary scale factor, and the critical limit is
obtained in the large-$b$ limit. To obtain more explicit scaling
relations, one can fix the arbitrariness of the scale parameter $b$,
such as
\begin{equation}
  b = u_t^{-1/z}\,,
  \label{bthlim}
  \end{equation}
which diverges in the critical limit $u_t\to 0$.  The following
asymptotic behavior of the free-energy density emerges:
\begin{equation}
  F = F_{\rm reg}(r,h^2) + u_t^{d/z+1} \, {\cal F} \big( u_t^{-y_r/z} u_r,
  u_t^{-y_h/z} u_h, \big\{ {u_t^{-y_i/z} v_i} \big\}
  \big)\,. \label{scalsingiv}
\end{equation}
For $h=0$ and $u_t\to 0$, since $y_i<0$ and $y_1=-\omega$,
we can expand it as
\begin{equation} 
  F \approx F_{\rm reg}(r,0) + u_t^{d/z+1} {\cal F}_0 \big( u_t^{-y_r/z} u_r \big)  
  + u_t^{d/z+1+\omega/z} {\cal F}_\omega\big( u_t^{-y_r/z} u_r \big) 
  + \ldots\,,
\label{Fsingiv-RG}
\end{equation}
where ${\cal F}_0$ and ${\cal F}_\omega$ are scaling functions.  Note that, to
eventually obtain the scaling relations in terms of the external
parameters $r, h$ and $T$ controlling the critical behavior, one needs
to expand the scaling fields, cf.~Eq.~\eqref{urhscal}, whose
subleading terms gives rise to {\em analytical} scaling corrections.
Therefore, the scaling relation~\eqref{Fsingiv-RG} predicts that the
free-energy density approaches the asymptotic behavior given by the
first term, with nonanalytic scaling corrections due to the irrelevant
RG perturbations and analytic contributions which are due to the
regular background associated with $F_{\rm reg}$ and to the expansions
of the nonlinear scaling fields in terms of the Hamiltonian
parameters.

Differentiating the Gibbs free energy, one can straightforwardly
derive analogous scaling ansatze for the energy density
\begin{equation}
  E(T,r,0) \equiv F - T{\partial F\over \partial T} \approx
  E_{\rm reg}(r,0) + 
  T^{d/z+1} \, \big[ {\cal E}_0 \big( T^{-y_r/z} r \big)  +
    T^{\omega/z} {\cal E}_\omega \big( T^{-y_r/z} r \big) 
    + \ldots \big]\,,
  \label{energydensitysca}
 \end{equation} 
and the specific heat 
\begin{equation}
  C_V \equiv {\partial E\over \partial T} \approx
  T^{d/z} \, \big[ {\cal C}_0 \big( T^{-y_r/z} r \big)  +
    T^{\omega/z} {\cal C}_\omega \big( T^{-y_r/z} r \big) 
    + \ldots \big]\,.
  \label{cviv}
\end{equation}
Notice that the specific heat does not get contributions from the
regular part of the free energy, since $F_{\rm reg}$, thus
$E_{\rm reg}$, is assumed to be independent of $T$.  At the critical point
$r = 0$, Eq.~\eqref{cviv} predicts
\begin{equation}
  C_V \sim T^{d/z} \big[ 1 + O(T^{\omega/z}) \big] \,.
  \label{cvnbeh}
\end{equation}

\subsection{Universality of the scaling functions}
\label{univscafu}

As already mentioned, the critical exponents characterizing the power
laws at continuous transitions are universal, i.e., they are shared by
all systems belonging to the given universality class, defined by a
few global features.  The universality extends to scaling functions
such as $\cal F$, related to the singular part of the free energy,
cf.~Eq.~\eqref{scalsing}, and more generally to the scaling functions
associated with other different quantities, which include the derivative
of the free energy, the correlation functions (see also below), etc.

Since scaling fields are arbitrarily normalized, universality holds
apart from a normalization of each argument and an overall
constant. Therefore, given two different models, if ${\cal A}_1$ and
${\cal A}_2$ are their scaling functions associated with a generic
observable, they are generally related as
\begin{equation}
  {\cal A}_1 \big( x_1,x_2, \ldots ) 
  = c_0 \, {\cal A}_2 \big( c_1 x_1,c_2 x_2, \ldots \big) \,,
\end{equation}
where $c_i$ are nonuniversal constants.  The above considerations
apply to generic scaling functions including those related to the
scaling corrections, extending also to the cases they allow for
further arguments, such as the system size and the spatial coordinates
for the correlation functions.

\subsection{Equilibrium finite-size scaling}
\label{fsssca}

The RG scaling theory developed in Sec.~\ref{freeen} holds in the
thermodynamic limit, that is expected to be well defined for any
$r\neq 0$ or $h\neq 0$, for which the correlation length $\xi$ is
finite.  Nonetheless, for practical purposes, both experimental and
numerical, one typically has to face with systems of finite
length. Such situations can be framed within the FSS framework, whose
scaling relations are obtained by also allowing for the dependence on
the size $L$.

Finite-size effects in critical phenomena have been the object of
theoretical studies for a long time~\cite{FB-72, Barber-83,
  Privman-book, PHA-91, Cardy-review, PV-02}. FSS describes the
critical behavior around a critical point, when the correlation length
$\xi$ of the critical modes becomes comparable to the size $L$ of the
system. For large sizes, this regime presents universal features,
shared by all systems whose transition belongs to the same
universality class.  Although originally formulated in the classical
framework, FSS also holds at zero-temperature QTs~\cite{SGCS-97,
  CPV-14} in which the critical behavior is driven by quantum
fluctuations.  Indeed, the RG approach~\cite{Wegner-76, Privman-book,
  PHA-91, PV-02} to FSS at classical finite-temperature transitions
(generally driven by thermal fluctuations) can be extended to QTs.
This allows us to characterize the asymptotic FSS behavior at QTs, and
also the nature of the corrections in systems of large, but finite,
size.

The FSS limit is essentially defined as the large-$L$ limit, keeping
the ratio $\xi/L$ fixed, where $\xi$ is a generic length scale of the
system, which diverges at the critical point. The FSS limit definitely
differs from the thermodynamic limit, which is generally obtained
by taking the large-$L$ limit, keeping $\xi$ fixed;
then the infinite-volume critical behavior is developed when
$\xi\to\infty$.  However, assuming an asymptotic matching of the FSS
and the thermodynamic limit, one may straightforwardly derive the
infinite-volume scaling behaviors from FSS, by taking the $\xi/L\to 0$
limit of the FSS ansatz.

The FSS approach is one of the most effective techniques for the
numerical determination of the critical quantities.  While
infinite-volume methods require that the condition $\xi \ll L$ is
satisfied, FSS applies to the less demanding regime $\xi\sim L$. More
precisely, the FSS theory provides the asymptotic scaling behavior when
both $L,\xi\to\infty$, keeping their ratio $\xi/L$ fixed.  However,
the knowledge of the asymptotic behavior may not be enough to estimate the
critical parameters, because data are generally available for limited
ranges of parameter values and system sizes, which are often
relatively small.  Under these circumstances, the asymptotic FSS
predictions are affected by sizable scaling corrections. Thus,
reliably accurate estimates of the critical parameters need a robust
control of the corrections to the asymptotic behavior.  This is also
important for a conclusive identification of the universality class of
the quantum critical behavior when it is {\em a priori} uncertain.
Moreover, an understanding of the finite-size effects is relevant
for the experiments, when relatively small systems are considered
(see, e.g., Ref.~\cite{GKMD-08}), or in particle systems trapped
by external (usually harmonic) forces, as in cold-atom setups
(see, e.g., Refs.~\cite{CW-02, Ketterle-02, Donner-etal-07, BDZ-08,
  CV-09, CV-10-tr}).

\subsubsection{The free-energy density}
\label{fsstra}

Within the FSS framework, the free-energy density can be written
as~\cite{Wegner-76, Diehl-86, Privman-book, PHA-91, SGCS-97, SS-00, CPV-14}
\begin{equation}
  F(L,T,r,h) = F_{\rm reg}(L,r,h^2) + F_{\rm sing}
  \big( u_l,u_t,u_r,u_h, \{v_i\}, \{ \widetilde{v}_{i} \} \big)\,,
  \label{Gsing-RG-1fss}
\end{equation}
where $F_{\rm reg}$ is again a nonuniversal function, and
$F_{\rm sing}$ bears the nonanalyticity of the critical behavior.
Notice that we extended the arguments of $F_{\rm sing}$ in
Eq.~\eqref{Gsing-RG-1}, also including the scaling field $u_l\sim L^{-1}$
associated with the system size $L$, and further irrelevant
surface scaling fields $\{\widetilde{v}_{i}\}$ with corresponding
negative RG scaling dimensions $\widetilde{y_i}<0$.~\footnote{See, e.g.,
  Ref.~\cite{CL-91} for a discussion of the boundary operators within CFT.}
The latter may arise from the presence of the boundaries, while they
are absent when the system of size $L$ has no boundaries,
such as for PBC or ABC.

Specifically, the scaling fields $u_l$ and $u_t$ are associated with
the size of the $(d+1)$-dimensional system. For classical systems in a
box of size $L^d$ with PBC or ABC and, more generally, for
translation-invariant BC, it is usually assumed that~\footnote{This
has been verified in many instances (for instance, in the $2d$
classical Ising model) and extensively discussed in Ref.~\cite{SS-00},
and can be justified as follows~\cite{CPV-14}. Consider a lattice
system and a decimation transformation which reduces the number of
lattice sites by a factor $2^d$.  In the absence of boundaries and for
short-range interactions, the new (translation-invariant) couplings
are only functions of the couplings of the original lattice and are
generally independent of $L$, while $L\to L/2$.  Therefore the flow of
$L$ is independent of the flow of the couplings, leading to the
conjecture that $u_l=1/L$ in the case of translation-invariant BC,
such as PBC or ABC.}
\begin{equation}
  u_l = 1/L\,. \label{ulsca}
\end{equation}
This condition does not generally hold for non translation-invariant
systems, such as OBC or FBC, where $u_l$ is generally expected to be
an arbitrary function of $1/L$, which can be expanded as
\begin{equation}
  u_l = L^{-1} + b L^{-2} + \ldots, \qquad \mbox{for } \;
  L \to \infty \,.~\footnote{A more detailed discussion of this expression
  and the effects of the subleading $O(L^{-1})$ terms 
  can be found in Ref.~\cite{CPV-14}.}
  \label{ull}
\end{equation}

At this point a comment is in order, to explain the
expression~\eqref{utscal} for the scaling field $u_t$ associated with
the temperature.  Let us assume that $z=1$, so that the quantum system
is equivalent to a classical $(d+1)$-dimensional system. The classical
system is, however, weakly anisotropic: couplings in the thermal
direction differ from those in the spatial one. Moreover, the
anisotropy depends on the model parameters. In classical weakly
anisotropic systems, universality is obtained only after transforming
them to an isotropic system by means of a scale transformation (see
Refs.~\cite{DC-09, Kastening-12} and references therein).  Therefore,
one generally obtains the nontrivial expression reported in
Eq.~\eqref{utscal}, even in the case $z=1$.
  
The scaling equation for the singular part $F_{\rm sing}$
in~\eqref{Gsing-RG-1fss} can be thus obtained from
Eq.~\eqref{Fsing-scaling}, by including the dependence on $L$:
\begin{equation}
  F_{\rm sing} \big( u_l,u_t,u_r,u_h,\{v_i\} \big)
  = b^{-(d+z)} \, F_{\rm sing} \big( b \,u_l, b^z u_t, b^{y_r} u_r , b^{y_h} u_h,
  \{b^{y_i} v_i\} \big) \,,
  \label{Fsing-scaling-fss}
\end{equation}
where $b$ is again arbitrary, and we have neglected the contributions
of the surface scaling $\{ \widetilde{v}_{i} \}$ (see
Ref.~\cite{CPV-14} for a more complete analysis including them).  To
derive the FSS behaviors, we fix $b = 1/u_l$, thus getting
\begin{equation}
  F_{\rm sing} =  
  u_l^{d+z} {\cal F} \big( u_l^{-z} u_t, u_l^{-y_r} u_r, u_l^{-y_h} u_h, 
  \{u_l^{-y_i} v_i\} \big) \,.
  \label{scalsing}
\end{equation}
The arguments $\{ u_l^{-y_i} v_i \}$, corresponding to the irrelevant
scaling fields, vanish for $L\to\infty$, since $y_i$ are negative.
Thus, provided that $F_{\rm sing}$ is finite and nonvanishing in this
limit, we can expand the singular part of the free energy as
\begin{equation} 
  F_{\rm sing} \approx
  u_l^{d+z} \, {\cal F}_0 \big( u_l^{-z} u_t, u_l^{-y_r} u_r, u_l^{-y_h} u_h \big)
  +v_1 u_l^{d+z+\omega} \,
  {\cal F}_\omega\big( u_l^{-z} u_t, u_l^{-y_r} u_r, u_l^{-y_h} u_h \big)
  + \ldots
  \label{Fsing-RG}
\end{equation}
where we only retain the contribution of the dominant (least)
irrelevant scaling fields, of RG dimension $-\omega$.~\footnote{The
  expansion~\eqref{Fsing-RG} is only possible below the upper critical
  dimension~\cite{Fisher-74}.  Above it, $F_{\rm sing}$ is singular
  and the failure of this expansion causes a breakdown of the
  hyperscaling relations, allowing one to recover the mean-field
  exponents.  For an analysis above the upper critical dimension
  see, e.g., Ref.~\cite{KK-10}.}

The dependence on $L$ of the regular function $F_{\rm reg}$ entering
Eq.~\eqref{Gsing-RG-1fss} can be generally assumed to be suppressed in
the case of translation-invariant systems~\cite{CPV-14}.
This extends analogous conjectures for classical systems,
where in the absence of boundaries, $F_{\rm reg}$ is assumed to be
independent of $L$, or, more plausibly, to depend on $L$ only through
exponentially small corrections~\cite{Privman-book, PHA-91,
  SS-00}. On the other hand, for generic spatial BC, a regular
expansion in powers of $1/L$ is assumed:
\begin{equation}
  F_{\rm reg}(L,r,h^2) \approx F_{{\rm reg},0}(r,h^2) + {1\over L} F_{{\rm
      reg},1}(r,h^2) + \ldots \,.
  \label{favb}
\end{equation}
Therefore the FSS behaviors for systems without boundaries are
substantially simpler than those with boundaries.  Indeed, in the former
situation, the FSS behavior of Eqs.~\eqref{Gsing-RG-1fss}
and~\eqref{Fsing-RG} simplifies to
\begin{align}
  F(L,T,r,h) & = F_{\rm reg}(r,h^2) + F_{\rm sing}( L,T,r,h)\,,
  \nonumber \\
  F_{\rm sing}(L,T,r,h) & \approx
  L^{-(d+z)} \left[ {\cal F}_0 \big( L^z u_t, L^{y_r} u_r , 
L^{y_h} u_h \big)
    + L^{-\omega} {\cal F}_\omega \big( L^z u_t, L^{y_r} u_r, 
L^{y_h} u_h \big)
    + \ldots \right]  \,.
  \label{Gsing-RG-1fss-ti}
\end{align}
Note that the scaling behavior in the thermodynamic limit (see
Sec.~\ref{freeen}) can be recovered from the FSS
law~\eqref{Fsing-scaling-fss}, by choosing $b=u_t^{-1/z}$ and taking
the limit $L/b \to \infty$.

The universality properties of the scaling function $\cal F$ discussed
at the end of Sec.~\ref{freeen} extend to the FSS framework.
However, the universal scaling functions depend on the shape of the
finite systems that are considered and on the type of BC.  On the
basis of the above general analyses, the asymptotic critical behavior
is subject to several different sources of scaling corrections:

\begin{itemize}
\item The irrelevant RG perturbations which generally give rise to
  $O(L^{-\omega})$ corrections, where $\omega$ is the universal
  exponent associated with the leading irrelevant RG perturbation.

\item Corrections arising from the expansion of the scaling fields
  $u_r$, $u_h$, and $u_t$ in terms of the Hamiltonian parameters $t$,
  $h$, and also the temperature $T$.  For example, they give rise to
  corrections of order $L^{-1/\nu}$, when $r\neq 0$.

\item Corrections arising from the analytic background term of the
  free energy, i.e., from $F_{\rm reg}$ in Eqs.~\eqref{Gsing-RG-1}
  and~\eqref{Gsing-RG-1fss}.

\item The irrelevant RG perturbations associated with the BC,
  which are of order $L^{-\omega_s}$, where $\omega_s$ is related to
  leading perturbation arising from the boundaries.  They are absent
  in finite systems without boundaries.

\item The $O(1/L)$ boundary corrections arising from the nontrivial
  analytic $L$-dependence of the scaling field $u_l$, Eq.~\eqref{ull}.
  The leading correction can be taken into account by simply
  redefining the length scale $L$, i.e., by using $L_{e}= L+c$ (where
  $c$ is an appropriate constant) instead of $L$~\cite{Hasenbusch-09a,
    Hasenbusch-12, Diehl-etal-12, CPV-14}.  These corrections are not
  present in the absence of boundaries, such as with PBC or ABC.
\end{itemize}

Equations~\eqref{scalsing} and~\eqref{Fsing-RG} provide the generic
FSS form of the free-energy density.  However, in certain cases the
behavior is more complex, due to the appearance of logarithmic
terms~\cite{Wegner-76}. They may be induced by the presence of
marginal RG perturbations, as happening in BKT transitions in
U(1)-symmetric systems~\cite{B-72, KT-73, AGG-80, Hasenbusch-05,
  PV-13}, or by resonances between the RG eigenvalues, as it occurs in
transitions belonging to the $2d$ Ising universality
class~\cite{Wegner-76, CHPV-02}~\footnote{In classical $2d$ systems,
  the logarithmic behavior of the specific heat is essentially related
  to a resonance between the identity operator and the thermal
  operator~\cite{Wegner-76}.} or to the $3d$ O$(N)$-vector
universality class in the large-$N$ limit~\cite{Diehl-etal-12, PV-98,
  PV-99}. We also mention that quantum particle systems at fixed
chemical potential may show further peculiar features when an infinite
number of level crossings occurs as the system size increases.  This
emerges in the BH models~\eqref{bhm} and XX models~\eqref{XXH} within
the zero-temperature critical superfluid phase~\cite{CV-10-XX,
  CMV-11-jstat, OYNR-15} (see also Sec.~\ref{modulations}).

Some interesting quantities can be obtained by taking derivatives of
the free energy. For example, in particle systems whose relevant
parameter $r$ is a linear function of the chemical potential $\mu$
(i.e., $r \sim \mu-\mu_c$ where $\mu_c$ is the critical point), the
FSS of the particle density is obtained by differentiating
Eq.~\eqref{Gsing-RG-1fss-ti} with respect to $\mu$ and therefore to
$r$. For $h=0$, we obtain
\begin{equation}
  \rho(L,r) \equiv {\partial F\over \partial r} \approx \rho_{\rm reg}(r)
  + {1\over L} \rho_{\rm reg,1}(r) + \ldots  + L^{-(d+z-y_r)} \,
  {\cal D}_p(L^z u_t, L^{y_r} u_r) + \ldots \,,
  \label{pade}
\end{equation}
where ${\cal D}_p$ is a universal scaling function apart from trivial
factors, such as a multiplicative one and the normalizations of its
arguments.  We note that the regular term represents the leading one
when $d+z-y_r>0$, as in most physically interesting systems.  In
particular at $T=0$ and for translation-invariant systems, the above
formula simplifies to
\begin{equation}
  \rho(L,r) \approx \rho_{\rm reg}(r) + L^{-(d+z-y_r)} \,
  \widetilde{\cal D}_p(L^{y_r}r)+ \ldots  \,.
\label{pade2}
\end{equation}
The compressibility can be obtained by taking an additional
derivative with respect to the chemical potential.

\subsubsection{The low-energy scales}
\label{fsslowene}

The singular part of the free energy is essentially determined by the
behavior of the low-energy levels at the CQT.  Therefore, the
low-energy scales, and in particular the energy difference $\Delta$ of
the lowest-energy levels, should show an analogous asymptotic behavior,
beside the leading term $\Delta \sim L^{-z}$~\cite{Sachdev-book}.
Thus, at $T=0$ and $h=0$, and for translation-invariant systems,
they are expected to show the asymptotic FSS behavior
\begin{equation}
  \Delta(L,h=0) \approx  L^{-z} \, \big[ {\cal D}_0( L^{y_r} u_r ) 
    + L^{-\omega} {\cal D}_\omega( L^{y_r} u_r ) + \ldots \big] \,.
  \label{Deltasca}
\end{equation}
Note that ${\cal D}_0(x)\sim x^{z\nu}$ for $x\to\infty$, to ensure
$\Delta\sim r^{z\nu}$ for $r>0$ (paramagnetic phase) in the
infinite-volume limit.
A more general discussion of the behavior of $\Delta$, allowing for
the existence of boundaries, can be found in Ref.~\cite{CPV-14}.

\subsubsection{The correlation function of the order-parameter field}
\label{fsstwopc}

We now consider the correlation function of the order-parameter field
$\phi({\bm x},t)$, as for example, the equal-time two-point function,
\begin{equation}
  G({\bm x},{\bm y}) \equiv \langle \phi({\bm x},t)
  \, \phi({\bm y},t) \rangle\,.
  \label{gxy}
\end{equation}
For vanishing magnetic field, the leading scaling behavior for
$L\to \infty$, $|{\bm x}-{\bm y}|
\to \infty$ with $|{\bm x} - {\bm y}|/L$ fixed,  
is given by
\begin{equation}
  G({\bm x},{\bm y};L,T,r) \approx u_l^{d+z-2+\eta} \, {\cal G}_0(u_l
  {\bm x}, \, u_l {\bm y}, \, u_l^{-z} u_t, u_l^{-y_r} u_r)\,,
  \qquad\qquad u_l = L^{-1} + O(L^{-2})\,.
  \label{gxysca}
\end{equation}
Corrections to Eq.~\eqref{gxysca} arise from two different sources.
First of all, they can be due to the scaling fields with negative RG
dimensions.  Moreover, there are also corrections due to {\em
  field-mixing} terms~\cite{CPV-14}, related to the fact that the
order-parameter field $\phi$ may in general be a linear combination
containing other odd operators ${\cal O}_{h,i}$, i.e.,  $\phi =
\sum_{i} a_i {\cal O}_{h,i}$.  Equation~\eqref{gxysca} represents
the contribution of the leading operator ${\cal O}_h$ since
\begin{equation}
  d + z - y_h = \tfrac12 (d +z - 2 + \eta) \, . 
  \label{dzeta}
\end{equation}

One may consider the space integral of the correlation
function~\eqref{gxy}, defined as
\begin{equation}
  \chi_{\bm y} \equiv  \sum_{{\bm x}} G({\bm y},{\bm x})\,.
  \label{chidef}
\end{equation}
In the presence of a boundary, as long as ${\bm y}$ is fixed in the
FSS limit, the leading scaling behavior is always the same, while
scaling corrections are expected to depend on ${\bm y}$.  When
translation invariance holds, $\chi\equiv \chi_{\bm y}$ is independent
of ${\bm y}$.  In the latter case and for $h=0$ and $T=0$, the
asymptotic FSS expansion is expected to be~\cite{CPV-14}
\begin{equation}
  \chi(L,r) \approx  L^{2-z-\eta} \big[ {\cal X}_0( L^{y_r} u_r) + 
    L^{-\omega} {\cal X}_\omega(L^{y_r} u_r) + \ldots \big] + B_\chi(r) ,
  \label{chisclaw}
\end{equation}
where the scaling functions ${\cal X}_\#$ are universal, apart from
multiplicative factors and a normalization of the argument. The
function $B_\chi$ is an analytic background term which represents the
contribution to the integral of points $\bm x$ such that
$|{\bm x}-{\bm y}|\ll L$.

One can also consider the length scale $\xi$ associated with the
critical modes, which diverges in the thermodynamic limit, as
described by Eqs.~\eqref{xidiv1} and~\eqref{xidiv2}.  Various
definitions of correlation lengths are usually considered.  One may
define a correlation length $\xi_e$ from the large-distance
exponential decay of the two-point correlation function~\eqref{gxy},
i.e.,
\begin{equation}
  G({\bm x},{\bm y})\sim \exp(-|{\bm x}-{\bm y}|/\xi_e) \,,
      \label{xiegxy}
\end{equation}
provided the system is not at a critical point, or assuming that at
least one of the spatial dimensions is infinite.  An alternative
definition, particularly useful for the analysis of finite systems,
can be extracted from the second moment of the correlation
function~\eqref{gxy}, i.e.,
\begin{equation}
\xi^2 \equiv {1\over 2d\chi_{\bm 0}} \sum_{\bm x} {\bm x}^2 G({\bm 0},{\bm x})
\label{xidef} 
\end{equation}
where the point ${\bm y}=0$ is at the center of the system.  In the
case of PBC or ABC, one may consider the more convenient definition
\begin{equation}
  \xi^2 \equiv  {1\over 4 \sin^2 (p_{\rm min}/2)} 
     {\widetilde{G}({\bm 0}) - \widetilde{G}({\bm p})\over 
       \widetilde{G}({\bm p})},
     \label{xidefpb}
\end{equation}
where $p_{\rm min} \equiv 2 \pi/L$, while ${\bm p}$ is a vector with
only one nonvanishing component equal to $p_{\rm min}$, and
$\widetilde{G}({\bm p})$ is the Fourier transform of $G({\bm x})$.
Leading scaling corrections turn out to be analogous to those for
$\chi$~\cite{CPV-14}.  Since the above length scales have RG dimension
1, their FSS behavior turns out to be
\begin{equation}
  {\xi(L,r)\over L}  = {\cal Y}(L^{y_r} u_r) + L^{-\omega} {\cal
      Y}_\omega( L^{y_r} u_r) + \ldots 
  \label{xisca}
\end{equation}
Note that, depending on the definition of the correlation length, we
may also have further subleading contributions arising from background
regular terms. They are absent for correlation lengths extracted from
asymptotic exponential behaviors in anisotropic lattices, such as
slab-like geometries where only one of the spatial directions is
finite. However, they are present in the case of the second-moment
definition~\eqref{xidefpb}, essentially related to the background term
in the susceptibility $\chi$, cf.~Eq.~\eqref{chisclaw}~\cite{CPV-14},
giving rise to further $O(L^{-(2-z-\eta)})$ scaling corrections.

\subsubsection{Renormalization-group invariant quantities}
\label{fssrginv}

Dimensionless RG invariant quantities are particularly useful to
investigate the critical region. Examples of such quantities are the
ratio
\begin{equation}
  R_\xi\equiv \xi/L\,,
  \label{rxidef}
\end{equation}
where $\xi$ is any length scale related to the critical modes, for
example the one defined in Eq.~\eqref{xidef}, and the ratios of the
correlation function $G$ at different distances, e.g.,
\begin{equation}
  R_g({\bm X},{\bm Y}) = \ln \left[ \frac{G({\bm 0},{\bm x})}
    {G({\bm 0},{\bm y})} \right] \,, \qquad {\bm X} \equiv \frac{\bm x}{L}\,,
  \quad {\bm Y} \equiv \frac{\bm y}{L}\,,
  \label{rgxy}
\end{equation}
where the point ${\bm 0}$ is located at the center of the system.
We denote them generically by $R$.  According to FSS, for
translation-invariant systems at $T=0$ and $h=0$, they must behave as
\begin{equation}
  R(L,r) \approx {\cal R}_0 (L^{y_r} u_r)  +
  L^{-\omega} \,{\cal R}_\omega( L^{y_r} u_r) + \ldots \,.
  \label{rscaf}
\end{equation}
Note that the expansion of the scaling field $u_r = r + c_r r^2 + \ldots$
gives rise to $O(L^{-y_r})$ corrections. Further corrections would
arise in the presence of boundaries~\cite{CPV-14}.

The scaling function ${\cal R}_0(x)$ is universal, apart from a trivial
normalization of the argument.  In particular, the limit
\begin{equation}
  \lim_{L\to\infty} R(L,0) =  {\cal R}_0(0)
  \label{rzero}
\end{equation}
is universal within the given universality class, i.e., it is
independent of the microscopic details of the model, although it
depends on the shape of the finite volume and on the BC.

The FSS behavior of the RG-invariant quantities $R$ can be exploited
to determine the critical point~\cite{Binder-83, PV-02}. Indeed, when
\begin{equation}
  \lim_{r\to0^-}\lim_{L\to\infty}R(L,r) > \lim_{L\to\infty}R(L,0) >
  \lim_{r\to0^+}\lim_{L\to\infty}R(L,r)
  \label{monbeh}
\end{equation}
or viceversa, one can define $r_{\rm cross}$ by requiring
\begin{equation}
  R(L,r_{\rm cross}) = R(2L,r_{\rm cross})\,.
  \label{rcross}
\end{equation}
The crossing point $r_{\rm cross}$ converges to $r=0$ with corrections
that are typically of order $L^{-1/\nu-\omega}$~\cite{CPV-14}.

We finally note that 
taking a RG invariant quantity $\widetilde{R}$ that is monotonic with
respect to the relevant parameter $r$, as is generally the case of
$R_\xi$, one may write another generic RG invariant quantity $R_i$ as
\begin{equation}
  \label{uvsrxi}
  R_i(L,r) = {\cal R}_i(\widetilde{R}) + O(L^{-\omega})\,,
\end{equation}
where ${\cal R}_i$ now depends on the universality class only, without
any nonuniversal multiplicative factor. This is true once the BC and
the shape of the lattice have been fixed, provided one uses
corresponding quantities in the different models. The
scaling~\eqref{uvsrxi} is particularly convenient to test
universality-class predictions, since it permits easy comparisons
between different models without any tuning of nonuniversal
parameters.~\footnote{See for example Refs.~\cite{PV-19-AH3d, BPV-19,
    BPV-20} for some applications of this strategy to identify the
  nature and universality class of classical finite-temperature
  transitions.}

\subsection{Modulated finite-size effects in some particle systems}
\label{modulations}

Some peculiar finite-size behaviors are observed in quantum particle
systems at fixed chemical potential, when an infinite number of level
crossings occurs as the system size increases. They are characterized
by modulations of the leading amplitudes of the power-law behaviors.
These phenomena are observed within the zero-temperature superfluid
phase of the BH and XX models~\cite{CV-10-XX, OYNR-15}.
They arise because the particle number is conserved, i.e., the
operator $\hat{N} = \sum_{\bm x} \hat n_{\bm x}$ commutes
with the BH Hamiltonian~\eqref{bhm} and the XX
Hamiltonian~\eqref{XXH}:
$[\hat{N}, \hat H_{\rm BH}] = [\hat{N}, \hat H_{\rm XX}] = 0$.
Thus the eigenvectors do not depend on $\mu$, even though the
eigenvalues do.  In a system of size $L$, the particle number
$N\equiv\langle\hat N\rangle$ is generally finite and increases as $N
\sim L^d$ with increasing the size, keeping the chemical potential
$\mu$ fixed.  Therefore, as $L\to\infty$, the system may generally
show an infinite number of ground-state level crossings where $N$
jumps by 1 and the gap $\Delta$ vanishes.

Some simple expressions can be obtained for the $1d$ BH model in the
hard-core limit, coinciding with the $1d$ XX chain, in the superfluid
region $|\mu|\le 2$. Within the superfluid phase, the system is
critical, with vanishing gap in the thermodynamic limit.  In the
infinite-size limit $L\to\infty$, the filling $f$ is given
by~\cite{Sachdev-book}
\begin{equation}
  f \equiv \langle n_i \rangle =  1 - \frac{1}{\pi} \arccos \,
  \bigg( \frac{\mu}{2} \bigg) \,, \qquad -2\le \mu \le 2\,.
  \label{fmu}
\end{equation}
Thus, in the range $-2<\mu<2$, we have $0<f<1$.  In the limit $\mu\to 2$
($f\to 1$), Eq.~\eqref{fmu} gives $N=L$ without vacuum degeneration,
which is the expected result for $\mu\ge 2$.  Analogously,
for $\mu\to -2$ ($f\to 0$), one obtains $N=0$ without vacuum
degeneration, which is the expected result for $\mu\le -2$.

Let us now consider a homogeneous system of finite size $L$ with OBC,
reporting some results from Ref.~\cite{CV-10-XX}.  The excitation
number $N$ for $|\mu|<2$ is exactly given by $N =
\lfloor(L+1)f\rfloor$, without any finite-$L$ correction.  For integer
$(L+1)f$, the ground state is degenerate ($\Delta=0$); the
lowest-energy simultaneous eigenvectors of the Hamiltonian and the
particle number give $N=(L+1)f$ and $N=(L+1)f-1$. For $f=1/s$ with
integer $s$, for every value of $N$ we have a vacuum degeneracy when
$L+1=Ns$, i.e., $\Delta=0$ for $L+1\equiv0 \, (\mbox{mod } s)$.  For
$f=r/s$ with integer $r$ and $s$, again $\Delta=0$ for $L+1\equiv0 \,
(\mbox{mod } s)$, but we can satisfy $L+1=Ns/r$ only for $N\equiv0 \,
(\mbox{mod } r)$.  For irrational $f$, $\Delta$ never vanishes for
integer $L$.

It is useful to define
\begin{equation}
  \varphi \equiv \{(L+1)f\}\,, \quad {\rm where}\;\; \{x\} \equiv
  x - \lfloor x\rfloor\,,
  \label{varphidef}
\end{equation}
which is the fractional part of $x$ (i.e., the
sawtooth function).  For integer $(L+1)f$, one can label the two
degenerate vacua with $\varphi=0\,,1$ according to $N+\varphi = (L+1)f$.
For each value of $\mu$, we have the asymptotic
behavior~\cite{CV-10-XX}
\begin{equation}
  L \,\Delta = \pi \sqrt{1-\mu^2} \, \big( 1/2 - |\varphi-1/2| \big) +
  O(L^{-1})\,.
%\qquad \varphi \equiv \{(L+1)f\}\,.
  \label{deltafss}
\end{equation}
Therefore, at fixed $\mu$, the amplitude of the leading $L^{-1}$ power
law turns out to be a periodic function of the size.  This corresponds
to an asymptotic periodicity of the $L$-dependence of $L\,\Delta$ with
period $1/f$.  Note that the amplitude vanishes for $\mu=\pm 2$,
indeed $\Delta=O(1/L^2)$ at $\mu=\pm 2$ without level crossings,
consistently with the fact that the corresponding continuum theory has
$z=2$.  Instead, for $|\mu|>2$ we have $\Delta = 2(|\mu|-1) + O(L^{-2})$.

At fixed chemical potential, the particle density in the middle of the
chain $\langle n_0\rangle$ for odd $L$, behaves as
\begin{equation}
  \langle n_0\rangle-f = {1-\bar\varphi\over L+1}\,,\qquad
  \bar\varphi \equiv 2\{[(L+1)f+1]/2\} \,,
  \label{eq:n0,allmus,pinf}
\end{equation}
without any finite-$L$ correction. Note that $0\le \bar\varphi < 2$
(thus either $\bar\varphi=\varphi$ or $\bar\varphi=\varphi+1$).
Moreover all the above formulas are invariant under the particle-hole
exchange $n_i\to1-n_i$, which implies $N\to L-N$, $f\to1-f$, and
$\mu\to-\mu$.

\subsection{Critical behavior in the presence of
  an external inhomogeneous field}
\label{inhomsyst}

Statistical systems are generally inhomogeneous in nature, while
homogeneous systems are often an ideal limit of experimental
conditions.  Thus, in the study of critical phenomena, an important
issue is how critical behaviors develop in inhomogeneous systems.
Particularly interesting situations arise when interacting particles
are constrained within a limited region of space by an external force.
This is a common feature of recent experiments with dilute atomic
vapors~\cite{CW-02, Ketterle-02} and cold atoms in optical
lattices~\cite{BDZ-08}, which have provided a great opportunity to
investigate the interplay between quantum and statistical behaviors in
particle systems.  In experiments of trapped particle systems aimed at
investigating their many-body critical behaviors at quantum and
thermal transitions, an accurate determination of the critical
parameters, such as the critical temperature, critical exponents,
etc., calls for a quantitative analysis of the trap effects.  This
issue has been much discussed within theoretical and experimental
investigations (see, e.g., Refs.~\cite{DSMS-96, vDK-97, WATB-04,
  RM-04, FWMGB-06, NCK-06, DZZH-07, HK-08, GZHC-09, BB-09, Taylor-09,
  CV-09, BDSB-09, ZKKT-09, RBRS-09, HR-10, Trotzky-etal-10, CV-10-tr,
  ZH-10, HZ-10, PPS-10, PPS-10b, NNCS-10, ZKKT-10, CV-10-XX, QSS-10,
  CV-10-ent, CV-10-offeq, FCMCW-11, ZHTGC-11, CV-11, Mahmud-etal-11,
  HM-11, CoV-11, CTV-12, Pollet-12, CT-12, KLS-12, CR-12, CTV-13,
  PV-13, DV-17a}.

\begin{figure}
  \begin{center}
    \includegraphics[width=0.5\columnwidth]{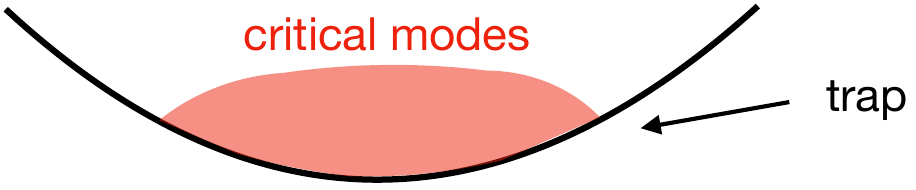}
    \caption{Sketch which highlights the emergence of the critical modes
      within a trap.}
    \label{trapsketch}
  \end{center}
\end{figure}

The critical behavior arising from the formation of BEC has been
investigated experimentally in a trapped atomic
system~\cite{Donner-etal-07}, observing an increasing correlation
length compatible with what is expected at a continuous transition.
However, the inhomogeneity due to the trapping potential drastically
changes, even qualitatively, the general features of the behavior at a
phase transition.  For example, the correlation functions of the
critical modes do not develop a diverging length scale in a trap.
Nevertheless, even in the presence of a trap, and in particular when
the trap is large, we may still observe a critical regime, although
distorted. Around the transition point, the critical behavior is
expected to show a power-law scaling with respect to the trap
size~\cite{CV-09, CV-10-tr}, which is called trap-size scaling (TSS),
controlled by the universality class of the phase transition of the
homogeneous system. TSS has some analogies with the FSS theory for
homogeneous systems, with two main differences: the inhomogeneity due
to the space dependence of the external field, and a nontrivial
power-law dependence of the correlation length $\xi$ of the critical
modes around the center of the trap, i.e., $\xi\sim \ell^\theta$, when
increasing the trap size $\ell$ at the critical point, where $\theta$
is the universal trap exponent. See Fig.~\ref{trapsketch} for a sketch
which suggests the critical behavior around the center of the trap.

The above considerations apply to general systems of interacting
bosonic particles trapped by an external harmonic potential. In
particular, we mention bosonic cold atoms in optical
lattices~\cite{BDZ-08}, which can be effectively
described~\cite{JBCGZ-98} by the BH model~\cite{FWGF-89} defined by
the Hamiltonian~\eqref{bhm}.  Experiments with cold atoms~\cite{CW-02,
  Ketterle-02, BDZ-08, CV-10-XX} are usually performed in the presence
of a trapping potential, which can be taken into account by adding a
corresponding term in the Hamiltonian:
\begin{equation}
\hat H_{\rm tBH} = \hat H_{\rm BH} + \sum_{\bm x} V_{\bm x} \hat n_{\bm x}\,, 
\qquad\qquad V_{\bm x} = v^p |{\bm x}|^p\,,  \label{bhmt}
\end{equation}
where $\hat H_{\rm BH}$ is given by Eq.~\eqref{bhm}, the trap is
assumed to be spherical around a center located at ${\bm x}=0$, $v$ is
a positive constant, and $p$ is a positive exponent.  Far from
the origin, the potential $V_{\bm x}$ diverges, therefore
$\langle \hat n_{\bm x}\rangle$ vanishes and the particles are trapped.
In the experiments, one usually has harmonic trapping potentials,
that is, $p = 2$.

A natural definition of trap size is provided by~\cite{BDZ-08, RM-04,
  DSMS-96, PGS-04, CV-10-XX, CV-10-offeq}
\begin{equation}
  \ell \equiv {t^{1/p}\over v}\,,\qquad {\rm so} \;\;{\rm that}\quad
  V_{\bm x}= t^{-1} \,\left({|{\bm x}|\over \ell}\right)^p\,,
  \label{trapsize}
\end{equation}
where $t$ is the kinetic constant.  In the following we set $t = 1$.
In the limit $p\to\infty$, one recovers the case of a homogeneous
system within a $d$-dimensional sphere of radius $\ell$ and OBC, whose
scaling behavior is described by the FSS theory discussed previously.
The definition~\eqref{trapsize} of trap size naturally arises when
considering the thermodynamic limit for a large number $N$ of
particles, which is generally defined by the limit $N,\,\ell\to\infty$
keeping the ratio $N/\ell^d$ fixed~\cite{PGS-04, BDZ-08}. This limit
turns out to be equivalent to introducing the chemical potential
$\mu$, as in Eqs.~\eqref{bhmt} and~\eqref{bhm}, i.e., to the limit
$\ell\to\infty$ keeping $\mu$ fixed.

\subsubsection{Local density approximation of the particle density}
\label{lodeapprox}

The inhomogeneous confining potential gives rise to a particle density
that depends on the space, in particular on the distance from the
center of the trap, in the case of spherical symmetry.  In such cases,
theoretical and experimental results have witnessed the coexistence
of MI and superfluid regions when varying the total occupancy
of the lattice (see, e.g.,
Refs.~\cite{FWMGB-06, JBCGZ-98, DLO-01, Batrouni-etal-02, KPS-02, KSDZ-04,
  PRHD-04, WATB-04, RM-04, RM-05, DLVW-05, GKTWB-06, ULR-06, RBRS-09, SBE-11}).

In the presence of a space-dependent external potential, the so-called
local density approximation (LDA) estimates the spatial dependence of
the particle density by taking the corresponding value for
the homogeneous system at the effective chemical potential
\begin{equation}
  \mu_{\rm eff}({\bm x}) \equiv \mu - V({\bm x})\,.
  \label{mueff}
\end{equation}
The LDA has been widely used to get quantitative information on the
behavior of BH models in a confining potential, and, more generally,
of inhomogeneous systems (see, e.g., Refs.~\cite{BHR-04, PBS-08,
  Batrouni-etal-02, PRHD-04, WATB-04}).

\begin{figure}
  \begin{center}
    \includegraphics[width=0.5\columnwidth]{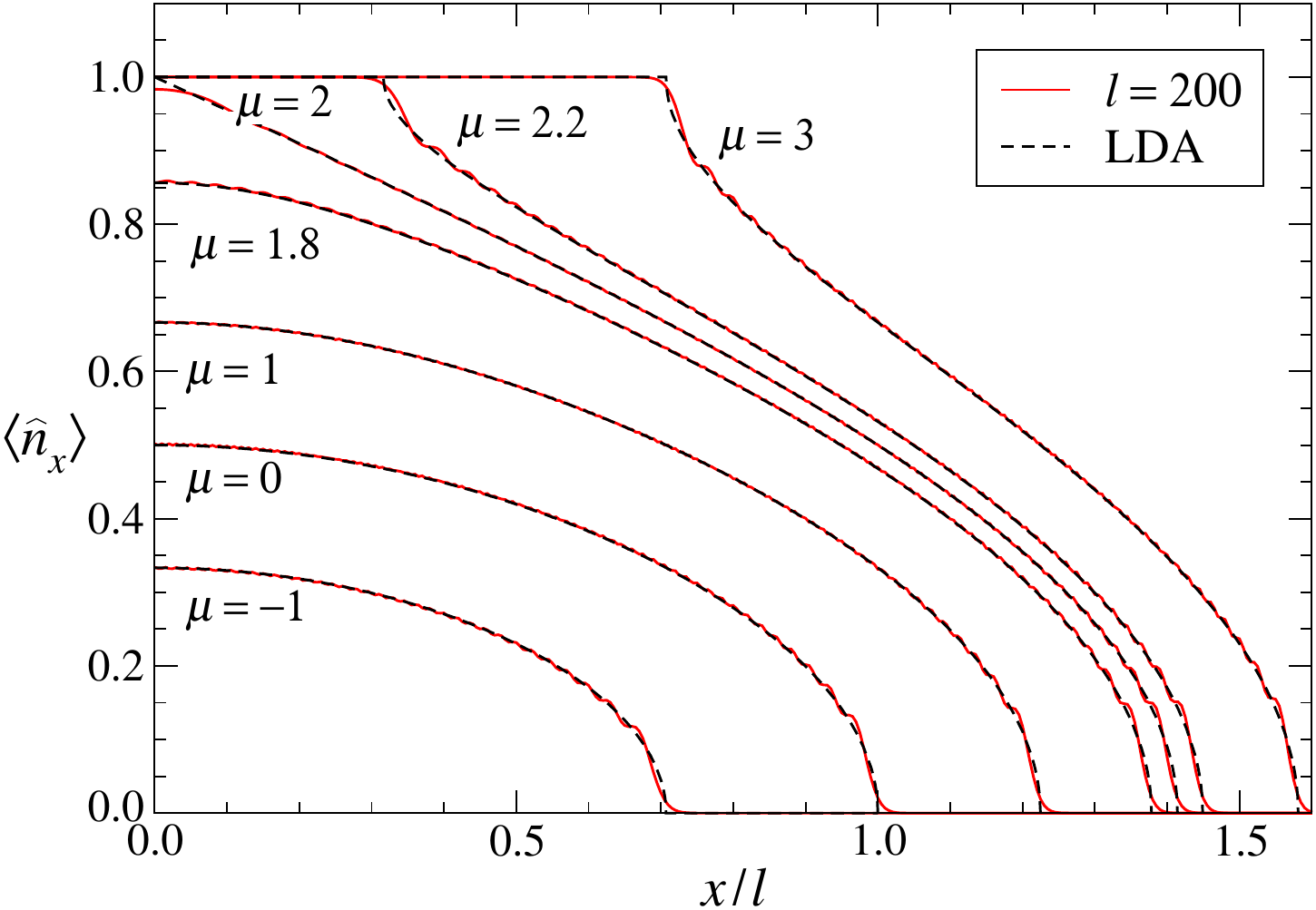}
    \caption{The particle density vs.\ $x/\ell$ for $p=2$ and several
      values of the chemical potential $\mu$, as obtained from the
      LDA, cf.\ Eq.~\eqref{nxlda}, and numerical calculations on a
      large chain with $\ell=200$.  The differences are hardly visible
      in the figure.  Adapted from Ref.~\cite{CV-10-XX}.}
    \label{XXpartdensnum1}
  \end{center}
\end{figure}

For example let us consider the $1d$ BH model in the $U \to \infty$
limit.  Such model has a nonzero filling above the low-density Mott
transition (i.e., for $\mu>-2$).  The filling $f$ in the
infinite-chain limit $L\to\infty$ is reported in Eq.~\eqref{fmu}.  The
corresponding Fermi momentum is $k_F=\pi f$.  In the following, $f$
will always denote the value for the infinite homogeneous chain.  The
LDA of the particle density reads
\begin{equation}
  \langle \hat n_x \rangle_{\rm lda} \equiv \rho_{\rm lda}(x/\ell) = 
  \left\{
  \begin{array}{lll}
    0 & \quad {\rm for} & \; \mu_{\rm eff}(x) < -2\,, \\
    1 - (1/\pi)\arccos[\mu_{\rm eff}(x)/2] &
    \quad {\rm for} & \; |\mu_{\rm eff}(x)| \le 2\,, \\
    1 & \quad {\rm for} & \; \mu_{\rm eff}(x) > 2\,. 
  \end{array} \right. 
  \label{nxlda} 
\end{equation}
This would imply the presence of a plateau at $n=1$ when $\mu_{\rm
  eff} \ge 2$, for
\begin{equation}
  x/\ell \le (-2+\mu)^{1/p}\,,
  \label{mott1ph}
\end{equation}
and a vanishing particle density when $\mu_{\rm eff} \le -2$, for
\begin{equation}
  x/\ell \ge (2+\mu)^{1/p}\,.
  \label{mott0ph}
\end{equation}
In Fig.~\ref{XXpartdensnum1}, the LDA of the particle density is
compared with numerical results for $p=2$ and the trap size
$\ell=200$~\cite{CV-10-XX}.  Note the flat regions related to the
$n=0,1$ Mott phases, have been already observed in experimental and
numerical works (see, e.g., Refs.~\cite{JBCGZ-98, Batrouni-etal-02,
  FWMGB-06}).  Analogous results are found for other powers of the
confining potential.  The LDA provides a good approximation of the
local particle density, which improves with increasing trap size.  The
differences of the trap-size dependence from the LDA results are
generally suppressed by powers of $\ell$ when increasing $\ell$, and
show a nontrivial scaling behavior, also characterized by peculiar
modulations~\cite{CV-10-XX} (see also below).  On the other hand,
critical correlation functions between different spatial points,
subject to different external forces, require a different framework,
which goes beyond LDA.

\subsubsection{Trap-size scaling}
\label{tsssca}

To investigate the critical behaviors within a trap, for example when
the chemical potential is closed to its Mott transition values, a
different approach is called for, which allows one to treat the
critical fluctuations neglected by the LDA. This issue can be studied
within the so-called TSS framework.

We again assume that the zero-temperature quantum critical behavior of
the homogeneous system is controlled by one relevant parameter
$r\equiv \mu-\mu_c$, as described in Sec.~\ref{cqte}. We suppose that
the inhomogeneity arises from an external potential coupled to the
particle density, as for the Hamiltonian~\eqref{bhmt}.  The presence
of the external trapping field significantly affects the critical
modes, introducing another length scale, the trap
size $\ell$ defined in Eq.~\eqref{trapsize}.  Within the TSS
framework~\cite{CV-09, CV-10-tr}, the scaling of the singular part of
the free-energy density around the center of the trap is generally
written as~\footnote{Hereafter we will always omit the subscript ${}_0$
  to indicate the leading contribution in all the scaling functions,
  thus neglecting the contribution of the irrelevant scaling fields,
  cf.~Eq.~\eqref{Fsing-RG}.}
\begin{equation}
  F({\bm x};v,T,r) \approx \ell^{-(d+z) \theta} \, {\cal F}(\ell^{-\theta} |{\bm
    x}|, \ell^{\theta z} T, \ell^{\theta y_r}r )\,,
  \label{freee}
\end{equation}
where $\theta$ is the {\em trap} exponent.  TSS implies that at the
critical point ($r=0$) the correlation length $\xi$ of the critical
modes behaves as
\begin{equation}
  \xi \sim \ell^{\theta}
  \label{xitheta}
\end{equation}
with increasing the trap size $\ell$.

The trap exponent $\theta$ depends on the universality class of the
transition, on its space dependence, and on the way it couples to the
particles. Its value can be inferred by a RG analysis of the
perturbation $P_V$ associated with the external trapping potential
coupled to the particle density~\cite{CV-09, CV-10-tr}.  Since the
particle density corresponds to the energy operator $|\phi|^2$ of the
corresponding field-theoretical description (see
Sec.~\ref{qtbhmodel}), we write the perturbation $P_V$ as
\begin{equation}
  P_V=\int {\rm d}^d x\, {\rm d}t \, V({\bm x}) \, |\phi({\bm x},t)|^2 = 
  \int {\rm d}^d x\, {\rm d}t \, v^p |{\bm x}|^p \, |\phi({\bm x},t)|^2\,.
  \label{pertu}
\end{equation}
The exponent $\theta$ is related to the RG dimension $y_v$ of the
coupling $v$ of the external field $V=(v r)^p$ by $\theta=1/y_v$.
Then, since the RG dimensions of the terms within space-time integral
representing $P_V$ must be equal to $d+z$ (that is, the RG dimension of the
space and time), we have
\begin{equation}
  p\,y_v - p + y_n = d+z\,,\label{rg1}
\end{equation} 
where $y_n=d+z-y_r$ is the RG dimension of the density/energy operator
$|\phi|^2$. We eventually obtain~\cite{CV-10-tr}
\begin{equation}
  \theta = {1\over y_v} = {p \over p + y_r}\,.
  \label{thetap}
\end{equation}
Multiplicative logarithms are typically expected at the upper
dimension of the given universality class, and also at the BKT
transition~\footnote{This should not be surprising, because they are
  already present in the scaling behavior of homogeneous
  systems~\cite{PV-13}.}.  Note that in the limit $p\to\infty$, when
the trap tends to become equivalent to a system of size $\ell$ with
OBC, the exponent $\theta\to 1$, as it should, since in this limit TSS
must become equivalent to FSS.  Analogous scaling ansatze for other
systems in the presence of inhomogeneous external forces have been put
forward in Refs.~\cite{Burkhardt-82, PKT-07, ZD-08, EIP-09, CKT-09,
  BDV-14}.

TSS equations can be derived for the correlation functions of the
critical modes. For example, the correlation function of the
fundamental complex field $\phi({\bm x})$ (the quantum field $\hat b$
in the BH model) is expected to behave as
\begin{equation}
  G({\bm x},{\bm y}) \equiv \langle \phi({\bm x}) \, \phi({\bm y})
  \rangle_c \approx \ell^{-\theta(d+z-2+\eta)} \, {\cal G}( \ell^{-\theta} {\bm x},
  \, \ell^{-\theta}{\bm y}, \, \ell^{\theta z} T, \, \ell^{\theta y_r} r)\,,
  \label{twopfpl}
\end{equation}
where ${\cal G}$ is a scaling function (the subscript c denotes the
connected component~\footnote{In this review, we will always use the
subscript $c$ to denote the connected component, $\langle \phi \, \psi
\rangle_c \equiv \langle \phi \, \psi \rangle - \langle \phi \rangle
\langle \psi \rangle$.}).

Other interesting observables are related to the particle density.  At
the vacuum-to-superfluid transition, $\mu_c=-2d$, the spatial
dependence of the particle density is described by the asymptotic
TSS~\cite{CV-10-XX}
\begin{equation}
  \rho_{\bm x}\equiv \langle \hat n_{\bm x}\rangle \approx \ell^{-y_n\theta}
      \, {\cal D}(\ell^{-\theta} {\bm x}, \, \ell^{\theta z} T)\,,
      \label{nxscalm}
\end{equation}
where $y_n$ is the RG dimension of the particle density operator
$\hat{n}_{\bm x}$, which is given by $y_n=d + z - y_r$.  Its behavior
behaviors turn out to be more complicated at the Mott $n\ge 1$
transitions, where modulation phenomena emerge, similar to those
appearing in the finite-size behavior within the superfluid phase.
The spatial dependence of the particle density turns out to be
described by the following scaling behavior:
\begin{equation}
  \rho({\bm x}) \approx \rho_{\rm lda}({\bm x}/\ell) +
  \ell^{-y_n\theta} \, {\cal D}(\ell^{-\theta} {\bm x},\varphi)\,.
  \label{nxscalm-1}
\end{equation}
The dominant term in the large trap-size limit is just given by the
LDA, which plays the role of an analytical contribution that must be
subtracted to observe scaling in the expectation value of the density
operators at phase transitions~\cite{CV-09, CV-10-tr}.  As noticed in
Refs.~\cite{CV-10-XX, CV-10-ent}, the scaling function is
characterized by a further dependence on a phase (which somehow
measures the distance from the closest level crossing), modulating the
TSS, arising from periodic level crossings with increasing the system
size, analogously to the phenomenon discussed in
Sec.~\ref{modulations}.

The above scaling relations tell us that the LDA is approached in the
large trap-size limit, with $O(\ell^{-y_n\theta})$ corrections.  However,
this asymptotic behavior changes close to the boundary of the trap
(i.e., when $\rho_{\rm lda}\approx 0$). In this case, the asymptotic
approach is controlled by the critical point associated with the
vacuum-to-superfluid transition in an effective external linear
potential~\cite{CV-10-XX, DV-17}.

The trap-size dependence predicted by TSS has been verified at various
thermal and quantum transitions: at the Ising transition of lattice
gas models~\cite{CV-09, QSS-10, CoV-11}, at the BKT transitions of
spin models~\cite{PV-13, CV-11}, at QTs of Ising-like and XY systems
(considering in particular their equivalent fermionic
systems)~\cite{CV-10-tr, CV-10-ent}, at the finite-temperature BEC
transition of bosonic particle systems such as those described by the
$3d$ BH model~\cite{CTV-13, CNPV-15-bec, DV-17a}, at the
finite-temperature BKT of $2d$ BH models~\cite{CNPV-13, DV-17}, at the
quantum Mott transitions of BH models~\cite{CV-10-tr, CV-10-XX, CTV-12,
  CT-12}, in fermionic gases~\cite{Vicari-12-ent, ACV-14, Nigro-17}.
In particle gases with a definite number $N$ of particles, the TSS can
be expressed in terms of $N$, being related to the scaling behavior
close to the QT separating the vacuum from the superfluid phase, at
$\mu_{\textrm{vs}}=-2d$ (see, e.g., Refs.~\cite{CMV-11, CV-10-offeq,
  Vicari-12-ent, Vicari-12-dyn, ACV-14, Vicari-19}).

\section{Equilibrium scaling behavior at first-order quantum transitions}
\label{escalingfoqt}

Quantum phase transitions are of the first order when ground-state
properties are discontinuous across the transition point. However,
singularities develop only in the infinite-volume limit.  If the system
size $L$ is finite, all the properties are analytic as a function
of the external parameter driving the transition. Nevertheless, around
the transition point, thermodynamic quantities and large-scale
properties develop peculiar scaling behaviors essentially depending on
the general features of the transition.  Their understanding is
essential to correctly interpret experimental or numerical data, which
implement relatively small systems.

These issues are also important for FOQTs, essentially for two good
reasons.  The first one is that FOQTs are phenomenologically relevant,
as they occur in a large number of quantum many-body systems,
including quantum Hall samples~\cite{PPBWJ-99}, itinerant
ferromagnets~\cite{VBKN-99, BKV-99}, heavy fermion
metals~\cite{UPH-04, Pfleiderer-05, KRLF-09}, SU($N$)
magnets~\cite{DK-16, DK-20}, quantum spin
systems~\cite{LMSS-12,CNPV-14,CNPV-15-fo, LZW-19}, etc.  The second
reason is that the low-energy properties at FOQTs are particularly
sensitive to the BC, giving rise to a variety of possible scenarios,
wider than those at CQTs.  Indeed, depending on the type of BC, for
example whether they are neutral or they favor one of the phases, the
behavior at FOQTs may be characterized by qualitatively different
dynamic properties~\cite{CNPV-14, CNPV-15-fo, CPV-15-bf, PRV-18-fo,
  YCDS-18, RV-18}, associated with exponential or power-law scaling.
Actually, in finite-size systems, this peculiar sensitivity of FOQTs
to the BC likely represents the most striking difference with CQTs.

Classical first-order transitions, driven by thermal
fluctuations, display FSS behaviors~\cite{NN-75, FB-82, PF-83, FP-85,
  Privman-book, CLB-86, Binder-87, BK-90, VRSB-93, CPPV-04}, similarly
to continuous transitions.  On the basis of the general
quantum-to-classical mapping of $d$-dimensional quantum systems onto
classical anisotropic $(d+1)$-dimensional systems, one expects that
the FSS theory of classical first-order transitions can be extended to
FOQTs, as put forward in Refs.~\cite{CNPV-14, CNPV-15-fo}.  However,
we recall the comment reported in Sec.~\ref{qtocl} on the
quantum-to-classical mapping, and the intrinsic anisotropy of the
resulting classical system.  In this respect, FOQTs may lead to
extremely anisotropic classical systems, such as for quantum
Ising-like systems, characterized by an exponentially larger length
scale along the imaginary time, analogously to the classical
models considered in Ref.~\cite{PF-83}.  Therefore, especially at
FOQTs, while the quantum-to-classical mapping remains a useful guiding
idea, a direct approach using quantum settings is simpler and likely
more illuminating than going through the former.

In the following we present the main features of the FSS behavior at FOQTs,
describing the different scenarios emerging when various types of BC
are considered.  We discuss this issue within the paradigmatic quantum
Ising models along their FOQT line, separating differently magnetized
phases. However most results apply to generic FOQTs as well.  We first
focus on FSS with neutral BC, where a quasi-level-crossing scenario
emerges (Sec.~\ref{fsspbcobc}), and then address BC giving rise to
domain walls, leading to a substantially different picture
(Sec.~\ref{fssabcfobc}).  In Sec.~\ref{fssfebc} we describe the more
complex situation arising from BC favoring one of the two magnetized
phases.  Finally, in Sec.~\ref{qutrdef} we discuss the possibility of
observing CQTs along FOQT lines driven by defects only.

Here we only mention, without further addressing it, that peculiar
scaling behaviors can be also observed in the presence of
inhomogeneous conditions, such as those arising from space-dependent
external fields, at both FOQTs~\cite{CNPV-15-inho} and classical
first-order transitions~\cite{BDV-14}.

\subsection{Finite-size scaling at first-order quantum transitions}
\label{fssfoqt}

We consider a generic finite $d$-dimensional system of size $L$ (as
usual, we may assume a cubic-like shape with a volume $L^d$), 
undergoing a FOQT driven by a magnetic field $h$, such as quantum Ising
systems defined by the Hamiltonian~\eqref{hisdef}, along the FOQT line
for $|g|<g_c$. We assume that the BC are such to preserve the parity
symmetry at $h=0$. This neutrality of BC is guaranteed by PBC, ABC,
and OBC in quantum Ising systems, and also by OFBC in the Ising
chains. The preservation of the parity symmetry implies that the
minimum of the difference of the lowest states occurs at strictly
$h=0$.

At FOQTs, the interplay between the temperature $T$, the field $h$,
and the system size $L$ gives rise to an asymptotic FSS of the
low-energy properties. To write down the corresponding scaling ansatz,
it is necessary to identify the relevant scaling combinations
associated with $h$ and $T$.  The scaling variable controlling the
interplay between $h$ and the size $L$ is naturally provided by the
ratio $\Phi$ between the energy variation $\delta E_h$ due to addition
of the longitudinal field $h$, and the gap $\Delta(L)$ at
$h=0$~\cite{CNPV-14}. Thus, we define 
\begin{equation} 
  \Phi \equiv {\delta E_h(L,h)/\Delta(L)} \,,\qquad
  \Delta(L) \equiv \Delta(L,h=0)\,.
  \label{kappah}
\end{equation}
If the BC are such that the lowest-energy states are the magnetized
states, the energy related to the longitudinal field $h$ can be
quantified as~\cite{CNPV-14}
\begin{equation}
  \delta E_h(L,h) = 2\, m_0 \,h \,L^d\,,
  \label{deltah}
\end{equation}
for sufficiently small $h$, where $m_0$ is the spontaneous
magnetization, obtained approaching the transition point $h\to 0$
after the infinite-volume limit, such as in Eq.~(\ref{sigmasingexp})
(of course, the normalization of $\delta E_h(L,h)$, and in particular
the factor two in its definition, is conventional).  The scaling
variable controlling the interplay between $T$ and $L$ is naturally
defined by the ratio
\begin{equation}
  \Xi \equiv {T \over \Delta(L)} \,.
  \label{kappat}
\end{equation}

The FSS limit corresponds to $L\to \infty$ and $h\to 0$, keeping
$\Phi$ and $\Xi$ fixed.  In this limit, the gap $\Delta$ is expected
to behave as~\cite{CNPV-14}
\begin{equation}
  \Delta(L,h) \approx \Delta(L) \, {\cal E}(\Phi) \,.
  \label{fde}
\end{equation}  
By definition ${\cal E}(0)=1$, and ${\cal E}(\Phi)\sim |\Phi|$ for
$\Phi\to\pm \infty$, in order to reproduce the expected linear behavior
$\Delta(L,h) \sim |h|L^d$ for sufficiently large $|h|$.  Moreover, the
local and global magnetization [such as those defined in
  Eq.~\eqref{longmagn}], are expected to scale as
\begin{equation}
  m_{\bm x}(L,T,h) \approx m_0 \, {\cal M}_x({\bm x}/L,\Xi,\Phi)\,,\qquad
  M(L,T,h) \approx m_0 \, {\cal M}(\Xi,\Phi)\,.
  \label{efssm}
\end{equation}
Because of the definition of $m_0$, cf. Eq.~\eqref{m0def}, we have
${\cal M}(0,\Phi)\to \pm 1$ for $\Phi \to \pm \infty$. Moreover ${\cal
  M}(T,0)=0$ by parity symmetry. For systems without boundaries,
$m_{\bm x}(L,T,h) = M(L,T,h)$ by translation invariance, thus ${\cal
  M}_x({\bm x}/L,\Xi,\Phi)={\cal M}(\Xi,\Phi)$.  The above FSS ansatze
represent the simplest scaling behaviors compatible with the
discontinuities arising in the thermodynamic limit.
The scaling variables $\Phi$ and $\Xi$, associated with $h$ and $T$
respectively, are related to the finite-size energy gap at the
transition, whose size behavior depends crucially on the BC
considered, disclosing exponential and power-law behaviors, as stated in
Sec.~\ref{foqtisi}.  Hence, once they are expressed in terms of the
size and of the parameter driving the transition, their $L$-dependence
changes according to the chosen BC.  As a consequence, the FSS
behaviors may be characterized by power laws, as in CQTs, or
exponential laws, which are peculiar to FOQTs only.

\subsection{Unified scaling picture at first-order and continuous
  quantum transitions}
\label{CQTanalog}

It is worth noticing that the definitions of $\Phi$ and $\Xi$ in
Eqs.~\eqref{kappah} and~\eqref{kappat} also apply to CQTs, for example
at the end point of the FOQT line for quantum Ising models at $g=g_c$.
Indeed, as discussed in the previous section, the CQT scaling variable
associated with the external field $h$ is given by
\begin{equation}
  \Phi = L^{y_h} h\,,
  \label{kappahcqt}
\end{equation}
which can be obtained using the more general definition~\eqref{kappah}.
Indeed, at CQTs the energy variation $\delta E$ arising from the perturbation
\begin{equation}
  \hat{H}_h = - h \sum_{\bm x} \hat P_{\bm x} \,, \qquad
  \hat P_{\bm x} = \hat \sigma^{(1)}_{\bm x} \,,
  \label{hhpoert}
\end{equation}
behaves as
\begin{equation}
  \delta E_h(L,h) \sim h \, L^{d - y_p} \,,
  \label{ehl}
\end{equation}
where $y_p$ is the RG critical dimension of the local operators
$\hat P_{\bm x}$ at the fixed point describing the quantum critical
behavior.  Moreover, at $g=g_c$, we have $\Delta(L,h=0) \sim L^{-z}$ where
$z$ is the universal dynamic exponent.
Then, using the scaling relation among the critical
exponents~\cite{CPV-14, Sachdev-book}
\begin{equation}
  y_h + y_p = d + z \,,
  \label{scalrel}
\end{equation}
where $y_h$ is the RG dimension of the perturbation $h$, we end up
with the expression for $\Phi$ reported in Eq.~\eqref{kappahcqt}.
Analogously, using Eq.~\eqref{kappat} and $\Delta \sim L^{-z}$,
we recover the CQT scaling variable $\Xi = L^z \, T$.
Therefore, the definitions~\eqref{kappah} and~\eqref{kappat} for the
scaling variables provide a unified FSS framework for CQTs and FOQTs.
The differences are encoded within the size dependence of $\Phi$ and $\Xi$.

\subsection{Neutral boundary conditions
  giving rise to a quasi-level-crossing scenario}
\label{fsspbcobc}

We now consider a class of BC that are neutral with respect to the
different phases separated by the FOQTs. For quantum Ising models,
where FOQTs are driven by the longitudinal field $h$, such BC are
provided by PBC or OBC.  In these cases the expected scenario is that
of a quasi-level-crossing of the two lowest states, with exponentially
suppressed energy differences $\Delta \sim \exp(-c L^d)$ of the lowest
states, such as Eqs.~\eqref{del}, \eqref{deltaobc}
and~\eqref{deltapbc}, while the energy differences $\Delta_n$
associated with higher excited states remain finite (or more generally
$\Delta_n/\Delta\to\infty$ exponentially, apart from power laws) in
the infinite-volume limit.  The general FSS theory outlined in
Sec.~\ref{fssfoqt} implies scaling variables $\Phi$ and $\Xi$ with an
exponential dependence on the lattice size, through the exponential
suppression of the gap in the large-$L$ limit. The corresponding FSS
functions, such as those introduced in Eqs.~\eqref{fde}
and~\eqref{efssm}, can be computed by exploiting a phenomenological
two-level theory, as explained below.

For sufficiently large $L$, the low-energy properties in the crossover
region $|h|\ll 1$ can be obtained by restricting the theory to the two
lowest-energy states $|0\rangle$ and $|1\rangle$, or equivalently to
the globally magnetized states $| + \rangle$ and $| - \rangle$,
exploiting a two-level truncation of the
spectrum~\cite{CNPV-14, PRV-18-loc}.  The effective evolution is ruled
by the Schr\"odinger equation
\begin{equation}
  {{\rm d \, |\Psi_2(t)\rangle} \over {\rm d}t}
  = -i \, \hat{H}_{2}(h) \, |\Psi_2(t)\rangle \,,
  \label{sceq}
\end{equation}
where $|\Psi_2(t)\rangle$ is a two-component wave function, whose
components correspond to the states $|+ \rangle$ and $|-\rangle$, and
\begin{equation}
  \hat{H}_{2}(h) = \varepsilon \, \hat \sigma^{(3)} + \zeta
  \, \hat \sigma^{(1)}\,.
  \label{hrtds}
\end{equation}
Here $\varepsilon$ is related to the perturbation induced by the
magnetic field $h$, while $\zeta$ is related to the small gap
$\Delta(L)$ at $h=0$~\cite{CNPV-14},
\begin{equation}
  \varepsilon =  m_0 h L^d\,,\qquad 
  \zeta = \langle - | \hat H_{\rm Is} | + \rangle = \tfrac12 \, \Delta(L,h=0) \,,
  \label{deltavarfo}
\end{equation}
so that one may easily see the correspondence between the scaling
variable $\Phi$ of the Ising system and the ratio $\varepsilon/\zeta$,
\begin{equation}
  \varepsilon / \zeta \to \Phi\,.
  \label{kappahverde}
\end{equation}
By diagonalizing the restricted Hamiltonian $\hat H_2$, we obtain the
energy difference corresponding to $\Delta(L,h)$. Moreover, by taking
the matrix element of the operator $\hat \sigma^{(3)}$, we get the
quantity corresponding to the magnetization of the quantum Ising
model.  Then, by matching with the scaling equations~\eqref{fde}
and~\eqref{efssm}, one easily obtains~\cite{CNPV-14}
\begin{equation}
  {\cal E}(\Phi) = \sqrt{1 + \Phi^2}\,, \qquad\qquad
  {\cal M}_{\bm x}({\bm x}/L,0,\Phi) = {\cal M}(0,\Phi) =
  {\Phi \over \sqrt{1+ \Phi^2}}\,. \label{fdemm}
\end{equation}
Note that in the case of OBC, the above behavior for the local
magnetization should hold for lattice sites ${\bm x}$
sufficiently far from the boundaries.  For sufficiently small
temperatures (i.e., $T \sim \Delta$, thus finite $\Xi$), the extension
to finite-temperature results can be obtained by averaging over the
two levels with the Gibbs weight, i.e.,
\begin{equation}
  M = Z_2^{-1}\, {\rm Tr} \big[ \hat \sigma^{(3)} e^{- \hat H_2/T} \big]\,,\qquad
  Z_2 = {\rm Tr} \big[ e^{-\hat H_2/T}\big] \,.
  \label{twolepa}
\end{equation}

The results of the above phenomenological two-level theory
are expected to be exact for quantum Ising models with neutral BC such
as PBC and OBC, i.e., they are expected to be exactly approached in the
large-$L$ limit.  This has been confirmed by means of computations
within quantum Ising chains with OBC~\cite{CNPV-14}.  The above approach
can be easily extended to the case of a larger finite (quasi) degeneracy
of the lowest states, such as quantum Potts models~\cite{CNPV-15-fo}.

Low-dimensional systems at FOQTs react rigidly to external local
fields $h_{\bm x}$~\cite{CNPV-14}. This property was checked for Ising
chains in which the external magnetic field is nonvanishing only at
one lattice site $x$, for example at the center of the chain.  The
general arguments should apply to this case as well, with $\varepsilon
\sim 2 m_0 h_x$ instead of $\varepsilon = 2 m_0 h L$.  Therefore, we
expect a FSS behavior analogous to that valid for the homogeneous
parallel field, the corresponding scaling variable being $\Phi_x = 2
m_0 h_x/\Delta(L)$.  Numerical results for the Ising chain are in full
agreement with these scaling predictions~\cite{CNPV-14}.

We finally mention that, in principle, the above scaling ansatze can be
derived by exploiting the general quantum-to-classical mapping, which
allows one to map the quantum system onto a classical one defined in
an anisotropic box of volume $V = L^d \, L_T$, with $L_T\sim 1/T$ and
$L_T\gg L$.  Then the above results can be also obtained from those
reported in Ref.~\cite{PF-83} for $(d+1)$-dimensional Ising-like
systems at a classical first-order transition with PBC along a
transverse dimension.~\footnote{In the classical case, FSS is
characterized by the scaling variables $u \sim hV$ and $ v \sim
\xi_T/L_T$, where $\xi_T$ is the characteristic transverse length at
the transition: $\xi_T\sim \exp (L^d \sigma)$, where $\sigma$ is the
interfacial tension.  Identifying $\Delta$ with $1/\xi_T$, we have the
correspondence $u/v\to\Phi$ and $v\to\Xi$ between the variables
$\Phi\sim hL^d/\Delta$ and $\Xi \sim T/\Delta$ and the classical ones
$u$ and $v$.}

\subsection{Boundary conditions giving rise to domain walls}
\label{fssabcfobc}

The general FSS behaviors reported in Sec.~\ref{fssfoqt} also hold in
systems at FOQTs with an infinite degeneracy of the ground state at
the transition, such as those with BC giving rise to domain walls.
For example, let us consider the Ising chain with ABC or OFBC.  In
these two cases, the lowest-energy states are associated with domain
walls (kinks), i.e., nearest-neighbor pairs of antiparallel spins,
which can be considered as one-particle states with $O(L^{-1})$
momenta. Hence, there is an infinite number of excitations with a gap
of order $L^{-2}$, cf. Eq.~\eqref{deltaabcfobc}.

\begin{figure}[tbp]
  \begin{center}
    \includegraphics[width=0.5\columnwidth]{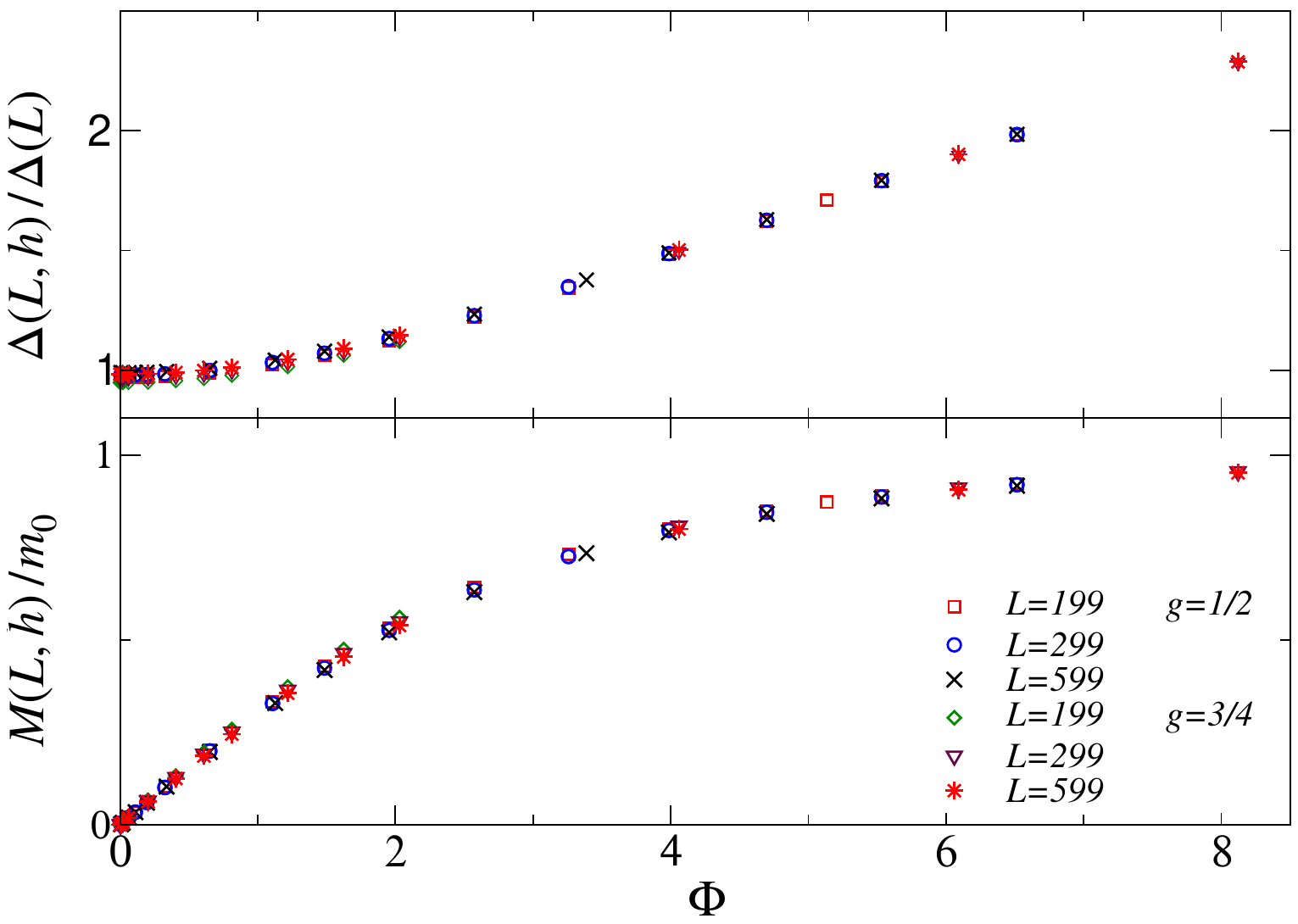}
    \caption{FSS of the energy difference of the lowest states and
      magnetization of the quantum Ising chain with OFBC, at $g=1/2$
      and $g=3/4$.  The two panels show data for the ratios
      $\Delta(L,h)/\Delta(L)$ (top) and $M(L,h)/m_0$ (bottom) versus
      $\Phi = 2 m_0 h L/\Delta(L)$.  The data approach nontrivial
      scaling curves with increasing $L$, which are independent of
      $g$, supporting the FSS Eqs.~\eqref{fde} and~\eqref{efssm}.
      Adapted from Ref.~\cite{CNPV-14}.}
    \label{deosx}
  \end{center}
\end{figure}

Applying the general FSS theory of Sec.~\ref{fssfoqt} to this case, we
obtain scaling variables $\Phi$ and $\Xi$ with a power-law dependence
on $L$.  Indeed, we have
\begin{equation}
  \Phi \sim L^3 \,h \,,\qquad \Xi \sim L^2 \,T\,.
  \label{kappahtabc}
\end{equation}
The scaling ansatze~\eqref{fde} and~\eqref{efssm} also hold in this
case~\cite{CNPV-14, CPV-15-bf}. Indeed, as shown in Fig.~\ref{deosx},
the ratios $\Delta(L,h)/\Delta(L)$ and $M(L,h)/m_0$ approach universal
(independent of $g$) scaling curves, when plotted versus $\Phi$.
However, in this case, scaling functions cannot be obtained by
performing a two-level truncation, because the low-energy spectrum at
the transition point presents a tower of excited states with
$\Delta_{n}=E_n-E_0=O(L^{-2})$, at variance with the OBC and PBC case, where
only two levels matter, close to the transition point.

These results reveal that the FSS at FOQTs may significantly depend on
the BC, being strictly connected with the large-volume low-energy
structure of the eigenstates.  This implies the possibility of
observing FSS with either exponential or power-law size dependence,
due to the fact that $\Phi$ may show either exponential or power-law
dependences with $L$, for different BC.  For instance, for the Ising
chain one has: (i)~$\Phi \sim h L e^{\alpha L}$ for OBC;
(ii)~$\Phi \sim h L^{3/2} e^{\alpha L}$ for PBC ($\alpha$ is a nonuniversal
positive constant); (iii)~$\Phi \sim h L^3$ for ABC or OFBC.
This sensitivity to the BC is likely the main feature distinguishing
FOQT and CQT behaviors in finite-size systems.

\subsection{Boundary conditions favoring one of the two phases}
\label{fssfebc}

\begin{figure}
  \begin{center}
    \includegraphics[width=0.5\columnwidth]{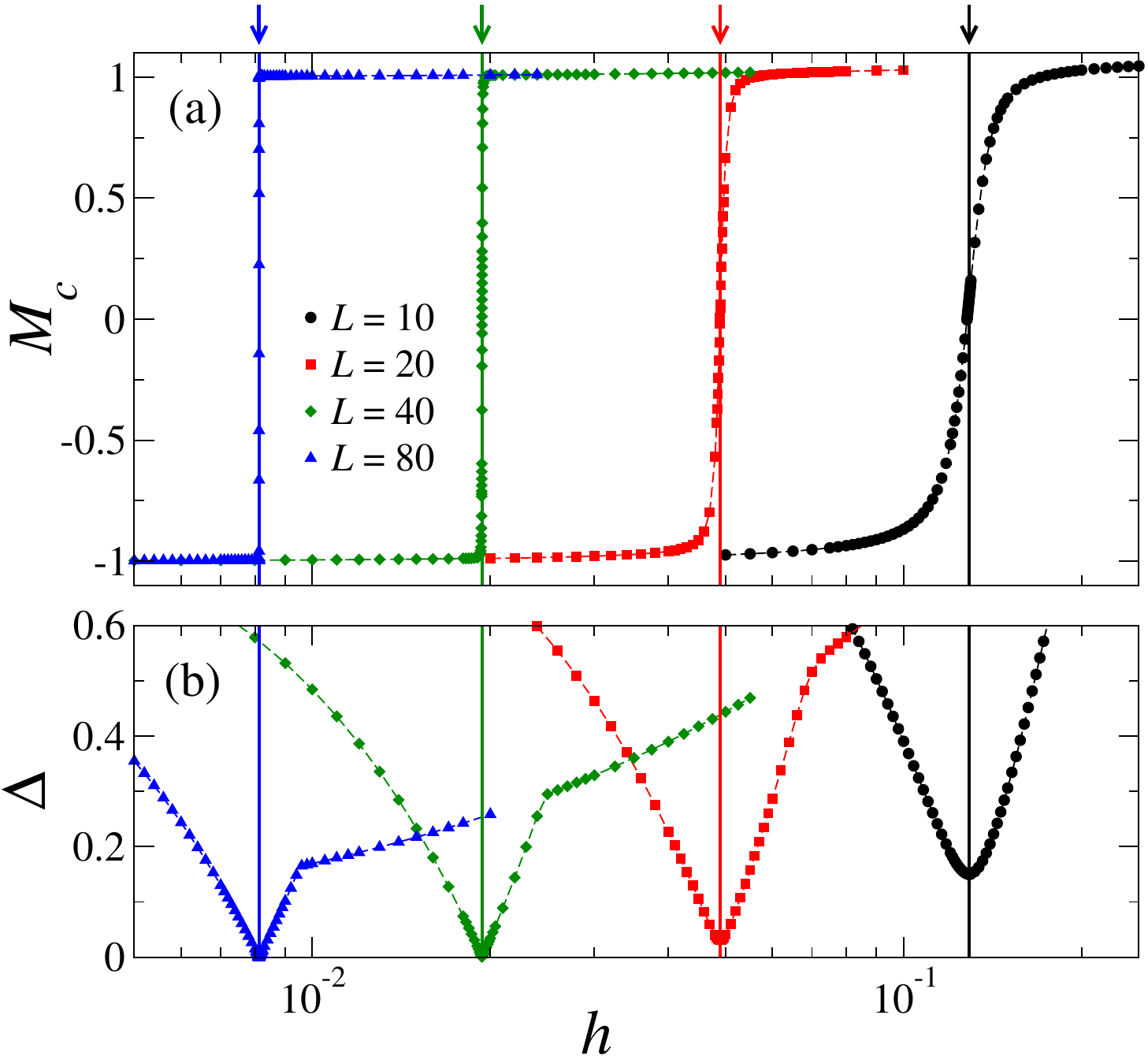}
    \caption{Central magnetization $M_c \equiv m_{L/2}/m_0$
      (normalized so that it ranges from -1 to 1) [panel (a)] and
      energy gap $\Delta$ [panel (b)] in the quantum Ising chain with
      EFBC, as a function of the longitudinal field $h$, for fixed
      transverse field $g=0.8$.  The various data sets correspond to
      different system sizes, as indicated in the legend.  The
      continuous vertical lines and arrows denote the magnetic fields
      $h_{tr}(L)$ corresponding to the minimum of the energy gap, as
      displayed in panel (b). From Ref.~\cite{PRV-18-fo}.}
    \label{fig:Mc}
  \end{center}
\end{figure}

Another interesting case is provided by quantum Ising chains with
equal and fixed BC (i.e., EFBC), favoring one of the two magnetized
phases.  They can be selected by restricting the Hilbert space to
states $|s\rangle$ at the boundaries $x=0$ and $x=L+1$ of an
open-ended chain, such that $\hat \sigma^{(1)}_0 |s\rangle = -
|s\rangle$ and $\hat \sigma^{(1)}_{L+1} |s\rangle = - |s\rangle$.
As numerically shown in Ref.~\cite{PRV-18-fo}, the interplay between
the size $L$ and the bulk longitudinal field $h$ is more complex than
that observed with neutral BC.  In the case of EFBC, for small values
of $h$, observables depend smoothly on $h$, up to $h_{tr}(L)\approx
c/L$ ($c$ is a $g$-dependent constant), where a sharp transition to
the oppositely magnetized phase occurs, see Fig.~\ref{fig:Mc}. The
field $h_{tr}(L)$ can be identified as the value of $h$ where the
energy difference $\Delta(L,h)$ of the lowest states is minimum.  Such
a minimum $\Delta_m(L)$ turns out to decrease exponentially with
increasing $L$, i.e., $\Delta_m(L)\sim e^{-bL}$.

\begin{figure}
  \begin{center}
    \includegraphics[width=0.47\columnwidth]{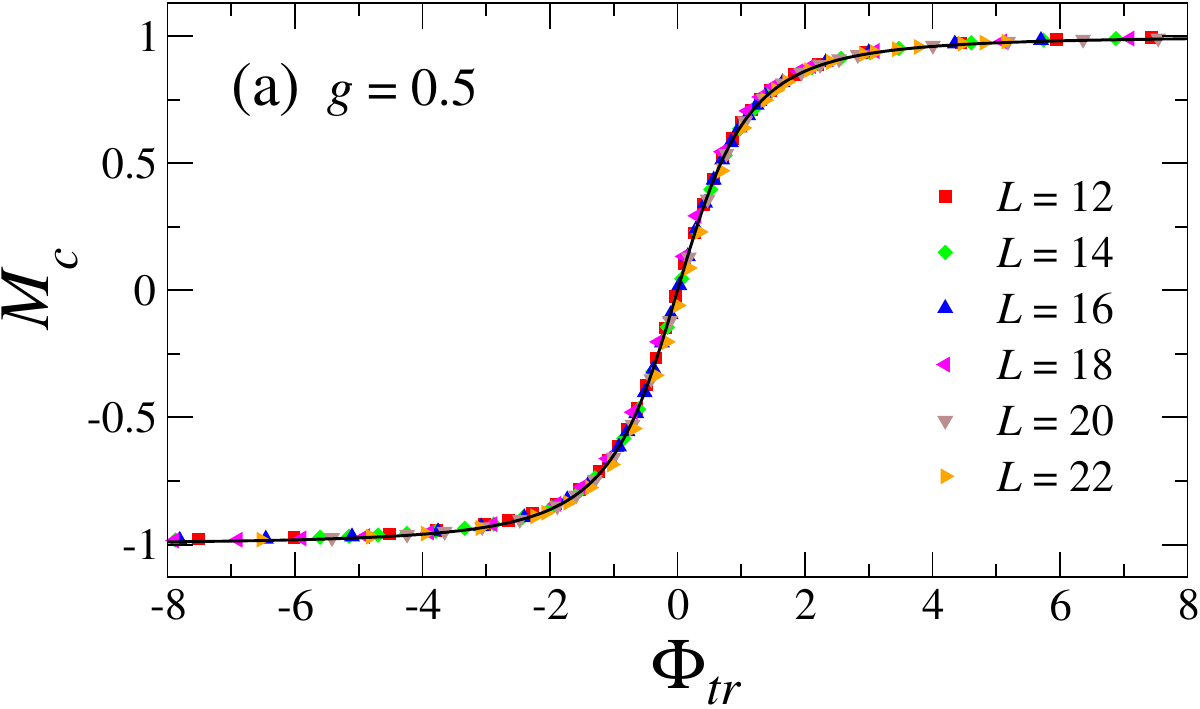}
    \hspace*{5mm}
    \includegraphics[width=0.47\columnwidth]{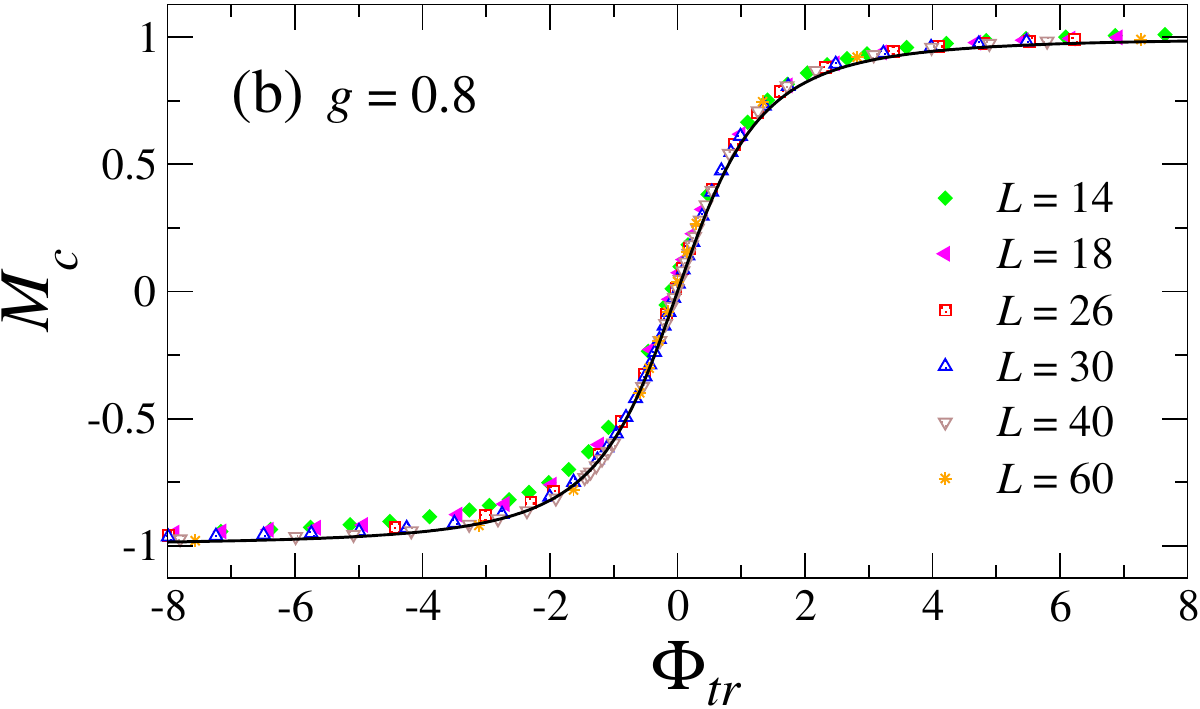}
    \caption{The central magnetization $M_c$, defined in
      Fig.~\ref{fig:Mc}, versus the scaling variable $\Phi_{tr}$, for
      several system sizes (see legend). Panel (a) is for $g=0.5$,
      while panel (b) is for $g=0.8$.  Even though the scenario is
      more complex than that for OBC (see Sec.~\ref{fsspbcobc}), a
      two-level model turns out to reproduce the scaling functions
      (continuous black curve).  Adapted from Ref.~\cite{PRV-18-fo}.}
    \label{fig:Mc_2lev}
  \end{center}
\end{figure}

In proximity of $h_{tr}(L)$, a universal FSS behavior emerges from the
competition of the two lowest-energy states, separated by a gap which
vanishes exponentially with $L$.  For $h$ close to $h_{tr}(L)$, it is
still possible to define a FSS behavior in terms of the scaling
variable
\begin{equation}
  \Phi_{tr} \equiv {2 m_0 L \, [h - h_{tr}(L)]\over \Delta_m(L)} \, ,
  \label{wdef}
\end{equation}
where $\Delta_m(L)$ is the minimum of the energy difference between
the two lowest states, defining the location of the pseudotransition
point $h_{tr}(L)$ at finite $L$.  The scaling variable $\Phi_{tr}$ is
the analogue of $\Phi$ defined in Eq.~\eqref{kappah}, with the
essential difference that the finite-size pseudotransition now occurs
at $h = h_{tr}(L)$, and not at $h = 0$.  Therefore, the relevant
magnetic energy scale is the difference between the magnetic energy at
$h$ and that at $h_{tr}(L)$, while the relevant gap is the one at
$h_{tr}(L)$.  For $h\approx h_{tr}(L)$, observables are expected to
develop a FSS behavior, such as
\begin{equation}
  \Delta(L,h) \approx \Delta_m(L)\, {\cal E}_f(\Phi_{tr}) \,, \qquad
  M_c(L,h) \approx {\cal M}_{cf}(\Phi_{tr}) \,,
  \label{febcsca}
\end{equation}
where $M_c(L,h) \equiv m_{L/2}(L,h) / m_0$ is the normalized
magnetization at the center of the chain. These predictions are nicely
supported by the data, as visible in Fig.~\ref{fig:Mc_2lev}. Further
details can be found in Ref.~\cite{PRV-18-fo}.

The infinite-volume critical point $h=0$ lies outside the region in
which FSS holds. Note that a crucial issue in the definition of the
scaling variable $\Phi_{tr}$ is that the values of $h_{tr}(L)$ and
$\Delta_m(L)$ must be those associated with the minimum of the gap for
the given size $L$, i.e., they cannot be replaced with their
asymptotic behaviors.

Analogous behaviors are expected in other FOQTs, for example in higher
dimensions, when BC favor one of the two phases. 

\subsection{Quantum transitions driven by defects}
\label{qutrdef}

The driving parameters of QTs are usually bulk quantities, such as the
chemical potential in particle systems, or external magnetic fields in
spin systems. In the presence of FOQTs, the bulk behavior is
particularly sensitive to the BC or to localized defects.  This
property makes it possible to induce a related CQT by
changing only the parameters associated with localized defects or
parameters controlling the boundaries~\cite{CPV-15-def, CPV-15-bf}.
Examples of this type of QTs can be observed in quantum Ising rings
with a localized defect.

We consider Ising rings of size $L=2\ell+1$
with a transverse magnetic field and one bond defect:
\begin{equation}
  \hat H_r = - J \sum_{x=-\ell}^{\ell-1} \hat \sigma^{(1)}_x
  \hat \sigma^{(1)}_{x+1} 
  - g \sum_{x=-\ell}^\ell \hat \sigma^{(3)}_x  
  - \zeta \; \hat \sigma_{-\ell}^{(1)} \hat \sigma_\ell^{(1)} \,.
  \label{hedef}
\end{equation}
We again set $J=1$ and assume $g\ge 0$.  Varying $\zeta$, it is
possible to recover models with PBC, OBC, and ABC, for $\zeta=1$, 0,
and $-1$, respectively.  The bond defect generally breaks translation
invariance, unless $\zeta=\pm 1$.  In the presence of an additional
parallel field $h$ coupled to $\hat \sigma^{(1)}_x$, FOQTs occur for
any $g < 1$.

For $g < 1$, we distinguish two phases: a {\em magnet} phase ($\zeta > -1$)
and a {\em kink} phase ($\zeta \le -1$).  The lowest states of
the magnet phase are superpositions of states $|\pm\rangle$ with
opposite nonzero magnetization $\langle\pm | \hat
\sigma_x^{(1)}|\pm\rangle = \pm m_0$ (neglecting local effects at the
defect), where${}^{~\ref{foot1}}$ $m_0=(1-g^2)^{1/8}$.  For a finite
chain, tunneling effects between the states $| + \rangle$ and $|
-\rangle$ lift the degeneracy, giving rise to an exponentially small
gap $\Delta(L,\zeta>-1)$~\cite{ZinnJustin-86, BC-87}~\footnote{See,
for example, Eq.~\eqref{deltaobc} for $\zeta=0$ (OBC).}.  An analytic
calculation for $\zeta\to -1^+$ gives~\cite{CPV-15-def, CPV-15-bf}
\begin{equation}
  \Delta(L,\zeta\to -1^+) \approx {8g\over 1-g} w^2 e^{-wL},
  \qquad  w={1-g\over g} \;(1+\zeta) \,.
  \label{defde}
\end{equation}
The large-$L$ two-point function is trivial,
\begin{equation}
  G(x_1,x_2) \equiv \big\langle \hat \sigma_{x_1}^{(1)} \,
  \hat \sigma_{x_2}^{(1)} \big\rangle \to m_0^2
  \label{gcmagnet}
\end{equation}
for $x_1\neq x_2$, keeping $X_i\equiv x_i/\ell$ fixed (but $X_i\neq \pm 1$).

The low-energy behavior drastically changes for $\zeta\le -1$, in
which the low-energy states are one-kink states (made of a
nearest-neighbor pair of antiparallel spins), behaving as
one-particle states with $O(L^{-1})$ momenta. In particular, for
$\zeta=-1$, corresponding to a system with ABC, the gap decreases as
$\Delta(L,-1) \sim L^{-2}$, cf.~Eq.~\eqref{deltaabcfobc}.  The first
two excited states are degenerate, thus $\Delta_{2}(L,-1)=
\Delta_{1}(L,-1)\equiv \Delta(L,-1)$.  For $\zeta<-1$, the ground
state and the first excited state are superpositions with definite
parity of the lowest kink $|\!\downarrow\uparrow\rangle$ and antikink
$|\!\uparrow\downarrow\rangle$ states.  The gap in this regime
decreases as $L^{-3}$~\cite{CJ-87, BC-87, CPV-15-def}, more precisely
\begin{equation}
\Delta(L,\zeta<-1) = {8 \zeta g^2 \over (1-\zeta^2)
  (1-g)^2}\,{\pi^2\over L^3} + O(L^{-4}) \,.
\label{deltallm3}
\end{equation}
On the other hand, $\Delta_{n}\equiv E_n-E_0$ for $n\ge 2$ behaves as $L^{-2}$,
\begin{equation} 
  \Delta_{2} = { 3 g \over (1-g)}\,{\pi^2\over L^2} 
  + {6(1-\zeta)g^2\over (1+\zeta)(1-g)^2} \,{\pi^2\over L^3} + O(L^{-4}) \,.
  \label{deltallm32}
\end{equation}
Corresponding changes are found in the asymptotic behavior of the
two-point function~\cite{CPV-15-def, CPV-15-bf}
\begin{equation}
  {G(x_1,x_2)\over m_0^2} = \left\{ \begin{array}{ll}
    1 - |x_1-x_2|/\ell & \; {\rm for}\;\; \zeta=-1\,, \vspace*{1mm} \\
    1 - |x_1-x_2|/\ell - |\sin(\pi x_1/\ell)-\sin(\pi x_2/\ell)|/\pi
    & \; {\rm for}\;\; \zeta<-1\,.
  \end{array} \right.
\end{equation}

\begin{figure}
  \begin{center}
    \includegraphics[width=0.5\columnwidth]{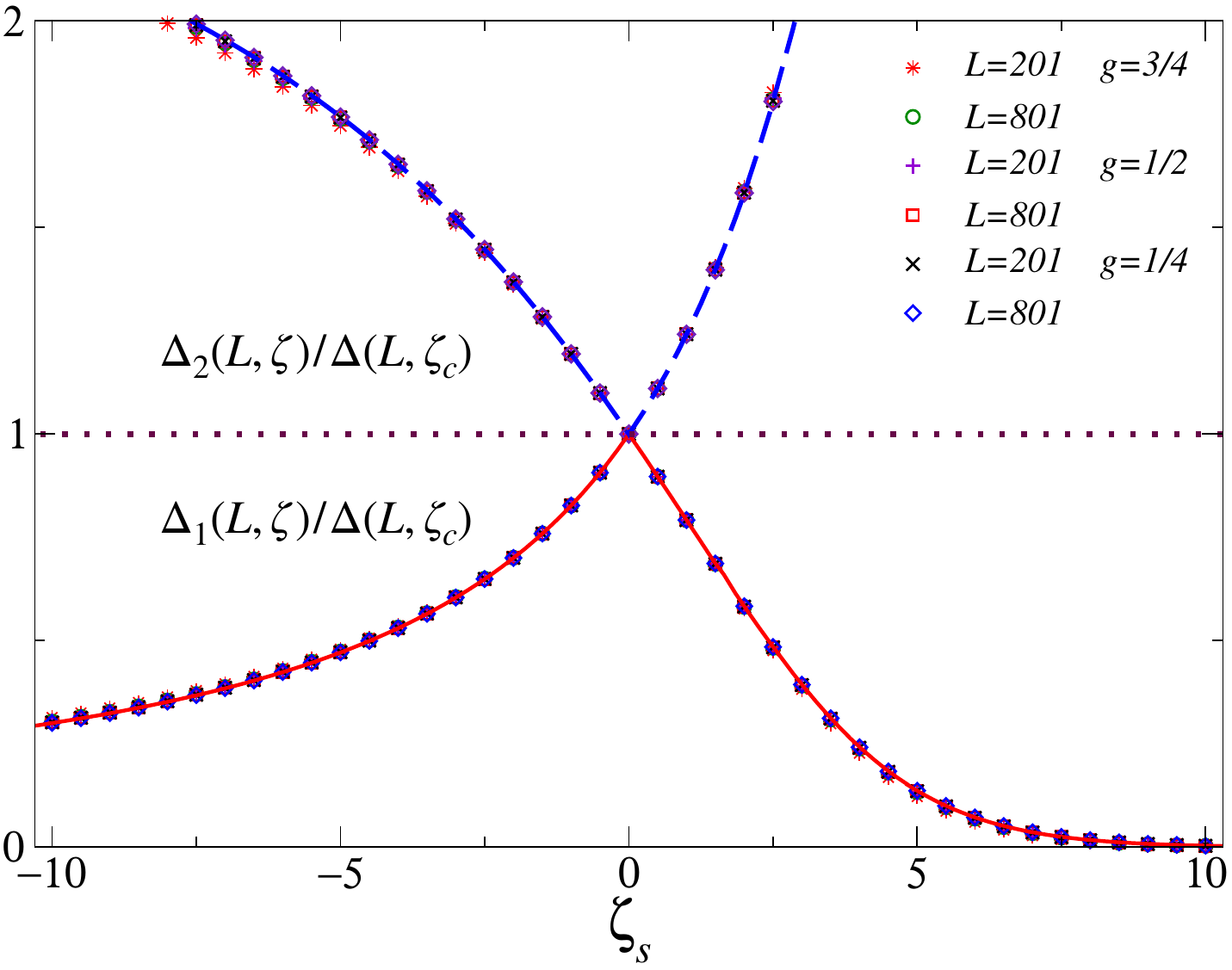}
    \caption{The scaling functions ${\cal D}_n(\zeta_s)$ of
      Eq.~\eqref{deltasca} (dashed lines) and numerical data (symbols)
      for the ratio $\Delta_{n}(L,\zeta)/\Delta(L,\zeta_c)$ as functions
      of the scaling variable $\zeta_s$, for $n=1$ (bottom) and $n=2$
      (top), separated by the dotted line.  Numerical data clearly
      approach the $g$-independent scaling curves ${\cal
        D}_n(\zeta_s)$ (differences are hardly visible).
      Adapted from Ref.~\cite{CPV-15-def}.}
    \label{dezetac}
  \end{center}
\end{figure}

The above behaviors show that the defect parameter $\zeta$ drives
a QT at $\zeta_c=-1$, separating the magnet and kink phases.
Moreover the system develops a universal $g$-independent scaling
behavior~\cite{CPV-15-def}, such as
\begin{equation}
  \Delta_{n}(L,\zeta) \equiv E_n(\zeta)-E_0(\zeta)
  \approx  \Delta(L,\zeta_c)\:{\cal D}_{n}(\zeta_s)\,,\qquad
  \zeta_s \equiv {1-g\over g} (\zeta - \zeta_c)\,L\,,
  \label{deltasca}
\end{equation}
for $L\to \infty$, keeping the scaling variable $\zeta_s$ fixed.  The
scaling functions ${\cal D}_1$ and ${\cal D}_2$, which have been
computed analytically, are shown in Fig.~\ref{dezetac}, where they are
also compared with some numerical results that support the scaling
ansatz~\eqref{deltasca}. The cusp-like behavior at $\zeta_s=0$ is the
consequence of the crossing of the first two excited states at
$\zeta=-1$.  The asymptotic large-$L$ behavior is generally approached
with $O(L^{-1})$ corrections.  Other observables satisfy analogous
scaling relations, such as the two-point function, and corresponding
susceptibility and correlation lengths. An analogous magnet-to-kink QT
can be observed in the Ising chain by appropriately tuning a magnetic
field coupled to $\hat \sigma^{(1)}$, localized at the
boundaries.~\footnote{These results can be reinterpreted in
  the equivalent fermionic picture of models~\eqref{hedef}.  In the
  magnet phase ($\zeta>\zeta_c$), the lowest eigenstates are
  superpositions of Majorana fermionic states localized at the
  boundaries or on the defect~\cite{Kitaev-01, Alicea-12}.  In finite
  systems, their overlap does not vanish, giving rise to the splitting
  $\Delta \sim e^{-L/l_0}$.  The coherence length $l_0$ diverges at
  the kink-to-magnet transitions as $l_0^{-1}\sim \zeta-\zeta_c$, 
  analogously to the behavior observed at the order-disorder transition
  $g\to 1^-$, where $l_0^{-1}\sim |\ln g|$.}

The above results reflect a QT with critical exponents given by the RG
dimension $y_\zeta=1$ of the relevant defect parameter, and dynamic
exponent $z=2$ due to the $L^{-2}$ behavior of the gap at $\zeta_c$,
see Eq.~\eqref{deltaabcfobc}.  This defect-driven CQT characterizes the
FOQT line of the Ising chain.  It is worth noting that, at the
critical value $g=1$, corresponding to the order-disorder continuous
transition, the bulk behavior is independent of the BC or of the
presence of defects, hence the magnet-to-kink transition only occurs
for $g$ strictly less than 1.  For instance, the gap at $g=1$ behaves
as $\Delta\sim L^{-1}$ for any $\zeta$~\footnote{Of course, the
  prefactor depends on the boundary fields~\cite{BP-91}.  Results for
  PBC, OBC, and ABC are reported in Eq.~\eqref{gapiscqt}.}

In conclusion, peculiar QTs can be induced by tuning the BC or by
changing lower-dimensional defect parameters, when the system is at a
FOQT. These scaling behaviors can be straightforwardly extended to
allow for a nonzero temperature $T$, by considering a further
dependence on the scaling variable $\Xi = L^z \, T$.  Even though we have
discussed the issue in $1d$, the same type of behavior is expected in
quantum $d$-dimensional Ising models defined in $L^{d-1}\times \bar L$
boxes with $L\gg \bar L$, in the presence of a $(d-1)$-dimensional
surface of defects or of opposite magnetic fields on the $L^{d-1}$
boundaries.

\section{Quantum information and many-body systems}
\label{qinfosec}

In this section we provide an overview of some concepts founded on the
recently developing {\em quantum information
  science}~\cite{Preskill-book, NC-book, BCRS-book}, which have been
proven useful to spotlight the presence of singularities at QTs in
many-body systems.  Among them, we mention the fidelity and its
susceptibility, the quantum Fisher information, as well as the
entanglement and other quantifiers of nonclassical correlations.  As
we shall see in the following, the net advantage of these approaches
is that they do not rely on the identification of an order parameter
with the corresponding symmetry-breaking pattern, and thus may be 
helpful also to detect topological transitions.  Other specific
reviews on these topics can be found in Refs.~\cite{AFOV-08, Gu-10,
  DS-18}.

\subsection{Fidelity and Loschmidt amplitude}
\label{sec:fidelity}

The strategy lying at the basis of the concept of fidelity in a
quantum many-body system is based on the comparison between two given
states of such a system, and quantifies the absolute value of the
overlap between them~\cite{GPSZ-06}.  To fix the ideas, consider for
example the ground state $|\Psi_0(g)\rangle$ of a given Hamiltonian
operator $\hat H(g)$, which depends on a control parameter $g$ (it can
be the intensity of an external magnetic field, the strength of
interactions, or some other parameter).  We define the fidelity
\begin{equation}
  A(g_1, g_2) \equiv
  \big| \langle \Psi_0(g_1) | \Psi_0(g_2) \rangle \big|
  \label{eq:fid1}
\end{equation}
as a measure of the distance between two ground states at different
values of $g$, say $g_1$ and $g_2$~\footnote{In the literature, the
following alternative definition for the fidelity is also encountered:
$\widetilde{A}(g_1, g_2) \equiv \big| \langle \Psi_0(g_1) | \Psi_0(g_2)
\rangle \big|^2$.  However all the reasonings made in this review are
independent of this choice.}.  Clearly $0 \leq A \leq 1$, where $A=1$
if $|\Psi_0(g_1)\rangle$ coincides with $|\Psi_0(g_2)\rangle$ up to a
global phase, while $A=0$ if such two states are orthogonal.
Now suppose to compare two infinitesimally close parameters, i.e.,
$g_1 = g_2 + \varepsilon$. It is rather intuitive that in proximity of
a QT, when the system observables may undergo singular behaviors, even
the slightest move in $g$ may result in a considerable deviation of
$A$ from the unit value, showing that the two ground states increase
their degree of orthogonality~\cite{ZP-06}.

The fidelity can be a convenient tool also to address the sensitivity
of the quantum motion under Hamiltonian perturbations.  In fact this
is the original context where it was first introduced~\cite{Peres-84},
in the framework of quantum chaos, and recently extended to the
many-body realm.  The setting can be put under the following general
grounds: From a given pure state $|\Psi_{\rm in}\rangle$ (not
necessarily the ground state), one can construct two state vectors
$|\Psi(t)\rangle$ and $|\Psi_\varepsilon(t)\rangle$ according to the
unitary evolutions ruled by $\hat H(g)$ and $\hat H(g+\varepsilon)$,
respectively. The fidelity is the absolute value of the overlap
between them, namely
\begin{equation}
  A_\varepsilon(t) = \big| \langle \Psi(t) | \Psi_\varepsilon(t) \rangle \big|
  \, ,
  \label{eq:FidLE}
\end{equation}
with
\begin{equation}
  |\Psi(t)\rangle = e^{-i \hat H(g) t} |\Psi_{\rm in}\rangle \qquad \mbox{and} \qquad
  |\Psi_\varepsilon(t)\rangle = e^{-i \hat H(g+\varepsilon) t} |\Psi_{\rm in}\rangle
  \,. 
\end{equation}
The above expression~\eqref{eq:FidLE}, without taking the absolute value,
is called the Loschmidt amplitude $\widetilde A_\varepsilon(t) =
\langle \Psi(t) | \Psi_\varepsilon(t) \rangle = \langle \Psi_{\rm in}(t)|
e^{i \hat H(g) t} \, e^{-i \hat H(g+\varepsilon) t} |\Psi_{\rm in}\rangle$,
since it can be equivalently seen as the overlap between a
given state and the one obtained by first evolving forward-in-time
with t<he unitary operator $e^{-i \hat H(g+\varepsilon) \, t}$,
followed by a backward-in-time perturbed evolution under $e^{-i \hat
  H(g) \, t}$.

We finally mention that the definition of the fidelity in
Eq.~\eqref{eq:fid1} can be generalized also to the cases in which one
or both (pure) states are substituted by mixed states.  Namely, the
fidelity $A \in [0,1]$ between a given pure state $|\psi\rangle$ and a
mixture $\rho$ is given by
\begin{equation}
  A(|\psi\rangle, \rho) \equiv \sqrt{\langle \psi | \rho | \psi \rangle} \,,
  \label{eq:fid2}
\end{equation}
while the fidelity between two mixtures $\rho$ and $\sigma$
is defined by~\cite{Uhlmann-76}
\begin{equation}
  A(\rho, \sigma) \equiv {\rm Tr} \left[
    \sqrt{\sigma^{1/2} \rho \, \sigma^{1/2}} \, \right] \,,
  \label{eq:fid3}
\end{equation}
where ${\rm Tr} [ \, \cdot \, ]$ indicates the trace operation.

\subsection{State discrimination and the quantum Fisher information}
\label{sec:QFI}

The concept of fidelity is intimately related to the {\em quantum
  estimation theory}, which can be put on formal grounds through the
quantum Cram{\'e}r-Rao bound~\cite{BC-94, BCM-96}. Indeed the latter
sets the limits on the precision that can be attained in the
estimation of an unknown parameter $\lambda$ from repeated independent
measurements of the quantum system. Namely, suppose to have a family
of quantum states $\{ \rho_\lambda \}$ that depend on
$\lambda$~\footnote{ The parametrization $\lambda \to \rho_\lambda$
can be static (e.g., for a given control parameter of the Hamiltonian)
or arise dynamically (e.g., for the time $t$ during a unitary
evolution).}.  Then, the variance of the estimated quantity $(\Delta
\lambda)^2$ should be limited by
\begin{equation}
  (\Delta \lambda)^2 \geq \frac{1}{M A_Q(\rho_\lambda)} \,,
  \label{eq:Cramer}
\end{equation}
where $M$ is the number of times the measurement is repeated and $A_Q$
is the so-called quantum Fisher information of the state
$\rho_\lambda$.
The quantum Fisher information can be expressed as
\begin{equation}
  A_Q(\rho_\lambda) = 8 \lim_{\delta \lambda \to 0}
    \frac{1 - A(\rho_\lambda, \rho_{\lambda + \delta \lambda})}
{\delta \lambda^2} \,,
    \label{eq:QFI}
\end{equation}
where $A(\rho_\lambda, \rho_{\lambda + \delta \lambda})$ is the
fidelity between two mixed states, defined in Eq.~\eqref{eq:fid3}.  In
fact, the right-hand side of Eq.~\eqref{eq:QFI} quantifies the
susceptibility of the quantum state to infinitesimal variations of the
parameter $\lambda$ and thus defines the fidelity susceptibility (see
also Sec.~\ref{setting}).~\footnote{Comparing Eq.~\eqref{eq:QFI} with
  Eq.~\eqref{expfide} in Sec.~\ref{setting}, one can realize that $A_Q
  = 4 \chi_A$, where $\chi_A$ is the fidelity susceptibility.}

The quantum Fisher information plays a crucial role in quantum
metrology, since it determines the reachable accuracy of the estimated
quantity~\cite{Paris-09}: through the above quantum Cram{\'e}r-Rao
bound~\eqref{eq:Cramer}, it may be possible to identify which
resources (e.g., entanglement or squeezing) are useful to improve the
measurement accuracy, as well as the proper procedures to adopt to
fully exploit them~\cite{GLM-11}.

\subsection{Entanglement}
\label{entangsec}

The concept of entanglement is one of the cornerstones of quantum
information theory and arises from the necessity to distinguish the
quantum correlations, that can occur in many-body quantum states, from
the classical ones~\cite{EPR-35, HHHH-09}.  Such distinction can be
made rigorous by first defining the amount of classical correlations,
a concept that relies on the possibility to perform local operations
supplemented by classical communication (LOCC) between partitions that
compose the whole system.  Any other correlations, that cannot be
simulated by the above means, are associated with quantum effects and
labelled as quantum correlations.  As described below, it is possible
to provide different types of quantifiers for the entanglement, which
treat it as an operational resource to perform tasks that are
impossible, if the quantum state is separable.

To be more precise, let us first consider a bipartite system (i.e., a
system made of two parts $\tt{A}$ and $\tt{B}$, typically called Alice
and Bob, to use the quantum information jargon) in a pure state
$|\psi\rangle_{\tt{AB}}$.  Such a state is called {\em separable} if
and only if it can be written as
\begin{equation}
  |\psi\rangle_{\tt{AB}} = |\alpha\rangle_{\tt A} \otimes
  |\beta\rangle_{\tt B}\,,
  \label{eq:Psi_separable}
\end{equation}
where $\otimes$ denotes the tensor product, while the (pure) states
$|\alpha\rangle_{\tt A}$ and $|\beta\rangle_{\tt B}$ respectively
describe the components of the two parts.  Any state
$|\psi\rangle_{\tt{AB}}$ that cannot be written as in
Eq.~\eqref{eq:Psi_separable} (that is, a non-separable state) is named
{\em entangled}.  The above definition of separability can be easily
generalized to the case of a bipartite mixed state:
\begin{equation}
  \rho_{\tt{AB}} = \sum_k p_k \, 
\rho_{\tt A}^{(k)} \otimes \rho_{\tt B}^{(k)}\,, \qquad
  \mbox{with } \;\; p_k \geq 0\,, \;\; \sum_k p_k = 1\,,
  \label{eq:Rho_separable}
\end{equation}
$\rho_{\tt A}^{(k)}$ and $\rho_{\tt B}^{(k)}$ being density matrices
for the corresponding two parts, and also to the more general case of
a $n$-partite system, made of parts $\tt{A, B, C,} \ldots$, according to
the following:
\begin{equation}
  \rho_{\tt{ABC \ldots}} = \sum_k p_k \, 
\rho_{\tt A}^{(k)} \otimes \rho_{\tt B}^{(k)} \otimes
  \rho_{\tt C}^{(k)} \otimes \ldots \,.
  \label{eq:Rho_separable2}
\end{equation}
Analogously as for pure states, mixed states that are not separable
are named entangled.

Finding quantifiers of nonclassical correlations is, in general, a
difficult task, however any good entanglement measure $\mathcal{E}$
shall fulfill a number of reasonable requirements.  For the sake of
simplicity, here we restrict the discussion to the case of the
bipartite entanglement $\mathcal{E}(\rho_{\tt{AB}})$, where the
situation is somewhat simpler to describe. The more involved cases of
$n$-partite scenarios will be briefly addressed below in
Sec.~\ref{sec:Multipartite}.

The quantity $\mathcal{E}$ can be defined according to the following:
\begin{enumerate}
\item
  It shall be a map from arbitrary density operators $\rho_{\tt{AB}}$
  of a bipartite quantum system to positive real numbers.  The measure
  should be normalized in such a way that there exist maximally
  entangled states (states, or a class of states, that are more
  entangled than all the others). These can be shown to be equivalent,
  up to unitary operations, to the generalized Einstein-Podolsky-Rosen
  (EPR)~\cite{EPR-35} pair
  \begin{equation}
    |\phi^+_d\rangle_{\tt{AB}} = \frac{1}{\sqrt{d}} \, \Big(
    |0\rangle_{\tt A} \otimes |0\rangle_{\tt B} + |1\rangle_{\tt A}
    |\otimes |1\rangle_{\tt B} + \; \ldots \; + | d-1\rangle_{\tt A}
    |\otimes |d-1\rangle_{\tt B} \Big)\,,
    \label{eq:EPR}
  \end{equation}
  where $d$ is the dimension of each of the two subsystems.
\item It should not increase under LOCC.
\item It should vanish for any separable state.
\end{enumerate}
Any function $\mathcal{E}(\rho_{\tt{AB}})$ satisfying these three
conditions is called an entanglement monotone.  Examples of such
monotones are the {\em entanglement cost} and the
{\em distillable entanglement}.
The first one is defined as the rate of maximally entangled states
needed to create $\rho_{\tt{AB}}$ using LOCC. The second one
quantifies the complementary optimal rate of maximally entangled
states that can be distilled from $\rho_{\tt{AB}}$ using LOCC.  We
also mention the {\em entanglement of formation}, being constructed
as the minimal possible average entanglement over all the pure-state
decompositions of $\rho_{\tt{AB}}$.

There are other requirements one should expect to enforce, even if
it could be complicated to construct a measure satisfying all of them.
In particular $\mathcal{E}$ should satisfy:
\begin{enumerate}
  \addtocounter{enumi}{3}
\item
  Continuity, in the sense that, for decreasing distance between two density
  matrices, the difference between their entanglement should tend to zero.
\item Convexity:
  $\mathcal{E}(\sum_i p_i \rho_i) \leq \sum_i p_i \, \mathcal{E}(\rho_i)$,
  where $p_i>0$, $\sum_i p_i=1$.
\item Additivity: $\mathcal{E}(\rho^{\otimes n}) = n \, \mathcal{E}(\rho)$.
\item Subadditivity: $\mathcal{E}(\rho \otimes \sigma)
  \leq \mathcal{E}(\rho) + \mathcal{E}(\sigma)$.
\end{enumerate}

\subsubsection{Bipartite entropies}
\label{sec:vN-Renyi}

For bipartite pure states $\rho_{\tt{AB}} = |\psi\rangle_{\tt{AB}}
\langle \psi|$, it is possible to define a unique measure of
entanglement $\mathcal{E} (|\psi\rangle_{\tt{AB}})$ satisfying all the
above conditions and coinciding with the above mentioned entanglement
monotones.  This is the von Neumann entropy $S$ of the reduced density
matrix of one of the two parts, say Alice: $\rho_{\tt{A}} = {\rm Tr}_{\tt B}
[|\psi\rangle_{\tt{AB}} \langle \psi|]$
(where ${\rm Tr}_{\tt{B}}[\,\cdot\,]$ denotes the partial trace over Bob),
also referred to as the entropy of entanglement,
\begin{equation}
  \mathcal{E} (|\psi\rangle_{\tt{AB}}) = S(\rho_{\tt A}) \equiv - {\rm
    Tr} \big[ \rho_{\tt A} \log \rho_{\tt A} \big]\,,
  \label{eq:EntEntro}
\end{equation}
which is thus a nonlinear function of the eigenvalues of
$\rho_{\tt A}$~\footnote{The bipartite entanglement
  $\mathcal{E}(|\psi\rangle_{\tt{AB}})$ can be equivalently obtained
  by calculating the von Neumann entropy on the reduced density matrix
  $\rho_{\tt B} = {\rm Tr}_{\tt A} [|\psi\rangle_{\tt{AB}} \langle \psi|]$
  of the complementary part, that is, $\mathcal{E}
  (|\psi\rangle_{\tt{AB}}) = S(\rho_{\tt A}) = - {\rm Tr} \big[
    \rho_{\tt B} \log \rho_{\tt B} \big]$.}.

Relaxing some of the requirements above, as for example the
subadditivity, one can define other bipartite entanglement measures
which are grouped under the name of R\'enyi entropies $S_n$,
defined as
\begin{equation}
  S_n(|\psi\rangle_{\tt{AB}}) \equiv
  \frac{1}{1-n} \log {\rm Tr} \big[ \rho_{\tt A}^n \big]\,,
\end{equation}
for any real number $n \geq 0$ and $n\neq 1$.  For $n\to 1$ one
recovers the von Neumann entropy $S$ in Eq.~\eqref{eq:EntEntro}.
These quantities, although not fulfilling all the expected properties
of entanglement measures, are often employed in the study of quantum
many-body systems, since they are typically easier to compute than the
von Neumann entropy.

\subsubsection{Concurrence}
\label{concurrencedef}

The situation is definitely more complex for bipartite mixed states
$\rho_{\tt{AB}} \neq |\psi\rangle_{\tt{AB}} \langle \psi|$, where the
above mentioned entanglement monotones do not coincide and, in
general, finding operative ways to define and calculate good
entanglement measures is cumbersome.  One of the scarce cases where
this is possible are mixed states of two spin-$1/2$ particles (also
named {\em qubits}), where there exists a measure called the 
concurrence~\cite{Wootters-98}. The concurrence $C$ is equivalent to
the entanglement of formation and is defined as
\begin{equation}
  C(\rho_{\tt{AB}}) \equiv \max \big\{ 0, \lambda_1 - \lambda_2 -
  \lambda_3 - \lambda_4 \big\}\,,
\end{equation}
where $\lambda_i$ are the square roots of the eigenvalues, in descending order,
of the Hermitian matrix
\begin{equation}
  R = \sqrt{\rho} \, \tilde \rho \, 
\sqrt{\rho} \qquad \mbox{with } \;
  \tilde \rho = \big( \hat \sigma^{(2)} \otimes 
\hat \sigma^{(2)} \big) \, \rho^\star \,
  \big( \hat \sigma^{(2)} \otimes \hat \sigma^{(2)} \big)\,,
\end{equation}
where the complex conjugation is taken in the basis of $\hat
\sigma^{(3)}$.

\subsubsection{Other entanglement indicators}
\label{sec:otherEnt}

There are other bipartite entanglement measures that can be useful in
different situations. Among them, we mention the negativity
$\mathcal{N}$, which is based on the partial transpose of the density
matrix $\rho_{\tt{AB}}$ with respect to one of the subsystems, namely
$\rho^{T_{\tt{A}}}_{\tt{AB}} = (\hat T_{\tt{A}} \otimes \hat
I_{\tt{B}}) \rho_{\tt{AB}}$, where $\hat T_{\tt{A}}$ is the transpose
operator acting on Alice, and $\hat I_{\tt{B}}$ the identity operator
acting on Bob.  The negativity is defined as~\cite{ZHSL-98}
\begin{equation}
  \mathcal{N} \equiv \sum_{\tilde \lambda_i <0} |\tilde \lambda_i|\,,
\end{equation}
with $\tilde \lambda_i$ indicating the negative eigenvalues of
$\rho^{T_{\tt{A}}}_{\tt{AB}}$.  Unfortunately, the negativity assigns
non-vanishing entanglement only to a subset of all the entangled
states, therefore $\mathcal{N} >0$ implies that $\rho_{\tt{AB}}$ is
entangled, but the opposite implication is not necessarily true.  More
specifically, it can be shown that the negativity is fully reliable
only for bipartite systems with $2 \times 2$ and $2 \times 3$
Hilbert-space dimensions.

Another quantity which is able to qualitatively disclose the degree of
correlations of a quantum state is the so-called purity, defined as
\begin{equation}
  {\cal P} (\rho_{\tt A}) \equiv {\rm Tr} \big[ \rho_{\tt A}^2 \big] \,.
  \label{eq:Purity}
\end{equation}
This quantity satisfies the constraint $d^{-1} \leq {\cal P} \leq 1$,
where $d$ is the dimension of Alice's Hilbert space.  The upper bound
is saturated for pure states, while the lower bound is obtained for
completely mixed states, given by multiples of the identity matrix,
$\hat I_d / d$. Also note that it is conserved under unitary
transformations (such as the unitary time evolution operator for a
non-dissipative system) acting on the density matrix $\rho_{\tt A}$.

In fact, for global pure states of a bipartite system, the purity can
be seen as another measure of entanglement.  For global mixed states
$\rho_{\tt AB}$, this quantifies the degree of local mixedness of
$\rho_{\tt AB}$ and can be interpreted as an indicator of the average lower
bound for the bipartite entanglement between ${\tt A}$ and ${\tt B}$.
For this reason, and due to its computational handiness, the purity is
often employed to discuss the coherence loss of quantum states during
the evolution, when they are coupled to some external environment.

\subsection{Quantum discord}

The concept of quantum correlations has been recently generalized
beyond the definition of entanglement.  Indeed, it is possible to
identify some separable mixed states as quantum correlated, yet not
entangled. To this purpose one can introduce the so-called quantum
discord, namely, a measure which takes into account any non-classical
source of correlations~\cite{OZ-01, HV-01}.

Focusing again on a system composed of two parts, Alice and Bob, the
bipartite quantum discord is defined as the difference between the
quantum mutual information
\begin{equation}
  \mathcal{I}_{\tt{AB}} \equiv 
  S(\rho_{\tt A}) + S(\rho_{\tt B}) - S(\rho_{\tt{AB}})\,,
  \label{eq:MutualInfo}
\end{equation}
and the one-way classical information
\begin{equation}
  \mathcal{J}_{\tt{B:A}}(\rho_{\tt{AB}}) \equiv S(\rho_{\tt{B}}) -
  S(\rho_{\tt{AB}} | \Pi_{\tt A})\,, \qquad \mbox{with} \quad
  S(\rho_{\tt{AB}} | \Pi_{\tt A}) = \sum_j p_j \, S(\rho_{{\tt
      B}|\Pi_j}) \,.
  \label{eq:1wayClass}
\end{equation}
The quantity $S(\rho_{\tt{AB}} | \Pi_{\tt A})$ denotes the average
amount of information that can be retrieved when measuring Alice
through a set of one-dimensional projectors $[\Pi_j]_{\tt A} =
|j\rangle\langle j|$, and $\{|j\rangle\}$ defines an orthonormal basis
in the Hilbert space of Alice.  The outcome $j$ is obtained with
probability $p_j$, collapsing Bob in the state $\rho_{\tt{B}|\Pi_j}$.

The quantum mutual information $\mathcal{I}_{\tt{AB}}$ in
Eq.~\eqref{eq:MutualInfo} measures the global amount of information
which is contained in the bipartite system $\rho_{\tt{AB}}$ and cannot
be accessed by looking at the reduced states $\rho_{\tt A}$ and
$\rho_{\tt B}$. Indeed one has $\mathcal{I}_{\tt{AB}} \geq 0$, and
such quantity is saturated to the null value only for tensor product
states.
Conversely, the one-way classical information measures the amount of
information that can be gained about Bob as a result of the
measurement $\{[\Pi_j]_{\tt A}\}$ on Alice.  Notice that, in general,
such quantity depends on which subsystem is measured, that is,
$\mathcal{J}_{\tt{A:B}}(\rho_{\tt{AB}}) \neq
\mathcal{J}_{\tt{B:A}}(\rho_{\tt{AB}})$.  Incidentally, for pure
states, the bipartite quantum discord coincides with the entropy of
entanglement in Eq.~\eqref{eq:EntEntro}.

Bayes rule ensures that, if $\tt{A}$ and $\tt{B}$ were classical
random variables, then $\mathcal{I}_{\tt{AB}} =
\mathcal{J}(\rho_{\tt{B:A}})$.  The quantum discord of a state
$\rho_{\tt{AB}}$ under an arbitrary projective measurement over Alice
is defined as the difference
\begin{equation}
  \mathcal{D}_{\tt{B:A}}(\rho_{\tt{AB}}) \equiv \mathcal{I}_{\tt{AB}}
  - \max_{\Pi_{\tt A}} \mathcal{J}_{\tt{B:A}}(\rho_{\tt{AB}})\,.
  \label{eq:Discord}
\end{equation}
We remark that, while the evaluation of $\mathcal{I}_{\tt{AB}}$ does
not alter the considered state, $\mathcal{J}_{\tt{B:A}}$ modifies the
quantum system, destroying the classical fraction of the initial
correlations, which cannot be acquired via a measurement on Alice.
This fact leads to an interpretation of the quantum discord in
Eq.~\eqref{eq:Discord} as a measure of genuinely quantum correlations.
Also note that the definition of
$\mathcal{D}_{\tt{B:A}}(\rho_{\tt{AB}})$ can be extended to
generalized measurements (positive operator-valued measurements),
although in most cases the minimization leading to optimal classical
correlations can be well approximated by just considering projective
measurements.

\subsection{Multipartite quantum correlations}
\label{sec:Multipartite}

Quantifying the amount of correlations beyond the bipartite case
is, in general, a cumbersome task, yet allowing to explore a scenario
that offers a much wider range of possibilities.
Consider, for example, a natural generalization of the EPR pair
in Eq.~\eqref{eq:EPR} to the $n$-partite case:
\begin{equation}
  |\Phi^+_d\rangle_{n} = \frac{1}{\sqrt{d}} \, \Big(
  |0\rangle^{(\otimes n)} + |1\rangle^{(\otimes n)} +
  \ldots +| d-1\rangle^{(\otimes n)} \Big) \,.
  \label{eq:EPRmulti}
\end{equation}
This state has the appealing property that its entropy of entanglement
across any bipartite cut assumes the largest possible value.  However,
there are entangled states that cannot be obtained from the
$|\Phi^+_d\rangle_{n}$ using LOCC alone. This argument shows that it
is not possible to establish a generic notion of a maximally entangled
state, and more involved processes need to be invoked.  While, in the
bipartite setting, such definition is univocal only for pure states,
here the situation is complex even in that case.

One of the most peculiar features of entanglement is the so-called
{\em monogamy}, which poses constraints on the maximally available
amount of correlations that can be established among several parts of
a quantum system~\cite{CKW-00}.  Consider the case of a system
composed of three parts, labelled as $\tt{A}$ (Alice), $\tt{B}$
(Bob), and $\tt{C}$ (Charlie). If Alice and Bob are very entangled, it
turns out that Charlie can only be very weakly entangled with either
Alice or Bob. For example, if each of the three parts is a qubit and
two of them are in an EPR state, then they cannot be entangled with
the other qubit at all.
This idea brings to the concept of geometric quantum frustration and
can be formalized according to the following statement.
Given a generic pure state of three qubits, the following inequality
must hold:
\begin{equation}
  C^2_{\tt{AB}} + C^2_{\tt{AC}} \leq C^2_{\tt{A|BC}}\,,
\end{equation}
where $C^2_{\tt{A|BC}}$ generalizes the notion of concurrence between
Alice's qubit $\tt{A}$ and the pair of qubits $\tt{BC}$ held by Bob
and Charlie.  This constraint allows us to define the three-tangle
\begin{equation}
  \tau_{\tt{ABC}} \equiv C^2_{\tt{A|BC}} - C^2_{\tt{AB}} - C^2_{\tt{AC}}
\end{equation}
as a measure of purely tripartite entanglement.

Generalizations to arbitrary $n$-partite systems are
also possible. We mention that other quantifiers of multipartite
entanglement have been put forward, such as the geometric
entanglement, the global entanglement, as well as the global quantum
discord (further details can be found in the review
papers~\cite{MBCPV-12, ABC-16}).

\section{Quantum information at quantum transitions}
\label{quinfoatqts}

QTs in many-body systems are related to significant qualitative
changes of the ground-state and low-excitation properties, induced by
small variations of a driving parameter.  As mentioned in the previous
Sec.~\ref{qinfosec}, alternative useful approaches for a proper
characterization of their main features may come from
quantum-information-based concepts, such as the ground-state fidelity,
its susceptibility related to the quantum Fisher information, the
entanglement, the concurrence, etc.~\cite{AFOV-08, Gu-10,
  Dutta-etal-book, Braun-etal-18}.  Below we give an overview of the
main results which have been obtained in this context.

\subsection{The ground-state fidelity and its susceptibility,  
  the quantum Fisher information}
\label{fidelity}

The fidelity constitutes a basic tool to analyze the variations of a
given quantum state in the Hilbert space.  In particular, the
ground-state fidelity quantifies the overlap between the ground states
of quantum systems sharing the same Hamiltonian, but associated with
different Hamiltonian parameters~\cite{AFOV-08, Gu-10, Braun-etal-18}.
Its relevance can be traced back to the Anderson's orthogonality
catastrophe~\cite{Anderson-67}: the overlap of two many-body ground
states corresponding to Hamiltonians differing by a small perturbation
vanishes in the thermodynamic limit. This paradigm is also realized in
many-body systems at QTs, however significantly different behaviors
emerge with respect to systems in normal conditions.  Besides that, as
outlined in Sec.~\ref{sec:QFI}, the fidelity susceptibility covers a
central role in quantum estimation theory, being proportional to the
quantum Fisher information.  The latter indeed quantifies the inverse
of the smallest variance in the estimation of the varying parameter,
such that, in proximity of QTs, metrological performances are believed
to drastically improve~\cite{ZPC-08, IKCP-08}.

Recently, there has been an intense theoretical activity focusing on
the behavior of the fidelity and of the corresponding susceptibility,
and more generally, of the geometric tensor~\cite{ZP-06, YLG-07, CZ-07},
at QTs (see, e.g., Ref.~\cite{Gu-10} for a review). In quantum many-body
systems, the establishment of a non-analytical behavior has been
exploited to evidence CQTs in several contexts, which have been deeply
scrutinized both analytically and numerically.  We quote, for example,
free-fermion models~\cite{CGZ-07, MPD-11, RD-11-tl, RD-11-xy,
  Damski-13, LZW-18}, interacting spin~\cite{CWHW-08, SAC-09, LLZ-09,
  AASC-10, Sirker-10, Nishiyama-13, SKV-15} and particle
models~\cite{BV-07, MHCFR-11, CMR-13, WLIMT-15, HWWW-16, Kettemann-16,
  SS-17}, as well as systems presenting peculiar
topological~\cite{YGSL-08, OS-14, KLS-16} and non-equilibrium
steady-state transitions~\cite{BGZ-14, MP-17}.  Recently, this issue
has been also investigated at FOQTs, showing that the fidelity and the
quantum Fisher information develop even sharper
behaviors~\cite{RV-18}.

\subsubsection{Setting of the problem}
\label{setting}

We define our setting by considering a $d$-dimensional quantum
many-body system of size $L^d$, with Hamiltonian
\begin{equation}
  \hat{H} = \hat{H}_{c} + w \hat{H}_{p} \,,
  \label{hdef}
\end{equation}
where $\hat{H}_c$ is a Hamiltonian at the QT point (i.e., its
parameters are tuned to their QT values). Moreover,
$[\hat{H}_c,\hat{H}_p]\neq 0$.~\footnote{We recall that if
  $[\hat{H}_c,\hat{H}_p]=0$, then the eigenstates of $\hat{H}$ do not
  depend on $w$. Therefore, when modifying $w$, the ground-state
  fidelity can only change in the presence of a level crossing
  involving the ground state.}  The tunable parameter $w$
drives the QT located at $w=0$. In the standard case of a CQT with
two relevant ($r$ and $h$) parameters, $w$ may be one of them and
$\hat{H}_p$ the corresponding perturbation (even or odd, respectively).
In particular, for the quantum Ising model in Eq.~\eqref{hisdef},
$\hat{H}_c$ can be identified with the critical Hamiltonian
at $g=g_c$ and $h=0$, while $w$ may correspond to the even
$r=g-g_c$ or the odd $h$ Hamiltonian parameters driving the CQT.
Alternatively, along the FOQT line, $\hat{H}_c$ may be the Ising
Hamiltonian for $g<g_c$ and $h=0$, while $w$ corresponds to the
parameter $h$ driving the FOQT.

We exploit the geometrical concept of the fidelity to monitor the changes
of the ground-state wave function $|\Psi_0(L,w)\rangle$ when varying
the control parameter $w$ by a small amount $\delta w$ around its
transition value.  Thus, we consider the fidelity between the ground
states associated with $w$ and $w+\delta w$,
\begin{equation}
  A(L,w,\delta w) \equiv \big| \langle \Psi_0(L,w) | \Psi_0(L,w+\delta w)
  \rangle \big| \,.
\label{fiddef}
\end{equation}
Assuming $\delta w$ sufficiently small, one can expand
Eq.~\eqref{fiddef} in powers of $\delta w$, as~\cite{Gu-10}
\begin{equation}
  A(L,w,\delta w) = 1 - \tfrac12 \delta w^2 \, \chi_A(L,w) + O(\delta w^3),
  \label{expfide}
\end{equation}
where $\chi_A$ defines the fidelity susceptibility.  As already noted
in Sec.~\ref{sec:fidelity}, the fidelity susceptibility is
proportional to the quantum Fisher information, defined in
Eq.~\eqref{eq:QFI}.  The cancellation of the linear term in the
expansion~\eqref{expfide} is related to the fact that the fidelity is
bounded ($A\le 1$).  The standard perturbation theory also allows us to
write $\chi_A$ as~\cite{Gu-10}
\begin{equation}
  \chi_A(L,w) = \sum_{n>0}
      {\big| \langle \Psi_n(L,w)| \hat H_p |\Psi_0(L,w)\rangle \big|^2 \over
        \big[ E_n(L,w) - E_0(L,w) \big]^2} \,,  \label{chifpert} 
\end{equation}
where $|\Psi_n(L,w)\rangle$ is the Hamiltonian eigenstate
corresponding to the eigenvalue $E_n(L,w)$.

In normal conditions, the fidelity susceptibility is expected to be
proportional to the volume, i.e.,
\begin{equation}
  \chi_A\sim L^d\,.
  \label{chianormal}
  \end{equation}
  At QTs, the leading behavior of $\chi_A$ can change: in CQTs it
  depends on their critical exponents, while at FOQT it develops an
  exponential divergence with increasing the size.  Therefore the
  divergence of the generally intensive quantity
\begin{equation}
  \widetilde{\chi}_A\equiv \chi_A / V \,, \qquad V=L^d\,,
  \label{tildechia}
\end{equation}
may be considered as a witness of a QT.

\subsubsection{Finite-size scaling at continuous and first-order
  quantum transitions}
\label{fssfisu}

The FSS framework provides a unified description of the behavior of
the fidelity susceptibility of finite-size systems, both at CQTs and at
FOQTs, which is particularly useful to distinguish them, and to obtain
correct interpretations of experimental and numerical results at
QTs. In particular, it allows us to determine its FSS behavior that
entails the expected power-law divergences associated with QTs.

As already discussed in Sec.~\ref{escalingcqt}, the FSS limit at CQTs
is generally obtained in the large-$L$ limit, keeping an appropriate
combination $\Phi$ of $w$ and $L$ fixed, i.e.,
\begin{equation}
  \Phi = w \, L^{y_w} \,,
  \label{lappaCQT}
\end{equation}
where $y_w$ is the RG dimension of $w$.  To derive the scaling
behavior of the fidelity $A(L,w,\delta w)$ and its susceptibility
$\chi_A(L,w)$, we assume that both $w$ and $w+\delta w$ are
sufficiently small to be in the transition region.  A natural
hypothesis for the scaling of the critical nonanalytic part of the
fidelity at the QT is given by~\cite{RV-18}~\footnote{Analogously to
the equilibrium free energy at CQTs (see Sec.~\ref{freeen}), we assume
that we can distinguish two different contributions: one of them is
related the standard analytical part that is also present far from the
QTs, and the other one is a nonanalytic term arising from the
critical modes at the transition point.}
\begin{equation}
  A(L,w,\delta w)_{\rm sing} \approx {\cal A}(\Phi,\delta\Phi) \,,
  \qquad \delta\Phi \equiv \delta w \, L^{y_w}\,.
  \label{fisca}
\end{equation}
The FSS of its susceptibility is then obtained from Eq.~\eqref{fisca},
by expanding ${\cal A}$ in powers of $\delta \Phi$, and matching it
with Eq.~\eqref{expfide}:
\begin{equation}
  \widetilde{\chi}_A = L^{-d} \, \chi_A(L,w) \approx L^{-d} \, \left(
            {\delta\Phi\over \delta w}\right)^2 \, {\cal A}_2(\Phi)\,.
\label{chifscal}
\end{equation}
This implies 
\begin{equation}
  \widetilde{\chi}_A(L,w) \approx L^{\zeta} {\cal A}_2(\Phi)\,, \qquad
  \zeta = 2y_w- d\,.
\label{cqtchil}
\end{equation}
On the basis of the above scaling ansatz, taking the thermodynamic
limit for $1\gg |w| > 0$, we also obtain
\begin{equation}
  \widetilde{\chi}_A \sim \lambda^{\zeta}\,,\qquad \lambda = |w|^{-1/y_w}\,,
  \label{chiathlim}
\end{equation}
where $\lambda$ plays the role of diverging length scale.  Of course,
the above scaling laws are in agreement with those obtained by other
scaling arguments reported in the literature (see, e.g.,
Refs.~\cite{Gu-10, Dutta-etal-book} and references therein).  Note that
the scaling behavior reported in Eq.~\eqref{cqtchil} should be
compared with the normal $O(1)$ contributions to $\widetilde{\chi}_A$,
arising from analytical behaviors, such as those far from the QT.
Therefore, the power law~\eqref{cqtchil} provides the leading
asymptotic behavior when $\zeta>0$, while it only gives rise to
subleading contributions when $\zeta<0$.

For example, let us consider the fidelity of the quantum Ising
models~\eqref{hisdef} associated with variations of the relevant
parameter $r=g-g_c$ and $h$.  According to the above scaling
arguments, one obtains the power laws
\begin{equation}
  \widetilde{\chi}_{A,r/h} \sim L^{\zeta_{r/h}}\,,\qquad
  \zeta_r = 2y_r-d =\frac{2}{\nu} - d\,,\qquad
  \zeta_h = 2 y_h-d = 2 + z -\eta,\,
\label{zetarh}
\end{equation}
respectively.  Therefore, inserting the values of the critical
exponents, see Sec.~\ref{cqtisi}, we find $\zeta_r=1$ and $\zeta_h
=11/4$ for 1$d$ Ising-like chains, $\zeta_r = 1.17575(2)$ and $\zeta_h
\approx 2.963702(2)$ for 2$d$ models (using the best estimates
reported in Table~\ref{isingres3d}), and $\zeta_r = 1$ and $\zeta_h =
3$ for 3$d$ models (where logarithms may also enter).

Several studies in the literature address the CQT of the 1$d$ quantum
Ising model, focussing on variations of the transverse field $g$
across the value $g=1$, while keeping the longitudinal field fixed at
$h~=~0$~\cite{ZP-06, CGZ-07, ZZL-08, GL-09, RD-11-tl, RD-11-xy,
  Damski-13, LZW-18, CWHW-08,DR-14}, reporting exact results.  In such
case, the FSS is controlled by the scaling variable $\Phi = r L^{y_r}$
with $r=g-1$ and $y_r=1$.  The results of these studies confirm the
power law $\widetilde{\chi}_A\sim L$, as predicted by
Eq.~\eqref{cqtchil}.  Indeed, at the critical point $r=0$ of the $1d$
Ising chain with PBC, the fidelity susceptibility with respect to
variations of $r$ is exactly given by~\cite{Damski-13}
\begin{equation}
  \widetilde{\chi}_A(L,r=0) = {L - 1\over 32}\,,\qquad {\rm thus}\quad
  {\cal A}_2(0) = \frac{1}{32}\,.
 \label{chiar0isi}
\end{equation}
On the other hand, in the normal disordered and ordered phases,
$\widetilde{\chi}_A(L,r) = O(1)$ in the large-size limit~\cite{Damski-13}.

The fidelity behavior has been also discussed at BKT transitions of
1$d$ quantum models~\cite{Yang-07, SKV-15, CRDZ-19}, where the length
scale diverges exponentially when approaching the critical point, as
$\xi\sim \exp(c/\sqrt{w-w_c})$~\cite{KT-73, Kosterlitz-74}.  At BKT
transitions, since the RG dimension $y_w$ formally corresponding to
the exponential behavior of the correlation length is $y_w=0$ (apart
from logarithms), the nonanalyticity of the fidelity susceptibility
does not provide the leading behavior, which remains
$\widetilde{\chi}_A= O(1)$ analogous to that far from the
QTs. Therefore, the fidelity susceptibility can be hardly considered
as an effective witness of BKT transitions.

The FSS of the fidelity can be extended to systems at FOQTs, where the
type of divergence is controlled by the closure of the gap
$\Delta(L,w)$ between the lowest energy levels, being exponential or
power law when increasing the system size, depending on the BC.  For
this purpose, we introduce a scaling variable $\Phi$ analogous to the
one reported in Eq.~\eqref{kappah}, which reduces to
Eq.~\eqref{lappaCQT} in the case of CQTs, as shown at the end of
Sec.~\ref{fssfoqt}. Therefore we define
\begin{equation}
  \Phi \equiv {\delta E_w(L,w) \over \Delta(L)} \,,\qquad \Delta(L) =
  \Delta(L,w=0)\,,
  \label{phideffo}
\end{equation}
where $\delta E_w(L,w)= E(L,w)-E(L,0)$ is the variation of energy
associated with changes of the parameter $w$.  For example, $w$ may
correspond to the longitudinal field $h$ at the FOQT line of quantum
Ising systems~\footnote{Note that here we are assuming that the
minimum of $\Delta(L,w)$ occurs at $w=h=0$, as it is the case of
Ising-like systems with PBC, OBC, ABC and OFBC.}.  We recall that, for
quantum Ising systems, the correspondence $w=h$, $\delta E_w(L,w) = 2
m_0 w L^d$ is assumed.~\footnote{One may extend the derivation to Ising-like
systems with EFBC (see Sec.~\ref{fssfebc}) by appropriate
redefinitions of $\Phi$ (details can be found in Ref.~\cite{RV-18}).}
Using scaling arguments analogous to those employed at CQTs, one
obtains the FSS behavior
\begin{equation}
  A(L,w,\delta w) \approx {\cal A}(\Phi,\delta\Phi) \,,
  \qquad
  \delta\Phi\equiv {\delta E_w(L,\delta_w) \over \Delta(L)}\,.  
  \label{fiscafo}
\end{equation}
The same scaling of Eq.~\eqref{chifscal} holds for the susceptibility.
However note that its prefactor $(\delta\Phi/\delta w)^2$ leads to
exponential laws for systems at FOQTs when the BC do not favor any of
the two phases separated by the transition, for which $\Delta(L)\sim
\exp(-c L^d)$.  In particular, for quantum Ising systems one obtains
\begin{equation}
  \widetilde{\chi}_A(L,w)  \approx {L^{d}\over \Delta(L)^2}
  \, {\cal A}_2(\Phi)\,.
  \label{chiafoqt}
\end{equation}
Therefore the asymptotic size dependence of the fidelity
susceptibility, or equivalently the quantum Fisher information,
significantly distinguishes CQTs from FOQTs, respectively
characterized by power laws and exponentials when boundaries are
neutral, such as the case of PBC and OBC in quantum Ising systems.
Moreover, as discussed in Sec.~\ref{fsspbcobc}, FOQTs in systems with
neutral boundaries can be effectively described exploiting a two-level
framework given by the Hamiltonian~\eqref{hrtds}. This allows one to
also compute the scaling function ${\cal A}_2(\Phi)$ as~\cite{RV-18}
\begin{equation}
  {\cal A}_{2}(\Phi) = {1\over 4 (1 + \Phi^2)^2} \,,
  \label{f2l}
\end{equation}
which is expected to hold for any FOQT effectively described by the
two-level framework, and therefore for $d$-dimensional quantum Ising
models with neutral BC along their FOQT lines.

\begin{figure}
  \begin{center}
    \includegraphics[width=0.47\columnwidth]{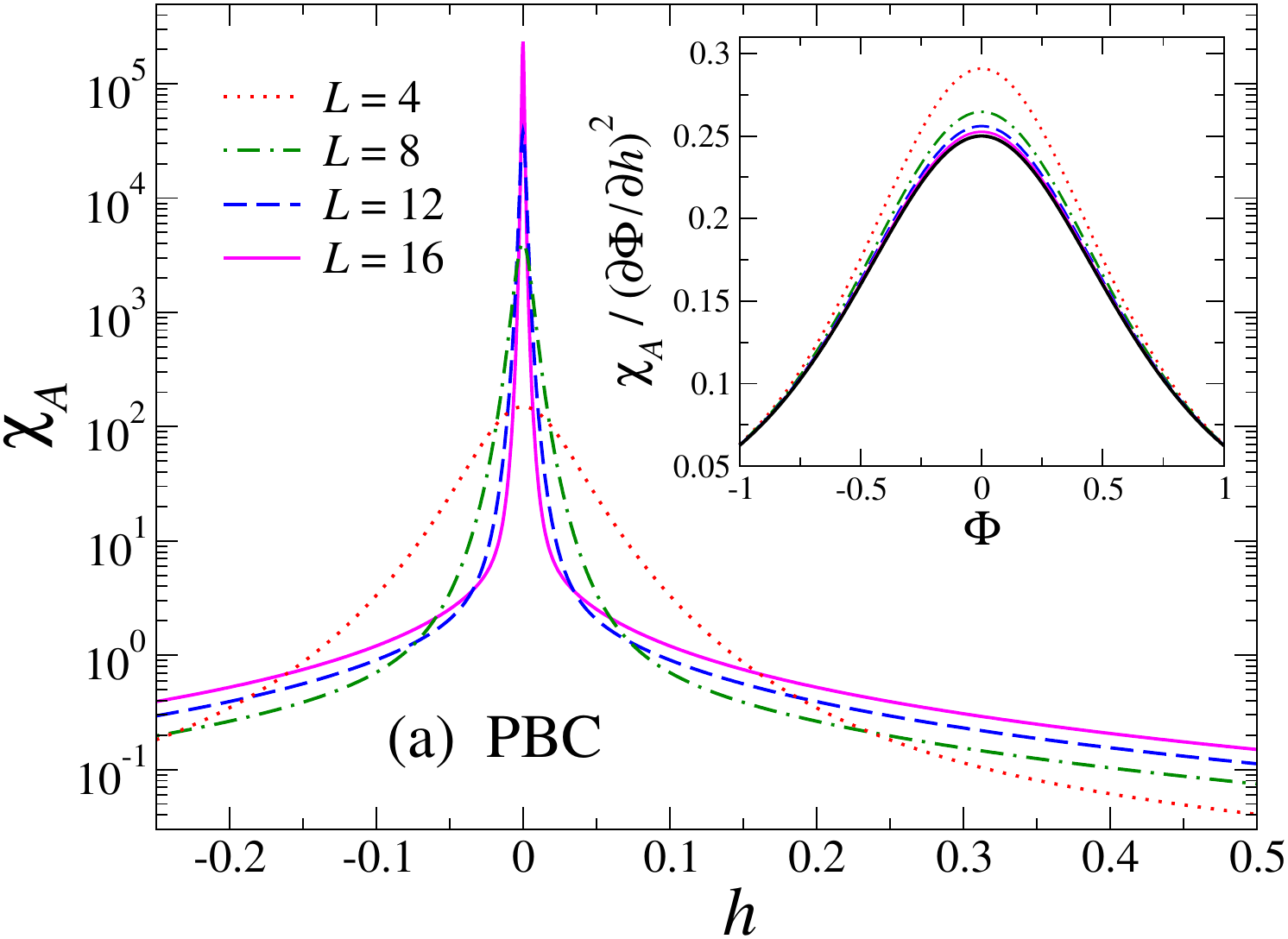}
    \hspace*{5mm}
    \includegraphics[width=0.47\columnwidth]{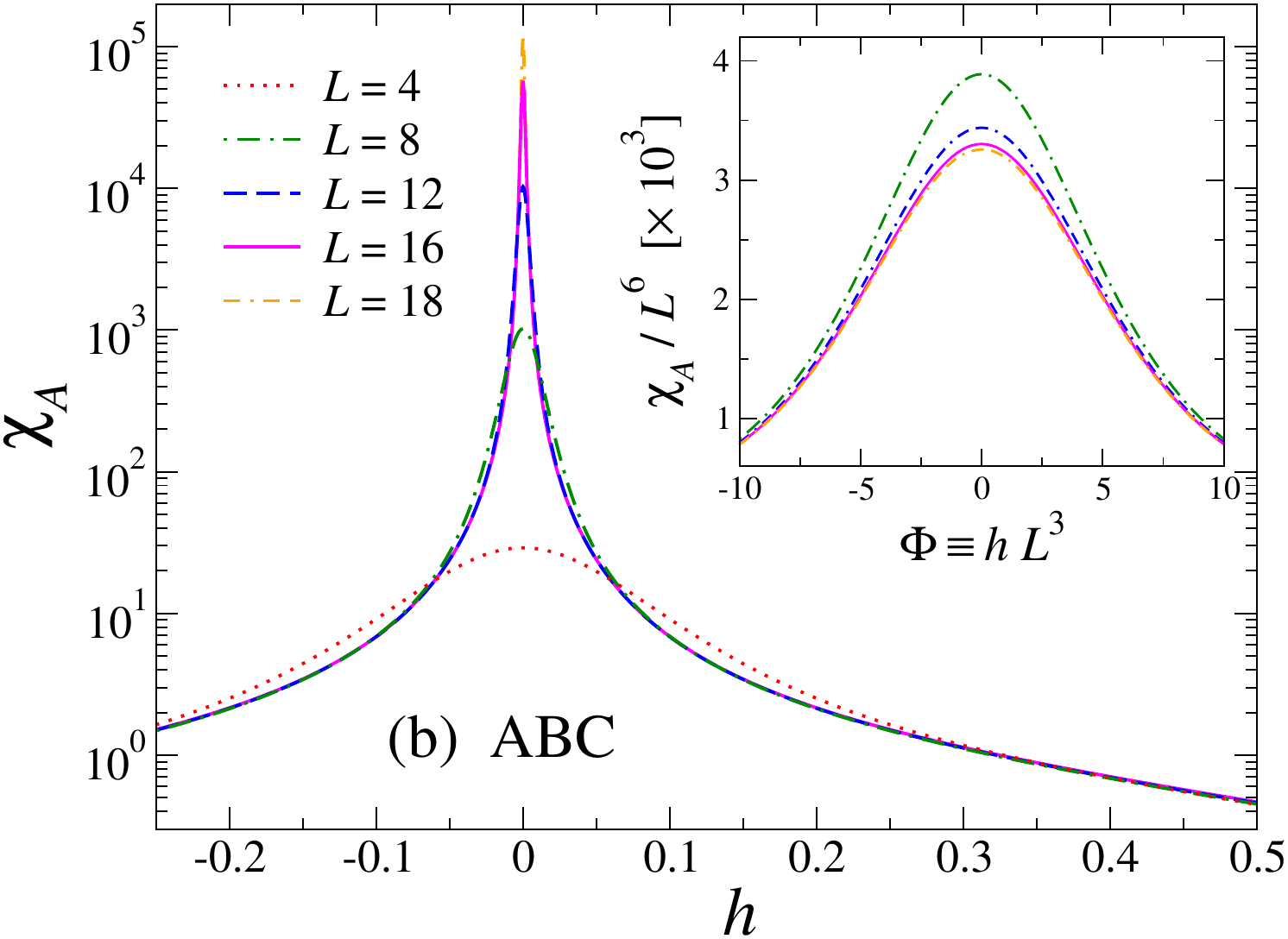}
    \caption{Fidelity susceptibility $\chi_A(L,h)$ for the quantum
      Ising ring~\eqref{hedef} at fixed $g$, associated with changes
      of the longitudinal parameter $h$, for some values of $L$.
      Panel $(a)$ is for $g=0.9$ and $\zeta=1$ (PBC), for which one
      obtains the scaling variable $\Phi = 2 m_0 h L /
      \Delta(L)$.  Panel $(b)$ is for $g=0.5$ and $\zeta=-1$ (ABC),
      for which the scaling variable $\Phi = h L^3$ can be used.
      The two insets display curves for $\chi_A/( \partial \Phi /
      \partial h)^2$, which clearly approach a scaling function of
      $\Phi$.  Note that curves in the inset of panel $(a)$ converge
      to the scaling function ${\cal A}_2(\Phi)$ (thick black line),
      cf.~Eq.~\eqref{f2l}.  Analogous results can be obtained for
      other values of $g<1$.  Adapted from Ref.~\cite{RV-18}.}
    \label{FidSusc_PBC_g09}
  \end{center}
\end{figure}

The scaling scenario of the fidelity and its susceptibility has been
confirmed by analytical and numerical results for the $1d$ quantum
Ising model across its FOQTs~\cite{RV-18}.
Figure~\ref{FidSusc_PBC_g09}, panel $(a)$, shows the fidelity
susceptibility across the FOQT of systems with PBC, obtained by means
of exact diagonalization and Lanczos algorithms.  In this case, using
the above correspondence, the scaling variable of Eq.~\eqref{phideffo}
is given by $\Phi = 2 m_0 h L / \Delta(L)$.  The results nicely
confirm the FSS at FOQTs reported in Eq.~\eqref{chiafoqt} and the
scaling function~\eqref{f2l} computed using the corresponding
two-level model~\eqref{hrtds}.

It is worth noting that the fidelity at FOQTs may also obey power-law
behaviors when the BC give rise to domain walls in the ground state,
such as ABC and OFBC. As discussed in Sec.~\ref{fssabcfobc}, in such
cases the gap behaves as $\Delta(L)\sim L^{-2}$, thus the scaling
variable of Eq.~\eqref{phideffo} is expected to scale as $\Phi \approx
h L^3$ and the scaling~\eqref{chiafoqt} of the fidelity susceptibility
predicts that $\chi_A\sim L^{6}$.  This implies
\begin{equation}
  \widetilde{\chi}_A\sim L^5\,,
  \label{chiaABC}
\end{equation}
which is anyway a quite large power-law divergence.  Expectations are
well confirmed by the numerical results shown in
Fig.~\ref{FidSusc_PBC_g09}, panel $(b)$.  In such case, the scaling
function is not the one reported in Eq.~\eqref{f2l}, since the
two-level framework does not apply (see
Sec.~\ref{fssabcfobc}).~\footnote{Other results for the case in which
the BC globally favor one of the phase, i.e., EFBC as discussed in
Sec.~\ref{fssfebc}, can be found in Ref.~\cite{RV-18}, showing in
particular that the peak of the fidelity susceptibility does not
strictly occur at $h=0$, but it is slightly shifted, being $O(1/L)$.}

We finally mention that the above FSS framework, both for CQTs and for
FOQTs, can be generalized to a finite
temperature $T$~\cite{Gu-10, Dutta-etal-book}.  In such case,
the quantum system is described by the density matrix
\begin{equation} 
  \rho_w\equiv \rho(L,T,w) = Z^{-1} \sum_n e^{-E_n/ k_B T} |
  \Psi_n\rangle \langle \Psi_n |\,,\qquad
  Z= \sum_n \langle \Psi_n | e^{-E_n / k_B T} | \Psi_n\rangle\,.
\end{equation}
One then needs Eq.~\eqref{eq:fid3} for the fidelity between mixed
states,
\begin{equation}
  A(L,T,w,\delta w) = {\rm Tr} \left[ \sqrt{\sqrt{\rho_w} \,
      \rho_{w+\delta w} \, \sqrt{\rho_w}} \right]\,,
  \label{altw}
\end{equation}
which reduces to
Eq.~\eqref{fiddef} for $T\to 0$.  The corresponding fidelity
susceptibility can be extracted analogously to Eq.~\eqref{expfide}.
At a QT, the $T=0$ scaling~\eqref{fisca} can be straightforwardly
extended to keep into account the temperature, by adding a further
scaling variable $\Xi = T/\Delta(L)$, cf.~Eq.~\eqref{kappat}.  For
example, at CQTs, the corresponding susceptibility should
asymptotically scale as
\begin{equation}
  \widetilde{\chi}_A(L,T,w) \approx L^{\zeta} \, {\cal A}_2(\Xi,\Phi)\,,
  \label{cqtchilT}
\end{equation}
when $\zeta>0$, cf.~Eq.~\eqref{cqtchil}.

\subsection{The bipartite entanglement entropy}
\label{entang}

In a quantum system, the reduced density matrices of subsystems, and,
in particular, the corresponding entanglement entropies and spectra,
provide effective probes of the nature of the quantum critical
behavior (see, e.g., Refs.~\cite{KP-06, AFOV-08, CCD-09, ECP-10,
  LW-06}).  Their dependence on the finite size of the system may be
exploited to determine the critical parameters of a QT~\cite{IL-08,
  MA-10, XA-11, XA-12, DLLS-12, DLS-13}.  The spatial entanglement of
systems near their QCP can be quantified by computing von Neumann or
R\'enyi entanglement entropies of the reduced density matrix of a
subsystem.  They generally satisfy an area law, with some notable
exceptions that present logarithmic corrections, such as free Fermi
gases in arbitrary dimensions~\cite{Wolf-06, GK-06, ECP-10, CMV-11,
  CMV-12, NV-13}, and 1$d$ many-body systems at CQTs described by 2$d$
CFT~\cite{HLW-94, VLRK-03, CC-04, JK-04, CCD-09}.  Recently, a
protocol to measure the second-order R{\'e}nyi entanglement entropy
$S_2$ has been also experimentally implemented in a trapped-ion
quantum simulator of spin chains~\cite{Brydges-etal-19}.

\subsubsection{Scaling behavior in
  one-dimensional continuous quantum transitions}
\label{d1systems}

We now discuss the scaling behavior of bipartite entanglement
entropies of 1$d$ systems with a QCP characterized by $z=1$, thus
realizing a CFT in two dimensions with central charge $c$.  For
example, one can consider $1d$ Ising-like quantum systems at their CQT for
$g=g_c=1$, whose corresponding central charge is $c=1/2$.  The lattice
system can be divided into two connected parts $\tt{A}$ and $\tt{B}$
of length $\ell_{\tt A}$ and $\ell_{\tt B} = L-\ell_{\tt A}$.  The
reduced density matrix of subsystem $\tt{A}$ is given by $\rho_{\tt A}
= {\rm Tr}_{\tt B} \, \rho$, where $\rho$ is the density matrix of the
ground state, and its corresponding von Neumann entropy is defined as
\begin{equation}
  S(L,\ell_{\tt A};r,h) = S(L,L-\ell_{\tt A};r,h)
  = -{\rm Tr}\,[ \rho_{\tt A} \log \rho_{\tt A}]\,.
  \label{vNen}
\end{equation}
The asymptotic behavior of bipartite entanglement entropies is known
at the critical point $r=h=0$~\cite{HLW-94, VLRK-03, CC-04, JK-04}:
\begin{equation}
  S(L,\ell_{\tt A};r=0,h=0)\approx q\, {c \over 6} \Big[ \log L 
    + \log \sin( \pi \ell_{\tt A}/L) + e_l \Big] \,,
  \label{ccfo}
\end{equation}
where $c$ is the central charge, $q$ counts the number of boundaries
between the two parts of the system (thus $q = 2$ for PBC, while $q=1$
for OBC), and $\log$ denotes the natural logarithm.  The constant
$e_l$ is nonuniversal and depends on the BC~\cite{CC-04, JK-04,
  IJ-08}.  The asymptotic behavior of the bipartite entanglement
entropies is also known in the thermodynamic limit close to the
transition point~\cite{CC-04, CCP-10}, i.e., for $L,\ell_A\ll \xi$,
where $\xi$ is the length scale of the critical modes, such as that
defined in Eq.~\eqref{xidef}.  One obtains~\cite{CC-04, FIK-07,
  CCP-10}
\begin{equation}
  S(L,\ell_{\tt A};r,h=0) \approx q \, {c \over 6}\log \xi + e_t\,,
  \qquad \xi\ll \ell_A,L\,,
  \label{salthl}
\end{equation}
where, again, $q = 2$ for PBC and $q=1$ for OBC, and $e_t$ is a
nonuniversal constant.

We now consider a critical system around its CQT point and consider
the dependence on the relevant parameters $r$ and $h$ associated with
the even and odd perturbations at the CQT, such as $r=g-1$ and $h$
entering the Hamiltonian~\eqref{hisdef} for the 1$d$ Ising chain.  In
the general FSS regime, using the notations of Sec.~\ref{freeen}, the
bipartite entanglement entropy has been conjectured to satisfy the
asymptotic FSS~\cite{CC-04, CC-09, CPV-14, WGK-17}
\begin{equation}
  S(L,\ell_{\tt A};r,h) -
  S(L,\ell_{\tt A};r=0,h=0) \approx \Sigma(\ell_{\tt A}/L, L^{y_r} r, L^{y_h}
  h)\,,
\label{asentsca}
\end{equation}
so that $\Sigma(x,0,0)=0$.  The corrections to this asymptotic
behavior may have various origins~\cite{CC-10, CCP-10, CE-10, CCEN-10,
  FC-11, XA-12, EEFR-12, DLF-12, CPV-14}.  Beside the FSS corrections
discussed in the previous sections, such as the $O(L^{-\omega})$
corrections due to the leading irrelevant operator, the $O(L^{-1})$
corrections due to the presence of boundaries, and the $O(L^{-1/\nu})$
corrections arising from the $O(r^2)$ terms of the scaling field
$u_r\approx r$, there are additional corrections. They are related to
the operators associated with the conical singularities at the
boundaries between the two parts~\cite{CC-10, CCP-10}. In the limit
$L,\ell_{\tt A}\to \infty$ at fixed $\ell_{\tt A}/L$, these new
operators give rise to terms of order $L^{-y_c}$ in the case of
OBC~\cite{CC-10, CCP-10} and of order $L^{-2 y_c}$ in the case of
PBC~\cite{CCEN-10, FC-11}. Here $y_c>0$ is the RG
dimension~\cite{CC-10, CCP-10} of the leading conical operator.  This
is conjectured to be essentially related to the energy
operator~\cite{CCEN-10, FC-11, XA-12, DLF-12}, hence $y_c = y_r$.
Moreover, the analysis of exactly solvable models shows the presence
of other corrections suppressed by integer powers of
$L$~\cite{CCEN-10}. Analogous results apply to the $n$-index R\'enyi
entropies, with the only difference that conical singularities lead to
$O(L^{-y_c/n})$ corrections in the case of OBC~\cite{CC-10, CCP-10}
and of order $L^{-2 y_c/n}$ in the case of PBC~\cite{CCEN-10, FC-11}.

A thorough analysis of the FSS behavior of the von Neumann and R\'enyi
entanglement entropies for the XY chain, confirming its asymptotic
behavior~\eqref{asentsca} with the predicted corrections, is reported
in Ref.~\cite{CPV-14}.

We finally mention that a number of works has also focused on the
entanglement spectrum, that is, the spectrum of the reduced
density matrix $\rho_{\tt A}$ of subsystem ${\tt A}$.
In particular, the behavior of the entanglement gap has been addressed
in proximity of QCPs of various models (see, e.g.,
Refs.~\cite{CL-08, PE-09, DLLS-12, DLS-13, CKS-14, DT-20, WAA-20}).

\subsubsection{Bipartite entanglement entropy in higher dimensions}
\label{d1systemshd}

In higher $d>1$ spatial dimensions, quantum spins or bosons develop an
entanglement entropy that scales as the boundary of the
bipartition~\cite{BKLS-86, Srednicki-93, AFOV-08, ECP-10}.  Some of the
subleading contributions to the area law provide important sources of
information for non-trivial quantum correlations, such as the
topological entanglement entropy in a gapped spin-liquid
phase~\cite{HIZ-05a, HIZ-05b, KP-06, LW-06}. At a QCP, subleading
terms contain information that identify the universality class (see,
e.g., Refs.~\cite{FrM-06, CH-07, MFS-09, SLRL-11, KHMS-11, MS-12,
  SMO-12, KPSS-12, IM-13, KHSM-13, Grover-14, KHIM-15, GK-15, WWS-17}).

The leading contribution to $S_{\tt A}$, associated with a bipartition of
2$d$ systems of bosonic particle or spin systems, displays the area law
\begin{equation}
  S_{\tt A} \approx b \,L_{\tt A} \,,
  \label{arealaw2d}
\end{equation}
where $L_{\tt A}$ is the linear size of the boundaries between the two
partitions, physically implying that the entanglement is local at the
boundary of the partitions even at the critical
point~\footnote{Notable exceptions to a strict area law are Fermi
gases in arbitrary dimensions~\cite{Wolf-06, GK-06, ECP-10, CMV-11,
  CMV-12, NV-13}, which present a multiplicative logarithmic
contribution.}.  The coefficient $b$ entering the area
law~\eqref{arealaw2d} is sensitive to the short-distance cut-off, and
is thus non-universal.  Therefore, unlike 1$d$ cases, the leading
behavior of the entanglement entropy in higher dimensions cannot be
used to characterize the critical behavior.  On the other hand, the
subleading $O(1)$, or equivalently the leading part of the subtracted
entanglement entropy
\begin{equation}
  \Delta S_{\tt A} \equiv S_{\tt A} - b\, L_{\tt A}\,,
  \label{deltasa}
\end{equation}
is expected to be universal in 2$d$ CQTs, thus providing information
on the nature of the transition.  Note that logarithmic behaviors of
$\Delta S_{\tt A}$ may arise from corner
contributions~\cite{MG-11, HSCWM-16}, while they are expected to be
absent for smooth boundaries between the partitions, where
\begin{equation}
  \Delta S_{\tt A}\approx \gamma \,,
  \label{deltasga}
\end{equation}
$\gamma$ being a universal constant.  Computations at the
Wilson-Fisher fixed point of O($N$) symmetry models,
cf.~Eq.~\eqref{HON}, are reported in Refs.~\cite{WWS-17, MFS-09}, for
various geometries, including those with sharp corners.  Concerning
the scaling properties of the bipartite entanglement entropy $S_{\tt A}$
around the QCP, we expect a behavior
analogous to that reported in Eq.~\eqref{asentsca}.

A strongly interacting quantum many-body system at zero temperature
can exhibit exotic order beyond the LGW paradigm, dubbed topological
order, whose defining property is that the ground-state degeneracy
depends on the topology of the space.  A class of models exhibiting
these features is provided by 2$d$ spin-liquid models.  Within these
models, the concept of topological entanglement entropy has been
proposed~\cite{LW-06, KP-06}, being proportional to the constant
$\gamma$ entering Eq.~\eqref{deltasga}, somehow replacing the role of
the order parameter in a topologically ordered system. Based on the
idea that the spin-liquid state is a type of collective paramagnet,
the topological entanglement entropy probes long-distance correlations
in the ground-state wave function that are not manifest as conventional
long-range order. Some computations supporting the above ideas can be
found, e.g., in Refs.~\cite{HIZ-05a, HIZ-05b} within the paradigmatic
toric code~\cite{Kitaev-03}, a quantum dimer model on the triangular
lattice~\cite{FM-07}, a BH on the Kagome lattice~\cite{IHM-11}.

\subsection{The concurrence between spins}
\label{concuspins}

Another aspect of the entanglement properties at CQTs is encoded into
the entanglement between two spins in quantum spin
models~\cite{AFOV-08}, such as the XY chain~\eqref{XYchain}.
In fact, the entanglement between two spin-1/2 systems has been
investigated within the XY chain~\cite{ON-02, OAFF-02}.
This can be quantified through the concurrence,
see Sec.~\ref{concurrencedef}.  The two-spin concurrence
turns out to vanish, unless the two sites are at most next-nearest
neighbors, therefore it is somehow related to 
short-range quantum correlations. However, the concurrence
$C_{\rm nn}$ between nearest-neighbor spins presents a critical behavior
at the critical point $g_c=1$: its derivative with respect to the
transverse-field parameter $g$ diverges logarithmically
in the thermodynamic limit~\cite{OAFF-02},
\begin{equation}
  \partial_g C_{\rm nn} \approx {8\over 3\pi^2} \log |g-g_c|\,.
  \label{dcnnbeh}
\end{equation}
Thus, from the scaling point of view, the concurrence $C_{\rm nn}$
behaves as the first derivative of the free energy with respect to the
relevant parameter $g$ (see, e.g., Ref.~\cite{CPV-14}), whose further
derivative diverges logarithmically at the critical point
(corresponding to the logarithmic divergence of the specific heat in
the corresponding classical 2$d$ Ising model).  A similar singular
behavior is found also in the quantum discord of two close
spins and other related quantities~\cite{Dillenschneider-08, Sarandy-09,
  TRHA-11, MCSS-12, Dutta-etal-book, HOG-14, CWSS-16}.
Results for other models have been also reported
(see, e.g., Refs.~\cite{GLL-03, GBF-03, LEB-04, RQJ-05, LL-11, MMJ-17},
the review article~\cite{AFOV-08} and references therein). We mention
that both the two-spin concurrence and the quantum discord are
discontinuous at FOQTs~\cite{GLL-03, GBF-03, Sarandy-09}.

\section{Out-of-equilibrium dynamics at continuous quantum transitions}
\label{dynqts}

The recent progress that has been achieved in the control and
manipulation of complex systems at the nano scale has enabled a wealth
of possibilities to address the unitary quantum evolution of many-body
objects~\cite{Bloch-08, GAN-14}.  These range from the (nearly)
adiabatic dynamics induced by a slow change in time of one of the
control parameters, to the deep out-of-equilibrium dynamics following
an abrupt quench.

The unitary dynamics across QTs can be induced by variations of
the Hamiltonian parameters, which generally give rise to significant
departures from adiabatic passages through equilibrium states.  At
both CQTs and FOQTs, a dynamic scaling behavior emerges when the
out-of-equilibrium dynamics arises from sudden quenches of the
Hamiltonian parameters, provided they remain within the critical
regime, i.e., sufficiently close to the QT point~\cite{PRV-18-dfss}.
It is worth stressing that at QTs the system is inevitably driven out
of equilibrium, even when the time dependence is very slow, because
large-scale modes are unable to equilibrate as the system changes
phase.  Off-equilibrium phenomena, as for example hysteresis and
coarsening, Kibble-Zurek defect production, aging, etc., have been
addressed in a variety of contexts, both experimentally and
theoretically, at classical and quantum phase transitions (see, e.g.,
Refs.~\cite{HH-77, Binder-87, Bray-94, CG-05, FM-06, BDS-06, WNSBD-08,
  Dziarmaga-10, PSSV-11, Ulm-etal-13, Pyka-etal-13, LDSDF-13, NGSH-15,
  Biroli-16} and references therein).  These studies have shown that
the time-dependent properties of systems evolving under such dynamics
obey out-of-equilibrium scaling behaviors, depending on the universal
static and dynamic exponents of the equilibrium
QT~\cite{Polkovnikov-05, ZDZ-05, Dziarmaga-05, DGP-10,
  GZHF-10, CEGS-12, PV-17-prl, PRV-18-loc, PRV-18-dfss}.

The so-called {\it quantum quench} represents one of the simplest
protocols in which a system can be naturally put in out-of-equilibrium
conditions~\cite{GMEHB-02, KWW-06, HLFSS-07, Trotzky-etal-12,
  Cheneau-etal-12, Gring-etal-12}, by suddenly changing a Hamiltonian
parameter.  Several interesting issues have been investigated in this
context.  They include the long-time relaxation and the consequent
spreading of quantum correlations and entanglement, the statistics of
the work, localization effects due to the mutual interplay of
interactions and disorder, dynamical phase transitions, the dynamic
scaling close to QTs, effects of dissipation or of measurements due to
interactions with an environment, to mention some of the most
representative ones (see, e.g., Refs.~\cite{Niemeijer-67, BMD-70,
  SPS-04, DMCF-06, SHLVS-06, RDYO-07, RDO-08, ZPP-08, PZ-09, IR-11,
  RI-11, GS-12, CEF-12a, CEF-12b, BRI-12, CE-13, HPK-13, FCEC-14,
  FHS-14, CTGM-16, CC-16, BD-16, NRVH-17, PRV-18-dfss, NRV-19-wo,
  NRV-19-cd, STT-20, RV-20-qu, KWW-06, HLFSS-07, CFFFI-09,
  Trotzky-etal-12, Cheneau-etal-12, Gring-etal-12, Schreiber-etal-15,
  Braun-etal-15, PCV-15, Kaufman-etal-16, Smith-etal-16, BLSKB-17,
  Zhang-etal-17, TNDTT-17, MCMJA-18, JJLM-19, Kohlert-etal-19,
  Maier-etal-19, CLSV-20, RCFG-21}).
All of them are eventually devoted to characterize the highly nonlinear
response of the system to the drive, where nonequilibrium fluctuation
relations may play a pivotal role~\cite{EHM-09, Jarzynski-11, CHT-11,
  Seifert-12}.

Several aspects related to the dynamics, and the asymptotic states,
arising from quantum quenches have been intensely investigated and
collected in a number of review articles, such as Refs.~\cite{NH-15,
  CC-16, VR-16, EF-16, LGS-16}.  The asymptotic behavior of the quench
dynamics in closed systems has been widely debated, in particular
whether the large-time stationary states arising from unitary dynamics
effectively behave as thermal states, thus leading to thermalization,
and under which conditions.  While the full density matrix can never
show thermalization under unitary time evolution, the reduced density
matrix related to sufficiently small subsystems can look effectively
thermal, with the remaining (integrated out) system acting like an
effective reservoir for it. The {\em eigenstate thermalization
  hypothesis} (ETH) has been argued to represent a key point to
understand thermalization, assuming that the relevant energy
eigenstates of a many-body Hamiltonian appear as thermal, i.e., that
they are practically indistinguishable from equilibrium thermal
states, as long as one focuses on macroscopic
observables~\cite{Deutsch-91, Srednicki-94, HZB-95, Tasaki-98, RDO-08,
  BKL-10, SR-10, RS-12, KIH-14, BMH-14, MFSR-16, Tang-etal-18,
  BLGR-20}.  Other ideas have been also proposed to characterize the
thermalization process in closed quantum systems, such as
typicality~\cite{Tasaki-16, PSW-06, Reimann-07} and quantum
correlation effects~\cite{GME-11, KH-13, FBC-17}.

Most quantum systems are expected to thermalize under a quench
associated with an extensive energy interchange, however there are
some notable exceptions, such as integrable systems~\cite{RDYO-07,
  Cazalilla-06, CEF-12a, CEF-12b, CNGS-13} (see, e.g.,
Refs.~\cite{Dziarmaga-10, EF-16, VR-16, IMPZ-16} for reviews on
quantum quenches in integrable spin chains and quadratic Hamiltonians,
also discussing the generalized Gibbs ensembles characterizing their
asymptotic states), and systems showing many-body
localization~\cite{BAA-06, OH-07, PH-10, SPA-14, VHA-15, KLSH-17,
  Agarwal-etal-17} (see Refs.~\cite{NH-15, AV-15, VM-16, AP-17, AL-18,
  AABS-19,Mitra-18} for reviews). We also mention the review
article~\cite{CC-16} focussing on quantum quenches in (1+1)-dimensional
CFTs~\cite{CC-06,CC-07a,CC-07b}. Another related and interesting issue
concerns the evidence of peculiar dynamical phase transitions and
prethermalization after quenches of O($N$) and other models~\cite{MMGS-13,
  SKDS-15, MCMG-15, CTGM-15, CTGM-16, CGDM-17, AF-17, LZMGS-19, JYS-19}.
We will not pursue the above-mentioned issues further, for which
we refer to the literature cited therein.

To stay within the focus of this review, in the following we discuss
issues closely related to the presence of CQTs (FOQTs are addressed in
Sec.~\ref{foqtdynamics}).  We primarily consider sudden soft quenches
probing the critical modes associated with equilibrium QTs, thus
involving sufficiently low energies to excite only long-range critical
fluctuations.  We characterize the scaling behaviors of the
out-of-equilibrium unitary dynamics at QTs arising from soft quenches.
As an exception, in Sec.~\ref{hardquenches} we also discuss hard
quenches to QCPs, reporting results that somehow signal the presence
of QCP, in particular in integrable systems, even though the energy
associated with the quench is relatively large.  The
out-of-equilibrium dynamics arising from slow changes of the
Hamiltonian parameters across QTs will be discussed in the forthcoming
section~\ref{KZdynamics}.

In the following we only consider homogeneous global quenches.
We mention that also local quenches have been studied in the
literature (see, e.g., Refs.~\cite{EP-07, CC-07b, EKPP-08, SD-11,
  Cardy-11}), for example to address issues related to the spread of
the quantum information after a sudden local change of the
system~\cite{BTC-18, BC-18, PBC-21}.

\subsection{Quench protocols}
\label{timedepproto}

A dynamic quench protocol is generally performed within a family of
Hamiltonians, that are written as the sum of two noncommuting terms:
\begin{equation}
  \hat{H}(w) = \hat{H}_{c} + w \hat{H}_{p} \,,
  \label{hlam}
\end{equation}
where $\hat{H}_c$ and $\hat{H}_p$ are independent of the parameter
$w$, and $[\hat{H}_c,\hat{H}_p]\neq 0$.~\footnote{In the case
$[\hat{H}_c,\hat{H}_p]=0$, the eigenstates of $\hat{H}$ do not depend
on $w$, therefore the time evolution can be written as a superposition
of single-eigenstate evolutions, making the problem relatively simple
and less interesting.}  The tunable parameter $w$ enables to modify
the strength of the {\em perturbation} $\hat{H}_{p}$, e.g., a
magnetic-field term in a system of interacting spins, with respect to
the {\em unperturbed} Hamiltonian $\hat{H}_c$. In the following we
assume that $\hat{H}_c$ represents a {\em critical} Hamiltonian, such
as the quantum Ising model~\eqref{hisdef} at $g=g_c$ and $h=0$, while
$w$ is a relevant parameter driving the QT. Thus, $w$ may correspond
to the transverse (t) field $r=g-g_c$ or the longitudinal
(l) field $h$, associated with the transverse and longitudinal spin
terms $\hat{H}_{p,{\rm t}}=-\sum_{\bm x} \hat \sigma_{\bm x}^{(3)}$
and $\hat{H}_{p,{\rm l}}=-\sum_{\bm x} \hat \sigma_{\bm x}^{(1)}$,
respectively. The critical point corresponds to $w=w_c=0$.

A quantum quench protocol is performed as follows:

\begin{itemize}
\item[$\bullet$] The system is prepared in the ground state of the
  Hamiltonian~\eqref{hlam} associated with an initial value $w_i$,
  that is, at $t=0$ the system starts from $|\Psi(t=0)\rangle \equiv
  |\Psi_0(w_i)\rangle$.  Alternatively, one may consider an initial
  state at temperature $T$, described by the Gibbs distribution
  $\rho_0(w_i) \propto \exp[-\hat H(w_i)/T]$.
  
\item[$\bullet$] Then the Hamiltonian parameter is suddenly changed to $w
  \neq w_i$.  The resulting dynamic problem corresponds to that of the
  unitary quantum evolution driven by the Hamiltonian $\hat{H}(w)$, i.e.,
  \begin{equation}
    |\Psi(t)\rangle = e^{-i \hat{H}(w) t}|\Psi(0)\rangle \,, \qquad
    \qquad |\Psi(0)\rangle \equiv |\Psi_0(w_i)\rangle \,.
    \label{afterque}
  \end{equation}
\end{itemize}

The out-of-equilibrium evolution of the system can be investigated by
monitoring observables and correlations obtained by taking expectation
values at fixed time.  For example, in the case of lattice spin
models, one may consider the longitudinal magnetization, the two-point
function of local operators related to the order parameter, etc.  In
particular, for quantum Ising models (see Sec.~\ref{isingmodels}), one
may consider the evolution of the local and global average
magnetization
\begin{equation}
  m_{\bm x}(t) \equiv \big\langle 
  \hat \sigma_{\bm x}^{(1)} \big\rangle_t \,,
  \qquad M(t) \equiv L^{-d} \sum_{\bm x} m_{\bm x}(t)\,,
  \label{magnt}
\end{equation}
as well as the fixed-time correlation function of the order-parameter
operator and its space integral,
\begin{equation}
  G(t,{\bm x}_1,{\bm x}_2) \equiv \big\langle \hat \sigma_{{\bm x}_1}^{(1)}
  \, \hat \sigma_{{\bm x}_2}^{(1)} \big\rangle_t\,,\qquad
  \chi(t,{\bm x}_0) = \sum_{\bm x} G({t,\bm x}_0,{\bm x})\,,
  \label{twopointt}
  \end{equation}
together with their connected contribution,
\begin{equation}
  G_c(t,{\bm x}_1,{\bm x}_2) \equiv G(t,{\bm x}_1,{\bm x}_2) -
  m_{{\bm x}_1}(t) \: m_{{\bm x}_2}(t)\,.
  \label{gct}
\end{equation}
In the above formulas~\eqref{magnt} and~\eqref{twopointt}, $\langle
\,\hat O \,\rangle_t$ denotes the expectation value of the operator
$\hat O$ at time $t$. In the case the system is in a pure state
$|\Psi(t)\rangle$, as for the quench protocol following
Eq.~\eqref{afterque}, this is given by $\langle \hat O \rangle_t
\equiv \langle \Psi(t)| \hat O |\Psi(t) \rangle$.  In the case the
system is in a mixed state $\rho(t)$, as when starting at finite
temperature or in the presence of dissipation (see Sec.~\ref{dissQT}),
this is given by $\langle \hat O \rangle_t \equiv {\rm Tr} \, \big[
  \hat O \, \rho(t) \big]$.

In the following, we distinguish between {\em soft} and {\em hard}
quenches around or to QTs.  Their difference is essentially related
to the extension of the energy levels they involve in the post-quench
dynamics. Soft quenches are related to changes of the Hamiltonian
parameters whose initial and final values stay close to the QT point,
thus only exciting the low-energy critical modes.
As we shall see, soft quenches can be effectively described by dynamic
scaling frameworks, within which the validity of the soft-quench
regime can be quantitatively addressed.  On the other hand, the more
general hard quenches are not limited by the above criticality
condition. For example they occur when the initial values of the
Hamiltonian parameters are far from the QT point, thus the post-quench
dynamics also involves high-energy levels that are not critical at
QTs.  In this case, the resulting out-of-equilibrium dynamics
is not generally expected to develop scaling behaviors.

\subsection{Homogeneous scaling laws for the out-of-equilibrium dynamics}
\label{homscaloaws}

As largely discussed in the literature, the unitary dynamics of closed
systems at CQTs develops dynamic scaling laws (see, e.g.,
Refs.~\cite{ZDZ-05, CEGS-12, Dziarmaga-05, DOV-08, DGP-10, GZHF-10,
  KCH-12, FDGZ-16, PRV-18-dfss, PRV-18-loc, Biroli-16, CC-16,
  Vicari-18, NRV-19-wo, RV-19-dec, RDZ-19, Sadhukhan-etal-20, PRV-20,
  RV-20-kz}).  In the following we discuss the conditions in which
dynamic scaling laws emerge from out-of-equilibrium regimes, and their
main features.  We focus on the quantum dynamics arising from
instantaneous quenches of closed systems at CQTs, described by the
Hamiltonian~\eqref{hlam}.

As a working hypothesis, we assume homogeneous scaling laws that
extend those holding at equilibrium (see Sec.~\ref{escalingcqt}), such
as that reported in Eq.~\eqref{Fsing-scaling-fss}. The main point of
such extension to the out-of-equilibrium dynamics is that the time
dependence of observables in the dynamic scaling limit is obtained
through the dependence of the scaling functions on a scaling
variable associated with time,
\begin{equation}
  \Theta \sim \Delta\,t \,,
  \label{eq:Theta_dyn}
\end{equation}
which is obtained by assuming that the relevant time scale of the
critical modes is proportional to the inverse energy difference
$\Delta$ of the lowest states. This can be implemented by adding a
dependence on $b^{-z} t$ in the scaling functions of the homogeneous
scaling laws, where $z$ is the critical exponent controlling the
suppression of the gap at the critical point.  Therefore, the
evolution of a generic observable $O$, such as the expectation value
at time $t$ of a local operator $\hat{O}$, is expected to
asymptotically satisfy the homogeneous scaling relation~\footnote{Here
we assume translation invariance, for example by taking systems with
PBC or ABC.\label{eq:notequench1}}
\begin{subequations}
  \begin{equation}
    O(t;L,w_i,w) \equiv \big\langle \hat O \big\rangle_t
    \approx b^{-y_o} \, {\cal O}(b^{-z} t, b^{-1} L, b^{y_w} w_i , b^{y_w} w)\,,
    \label{Oscadyn}  
  \end{equation}
where $b$ is an arbitrary positive parameter, $y_{o}$ is the RG
dimension of the operator $\hat{O}$, $y_w$ is the RG dimension of the
Hamiltonian parameter $w$, $L$ is the size of the system, and ${\cal
  O}$ is a universal scaling function apart from normalizations, see
Sec.~\ref{univscafu}.  Equation~\eqref{Oscadyn} is expected to provide
the asymptotic power-law behavior in the large-$b$ limit.

Analogously, in the case of fixed-time correlation functions of two
generic local operators $\hat{O}_1$ and $\hat{O}_2$, two-point
correlation functions only depend on the difference ${\bm x}\equiv
{\bm x}_2- {\bm x}_1^{~\ref{eq:notequench1}}$.  Their dynamic scaling
is thus expected to satisfy the homogeneous law
\begin{equation}
  G_{12}(t,{\bm x};L,w_i,w) \equiv \big\langle 
  \hat{O}_1({\bm x}_1) \, \hat{O}_2({\bm x}_2) \big\rangle_t
  \approx b^{-\varphi_{12}} \, {\cal G}_{12}(b^{-z} t, b^{-1} {\bm x}, b^{-1} L,
  b^{y_w} w_i , b^{y_w} w)\,, \label{g12scadyn}
\end{equation}
\label{scadyn}
\end{subequations}
where $\varphi_{12}=y_1+y_2$ and $y_j$ are the RG dimensions of the
operators $\hat{O}_j$.~\footnote{In the presence of boundaries, such
as systems with OBC, translation invariance is only recovered in the
thermodynamic limit. Therefore, one should keep the separate
dependence on both ${\bm x}_1/L$ and ${\bm x}_2/L$, which may arise
from boundary effects.}

To simplify the presentation, in the dynamic homogeneous scaling
laws~\eqref{scadyn} we have not inserted the scaling fields within the
arguments of the scaling functions ${\cal O}$ and ${\cal G}_{12}$, as
was done in Sec.~\ref{escalingcqt}, but we have only reported their
leading approximation in terms of the Hamiltonian parameters. The
asymptotic behavior is not changed; the differences arise at the level
of some typically subleading scaling corrections, which will be
neglected in the following.

The dynamic scaling framework can be extended to situations where the
initial condition is given by a Gibbs ensemble at temperature $T$, by
adding a further dependence on the product $b^z T$ in the scaling
functions of Eqs.~\eqref{scadyn}.

We finally note that the above dynamic scaling behaviors, and in
particular the scaling of the time dependence, is analogous to that
arising in the critical dynamics of classical systems, within RG
frameworks~\cite{HH-77, FM-06}. We recall that the exponent $z$ of the
critical dynamics at classical phase transitions is a further
independent critical exponent that also depends on the type of
dynamics considered~\cite{HH-77}. In quantum many-body systems, the
time dependence arises from the unitary quantum dynamics, therefore
the time dependence of the corresponding critical dynamics is
controlled by the same dynamic exponent $z$ describing the suppression
of the energy difference of the lowest states at the transition point.

\subsection{Dynamic scaling arising from soft quenches at
  quantum transitions}
\label{dynscaqu}

\subsubsection{General scaling behaviors}
\label{genscabehqu}

The dynamic scaling in the infinite-volume thermodynamic limit can be
straightforwardly obtained from the scaling laws~\eqref{scadyn}, by
setting
\begin{equation}
  b = \lambda \equiv |w|^{-1/y_w}\,, 
  \label{bla1}
\end{equation}
and taking the limit $L/\lambda\to \infty$ (of course this limit is
meaningless at strictly $w=0$). Assuming that such limit is well
defined, one obtains
\begin{subequations}
  \begin{align}
    O(t; w_i, w) \,
    \approx \; & \lambda^{-y_o} \, {\cal O}_\infty(\lambda^{-z} t,
    \lambda^{y_w} w_i)\,,
    \label{Oscaqueil}\\  
    G_{12}(t, {\bm x}; w_i, w) \, \approx \; & \lambda^{-\varphi_{12}} \,
    {\cal G}_\infty(\lambda^{-z} t, \lambda^{-1} {\bm x}, \lambda^{y_w} w_i)
    \label{g12scaqueil}\,.
  \end{align}
  \end{subequations}
We should emphasize that the above dynamic scaling limit requires that
the pre- and post-quench Hamiltonians remain in the {\em critical} regime
of a QT, i.e., $w_i$ and $w$ are sufficiently small.

The asymptotic dynamics scaling allowing for the finite size $L$ of
the system can be obtained from the same scaling laws~\eqref{scadyn},
by simply setting $b = L$, while the parameter $w$ may take any value,
including $w=0$. The resulting scaling ansatze can be written as
\begin{subequations}
  \begin{align}
    O(t; L,w_i, w) \, \approx \; & L^{-y_o} \, {\cal O}(L^{-z} t, w_i/w,
    L^{y_w} w)\,,
    \label{Oscaquefss}\\
    G_{12}(t, {\bm x}; L, w_i, w) \, \approx \; & L^{-\varphi_{12}} \,
    {\cal G}_{12}(L^{-z} t, L^{-1} {\bm x}, w_i/w, L^{y_w} w)
    \label{g12scaquefss}\,.
  \end{align}
  \label{scaquefss}
\end{subequations}
Therefore, an asymptotic dynamic FSS is expected to emerge
in the large-$L$ and large-time limit, keeping the scaling variables
\begin{equation}
  \Theta = L^{-z} \, t\,, \qquad {\bm X} = \frac{\bm x}{L}\,,\qquad
  \Phi_{i} = L^{y_w} \, w_i \,, \qquad \Phi = L^{y_w} \,w\,,\qquad
  \delta_w = \frac{w}{w_i}-1\,,
  \label{scalvarque}
\end{equation}
fixed~\footnote{Note that the equilibrium (ground-state) FSS behavior
must be recovered in the limit $\delta_w \to 0$.}.  It is also
possible to include the effect of a small finite temperature, assuming
a Gibbs ensemble as initial condition, by adding the scaling variable
$\Xi = L^z \, T$ as a further argument of the dynamic FSS
functions~\eqref{scaquefss}.

The above considerations provide the appropriate conditions under
which we may observe dynamic scaling corresponding to the soft-quench
scaling regime. This is obtained by keeping the scaling
variables~\eqref{scalvarque} fixed in the large-size limit, thus
we need to tune the Hamiltonian parameters so that they stay close
to the QTs, according to $w_i \sim w \sim L^{-y_w}$, which is the
condition to ensure that only the low-energy critical modes effectively
get excited during the post-quench dynamics, for $t \sim L^z$.
Moreover the asymptotic dynamic FSS, keeping
the scaling variables~\eqref{scalvarque} fixed, is expected to be
approached with power-law suppressed corrections, analogously to the
approach to the asymptotic equilibrium FSS discussed in
Sec.~\ref{escalingcqt}.

A similar dynamic scaling behavior has been also put forward in the context
of trapped bosonic gases~\cite{CV-10-tr}, extending the equilibrium
TSS discussed in Sec.~\ref{inhomsyst} to a dynamical TSS. This can be
obtained by replacing $L$ with $\ell^\theta$ in the dynamic FSS equations
[we recall that $\ell$ is the trap size, cf.~Eq.~\eqref{trapsize},
  and $\theta\le 1$ is the trap exponent, cf.~Eq.~\eqref{thetap}].

A related important issue regards the {\em thermalization}, that is,
whether the system presents a local thermal-like behavior at an
asymptotically long time after the quench.  Understanding under which
circumstances this occurs is a highly debated issue~\cite{PSSV-11,
  EFG-15, GE-16, DKPR-16, MIKU-18,  NH-15}, being related to the
integrability properties of the
Hamiltonian $\hat H_c$, the mutual interplay of interactions and
inhomogeneities, and the nature of the spectrum.  An effective
thermalization may emerge in the large-volume limit (of nonintegrable
systems), keeping the parameters $w\neq w_i$ fixed, i.e., in the limit
$\Phi\to\infty$, when the quench protocol entails energy interchanges
growing as the volume. However, in this hard-quench regime also
noncritical modes get excited, therefore the dynamics is expected to
lose the typical features of the critical dynamics.  We will return
to this point in Sec.~\ref{hardquenches}.

\subsubsection{Quantum quenches at the continuous transition
  of the Ising chain}
\label{dynqueisi}

\begin{figure}[tbp]
  \begin{center}
    \includegraphics[width=0.5\columnwidth]{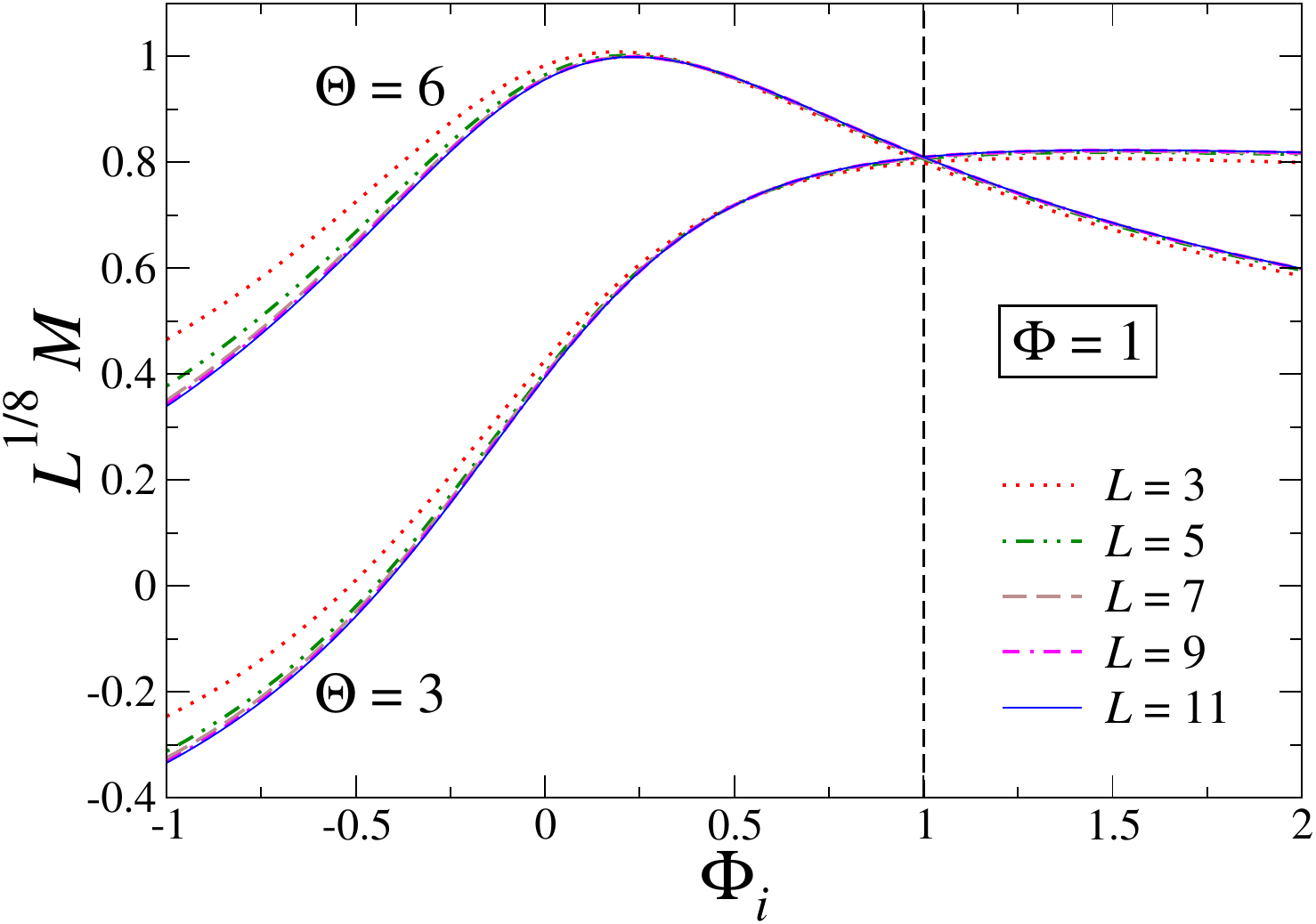}
    \caption{Magnetization for fixed $\Phi=1$ and for two different
      rescaled times $\Theta$.  The curves are plotted against the
      rescaled parameter $\Phi_i$, which is used to compute the
      initial state.  Notice that, at $\Phi_i=\Phi=1$, the equilibrium
      behavior is recovered (vertical dashed line). The curves at
      different size appear to approach an asymptotic function, in
      accordance with the dynamic FSS theory.  Adapted from
      Ref.~\cite{PRV-18-dfss}.}
    \label{fig:CQT_Kivar}
  \end{center}
\end{figure}

The above dynamic scaling ansatze have been confirmed by numerical
studies within the 1$d$ Ising chain~\eqref{hisdef} at its CQT point
$g_c=1$ and $h=0$. In particular, let us consider the case in which
the instantaneous quench is performed on the longitudinal field $h$,
keeping $g=g_c$ fixed.  Specializing to the Ising chain, the
magnetization $M$ and the two-point connected correlation function
$G_c$ of the order-parameter field,
cf. Eqs.~\eqref{magnt}-\eqref{gct}, show the asymptotic dynamic FSS
behaviors:
\begin{equation}
  M(t;L,w_i,w) \approx L^{-\beta/\nu} \, {\cal M}(\Theta,\Phi_i,\Phi)\,,
  \qquad 
  G_c(t,x;L,w_i,w) \approx L^{-\eta/\nu} \,
  {\cal G}(\Theta, X,\Phi_{i},\Phi)\,,
  \label{mgcheckisi}  
\end{equation}
where we assumed translation invariance, $\beta=1/8$ denotes the
magnetization critical exponent, $\nu=1$ and $\eta=1/4$, and thus
$\beta/\nu=1/8$ and $\eta/\nu=1/4$.  This out-of-equilibrium FSS
behavior has been checked by numerical computations based on exact
diagonalization~\cite{PRV-18-dfss}.  As an example,
Fig.~\ref{fig:CQT_Kivar} displays the magnetization after a quench at
fixed rescaled time $\Theta$ and $\Phi=1$, as a function of $\Phi_i$.
The various curves spotlight the emergence of a scaling behavior.
Further results can be found in the original work~\cite{PRV-18-dfss}.

\subsection{Scaling behavior of the Loschmidt echo}
\label{Loschecho}

The Loschmidt amplitude quantifies the deviation of the post-quench
state at time $t > 0$ from the initial state before the quench, and is
strictly connected with the concept of the fidelity (see
Sec.~\ref{sec:fidelity}). It is defined as the overlap
\begin{equation}
  \widetilde A(t) = \langle \Psi_0(w_i) | \Psi(t) \rangle = 
  \langle \Psi_0(w_i) | e^{-i \hat H(w) t} | \Psi_0(w_i) \rangle
  \,.~\footnote{The fidelity is given by
  $A(t) = \vert \widetilde A(t) \vert = \big\vert \langle \Psi_0(w_i) |
  e^{-i \hat H(w) t} | \Psi_0(w_i) \rangle \big\vert$.}  
  \label{ptdef}
\end{equation}
The so-called Loschmidt echo $Q(t)$ is identified with its rate function,
\begin{equation}
  Q(t) = - \ln \big| \widetilde A(t) \big|^2\,.
  \label{loschdef}
\end{equation}
Note that $Q(t)=0$ implies the restoration of the initial quantum
state.

\begin{figure}[!t]
  \begin{center}
    \includegraphics[width=0.47\columnwidth]{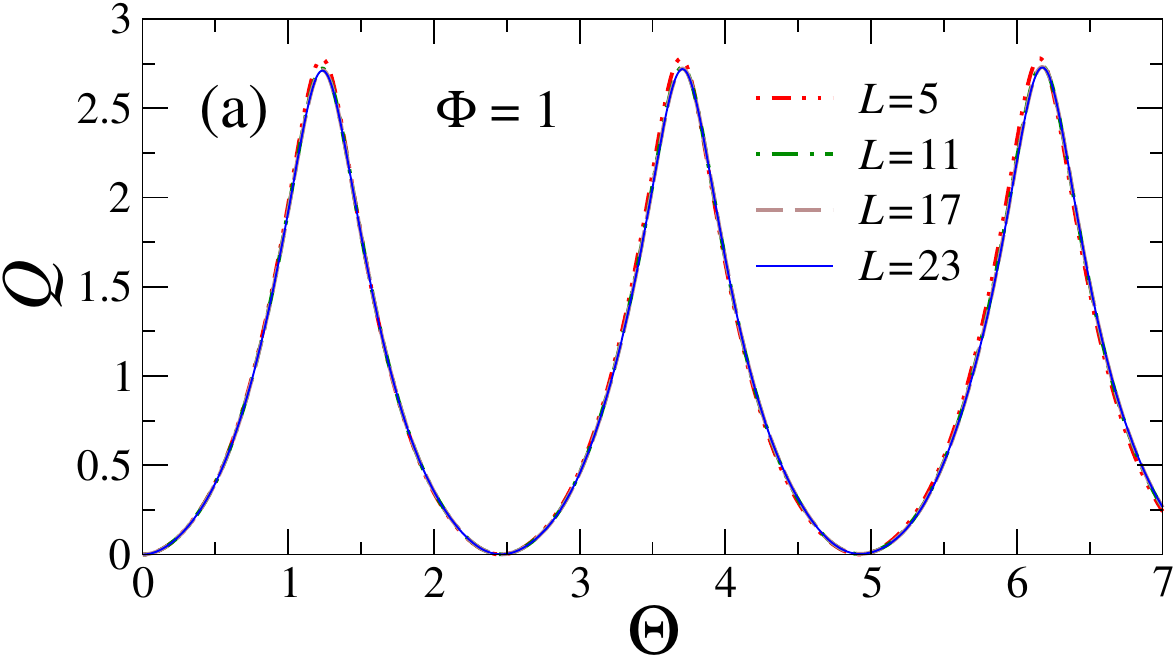}
    \hspace*{5mm}
    \includegraphics[width=0.47\columnwidth]{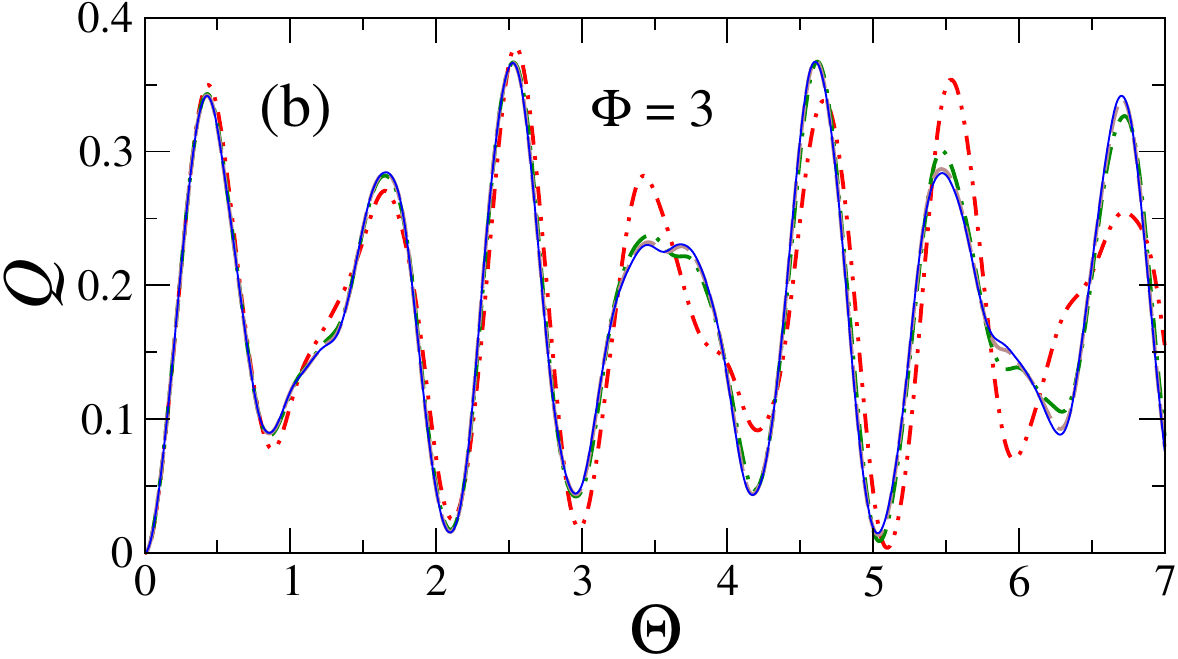}
    \caption{Temporal behavior of the Loschmidt echo $Q(t)$ defined in
      Eq.~\eqref{loschdef} for $\delta_w = -2$, and two different
      values of $\Phi = 1$ ($a$) and $\Phi = 3$ ($b$).
      Data are plotted against the rescaled time $\Theta = L^{-z} \,t$,
      so that the convergence to a scaling function, in the large-$L$
      limit, is clearly visible. Adapted from Ref.~\cite{PRV-18-dfss}.}
    \label{fig:LEcho_kvar}
  \end{center}
\end{figure}

The dynamic scaling behavior of $Q(t)$ at QTs was addressed in
Ref.~\cite{PRV-18-dfss}, focusing on quenches of the longitudinal
field $h$ in quantum Ising models, thus $w\equiv h$, and numerically
investigating it within 1$d$ systems. Its time
dependence after the quench is conjectured to obey the dynamic FSS
behavior
\begin{equation}
  Q(t;L, w_i, w) \approx {\cal Q} (\Theta, \Phi, \delta_w)\,.
  \label{Lasca}
\end{equation}
An illustrative example is reported in Fig.~\ref{fig:LEcho_kvar},
which shows that the numerical data fully support the dynamic FSS
predicted by the scaling equation~\eqref{Lasca}.  The emerging pattern
is similar to that of the magnetization.  We also note the presence of
quasi-complete revivals of the quantum states along the quantum
evolution, when $Q(t)\ll 1$.

\subsection{Scaling properties of work fluctuations after quenches
  near quantum transitions}
\label{workflu}

We now address issues related to the statistics of the work done on a
many-body system close to a QT, when this is driven out of equilibrium
by suddenly switching one of the control parameters~\cite{CHT-11,
  GPGS-18}, such as the quenching protocol described in
Sec.~\ref{timedepproto}.  Several issues related to this topic have
been already discussed in a variety of physical implementations,
including spin chains~\cite{Silva-08, DPK-08, DGCPV-12,
  Mascarenhas-etal-14, MS-14, ZT-15, SD-15, Bayat-etal-16}, fermionic
and bosonic systems~\cite{DL-08, GS-12, SRH-14, SGLP-14}, quantum
field theories~\cite{SGS-13, SS-13, PS-14, Palmai-15}, as well as in
different contexts like cyclically driven systems~\cite{BAKP-11} and
dynamic quantum phase transitions~\cite{HPK-13}.  It has also been 
shown that the work statistics can be experimentally measured in
present-day ultracold-atom systems, by means of ion
traps~\cite{HSDL-08} or Ramsey interferometry~\cite{Dorner-etal-13,
  MDP-13}.

The research done so far in this context has mostly addressed the
thermodynamic limit of systems close to criticality
(see, e.g., Ref.~\cite{GPGS-18} and references therein).
However, it is phenomenologically important to know the impact of having
a finite size, in order to achieve a deep understanding of any reliable
quantum-simulation experiment.  Indeed, for a global quench the work
is extensive.  Therefore in the thermodynamic (i.e., large-volume) limit,
keeping the Hamiltonian parameter fixed, one expects the
probability associated with the work density $W/V$ (where $V=L^d$ is the
volume of the system) to be sharply peaked around its average value,
with typical fluctuations suppressed as $V^{-1/2}$.  The work
distribution is expected to approach a quasi-Gaussian distribution in
the infinite volume limit, around the average value of the work
density $W/L^d$. One also expects $O(L^{-d/2})$ fluctuations having
the general form $P(W) \sim \exp[-L^d I(W/L^d)]$, with $I(x)\ge
0$~\cite{GS-12,GPGS-18}.  This suggests that fluctuations around the
work average, and in particular deviations from Gaussian behaviors,
may be only observable for relatively small systems.

\subsubsection{Work fluctuations associated with a quench}
\label{woque}

The quantum work $W$ associated with a quench protocol, i.e., the work
done on the system by quenching the control parameter $w$, does not
generally have a definite value.  More specifically, this quantity can
be defined as the difference of two projective energy
measurements~\cite{CHT-11}.  The first one, at $t=0$, projects onto the
eigenstates of the initial Hamiltonian $\hat{H}(w_i)$ with a
probability given by the equilibrium Gibbs distribution. Then the
system is driven by the unitary operator $e^{-i \hat{H}(w) t}$ and the
second energy measurement projects onto the eigenstates of the
post-quench Hamiltonian $\hat H(w)$.  The work probability
distribution can thus be written as~\cite{CHT-11, TH-16, TLH-07}
\begin{equation}
  P(W) = \sum_{n,m} \delta \big\{
    W-[ E_{n}(w)-E_{m}(w_i) ] \big\} \, \big| \langle \Psi_n(w) | \Psi_m(w_i)
  \rangle \big|^2 \, p_m\,,\quad
  \label{pwdefft}
\end{equation}
where $\{E_{n}(w) \}$ and $\{ |\Psi_n(w)\rangle \}$ denote eigenvalues
and eigenstates of the Hamiltonian $\hat{H}(w)$, and $p_m$ are the
probabilities of the initial eigenstates $| \Psi_m(w_i)\rangle$.  For
example, in the case of an initial Gibbs distribution with temperature
$T$, one has $p_m \propto \exp[-E_{m}(w_i)/T]$.  One may also
introduce a corresponding characteristic function~\cite{Silva-08,
  CHT-11}
\begin{equation}
  C(s)  = \int dW \, e^{i s W} \, P(W),
  \label{gudef}
\end{equation}
encoding full information of the work statistics. 

The work distribution~\eqref{pwdefft} satisfies the quantum version of
the Crooks fluctuation relation~\cite{CHT-11}
\begin{equation}
  \frac{P(W, T, w_i, w)}{P(-W, T, w_i, w)}
  = e^{W/T} e^{-\Delta F/T}\,,\qquad  \Delta F = F(w)-F(w_i)\,,
  \label{eq:Crooks}
\end{equation}
where the probability distribution in the denominator corresponds to
an inverted quench protocol, from $w$ to $w_i$, and where $F = - T
\,\ln Z$ is the free energy. It also satisfies the Jarzynski
equality~\cite{Jarzynski-11, CHT-11}:
\begin{equation}
  \langle e^{-W/T} \rangle \equiv \int dW\, e^{-W/T}\, P(W) 
  = e^{-\Delta F/T}\,.
  \label{jeq}
\end{equation}
One may also define the so-called dissipative work~\cite{CHT-11}
$W_d = W - \Delta F$, satisfying the inequality $\langle W_d \rangle \ge 0$.

The zero-temperature limit corresponds to a quench protocol starting
from the ground state $|\Psi_0(w_i)\rangle$ of $\hat{H}(w_i)$.
Assuming that such ground state is nondegenerate, the work
probability~\eqref{pwdefft} reduces to
\begin{equation}
  P(W) = \sum_{n} \delta \big\{
    W - [E_{n}(w)-E_{0}(w_i)] \big\} \, \big| \langle \Psi_n(w) | \Psi_0(w_i)
  \rangle \big|^2 \,.
  \label{pwdef}
\end{equation}
In this case, the dissipative work simplifies to $W_d= W - [E_0(w)-E_0(w_i)]$,
and the characteristic function $C(s)$ can be simply written as the amplitude
\begin{equation}
  C(s) = \langle \Psi_0(w_i) | e^{-i \hat{H}(w_i) s} \,e^{i\hat{H}(w) s}
  | \Psi_0(w_i) \rangle\,,
  \label{csamp}
\end{equation}
whose absolute value is related the Loschmidt echo $Q(t) = - \ln |
\langle \Psi_0(w_i) | \Psi(t)\rangle|^2$, providing information on the
overlap between the initial state $|\Psi_0(w_i)\rangle$ and the
evolved quantum state $|\Psi(t)\rangle$ at time $t$.

\subsubsection{Scaling of the work fluctuations}
\label{scawoque}

We now discuss the behavior of such quantities at CQTs.  To this
purpose, we distinguish two cases:
\begin{itemize}
  \item[(i)] the case in which $w$ is an {\em odd} Hamiltonian
    parameter, such as the longitudinal field $h$ in quantum Ising
    models, thus correspondingly $\langle \hat{H}_p \rangle=0$ at the
    critical point $w=0$;
  \item[(ii)] the case in which $w$ is an {\em even} Hamiltonian
    parameter, such as the transverse field $g$ in quantum Ising
    models (more precisely $w=r=g-g_c$), thus $\langle \hat{H}_p
    \rangle\neq 0$ at the critical point $w=0$.
\end{itemize}
In case (i) work fluctuations develop notable scaling
behaviors, while in case (ii) things get more complicated.

Assuming the existence of a nontrivial dynamic FSS limit for the work
distribution $P(W)$, as expected for the case (i), a natural working
hypothesis is that it is defined as the large-size limit keeping the
appropriate scaling variables fixed, such as~\cite{NRV-19-wo}
\begin{equation}
  P(W,L,w_i,w) \approx \Delta(L)^{-1} \: {\cal
    P}(\Omega,\Phi,\delta_w)\,,\qquad \Omega \equiv {W \over \Delta(L)} \,,
  \label{genpwsca}
\end{equation}
in the zero-temperature limit, where $\Delta(L)$ is the energy gap of
the lowest states at $w=0$, and we have introduced a further scaling
variable $\Omega$.  The dynamic FSS limit is defined as the large-size
$L\to\infty$ limit, keeping $\Omega$, $\Phi$, and
$\delta_w=w/w_i-1=\Phi/\Phi_i-1$ fixed.  The dynamic
FSS~\eqref{genpwsca} allows us to infer the scaling behavior of the
average of the work and its higher moments $\langle W^k \rangle \equiv
\int dW\, W^k\, P(W)$, as well.  Equation~\eqref{genpwsca} also
determines the scaling behaviors of the corresponding characteristic
function~\eqref{gudef},
\begin{equation}
  C(s;L,w_i,w) \approx {\cal C}(\Theta_s, \Phi, \delta_{w})\,,\qquad
  \Theta_s \equiv L^{-z} \, s\,.
  \label{gtsca}
\end{equation}
The above formulas can be straightforwardly extended to allow for a
finite temperature $T$ of the initial state.

Using the general dynamic FSS of the work probability, one can derive
the corresponding behavior of the zero-temperature average work, i.e.,
$\langle W \rangle= \Delta(L) \, {\cal W}_1(\Phi_i,\Phi)$.  Actually,
the same scaling behavior can be equivalently derived by noting that
\begin{equation}
  \langle W \rangle = \langle \Psi_0(w_i) | \hat{H}(w)
  - \hat{H}(w_i)| \Psi_0(w_i)
  \rangle = (w-w_i)\langle \Psi_0(w_i) | \hat{H}_p | \Psi_0(w_i) \rangle
  \approx L^{d-y_p} \, \delta_w w_i\, f_p(\Phi_i) \,,
  \label{wavsca}
\end{equation}
where $y_p$ and $f_p$ are, respectively, the RG dimension and the
equilibrium FSS function associated with the observable
$\hat{H}_p/L^{d}$.  Then, taking into account the hyperscaling
relation $y_p + y_w = d+z$ between the RG dimensions of $w$ and of the
associated perturbation $\hat{H}_p$~\cite{Sachdev-book, CPV-14}, we
have that
\begin{equation}
  \langle W\rangle \approx  L^{-z} {\cal W}_1(\Phi_i,\Phi)\,,
  \qquad 
      {\cal W}_1(\Phi_i,\Phi) \sim \delta_w \,\Phi_i\,f_p(\Phi_i)\,,
      \qquad \delta_w = {w\over w_i}-1\,,
      \label{calla3}
\end{equation}
in agreement with the scaling behavior of the average work $\langle W
\rangle = \int dW \, W \, P(W)$ that is obtained using the scaling
ansatz~\eqref{genpwsca}.  This provides a relation between the dynamic
FSS function of the average work and the equilibrium FSS function of
the expectation value of the Hamiltonian term $\hat{H}_p$ associated
with the driving parameter $w$. Taking the large-volume limit ($L \to
\infty$) of the above scaling behaviors, the average work is expected
to grow as the volume, which implies
\begin{equation}
  f_p(\Phi_i) \sim |\Phi_i|^{y_p/y_w}, \qquad |\Phi_i|
  \to \infty, \qquad \langle W\rangle \sim L^{d}\,
  (w-w_i) \,|w_i|^{y_p/y_w}\,.
\label{calla2li}
\end{equation}
This can be also written as
\begin{equation}
  {\langle W\rangle / L^{d} } \sim \delta_w\,\xi_i^{-(d+z)} \,,
  \qquad \xi_i \sim |w_i|^{-1/y_w}\,,
  \label{altivsca}
\end{equation}
where $\xi_i$ represents an infinite-volume correlation length
associated with the initial ground state of $\hat{H}(w_i)$.

\begin{figure}[!t]
  \begin{center}
    \includegraphics[width=0.47\columnwidth]{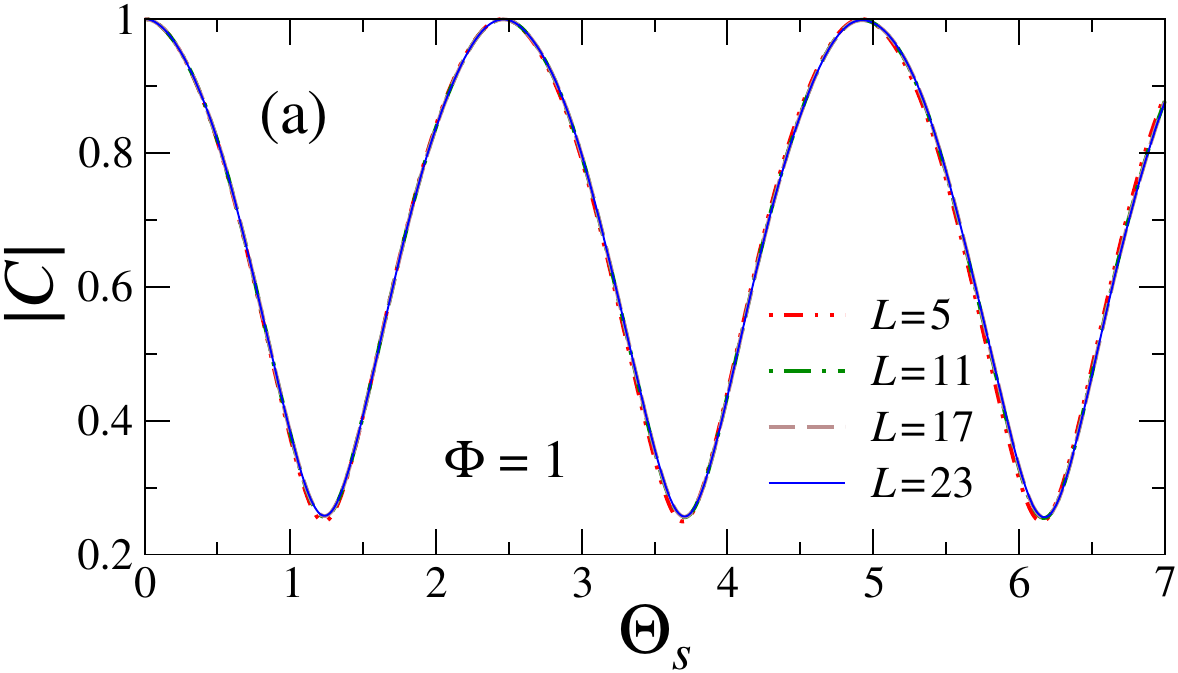}
    \hspace*{5mm}
    \includegraphics[width=0.47\columnwidth]{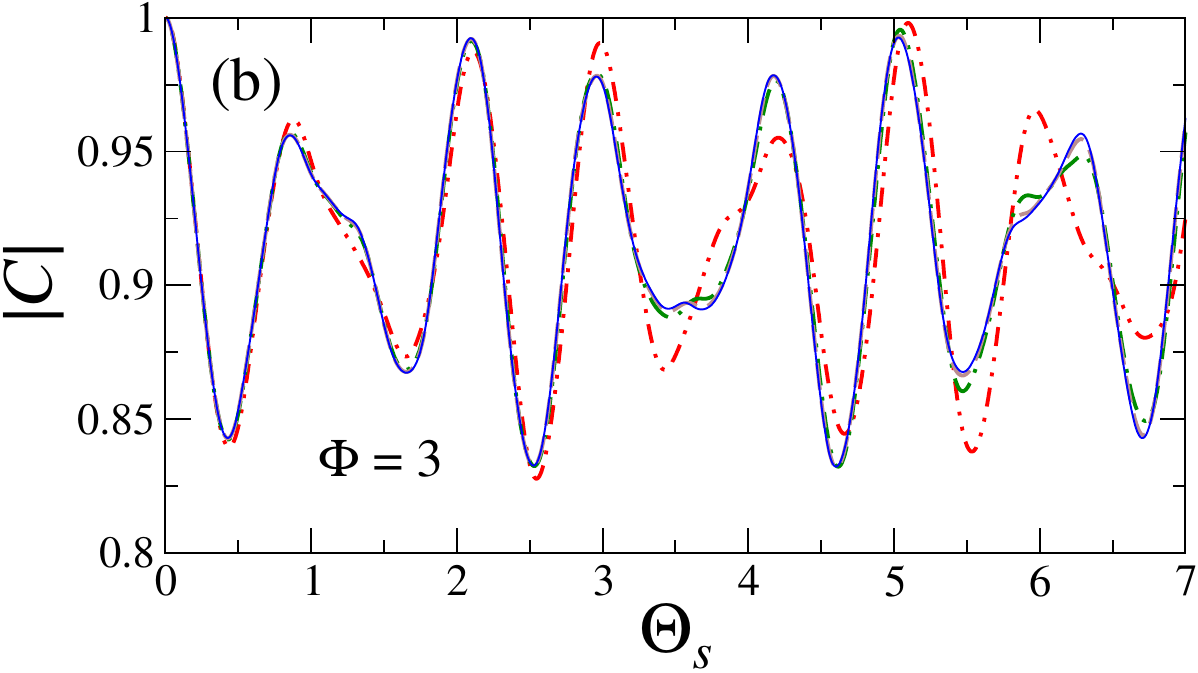}
    \caption{Scaling of the modulus of the characteristic function
      $|C(s)|$ associated to a quench of the Ising model for $\delta_w
      = -2$. The two panels refer to $\Phi=1$ ($a$) and $\Phi=3$
      ($b$).  Data are plotted against the rescaled time
      $\Theta_s = L^{-z}\,s$, so that the convergence to a scaling
      function, in the large-$L$ limit, is clearly visible.
      Adapted from Ref.~\cite{NRV-19-wo}.}
    \label{CharFun_CQT}
  \end{center}
\end{figure}

As already mentioned, the above scaling behaviors of the work
fluctuations are expected to apply when the driving parameter is odd
[case (i)], i.e., when the expectation value of $\hat{H}_p$ vanishes
on the ground state of the critical Hamiltonian (for $w=0$).  This is
demonstrated by the data in Fig.~\ref{CharFun_CQT} for the
characteristic work function, cf. Eq.~\eqref{gudef}, of the 1$d$ Ising
chain, associated with a quench of the longitudinal field $h$: they
clearly support the scaling behavior put forward in Eq.~\eqref{gtsca}.
In contrast, as discussed in Ref.~\cite{DRV-20}, when the driving
parameter is associated with an even perturbation [case (ii)], such as
for $w=r=g-g_c$ in the quantum Ising models, then work fluctuations
get dominated by analytical terms (mixing with the identity operator),
while the scaling terms remain subleading.

We finally mention that the behavior of work fluctuations arising
from soft quenches has been also investigated within the BH model in
the hard-core limit, driven by an external constant field globally
coupled to the bosonic modes, through the vacuum-to-superfluid CQT,
extending the analysis to particle systems confined by inhomogeneous
external potentials~\cite{NRV-19-wo}.

\subsection{Scaling of the bipartite entanglement entropy after
  soft quenches}
\label{dynscaentent}

\begin{figure}[!t]
  \begin{center}
    \includegraphics[width=0.5\columnwidth]{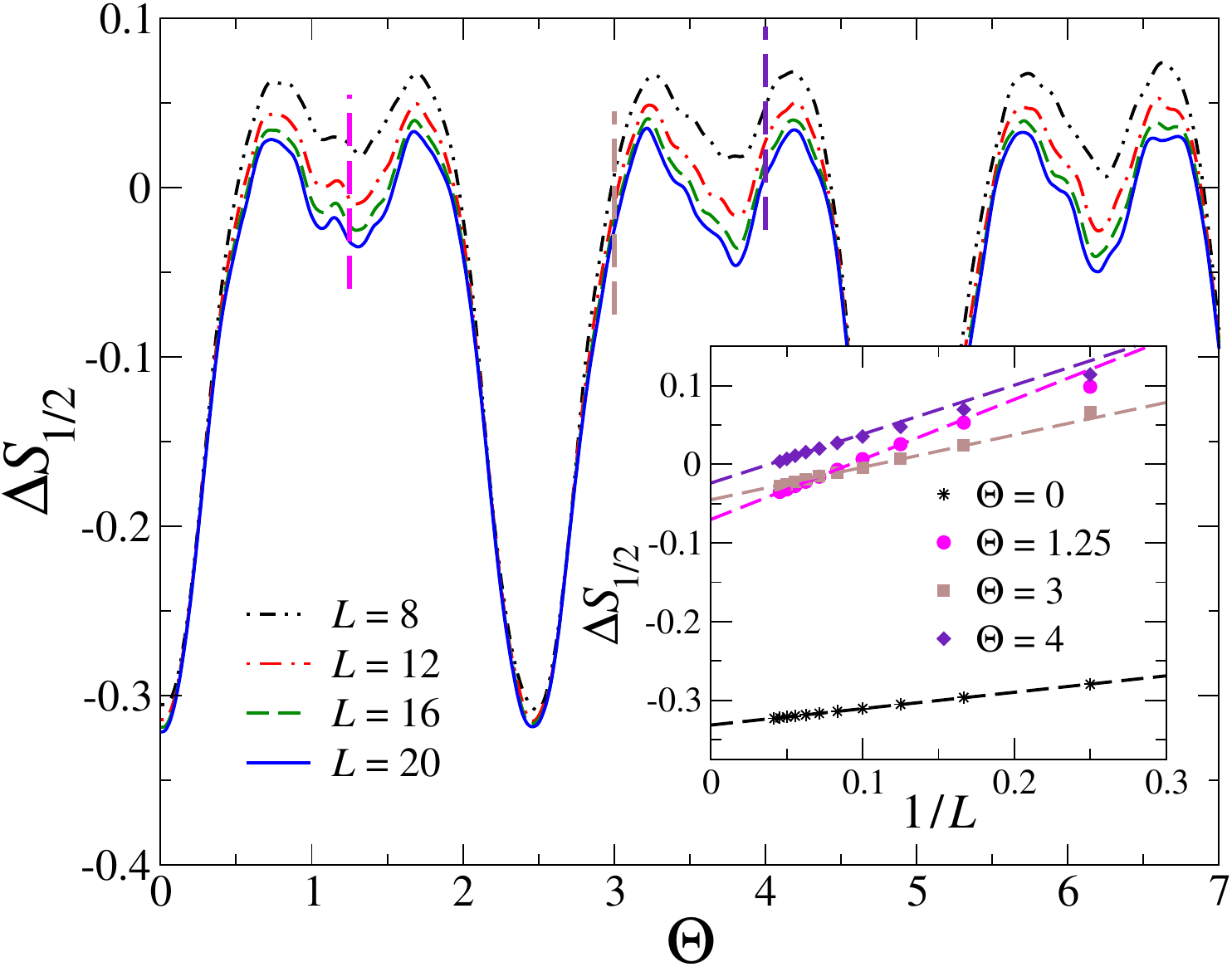}
    \caption{Temporal behavior of the entanglement entropy $\Delta
      S_{1/2}\equiv \Delta S(t;L,L/2;w_i,w)$ for a balanced
      bipartition in a quantum Ising ring, after quenching the
      longitudinal field $h=w$ from $w_i=0$. The inset displays the
      convergence with $L$ of the various curves (up to $L=22$), for
      four values of $\Theta$ (see the long-dash lines in the main
      panel), plotted against $1/L$. Black stars denote data
      corresponding to the equilibrium condition $\Theta = 0$.  Dashed
      lines are numerical fits of the data at the largest available
      $L$.  Adapted from Ref.~\cite{PRV-18-dfss}.}
    \label{fig:VNentro}
  \end{center}
\end{figure}

The time evolution of the entanglement entropy of bipartitions
$\tt{A}|\tt{B}$ of the system, cf.~Eq.~\eqref{vNen}, quantifies the
amount of quantum correlations that are present between the two parts
of the chain after soft quenches around a CQT.~\footnote{The time
dependence of the entanglement entropy has been also investigated for
generic quenches, see for example
\cite{CC-05,DMCF-06,CC-07b,FC-08,Cardy-11,CC-16,AC-17}, without any
particular connection with issues related to QTs.}  Here we focus on
1$d$ quantum spin systems, such as Ising-like chains, and consider a
bipartition into two parts of size $\ell_{\tt A}$ and $L-\ell_{\tt
  A}$.  Extending equilibrium scaling arguments (see,
Sec.~\ref{entang}), the following dynamic FSS behavior has been
conjectured for its behavior after quenches from $w_i$ to
$w$~\cite{PRV-18-dfss}
\begin{equation}
  \Delta S(t;L,\ell_{\tt A};w_i,w) \equiv S(t;L,\ell_{\tt A};w_i,w) -
  S_{c}(L,\ell_{\tt A}) \approx {\cal S}(\Theta,\ell_{\tt
    A}/L,\Phi,\delta_w)\,,
  \label{esca}
\end{equation}
where $S_c(L,\ell_{\tt A})$ accounts for the critical FSS
behavior~\eqref{ccfo}.  Numerical checks within the Ising
chain~\cite{PRV-18-dfss} support this asymptotic scaling behavior, as
shown in Fig.~\ref{fig:VNentro}, with corrections generally decaying
as $1/L$.

\subsection{Out-of-equilibrium dynamics after hard quenches
  to quantum critical points}
\label{hardquenches}

As discussed in the previous subsections, the out-of-equilibrium
dynamics arising from quantum quenches at CQTs develops scaling
behaviors controlled by their universality class. These require the
Hamiltonian parameters associated with the quench protocol, $w_i$ and
$w$, to be close to the QCP, at $w=0$.  However, in the case of hard
quenches across the CQT (i.e., when $w_i$ and $w$ differ
significantly), a dynamic scaling controlled by the universality class
of the equilibrium QT is generally not expected, essentially because
the instantaneous change of the Hamiltonian parameters entails a
significant amount of energy exchange. Indeed, the instantaneous
change from $w_i$ to $w$ gives rise to a relatively large amount
$\Delta E$ of energy above the ground level of the Hamiltonian $\hat
H(w)$,
\begin{equation}
  \Delta E = \sum_{n} \big[ E_{n}(w) - E_{0}(w) \big] \, 
  \big| \langle \Psi_n(w)| \Psi_0(w_i) \rangle \big|^2 \sim L^d\,,
  \label{deltae}
\end{equation}
Typically $\Delta E$ is much larger than the energy scale $E_c$ of the
low-energy excitations at criticality $w=0$, which is given by
$E_c\sim L^{-z}$ with $z=1$ for CQTs of Ising-like systems.  This
would naturally lead to the expectation that the unitary
energy-conserving dynamics, after quenching to the QCP, is not
significantly related to the quantum critical features of the
low-energy spectrum of the critical Hamiltonian.  However, as argued
in Refs.~\cite{BDD-15, RMD-17, TIGGG-19, HPD-18, Haldar-etal-20,
  RV-20-qu}, some signatures may emerge as well. In particular,
integrable many-body systems (such as those mappable into generic
noninteracting fermionic systems) develop peculiar
discontinuities even in the asymptotic stationary states arising from
the quantum quenches~\cite{BDD-15, RMD-17, RV-20-qu}.  On the other
hand, local observables are not expected to present singularities in
nonintegrable systems where generic quantum quenches lead to
thermalization, since they are generally smooth functions of the
temperature (exceptions to this general behavior are seen to emerge
in integrable models and in the presence of localization phenomena).
Nonetheless, it has been recently argued that it is possible to
recover signals from intermediate regimes of the post-quench
quantum evolution~\cite{Haldar-etal-20}.

\subsubsection{Singular behavior of the XY chain}
\label{tgermohardqu}

The above issue related to hard quenches has been thoroughly
investigated within the quantum XY model~\eqref{XYchain}, under
quenches of the Hamiltonian parameter $g$ driving the transition that
separates the quantum paramagnetic and ferromagnetic
phases~\cite{CEF-12a, CEF-12b, BDD-15, EF-16, RV-20-qu}, starting from
the ground states associated with initial values $g_i$.  In the
thermodynamic limit, the out-of-equilibrium evolution of the energy
density~\cite{BDD-15}, and the longitudinal and transverse
magnetizations~\cite{RV-20-qu}, develop a singular dependence on the
quench parameter $g$, or $w\equiv g-g_c$, around the critical value
$g=g_c$, for any anisotropy parameter $\gamma>0$ and starting point
$g_i$, including the extremal ones corresponding to fully disordered
and ordered initial states.

\begin{figure}[tbp]
  \begin{center}
    \includegraphics[width=0.5\columnwidth]{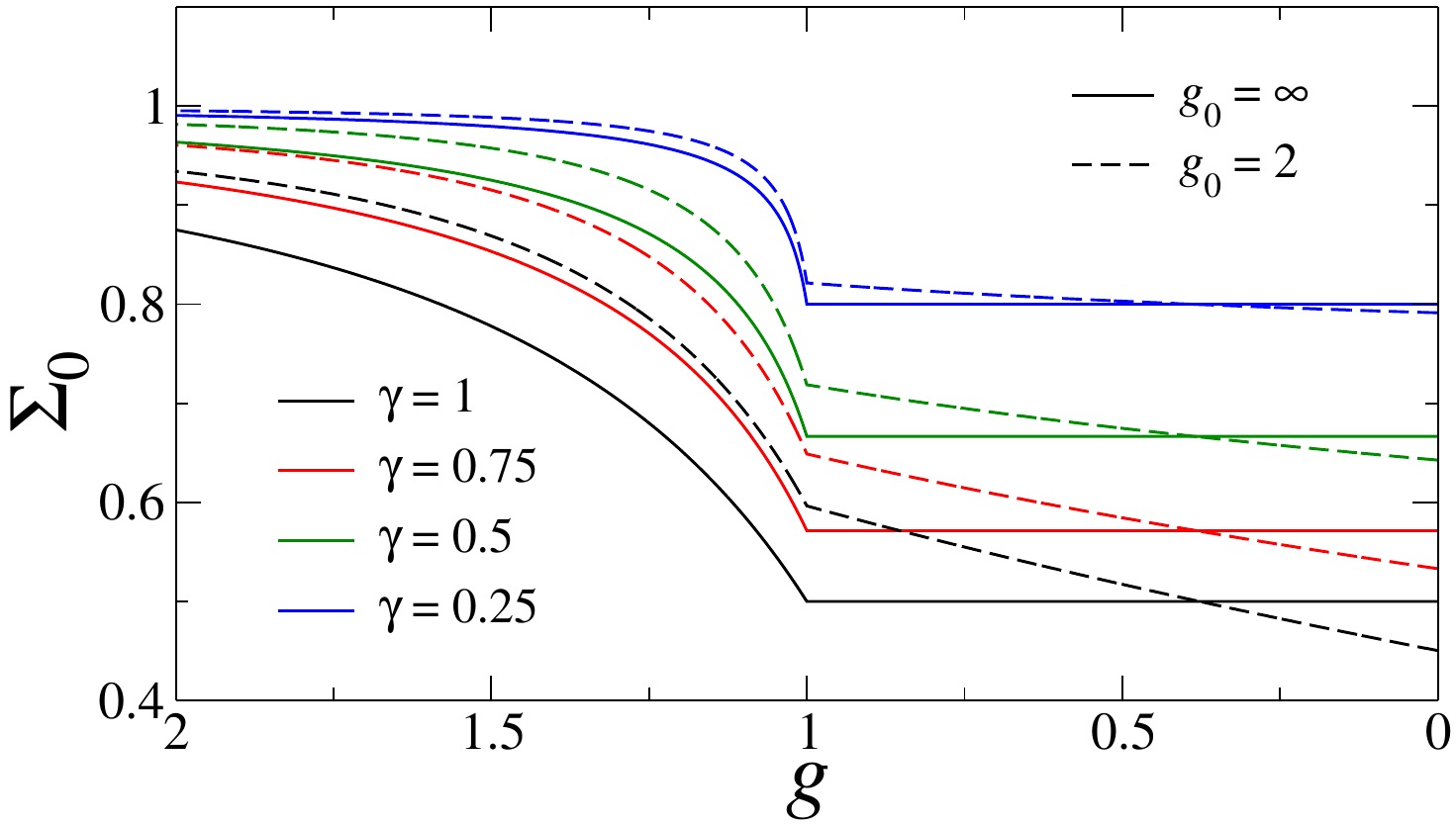}
    \caption{The large-time limit of the transverse magnetization for
      the quantum XY chain with different anisotropies $\gamma$ in the
      thermodynamic limit, as a function of $g$ and for two values of
      $g_i=\infty$ (continuous lines) or $g_i=2$ (dashed lines).  The
      curves for $\Sigma_0$ display a singular behavior (i.e., a
      discontinuity in the derivative) at $g=g_c=1$.  Adapted from
      Ref.~\cite{RV-20-qu}.}
    \label{Mt_asynt_an}
  \end{center}
\end{figure}

The behavior of the transverse (t) magnetization of the XY chain
after a quench from $g_i$ to $g$,
\begin{equation}
  M_{\rm t}(t) \equiv L^{-1} \sum_x m_{{\rm t},x}(t)\,, \qquad
  m_{{\rm t},x}(t) = \big\langle \hat \sigma^{(3)}_x \big\rangle_t\,,
  \label{tramag}
\end{equation}
has been analytically computed in the infinite-volume
limit~\cite{Niemeijer-67, Niemeijer-68, BMD-70, BM-71a, BM-71b,
  MBA-71}.  For convenience, we write it as
\begin{equation}
  M_{\rm t}(t;L\to\infty,g_i,g) \equiv \Sigma(t;g_i,g) \,,\qquad
  \Sigma(t;g_i,g) \equiv \Sigma_0(g_i,g) + \Sigma_t(t;g_i,g)\,,
  \label{ssigmadef}
\end{equation}
where $\Sigma_0(g_i,g)$ represents the asymptotic time-independent
term, i.e., $\Sigma_t$ vanishes in the large-time limit.  For
$g_i\to\infty$, the analytical expressions somehow simplify
into~\cite{Niemeijer-67,BMD-70}
\begin{equation}
  \Sigma_0(g_i\to\infty,g) = \int_0^\pi {dk\over \pi} \; {[g-\cos(k)]^2\over
    \Lambda(k,g)^2} \,,\qquad
  \Sigma_t(t;g_i\to\infty,g) =  \int_0^\pi {dk\over \pi} \;
        {\gamma^2 \sin(k)^2 \cos\left[ 4 \, \Lambda(k,g)\,t \right]
          \over \Lambda(k,g)^2} \,,
  \label{fagg0inf} 
\end{equation}
where $\Lambda(k,g) = \sqrt{[g - \cos(k)]^2 + \gamma^2 \sin(k)^2}$.
Figure~\ref{Mt_asynt_an} shows some curves for the transverse
magnetization in the large-time limit [i.e., of the function
  $\Sigma_0(g_i,g)$], for the initial couplings $g_i=\infty$ and
$g_i=2$ within the disordered phase. As clearly visible, the function
$\Sigma_0(g_i,g)$ presents a nonanalytic behavior in correspondence of
the QCP at $g_c = 1$. Indeed, we have that~\cite{RV-20-qu}
\begin{equation}
  \lim_{g\to g_c^+} {\partial \Sigma_0(g_i,g)\over \partial g} -
  \lim_{g\to g_c^-} {\partial \Sigma_0(g_i,g)\over \partial g} =
  \gamma^{-1}\,.
  \label{singdermt}
\end{equation}
Note that such a discontinuity is independent of $g_i>1$.  In
particular, for $\gamma=1$ and $g_i\to\infty$, the behavior around
$g_c$ turns out to be $\Sigma_0(\infty,g) = 1/2 + ( g- 1) + O[(g-1)^2]$
for $g\ge 1$, and $\Sigma_0(\infty,g)=1/2$ for $g\leq 1$.

Singular behaviors are also found for quenches starting from the
ordered magnetized phase ($g_i<1$), in particular when monitoring the
longitudinal (l) magnetization
\begin{equation}
  M_{\rm l}(t) \equiv L^{-1} \sum_x m_{{\rm l},x}(t)\,, \qquad
  m_{{\rm l},x}(t) = \big\langle \hat \sigma^{(1)}_x \big\rangle_t\,.
  \label{longmag}
\end{equation}
For example, one may consider quenches of the quantum Ising chain
($\gamma=1$) starting from a completely ordered state (i.e., a fully
magnetized state $|\Psi(0)\rangle = |\!\! \uparrow, \ldots , \uparrow
\rangle$).  This corresponds to one of the degenerate ground states in
the thermodynamic limit when $g_i\to 0$, obtainable by the ground
state $|\Psi_0(g_i,h)\rangle$ in the limit
\begin{equation}
  |\Psi(0)\rangle = \lim_{g_i\to 0} \lim_{h\to 0^+} \lim_{L\to\infty}|
  \Psi_0 (g_i,h)\rangle \,,
\end{equation}
where $h$ is an external homogeneous magnetic field coupled
to the longitudinal magnetization.
Since such initial state breaks the $\mathbb{Z}_2$ symmetry of the
model, one finds a nonzero longitudinal magnetization $M_{\rm l}$ along
the quantum unitary evolution after quenching the transverse-field
parameter $g$.  In the thermodynamic limit, it vanishes in the
large-time limit, showing an asymptotic exponential
decay~\cite{CEF-12a}
\begin{equation}
  M_{\rm l}(t;g) \approx {\cal M}_{{\rm l},a}(t,g) = {\rm A}(g)\,
  \exp[-\Gamma(g)\,t ]\,.
  \label{mlgg0}
\end{equation}
Similarly to the transverse magnetization, there is a singular
behavior at $g=g_c$. Indeed, around $g_c=1$, the decay function
$\Gamma(g)$ behaves as~\cite{CEF-12a, RV-20-qu}
\begin{equation}
  \Gamma(g) = \left\{ \begin{array}{ll}
    4/\pi - 2\sqrt{2(1 - g)} + O(1-g) & {\rm for} \; g<1 \,,  
    \vspace*{2mm} \\
    4/\pi & {\rm for} \; g\ge 1 \,. 
  \end{array} \right. \label{gammag1beh}
\end{equation}

\subsubsection{Revival phenomena in finite-size systems}
\label{revivalhardqu}

\begin{figure}[tb]
  \begin{center}
    \includegraphics[width=0.5\columnwidth]{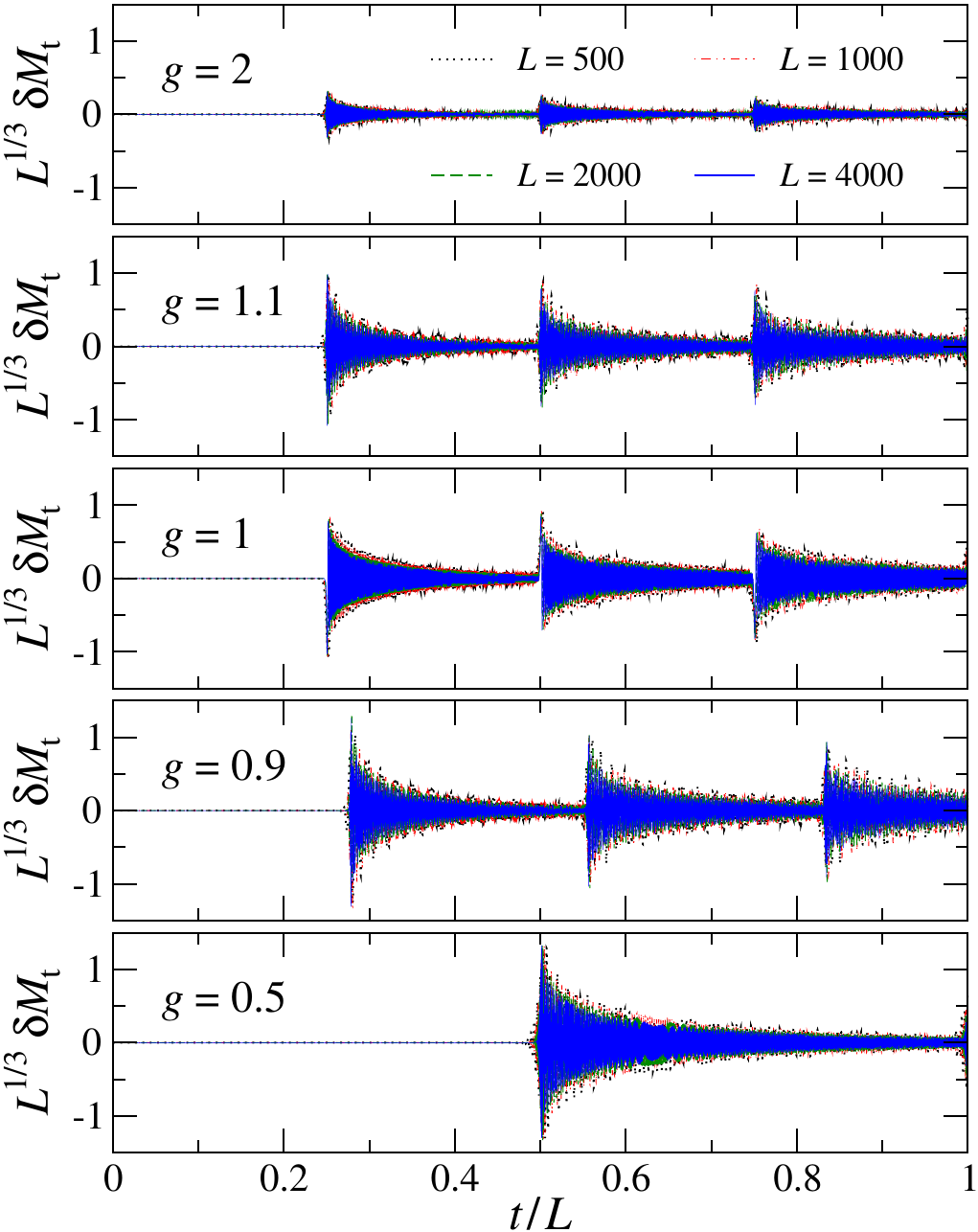}
    \caption{Finite-size features of the temporal behavior of the
      transverse magnetization after a quench from $g_i=\infty$.  The
      figure shows $L^{1/3} \, \delta M_{\rm t}$ versus the rescaled time
      $t_L\equiv t/L$, where $\delta M_{\rm t}$ in Eq.~\eqref{submat}
      quantifies the finite-size effects. Panels from top to bottom
      are for $g=2, \, 1.1, \,1, \, 0.9, \,0.5$.  Colored curves stand
      for various system sizes, as indicated in the legend. They
      support a power-law suppression of the finite-size revival
      effects as $L^{-1/3}$.  Adapted from Ref.~\cite{RV-20-qu}.}
    \label{fig:Rescal1_full_g}
  \end{center}
\end{figure}

In finite systems, quantum quenches from the disordered phase ($g_i>1$)
show the emergence of peculiar revival phenomena (see, e.g.,
Refs.~\cite{HHH-12, KLM-14, Cardy-14, JJ-17, JJLM-19, JA-20, MAC-20,
  RV-20-qu}).
This is evident from the data shown in Fig.~\ref{fig:Rescal1_full_g},
for the subtracted transverse magnetization
\begin{equation}
  \delta M_{\rm t}(t;L,g_i,g) \equiv M_{\rm t}(t;L,g_i,g)
  - \Sigma(t;g_i,g) \,,
  \label{submat}
\end{equation}
$\Sigma(t;g_i,g)$ being the infinite-size limit in
Eq.~\eqref{ssigmadef}, for finite systems of size $L$ with PBC and
$g_i\to\infty$.  The numerical results show the following behavior
\begin{equation}
  \delta M_{\rm t}(t;L) = L^{-a} f_e(t_L) \, f_o(t;L) + O(L^{-1})\,, \qquad
  t_L \equiv t/L \,,\qquad a \approx 1/3 \,,
  \label{asyd}
\end{equation}
where $f_o(t;L)$ is a rapidly oscillating function around zero
depending on both $t$ and $L$, while the envelope function $f_e(t_L)$ is a
(non-oscillating) function of $t_L$ with apparent discontinuities
located at $t_L=t_{L,k}$. This behavior of the function $f_e$ is
essentially related to revival phenomena, which appear at times
\begin{equation}
  t_{L,k} \equiv  {k\over 2v_m}\,,  \qquad k = 1,2, \ldots\,,
  \qquad v_m = 2 \, {\rm Min} [g,1]\,,
  \label{tk}
\end{equation}
where $v_m$ is the maximum velocity of the
quasi-particle modes~\cite{LR-72, CC-05, CEF-12a}.  The amplitude of
such discontinuities generally tends to decrease with increasing $k$,
as it can be seen from the various panels of
Fig.~\ref{fig:Rescal1_full_g}.  In particular, numerical results for
$g=g_c=1$ show that the first sharp dip is asymptotically located at
$t_{L,1} = 1/4 + O(L^{-(1-a)})$.  This can be related to the
interference between the signals traveling in the opposite direction
with velocity $v_{m} = 2$,~\cite{KLM-14} taking a time
$t = L/(2 v_m) = L/4$ to approach each other.  The scaling
behavior~\eqref{asyd} has been observed in numerical simulations
for any value of $g$; the functions $f_e$ and $f_o$ change with $g$, but
maintain a similar structure.  Therefore, the main features of the
scaling laws associated with the finite-size revival effects (in
particular its power law $L^{-a}$) are not related to the existence
of quantum criticality at $g_c$.

We finally mention that similar revival phenomena are also observed
when quenching from the ordered phase and looking at the longitudinal
magnetization in finite-size systems~\cite{RV-20-qu}.

\subsubsection{Moving away from integrability}
\label{nonintegrhq}

Numerical studies of the effects of moving Ising-like systems away
from integrability, for example by adding nonintegrable Hamiltonian
terms as those of the anisotropic next-to-nearest-neighbor Ising
(ANNNI) models~\cite{Dutta-etal-book}, have shown that some
qualitative features of the post-quench dynamics persist.  In
particular, the evolution of the longitudinal magnetization displays
qualitative differences when the quenches are performed within the
ordered phase or crossing the transition point toward the disordered
phase~\cite{RV-20-qu}.

As originally proposed in Ref.~\cite{Haldar-etal-20}, hard-quench
protocols may be used to find evidence of phase transitions even
in nonintegrable systems.  For this purpose, one can exploit some
features of the intermediate-time dynamics of local
observables and the entanglement entropy after quantum quenches
to the critical point of an underlying equilibrium QT.
Numerical investigations on the derivatives of such quantities
with respect to the quench parameter, in ANNNI chains,
develop strong dips/peaks in the vicinity of the QT.

\section{Out-of-equilibrium dynamics arising from slow Kibble-Zurek protocols}
\label{KZdynamics}

In this section we focus on the dynamics arising from
slow changes of the Hamiltonian parameters across CQTs.  Dynamic
KZ protocols involving such slow changes are generally
performed within systems described by the general Hamiltonian
\begin{equation}
  \hat{H}(t) = \hat{H}_{c} + w(t) \, \hat{H}_{p} \,,
  \label{hlamt}
\end{equation}
where $\hat{H}_c$ and $\hat{H}_p$ do not depend on time.  As assumed
in Eq.~\eqref{hlam}, also in this case we suppose that
$[\hat{H}_c,\hat{H}_p]\neq 0$ and that $\hat{H}_c$ represents a
critical Hamiltonian at its CQT point.  The tunable parameter $w$
controls the strength of the coupling with the perturbation
$\hat{H}_p$, and is again taken as a relevant parameter driving the
CQT, such as the longitudinal ($h$) or the transverse ($r=g-g_c$)
external field strength in the quantum Ising models~\eqref{hisdef}, or
the chemical potential $r=\mu-\mu_c$ in the Kitaev fermionic
models~\eqref{kitaev2}.  Therefore $w_c = 0$ corresponds to the
transition point.

Across a QT, the growth of an out-of-equilibrium dynamics is
inevitable in the thermodynamic limit, even for very slow changes of
the parameter $w$, because large-scale modes are unable to equilibrate
as the system changes phase. Indeed, when starting from the ground
state associated with the initial value $w_i$, the system cannot pass
through ground states associated with the time dependence of $w(t)$
across the transition point, thus departing from an adiabatic
dynamics. This gives rise to a residual abundance of defects after
crossing the transition point~\cite{Kibble-76, Kibble-80, Zurek-85,
  Zurek-96, ZDZ-05, Polkovnikov-05, Dziarmaga-05, PG-08, Dziarmaga-10,
  Dutta-etal-book, CEGS-12, DZ-14, PSSV-11, Damski-05, USF-07, USF-10}.

\subsection{Kibble-Zurek protocols}
\label{KZprot}

Slow (quasi-adiabatic) passages through QTs allow us to probe some
universal features of quantum fluctuations in such circumstances.  In
this respect, a special role is played by the KZ
problem~\cite{Kibble-76, Zurek-85}, related to the amount of final
defects that are generated when slowly moving the system from the
disordered to the ordered phase.  KZ-like protocols have been largely
employed to investigate the critical dynamics of closed systems,
subject to unitary time evolutions.  To this purpose, one can assume
that negative values $w < 0$ correspond to a gapped quantum disordered
phase. Quasi-adiabatic passages through the QT are obtained by slowly
varying $w$ across $w_c = 0$, following, e.g., the standard procedure:

\begin{itemize}
\item[$\bullet$] One starts from the ground state of the many-body system at
  $w_i < 0$, given by $|\Psi(t=0)\rangle \equiv |\Psi_0(w_i)\rangle$.
  
\item[$\bullet$] Then the out-equilibrium unitary dynamics, ruled by the
  Schr\"odinger equation
  \begin{equation}
    {{\rm d} \, |\Psi(t)\rangle \over {\rm d} t} =
    - i \, \hat H(t) \, |\Psi(t)\rangle \,,
    \label{unitdyn}
  \end{equation}
  arises from a linear dependence of the time-dependent parameter
  $w(t)$, such as
  \begin{equation}
    w(t) = c \,t\,, \qquad t_s \equiv c^{-1}>0\,,
    \label{wtkz}
  \end{equation}
  up to a final value $w_f>0$. Therefore the KZ protocol starts at
  time $t_i = t_s \, w_i<0$ and stops at $t_f= t_s \, w_f>0$.
  The parameter $t_s$ denotes the time scale of the slow variations
  of the Hamiltonian parameter $w$~\footnote{The linear time
  dependence~\eqref{wtkz} can be also extended to more general time
  dependencies, such as generic power laws
  $w(t) = {\rm sign}(t) \, |t/t_s|^a$, with $a>0$.}.
\end{itemize}

The resulting out-of-equilibrium evolution of the system can be
investigated by monitoring observables and correlations at fixed time,
such as those reported in Sec.~\ref{timedepproto}.  Other similar
protocols have been introduced (see, e.g., Refs.~\cite{Dziarmaga-10,
  CEGS-12}), for example crossing the QT with different power-law
dependencies of $w(t)$. The scaling laws reported below can be
straightforwardly adapted to these more general cases.

\subsection{The Kibble-Zurek mechanism}
\label{KZprob}

The so-called KZ mechanism, associated with slow passages through
critical points, was introduced by Kibble~\cite{Kibble-76, Kibble-80}
in the context of the expanding universe arising from the Big Bang,
and then recast in the language of critical phenomena by
Zurek~\cite{Zurek-85}.  Their proposal is a theory of the defects
generated in a system being slowly cooled across a continuous
phase transition. The system
inevitably goes out of equilibrium, ending up in the broken-symmetry
phase with different spatial regions realizing different orientations
of the broken symmetry, and topological defects as a result.

The KZ problem~\cite{Zurek-96} at CQTs is related to the abundance of
defects arising from the KZ protocol due to the inevitable loss of
adiabaticity when crossing a CQT, even in the limit of very slow
variations of the driven parameter. For asymptotically slow changes of
the parameters around the CQT, the dynamic behavior is expected to be
essentially controlled by the universal features associated with the
universality class of the equilibrium CQT.  Far away from the critical
point within the disordered phase [i.e., $w(t)\lesssim -1$], the
equilibrium relaxation time $t_r$ is very small with respect to the
time scale $t_s$ of the variation of $w$, thus enabling an {\em
  adiabatic} dynamics through the ground states at the instantaneous
values of $w(t)$.  On the other hand, close to the transition point
[i.e., $|w(t)|\ll 1$], the dynamics gets approximately {\em frozen},
due to the divergence of the equilibrium relaxation time (critical
slowing down), behaving as $t_r \sim |w|^{-z\nu}$.  The critical modes
cannot adjust to the change of the driving parameter $w(t)$, giving
rise to an out-of-equilibrium dynamics.

\begin{figure}
  \begin{center}
    \includegraphics[width=0.6\columnwidth]{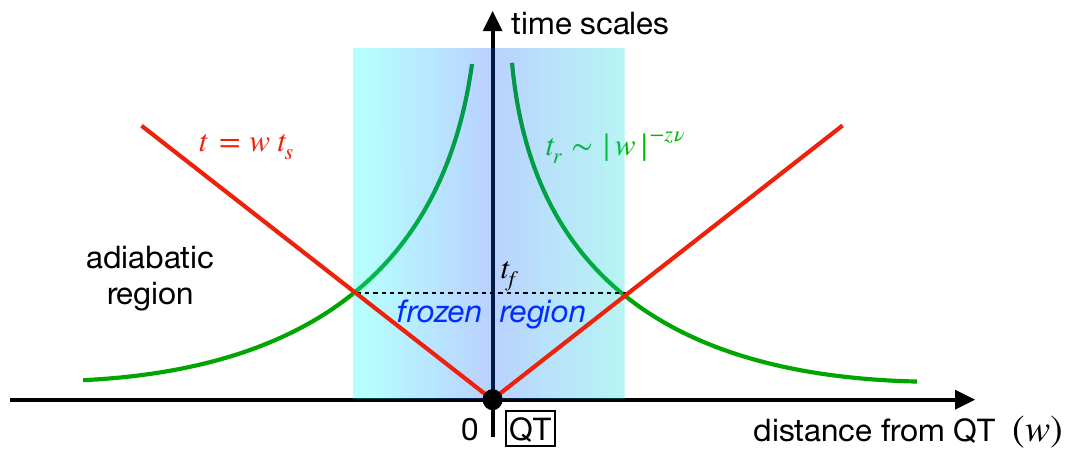}
    \caption{A sketch containing the idea beyond the KZ mechanism.
      The green and red curves respectively indicate the relaxation
      time $t_r$ and the time $t$ to reach the QT point (located at
      $w=0$), as a function of the parameter $w$.  The intersection
      between the two curves denotes the time $t_f$ at which the
      system ceases to behave adiabatically and enters a region where
      the dynamics remains approximately frozen.}
    \label{fig:Sketch_KZ}
  \end{center}
\end{figure}

A phenomenological derivation of the KZ mechanism has been reported in
the pioneering works~\cite{Kibble-76, Zurek-85, ZDZ-05} and is
sketched in Fig.~\ref{fig:Sketch_KZ}.  The turning point from the
adiabatic to the out-of-equilibrium frozen regime occurs at the time
$t_f$ when the relaxation time $t_r$ is of the order of the remaining
time to reach the value $w=0$,
\begin{equation}
  t_r\sim |w(t_f)|^{-z\nu} \approx |t_f|\,, \quad {\rm thus}\quad
  |t_f| \sim t_s^{z\nu/(1 + z\nu)}\,.
  \label{trf}
\end{equation}
Such {\em freezing} time $t_f$ allows us to define a corresponding
length scale $\lambda_f \equiv (t_f/t_s)^{-\nu}=
t_s^{\nu/(1+z\nu)}$. The number of defects eventually emerging from
the frozen region is thus expected to follow the power law
\begin{equation}
  \rho_{\rm def} \sim \lambda_f^{-d} = t_s^{-{d\nu\over 1+z\nu}}\,.
  \label{rhodefori}
\end{equation}
An analogous power law can be also derived by arguments based on the
adiabatic perturbation
theory~\cite{Polkovnikov-05, Dziarmaga-10, Dutta-etal-book, PSSV-11}.

We mention that, beside the average density of defects $\rho_{\rm def}$,
also the corresponding
full counting statistics can be shown to exhibit universal features.
In particular, all the cumulants of the defects number distribution
(including the variance, third-centered moment, etc.) inherit
a universal power-law scaling with the driving time, thus generalizing
the KZ prediction of Eq.~\eqref{rhodefori}~\cite{delCampo-18, GMD-20}.

\subsection{Dynamic scaling arising from Kibble-Zurek protocols}
\label{dynscaKZ}

Dynamic scaling laws are expected to develop in the limit of large
time scale $t_s$ of the driven parameter $w(t)$.  A phenomenological
scaling theory is obtained by assuming the homogeneous scaling laws
reported in Sec.~\ref{homscaloaws}, taking into account that the
parameter $w(t)=t/t_s$ has a linear time dependence, when
crossing the critical point $w=0$.  For example, for generic
observables and correlation functions defined using local operators
$\hat{O}$, we consider the homogeneous scaling laws
\begin{subequations}
  \begin{eqnarray}
    O(t,t_s;L,w_i) \!\! & \!\! \equiv \!\! & \!\!
    \big\langle \hat{O} \big\rangle_t
    \approx b^{-y_o} \, {\cal O}(b^{-z} t, b^{y_w} w(t), b^{-1} L, b^{y_w} w_i)\,,
    \label{Oscadynwt}  \\ 
    G_{12}(t,t_s,{\bm x};L,w_i) \!\! & \!\! \equiv \!\! & \!\! \big\langle 
    \hat{O}_1({\bm x}_1) \hat{O}_2({\bm x}_2) \big\rangle_t
    \approx b^{-\varphi_{12}} \, {\cal G}_{12}(b^{-z} t, b^{y_w} w(t),
    b^{-1}{\bm x}, b^{-1}L, b^{y_w} w_i)\,, \qquad \label{g12scadynwt}
  \end{eqnarray}
  as working hypotheses for the derivation of the dynamic
  scaling laws in the thermodynamic and FSS limits~\footnote{The
    spatial dependence on ${\bm x}={\bm x}_2-{\bm x}_1$ alone
    in Eq.~\eqref{g12scadynwt} reflects the assumption
    of translation invariance.}.
  \label{scadynwt}
\end{subequations}

\subsubsection{Dynamic scaling in the thermodynamic limit}
\label{dynkzinfL}

To derive a dynamic scaling theory for infinite-volume systems, it is
possible to exploit the arbitrariness of the scale parameter $b$ in
Eqs.~\eqref{scadynwt}.  For this purpose we set
\begin{equation}
  b = \lambda \equiv t_s^{1\over y_w+z} \,,
  \label{sbla}
\end{equation}
where $\lambda$ is the length scale associated with the KZ protocol,
and take the thermodynamic limit $L/\lambda\to\infty$.  By simple
manipulations, this leads to the dynamic KZ scaling ansatze
\begin{equation}
  O(t,t_s;w_i) \approx  \lambda^{-y_o}\, {\cal O}_\infty(\Omega_t,
  \lambda^{y_w} w_i)\,,\qquad
  G_{12}(t,t_s,{\bm x}; w_i)  \approx  \lambda^{-\varphi_{12}}
  \,{\cal G}_\infty(\Omega_t, {\bm x}/\lambda, \lambda^{y_w} w_i)\,,
  \label{kzG}
\end{equation}
where $\Omega_t$ is the rescaled time
\begin{equation}
  \Omega_t \equiv {t \over t_s^{\kappa}}\,,\qquad \kappa = {z \over y_w+z}\,.
  \label{tauvar}
\end{equation}
The dynamic KZ scaling limit, where the above asymptotic behaviors
apply, is obtained by taking $t_s\to\infty$, keeping the arguments of
the scaling functions ${\cal O}_\infty$ and ${\cal G}_\infty$ fixed.
Actually, introducing a temporal scaling variable related to the initial
time of the KZ protocol, defined as
\begin{equation}
  \Omega_{t_i} \equiv {t_i \over  t_s^{\kappa}}\,,\qquad t_i = w_i\, t_s\,,
  \label{tauidef}
\end{equation}
we may rewrite the scaling Eqs.~\eqref{kzG} as
\begin{equation}
 O(t,t_s;w_i) \approx \lambda^{-y_o}\, \widetilde {\cal
   O}_\infty(\Omega_{t}, \Omega_{t_i} )\,,\qquad
  G_{12}(t,t_s,{\bm x};w_i) \approx  \lambda^{-\varphi_{12}} \,
  \widetilde {\cal G}_\infty(\Omega_{t}, {\bm x}/\lambda, \Omega_{t_i})\,.
  \label{kzG2}
\end{equation}

Since the KZ protocol starts from $w_i<0$ corresponding to the gapped
phase, whose gap decreases as $\Delta\sim \xi^{-z}$ and the
ground-state length scale $\xi$ diverges only at the critical point
$w=0$, the emerging dynamic KZ scaling should be independent of the
actual finite value of $w_i< 0$, if this is kept fixed in the KZ
scaling limit.  This is due to the fact that, in a gapped phase, the
evolution arising from slow changes of the parameters is essentially
adiabatic, from $w_i$ to the relevant scaling interval $I_w$ around
$w=0$, which effectively decreases as
\begin{equation}
  I_{w} \sim t_s^{-1+\kappa} \;\to\; 0
  \label{deltamu}
\end{equation}
in the dynamic KZ scaling limit.  Therefore, when increasing $t_s$,
keeping $w_i<0$ constant and finite, the dynamic KZ scaling must be
independent of $w_i$, corresponding to the $\Omega_{t_i} \to -\infty$
limit of the relations~\eqref{kzG2}.  When keeping $w_i<0$ fixed, this
leads to the dynamic scaling ansatze
\begin{equation}
  O(t,t_s;w_i)\approx \lambda^{-y_o}\, {\cal O}_\infty (\Omega_t)\,,
  \qquad
  G_{12}(t,t_s,{\bm x};w_i) \approx \lambda^{-\varphi_{12}} \, {\cal
    G}_\infty (\Omega_t,{\bm x}/\lambda)\,,
  \label{sinflimG}
\end{equation}
for fixed-time observables. These scaling laws can be extended to
correlations at different times~\cite{CEGS-12}.

One may also define an out-of-equilibrium correlation length $\xi_{\rm
  KZ}(t)$ from the large-distance exponential decay of the two-point
equal-time correlation function of the site variables, or its second
moment.  Using the above scaling behaviors, in particular
Eq.~\eqref{sinflimG}, one gets this scaling ansatz (keeping $w_i$
fixed):
\begin{equation}
  \xi_{\rm KZ}(t,t_s) \approx t_s^{\kappa/z} \,{\cal L}(\Omega_t)\,.
  \label{xikzsca}
\end{equation}

The dynamic scaling functions introduced above are expected to be
universal with respect to changes of the microscopic details of the
Hamiltonian within the given universality class.  Of course, like any
scaling function at QTs, such a universality holds, apart from a
multiplicative overall constant and normalizations of the scaling
variables.  The approach to the asymptotic dynamic scaling behavior is
expected to be generally characterized by power-law suppressed
corrections, such as those discussed at equilibrium (see
Sec.~\ref{escalingcqt}).

As already mentioned, the so-called KZ problem genuinely addresses the
formation of defects when slowly crossing the QT, from the disordered
to the ordered phase. The defect number arising from the
out-of-equilibrium condition is expected to scale as the inverse of
the scaling volume, that is,
\begin{equation}
  \rho_{\rm def} \sim \xi_{\rm KZ}^{-d}\approx t_s^{-{d\over
      y_w+z}} \, {\cal R}_{\rm def}(\Omega_t)\,,
  \label{rhodef}
\end{equation}
in agreement with Eq.~\eqref{rhodefori} identifying $y_w=1/\nu$.
However, we should note that Eq.~\eqref{rhodef} is derived
in the KZ scaling limit, i.e., in the large-$t_s$ limit keeping
$\Omega_t = t/t_s^\kappa$ finite, or equivalently when the final value
$w_f$ of $w$ scales appropriately around $w=0$ (i.e., as $w_f \sim
t_s^{-1+\kappa}$ when increasing $t_s$).  In the large-$\Omega_t$
limit, important dynamic effects related to the ordered region and its
degenerate lowest states may set in, particularly when the global
symmetry is preserved by the KZ protocol and its initial
state. Therefore, the determination of the asymptotic power law of the
defect number would also require some knowledge of the asymptotic
behavior of the scaling function ${\cal R}_{\rm def}(\Omega_t)$ in
Eq.~\eqref{rhodef}. In particular, in order to recover the KZ
prediction~\eqref{rhodefori} from the scaling law~\eqref{rhodef},
it is required that $\lim_{\Omega_t \to \infty} {\cal R}_{\rm
  def}(\Omega_t)$ converges to a finite and constant value.  This
issue is relevant for possible comparisons of the KZ
prediction~\eqref{rhodefori} for the number of defects with
measurements taken in the very large time limit, deep in the ordered
phase.

For example, classical systems show coarsening phenomena when they are
quenched to an ordered phase with multiple degenerate low-energy
states. Local broken-symmetry regions grow in time and the system is
asymptotically self-similar on a characteristic coarsening length
scale $\ell_c$, which generally increases as a power law of the
time~\cite{Bray-94}.  The correlation functions approach their
equilibrium value on the scale $\xi_e\ll \ell_c$ within each
quasi-homogeneous domain, but they get exponentially suppressed
between different domains.  In the late-time regime, a dynamical
coarsening scaling is eventually realized when there are no growing
scales competing with $\ell_c$.  As argued in Ref.~\cite{CEGS-12},
coarsening dynamic phenomena are expected to become relevant in the
large-time regime of the KZ protocols, deep in the ordered phases,
determining the asymptotic $\Omega_t\to\infty$ behavior of the KZ
scaling functions.

The general features of the KZ dynamic scaling in the thermodynamic
limit, and in particular the KZ predictions for the abundance of
residual defects, have been confirmed by several analytical and
numerical studies (see, e.g., Refs.~\cite{Dziarmaga-10,
  Dutta-etal-book, PSSV-11, CEGS-12, NDP-13} and citing references).
In particular, the computations~\cite{Dziarmaga-05, CL-06,
  Dziarmaga-10} for Ising and XY chains using KZ protocols driven by
the transverse field $g$ confirm the scaling behavior predicted by
Eq.~\eqref{rhodef}, i.e.  $\rho_{\rm def}\sim t_s^{-1/2}$. Further
results supporting the KZ dynamic scaling will be presented in
Sec.~\ref{kitaevKZ} for the related fermionic Kitaev wire.  The KZ
defect production has been also investigated at multicritical
points~\cite{DMDS-09, DOV-09, MD-10}.  Results for quantum-information
quantities, such as the entanglement, the concurrence and the discord,
during KZ protocols can be found in
Refs.~\cite{CDRZ-07, SS-09, NPD-11, NDP-13, CENTV-14, TTD-14, HYZ-15}.
We finally mention that the scaling law of the abundance of residual
defects after crossing phase transitions has been also verified
in experiments for various physically interesting systems
(see, e.g., Refs.~\cite{DRGA-99, MMARK-06, SHLVS-06, WNSBD-08, CWBD-11,
  Griffin-etal-12, Ulm-etal-13, Pyka-etal-13, LDSDF-13, Braun-etal-15,
  Chomaz-etal-15, NGSH-15, Cui-etal-16, Gong-etal-16, Anquez-etal-16,
  CFC-16, Keesling-etal-19}.

\subsubsection{Dynamic finite-size scaling}
\label{dynkzfss}

The scaling Eqs.~\eqref{scadynwt} allow to derive dynamic FSS
relations, which are valid far from the thermodynamic limit,
and which extend those predicted by the FSS theory for systems
at equilibrium.  For example, by setting $b=L$ in Eq.~\eqref{g12scadyn},
we obtain
\begin{equation}
  G_{12}(t,t_s,{\bm x};L,w_i) \approx L^{-\varphi_{12}}\, {\cal
    G}_L(\Theta, {\bm X}, L^{y_w}t/t_s, \Omega_{t_i})\,,
  \label{kzfssG}
\end{equation}
where ${\bm X} = {\bm x}/L$, $\Theta = L^{-z} t$,
$\Omega_{t_i} = L^{y_w} w_i$, similarly to the case of quantum
quench protocols, cf.~Eq.~\eqref{scalvarque}.

This dynamic FSS behavior is expected to be obtained by taking
$L\to\infty$, while keeping the arguments of the scaling function
${\cal G}_L$ fixed.  One may introduce more convenient scaling
variables, which are combinations of those entering
Eq.~\eqref{kzfssG}.  For example, one can write it as
\begin{equation}
  G_{12}(t,t_s,{\bm x};L,w_i) \approx L^{-\varphi_{12}}\,
  \widetilde {\cal G}_L(\Omega_{t}, {\bm X}, \Upsilon, \Omega_{t_i})\,,
  \label{kzfssG2} 
\end{equation}
where 
\begin{equation}
  \Upsilon \equiv {t_s \over  L^{y_w+z}}\,, \label{iupsvar}
\end{equation}
and $\Omega_t,\,\Omega_{t_i}$ are defined in Eqs.~\eqref{tauvar}
and~\eqref{tauidef}, respectively.

Assuming again that the KZ protocol starts from the gapped disordered
phase and that the initial $w_i<0$ is kept fixed in the dynamic scaling
limit, the same dynamic FSS is expected to hold, irrespective of the
value of $w_i$.  Thus, the dynamic FSS behavior in Eq.~\eqref{kzfssG2}
simplifies into
\begin{equation}
  G_{12}(t,t_s,{\bm x};L,w_i) \approx L^{-\varphi_{12}} \, {\cal
    G}_{L,\infty} (\Omega_t, {\bm X}, \Upsilon)\,.
  \label{kzfssG2mufixed} 
\end{equation}
Indeed, with increasing $L$, the dynamic FSS occurs within a smaller
and smaller interval $I_w$ of values of $|w|$ around $w=0$: since the
time interval of the out-of-equilibrium process described by the
scaling laws scales as $t_{\rm KZ}\sim t_s^\kappa$, the relevant
interval $I_w$ of values of $|w|$ shrinks as
\begin{equation}
  I_w \sim {t_{\rm KZ} / t_s}\sim L^{-y_w}\,,
  \label{iwfss} 
\end{equation}
when keeping $\Upsilon$ fixed.

Note that, in the limit $\Upsilon\to\infty$, the evolution as a
function of $w(t)=t/t_s$ corresponds to an adiabatic dynamics. Indeed,
since the finite size $L$ guarantees the presence of a gap between the
lowest states, one may adiabatically cross the critical point if
$\Upsilon\to\infty$, passing through the ground states of the
finite-size system for $w(t)$. The adiabatic evolution across the
transition point is prevented only when $L\to\infty$ (before the limit
$t_s\to\infty$), i.e., when the time scale $t_{r}$ of the critical
correlations diverges, as $t_r\sim \Delta^{-1}\sim L^z$.

\subsection{Kibble-Zurek protocols within the 1$d$  Kitaev model}
\label{kitaevKZ}

\begin{figure}
  \begin{center}
    \includegraphics[width=0.47\columnwidth]{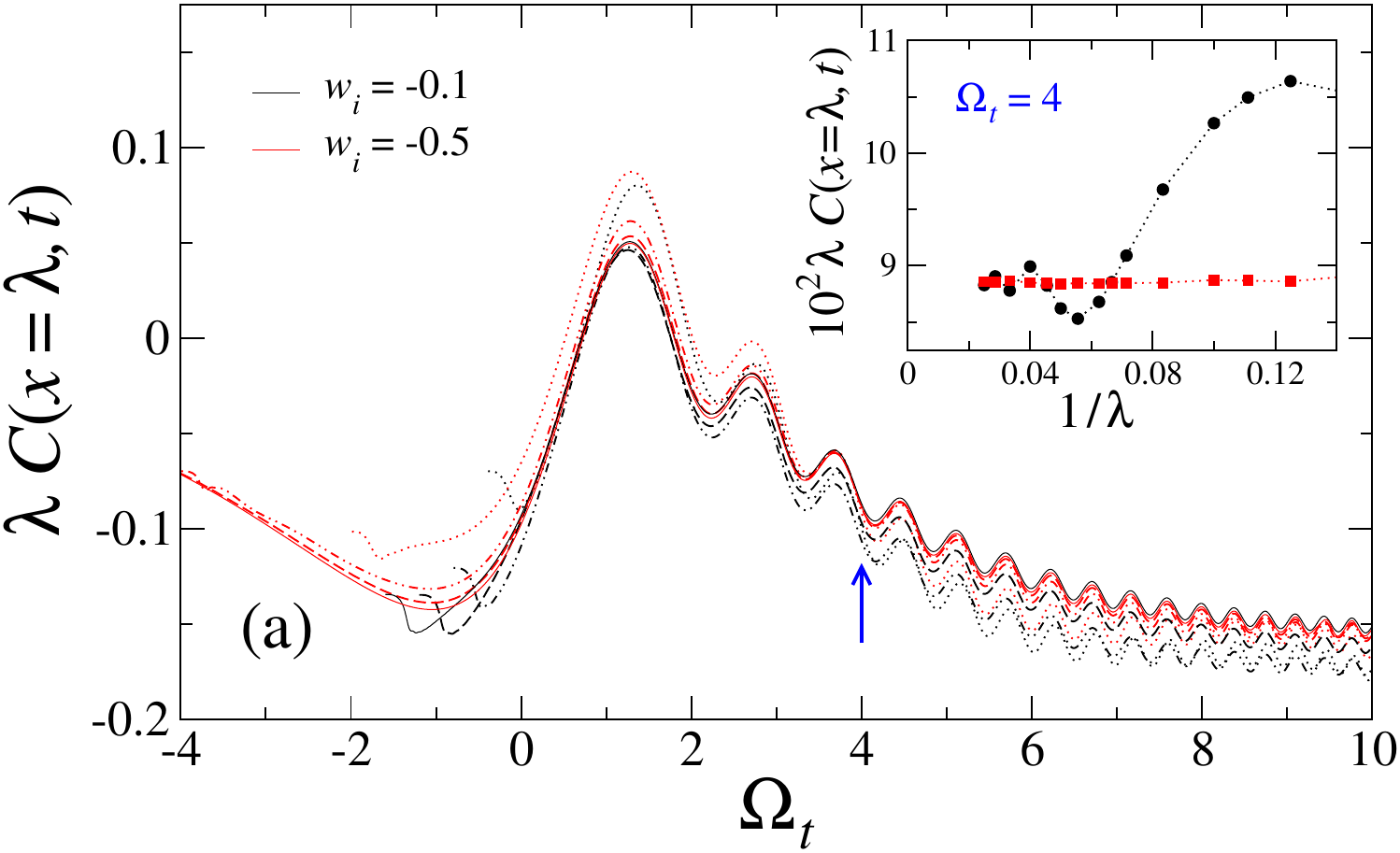}
    \hspace*{5mm}
    \includegraphics[width=0.47\columnwidth]{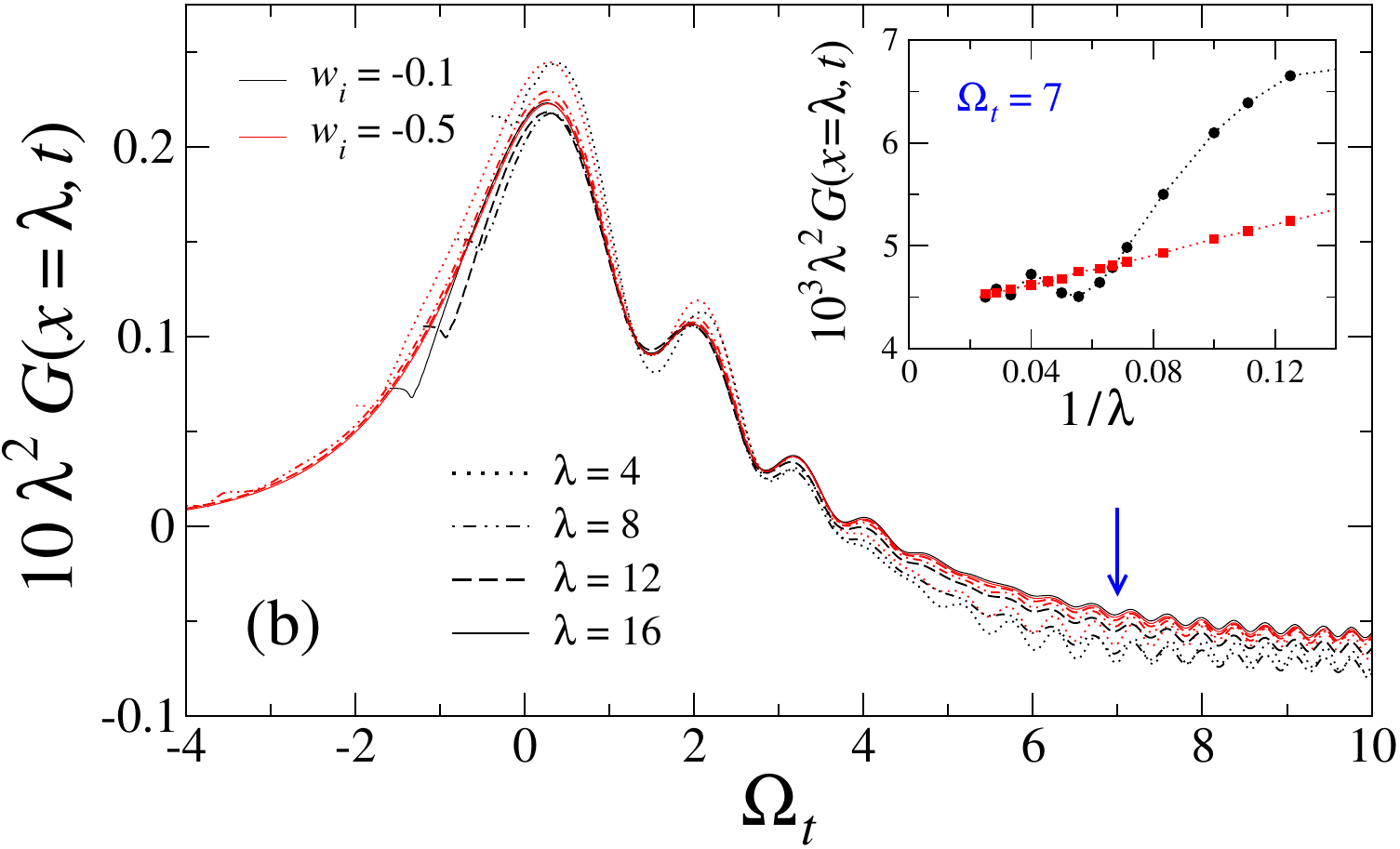}
    \caption{Rescaled correlations $\lambda \, C(x,t)$ $(a)$ and
      $\lambda^2 \, G(x,t)$ $(b)$, at fixed $x/\lambda = 1$ (results
      for other values of $x/\lambda$ show analogous behaviors), for
      the unitary dynamics of the Kitaev quantum wire in the
      thermodynamic limit, as a function of the rescaled time
      $\Omega_t$. The correlation functions $C$ and $G$ are defined in
      Eqs.~\eqref{eq:corrC} and~\eqref{eq:corrG}, respectively.
      Different line styles stand for various values of the length
      scale $\lambda$, from $4$ to $16$ [see legend in ($b$)].  Data
      belonging to one of the two color sets correspond to a given
      initial Hamiltonian parameter $w_i<0$, which is kept fixed and
      equal to either $w_i = -0.1$ (black curves and circles) or $w_i = -0.5$
      (red curves and squares).  The insets in the two panels display rescaled
      correlations as a function of $1/\lambda$, for both cases of
      $w_i$ presented in the main frames, at the $\Omega_t$ value
      indicated by the blue arrow.  In the figure, time evolutions up
      to length scales $\lambda \sim 50$ are shown [corresponding to
        time scales $t_s = \lambda^{y_w+z} = \lambda^2$ of the order
        $O(10^3)$], and have been obtained from much larger systems of
      size $L = 4096$.  Analogous results are obtained for the
      correlation function $P$, cf.~Eq.~\eqref{eq:corrP}.  Adapted
      from Ref.~\cite{RV-20-kz}. }
    \label{fig:PCG_TL_nodiss_mu}
  \end{center}
\end{figure}

The KZ scaling behaviors in the thermodynamic and FSS limits can be
accurately checked within the paradigmatic 1$d$ Kitaev fermionic
chain~\eqref{kitaev2}, for which results for very large systems can be
obtained by diagonalizing the Hamiltonian, as outlined in
Sec.~\ref{isingmodels}. We remark that, although the XY
chain~\eqref{XYchain} can be exactly mapped into the Kitaev quantum
wire in the thermodynamic limit, differences emerge when finite-size
systems are considered, because the nonlocal Jordan-Wigner
transformation does not preserve the BC (see the
discussion in Sec.~\ref{foqtisi}), with the exception of the simplest
OBC. In particular, Kitaev quantum wires with PBC or ABC do not map
into XY chains with analogous BC. This point is particularly relevant
for the dynamic FSS. For example, as already mentioned, the Hilbert
space of the Kitaev quantum wire with ABC is restricted with respect
to that of the quantum Ising chain, so that it is not possible to
restore the competition between the two vacua belonging to the
symmetric/antisymmetric sectors of the XY chain~\cite{Katsura-62,
  Kitaev-01, CPV-14}, controlled by the symmetry-breaking longitudinal
field $h$. 

The Kitaev quantum wire in Eq.~\eqref{kitaev2} undergoes a CQT
at $\mu_c=-2$, independently of $\gamma$, belonging to the 2$d$
Ising universality class.  Thus we
introduce the relevant parameter $w=\mu-\mu_c$, whose RG dimension
$y_w = 1$ determines the length-scale critical exponent $\nu = 1/y_w =
1$. The dynamic exponent associated with the unitary quantum dynamics
is $z = 1$.  To characterize the quantum evolution, various fixed-time
correlations of fermionic operators can be considered, such as
\begin{subequations}
\begin{eqnarray}
  P(x,t) \!\! & \!\! \equiv \!\! & \!\! \big\langle \hat c_j^\dagger 
  \hat c_{j+x}^\dagger + \hat c_{j+x} \hat c_{j} \big\rangle_t \, ,
  \label{eq:corrP} \\
  C(x,t) \!\! & \!\! \equiv \!\! & \!\! \big\langle
  \hat c_j^\dagger \hat c_{j+x} 
  + \hat c_{j+x}^\dagger \hat c_{j} \big\rangle_t \, ,
  \label{eq:corrC} \\
  G(x,t) \!\! & \!\! \equiv \!\! &
  \!\! \big\langle \hat n_j \hat n_{j+x} \big\rangle_t - 
  \big\langle \hat n_j \big\rangle_t \,
  \big\langle \hat n_{j+x} \big\rangle_t \,,
  \label{eq:corrG}
\end{eqnarray}
where $j,x \in [1,L/2]$, and averages are calculated over the state
of the system at time $t$, arising from the KZ protocol described in
Sec.~\ref{KZprot}, and starting from the ground state associated with
$w_i=\mu_i-\mu_c$.
\label{gpcntf}
\end{subequations}

According to the scaling arguments outlined in Sec.~\ref{kitaevKZ}, these
correlation functions are predicted to show dynamic scaling behaviors.
In particular, Eqs.~\eqref{sinflimG} and~\eqref{kzfssG2mufixed}
are expected to describe their asymptotic dynamic KZ scaling in the
thermodynamic and FSS limits, respectively, when the initial parameter
$w_i$ is kept fixed in the dynamic KZ limit, so that its dependence
asymptotically disappears.  The corresponding exponents are
$\varphi_{12} = 1$ for the correlations $P$ and $C$, and
$\varphi_{12} = 2$ for $G$, according to the RG dimensions $y_c=1/2$
of the fermionic operators $\hat{c}_x$, and $y_n=1$ of the density
operator $\hat{n}_x$~\cite{Sachdev-book}.

\begin{figure}
  \begin{center}
    \includegraphics[width=0.47\columnwidth]{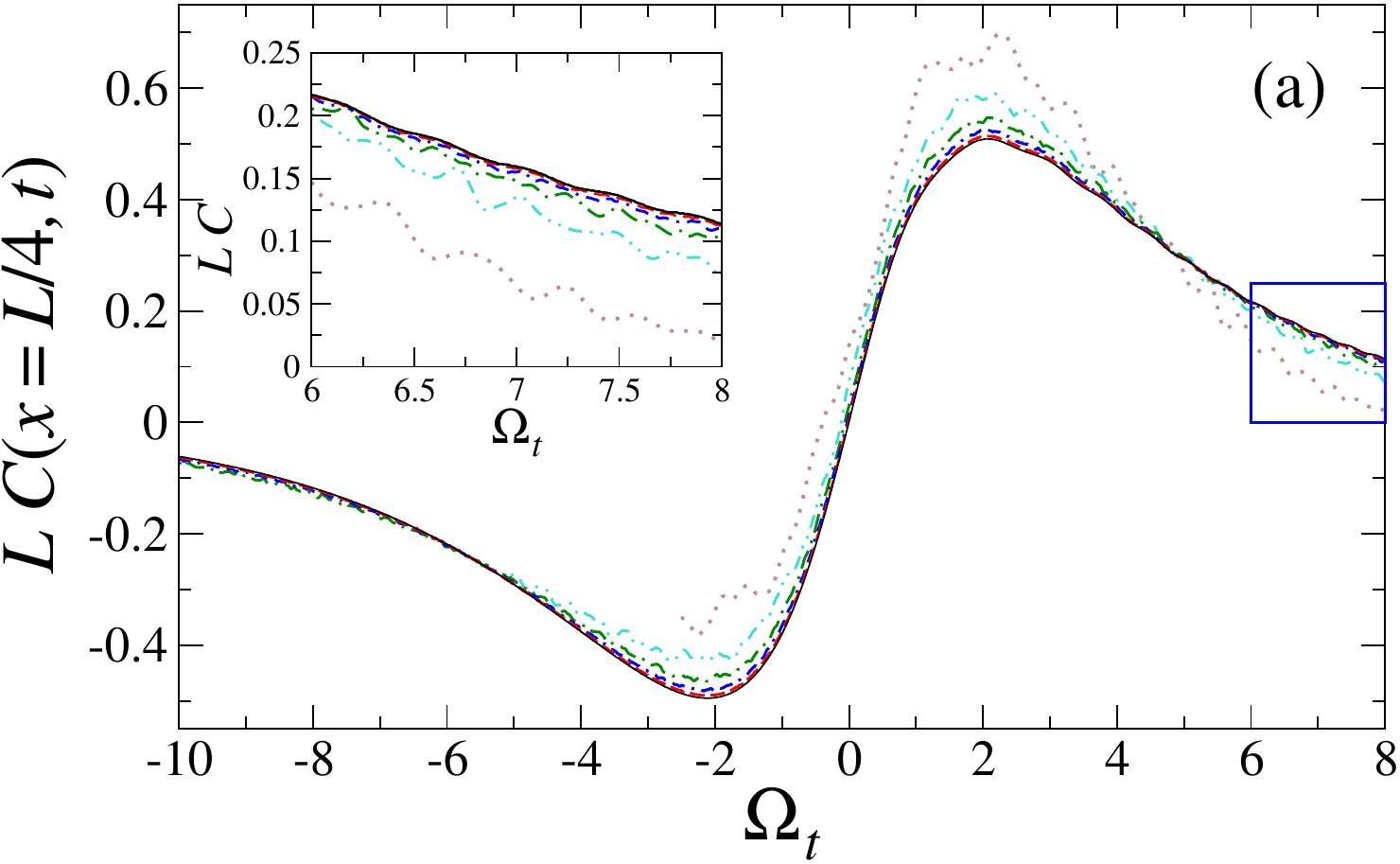}
    \hspace*{5mm}
    \includegraphics[width=0.47\columnwidth]{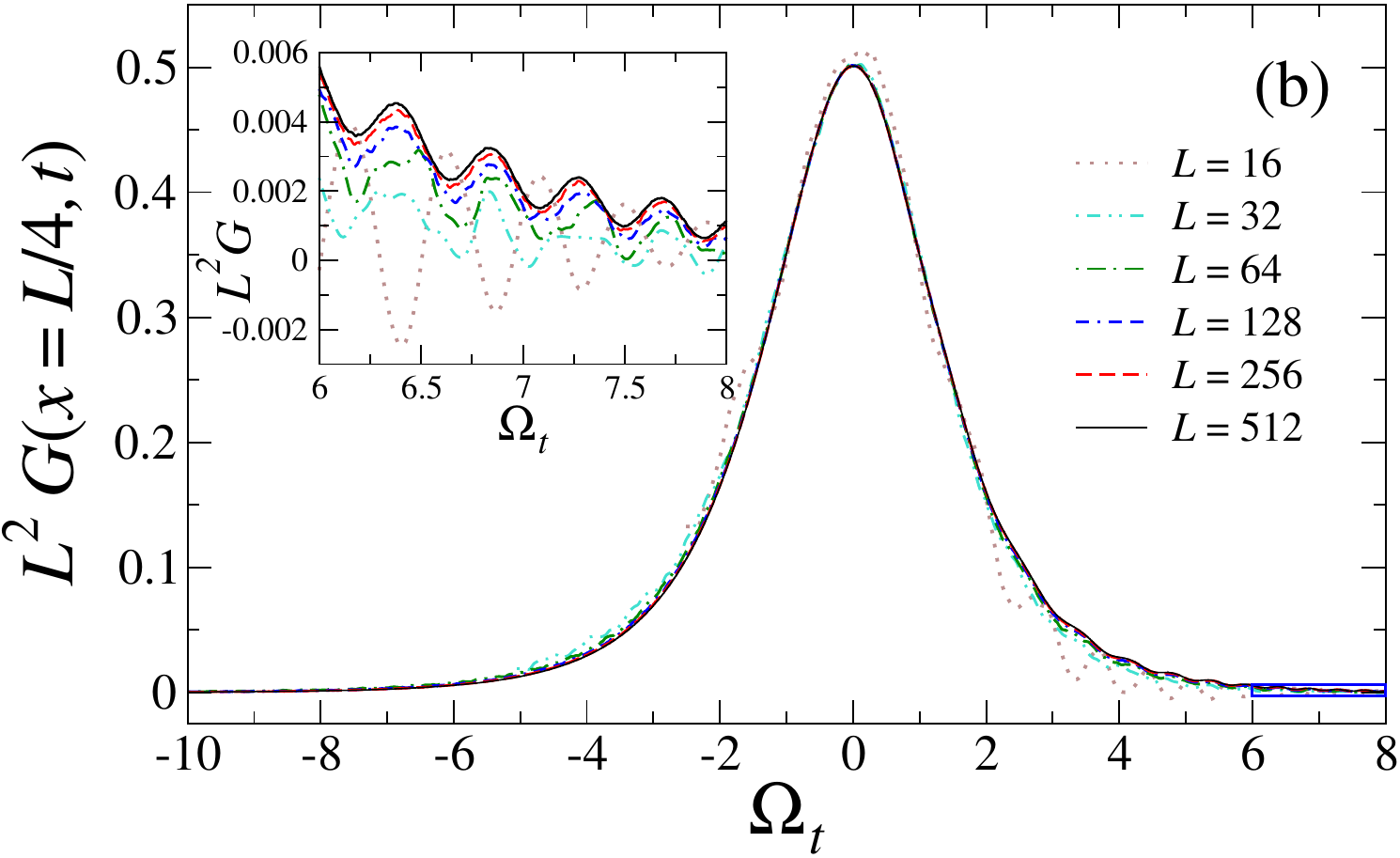}
    \caption{Rescaled correlations $L \, C(x,t)$ $(a)$ and $L^2 \,
      G(x,t)$ $(b)$, defined in Eqs.~\eqref{eq:corrC} and~\eqref{eq:corrG}
      respectively, as a function of the rescaled
      time $\Omega_t$, fixing the initial Hamiltonian parameter
      $w_i=-0.5$. We also fix $x/L = 1/4$ and $\Upsilon=0.1$.
      Different lines are for various sizes $L$, as indicated in the
      legend. The insets show magnifications of the main frames in the
      regions $6 \leq \Omega_t \leq 8$ enclosed in the blue boxes, to
      highlight discrepancies that reduce with increasing $L$.
      Analogous results can be obtained for the correlation function
      $P$, cf.~Eq.~\eqref{eq:corrP}.}
    \label{fig:CG_FSS_nodiss_mu}
  \end{center}
\end{figure}

Numerical results for the Kitaev wire with ABC (which are particularly
convenient for computations) confirm the dynamic KZ scaling
predictions for the above correlation functions. In particular,
Fig.~\ref{fig:PCG_TL_nodiss_mu} reports some results for the
correlations $C(x,t)$ and $G(x,t)$ in the thermodynamic limit,
keeping the initial
parameter $w_i$ fixed (two values of $w_i$ are considered in the
figure).  The results also show that the asymptotic scaling behaviors
match for different starting point $w_i$, thus confirming that the
asymptotic dynamic scaling is independent of $w_i$, when this is kept
fixed.  Fig.~\ref{fig:CG_FSS_nodiss_mu} shows some results for
the same correlation functions in the FSS limit.  Again the predicted
asymptotic scaling behavior is clearly approached, with scaling
corrections that are typically $O(1/L)$.

\subsection{Quantum annealing}
\label{sec:QAnnealing}

We finally discuss a somehow related issue, which deals with
the quantum dynamics arising from a gentle modulation in time
of one of the Hamiltonian control parameters.
The principle of adiabatic quantum dynamics has been adopted
in recently developed quantum devices for
quantum-computation purposes (see, e.g., Refs.~\cite{Johnson-etal-11,
  Boixo-etal-14} and citing references).  This leads to the so-called
{\em adiabatic quantum computing}, whose key principle is very simple:
evolving a quantum system from the ground state of a given Hamiltonian
$\hat H_i$ to that of another Hamiltonian $\hat H_f$, which encodes
the solution to the problem of interest.  The evolution should be
performed in such a way that the system is always in its instantaneous
ground state, by slowly varying the parameter $s(t): 0 \to 1$ in the
Hamiltonian
\begin{equation}
  \hat H[s(t)] = \big[ 1-s(t) \big] \, \hat H_i + s(t) \, \hat H_f \,,
  \qquad s(t_i) = 0\,, \quad s(t_f) = 1\,,
\end{equation}
and such protocol is usually referred to as quantum
annealing~\cite{KN-98, Fahri-etal-01, ALi-18}.

Among the possible quantum annealing protocols, a particularly relevant
case is found when crossing a QT, an occurrence giving rise to
the KZ mechanism, which also emerges when hard problems are addressed.
This kind of situation has been already implemented on physical
systems made of superconducting qubits~\cite{GDZZ-18} and of trapped
ions~\cite{Cui-etal-20}, with the purpose to measure the mean number
of topological defects (and the cumulants of the distribution)
generated by the (quasi-)adiabatic passage across
the QCP of prototypical many-body models, as the quantum
Ising chain in a transverse field.
However, it was soon realized that dissipation in certain physical
devices might play an important role (see also Sec.~\ref{KZdiss}).

For practical purposes, it can be shown that the solution to
computationally ``hard'' (NP) problems, encoded into the ground state
of $\hat H_f = \hat H[s(t_f)]$, cannot be adiabatically connected with
the ground state of an ``easy'' Hamiltonian $\hat H_i = \hat H[s(t_i)]$.
Indeed such connection typically reflects into the passage through a FOQT,
where the gap may close exponentially with the system size,
thus requiring an exponentially large amount of time for the annealing
process to reach the target state~\cite{JKKM-08, JLSZ-10, BFKSZ-13}.

\section{Dynamic finite-size scaling at first-order quantum transitions}
\label{foqtdynamics}

In this section we focus on systems at FOQTs, showing that they
develop a dynamic FSS besides the equilibrium FSS outlined in
Sec.~\ref{escalingfoqt}. In particular, even the dynamic behavior
across a FOQT is dramatically sensitive to the BC, giving rise to
out-of-equilibrium evolutions with exponential or power-law time
scales, unlike CQTs where the power-law time scaling is generally
independent of the type of boundaries.  Again, this is essentially
related to the sensitivity of the energy difference of the lowest
states to the boundaries, whether they are neutral or favor one of the
two phases separated by the FOQT. In particular, exponentially large
time scales are expected when the boundaries are such to give rise to
a quasi-level-crossing scenario, with exponentially suppressed energy
differences of the two lowest levels with respect to the rest of the
spectrum at larger energy, while power-law time scales are expected
for boundaries favoring domain walls, for which the low-energy
spectrum is dominated by their dynamics.

In the following, we consider again dynamic protocols entailing
instantaneous or slow changes of the Hamiltonian parameter, as
outlined in Secs.~\ref{timedepproto} and~\ref{KZprot}, respectively.
Specifically, the Hamiltonian is supposed to be written as in
Eq.~\eqref{hlamt}, being the sum of two terms, $\hat{H}(t) =
\hat{H}_{c} + w(t) \hat{H}_{p}$, where the unperturbed term
$\hat{H}_c$ assumes the parameter values corresponding to a FOQT, and
the other one $\hat{H}_p$ may be associated with an external varying
magnetic field.  For the quantum Ising models~\eqref{hisdef},
$\hat{H}_c$ may correspond to $h=0$ and any point along the FOQT line
for $g<g_c$. Therefore, $w$ corresponds to the magnetic field $h$.

As we shall see, the equilibrium FSS theory at FOQTs outlined in
Sec.~\ref{escalingfoqt} can be straightforwardly extended, essentially
by allowing for a time dependence associated with the further scaling
variable
\begin{equation}
  \Theta \equiv \Delta(L)\:t \,,
  \label{thetadeffo}
  \end{equation}
besides those already introduced at equilibrium, analogously to the
case of the dynamic FSS at CQTs [see Eq.~\eqref{eq:Theta_dyn} in
  Sec.~\ref{homscaloaws}].  Again, note that the size dependence of
the time scaling variable $\Theta$ may be exponential or power law,
depending on the boundaries and the corresponding size dependence of
the gap $\Delta(L)$.  For the sake of concreteness, we will focus on
the 1$d$ Ising chain, thus along its FOQT line for $g<g_c=1$. However,
analogous behaviors are expected for larger dimensions, with the
appropriate correspondence of the BC.

We mention that FOQTs have been studied in the context of adiabatic
quantum computation as well (see, e.g., Refs.~\cite{AC-09, YKS-10,
  LMSS-12}), due to their peculiar features of realizing an
exponentially suppressed energy difference of the lowest eigenstates.
Moreover, the out-of-equilibrium dynamical behavior has been also
investigated at classical first-order transitions, driven by the
temperature or by external magnetic fields (see, e.g.,
Refs.~\cite{Binder-87, PV-15, PV-16, PV-17-prl, PV-17-pre, PPV-18,
  SW-18, Fontana-19}).

\subsection{Quantum quenches at first-order quantum transitions}
\label{neutraldyn}

As described in Sec.~\ref{timedepproto}, the quench protocol
introduces the parameter $w_i$ determining the initial ground state,
and then the parameter $w\neq w_i$ determining the quantum evolution,
according to Eq.~\eqref{afterque}.  The dynamic FSS of observables can
be obtained by extending the one at equilibrium,
cf. Eq.~\eqref{efssm}, allowing for a time dependence through the
scaling variable $\Theta$ defined in Eq.~\eqref{thetadeffo}.  We also
need the scaling variables $\Phi_i$ and $\Phi$ associated with the
values of $w_i$ and $w$ involved in the quench protocol, which can be
defined as in Eq.~\eqref{kappah}, replacing $h$ with $w_i$ and $w$,
respectively.  The dynamic FSS limit is defined as the infinite-size
limit, keeping the scaling variables $\Phi_i$, $\Phi$ and $\Theta$
fixed.  In this limit, the central-lattice and global magnetizations
are expected to behave as
\begin{equation}
  M(t;L,w_i,w) \approx m_0 \, {\cal M}(\Theta,\Phi_i,\Phi)\,,
  \label{mcheckfoqt}
\end{equation}
where $m_0$ is the infinite-size magnetization approaching the FOQT in
the limit $h\to 0^+$, cf.~Eq.~\eqref{sigmasingexp}.  This scaling
behavior is expected to hold for any $g<1$, and the scaling function
${\cal M}$ should be independent of $g$, apart from trivial
normalizations of the arguments.  However, analogously to equilibrium
FSS at FOQTs, the size-dependence of the scaling variables and the
form of ${\cal M}$ significantly depend on the BC.

The above dynamic scaling relations can be straightforwardly extended
to any FOQT, by identifying the scaling variable $\Phi$ as the ratio
$\delta E_p(L)/\Delta(L)$, where $\delta E_p(L)$ is the energy
variation associated with the perturbation $w\hat{H}_p$ and
$\Delta(L)$ is the energy difference between the two lowest states at
the transition point. The scaling variable $\Theta$ associated with
time is always defined as in Eq.~\eqref{thetadeffo}.

\subsubsection{Neutral boundary conditions}
\label{quneubou}

In the case of neutral boundary conditions, such as OBC or PBC, one
may again exploit the two-level truncation to compute the scaling
function ${\cal M}$ associated with the magnetization, extending the
equilibrium computation reported in Sec.~\ref{fsspbcobc}.  As shown in
Ref.~\cite{PRV-18-dfss}, in the long-time limit and for large systems,
the scaling properties in a small interval around $w=0$ are captured
by a two-level truncation, restricting the problem to the two
nearly-degenerate lowest-energy states.  More precisely, this
approximation is appropriate when
\begin{equation}
  \delta E_w(L,w) = 2 m_0 |w| L^d = |\Phi| \, \Delta(L) \ll E_2(L)-E_0(L)\,.
  \label{twolappco}
\end{equation}
In the case at hand (i.e., for systems with PBC or OBC),
$E_2(L)-E_0(L)=O(1)$ in the large-$L$ limit.

The effective quantum evolution with the reduced two-dimensional
Hilbert space is determined by the Schr\"odinger equation with the
truncated Hamiltonian $\hat{H}_2(w)$, cf.~Eq.~\eqref{sceq}, using the
two-component ground state corresponding to $w_i$ as initial condition
at $t=0$. Simple calculations show that, apart from an irrelevant
phase, the time-dependent state $|\Psi_2(t)\rangle$ evolves
as~\cite{PRV-18-dfss},
\begin{equation}
  |\Psi_2(t)\rangle = \cos\left({\alpha_i-\alpha\over 2}\right) |0\rangle
  + e^{-i \, \Theta \sqrt{1 + \Phi^2}} \sin\left({\alpha_i-\alpha\over 2}\right)
  |1\rangle\,,
  \label{psitfo}
\end{equation}
with $\tan \alpha_i = \Phi_i^{-1}$, $\tan \alpha = \Phi^{-1}$, and
\begin{equation}
  |0\rangle = \sin(\alpha/2) \, |-\rangle +  
  \cos(\alpha/2) \, |+\rangle\,, \qquad 
  |1\rangle =  \cos(\alpha/2) |-\rangle -
  \sin(\alpha/2) \, |+\rangle\,,
\end{equation}
$|\pm \rangle$ being the eigenstates of the operator $\hat \sigma^{(3)}$.
Then, the expectation value $\langle \Psi_2(t) | \hat \sigma^{(3)}
|\Psi_2(t) \rangle$ gives the magnetization, which
produces the following dynamic scaling function defined in
Eq.~\eqref{mcheckfoqt}:
\begin{equation}
  {\cal M} = \cos (\alpha-\alpha_i)
  \cos \alpha + \cos \big(\Theta \sqrt{1+\Phi^2}
  \big) \sin(\alpha-\alpha_i) \sin \alpha\,.
  \label{m2lsca}
\end{equation}
Numerical results for the Ising chain with PBC~\cite{PRV-18-dfss} have
shown that the above scaling behavior is rapidly approached with
increasing $L$ (corrections in $L$ appear exponentially suppressed),
confirming that, when the energy scale of the dynamic process is of
the order of the energy difference of the lowest states, the dynamics
is effectively described by restricting the problem to the two
lowest-energy states of the Hilbert space. The above
behavior is expected at FOQTs where the two-level scenario applies,
therefore it holds for generic quantum Ising models in any
dimension, along their FOQT line.  In a sense, the two-level model
thus represents a large universality class for the asymptotic dynamic
FSS at FOQTs.

An analysis of the work fluctuations in quenches at FOQT is reported
in Ref.~\cite{NRV-19-wo}, where the scaling predictions are
compared with computations within the two-level framework.  Again,
numerical results nicely support them.

We finally mention that in the case of EFBC, favoring a magnetized
phase, such as those discussed in Sec.~\ref{fssfebc}, the situation is
subtler, due to the fact that the pseudo critical point is shifted.
However, the two-level scenario applies as well, in the appropriate
limit~\cite{PRV-20}.

\subsubsection{Boundary conditions giving rise to domain walls}
\label{qudomwall}

\begin{figure}
  \begin{center}
    \includegraphics[width=0.47\columnwidth]{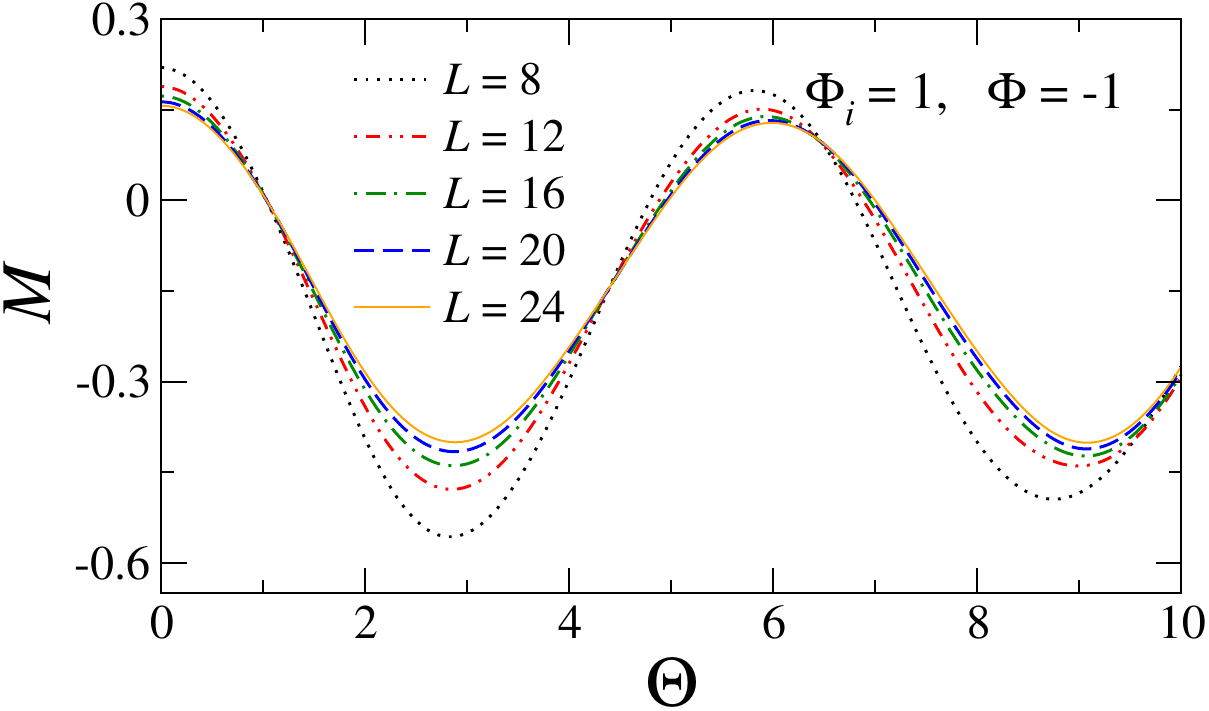}
    \hspace*{5mm}
    \includegraphics[width=0.47\columnwidth]{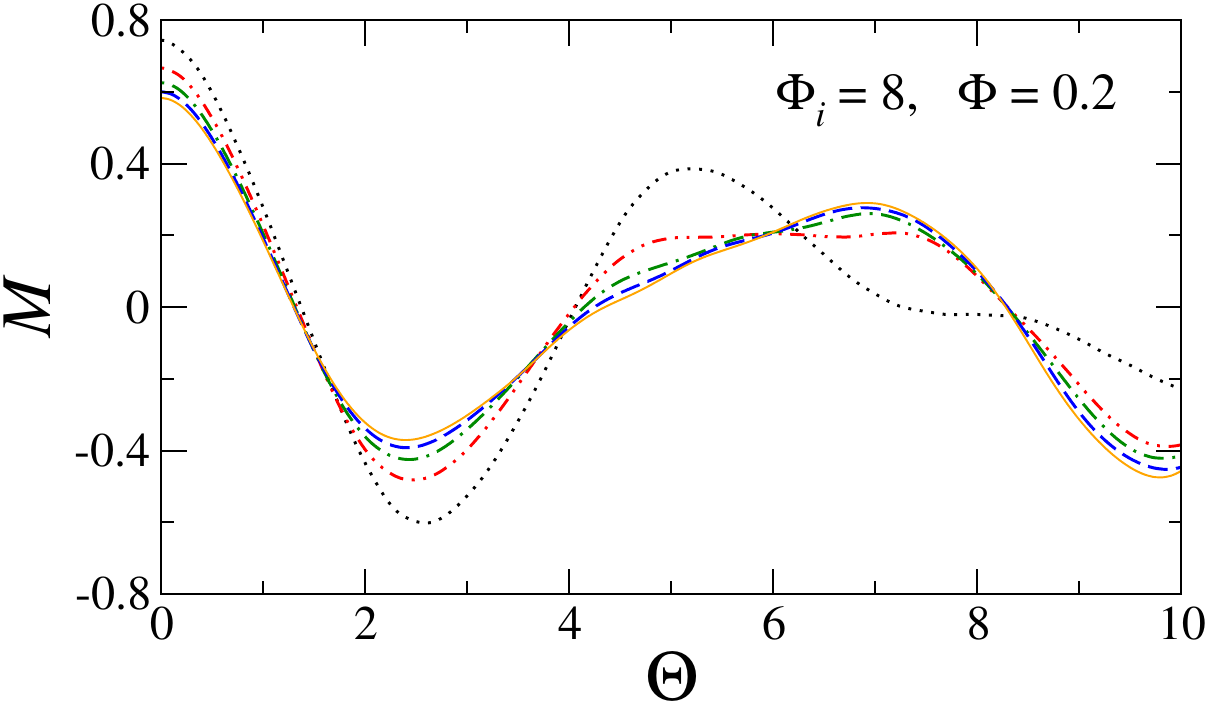}
    \caption{Average magnetization $M$ for the quantum Ising chain
      with OFBC, after a sudden quench of the longitudinal field close
      to the FOQT, as a function of the rescaled time variable
      $\Theta\sim L^{-2}\, t$.  We fix $g=0.5$ and the values of the
      rescaled variables $\Phi_i$ and $\Phi$ (see legends in the two
      panels).  Different data sets are for various chain lengths $L$,
      as indicated in the legend.  In both cases, as $L$ increases,
      the data nicely approach an asymptotic scaling function. Adapted
      from Ref.~\cite{PRV-20}.}
    \label{fig:Magnet_Quench_KL}
  \end{center}
\end{figure}

The two-level scenario does not apply to BC giving rise to domain
walls, such as Ising chains with ABC and OFBC, see
Sec.~\ref{fssabcfobc}, where the gap turns out to behave as
$\Delta(L)\sim L^{-2}$.  However the dynamic FSS behavior~\eqref{mcheckfoqt}
holds as well, with the appropriate scaling variables~\cite{PRV-20},
as shown in Fig.~\ref{fig:Magnet_Quench_KL}.  Analogous scaling
behaviors are expected in higher dimensions, for example in the
presence of more involved situations as for mixed boundaries, such as
ABC or OFBC along one direction, and PBC or OBC along the others.

\subsection{Dynamic scaling arising from  Kibble-Zurek protocols}
\label{dynkzfo}

We now focus on the dynamic behavior arising from show changes of the
Hamiltonian parameter $w(t)$, such as the KZ protocol outlined in
Sec.~\ref{KZdynamics}. In particular, we discuss the case in which the
initial value $w_i<0$ is kept fixed in the large-$L$ limit.  Since the
equilibrium FSS should be recovered in the appropriate limit, one of
the scaling variables can be obtained from the equilibrium scaling
variable introduced in Eq.~\eqref{kappah}, by replacing $h$ with
$w(t)$, i.e., we consider
\begin{equation}
  \Phi_{\rm KZ} \equiv {2 m_0 \, w(t) \, L^d \over \Delta(L)} = {2 m_0 t L^d
    \over \Delta(L) \: t_s}\,.
  \label{katdef}
\end{equation}
As a further scaling variable associated to the time, we may take
again $\Theta=\Delta(L) \,t$.
Let us also define the related useful scaling variable
\begin{equation}
  \Upsilon \equiv {\Theta\over \Phi_{\rm KZ}} =
           {\Delta^2(L) \: t_s\over 2 m_0 L^d}\,.
  \label{taudef}
\end{equation}
The dynamic FSS limit corresponds to $t, t_s, L\,\to\infty$, keeping the
above scaling variables fixed. In this limit the magnetization is
expected to show the dynamic FSS behavior
\begin{equation}
  M(t,t_s;w_i,L) \approx m_0\,{\cal M}(\Upsilon, \Phi_{\rm KZ}) \,,
  \label{mtsl}
\end{equation}
independently of $w_i<0$.  In the adiabatic limit ($t,t_s\to \infty$
at fixed $L$ and $t/t_s$), ${\cal M}(\Upsilon\to\infty, \Phi_{\rm
  KZ})$ must reproduce the equilibrium FSS
function~\eqref{efssm}. The scaling functions are expected
to be universal (i.e., independent of $g$ along the FOQT line, for a
given class of BC).  Note that the dynamic FSS behavior develops in a
narrow range of $w(t)\approx 0$; indeed, since $\Phi_{\rm KZ}$ is kept
fixed in the dynamic FSS limit, the relevant scaling behavior develops
within a range $I_w \sim \Delta/L^d$, thus the interval $I_w$ rapidly
shrinks when increasing $L$.  This implies that the dynamic FSS
behavior is independent of the initial value ($w_i$) of $w$.

The above scaling behaviors are expected to hold both for neutral
boundaries, such as PBC and OBC, and for BC giving rise to domain
walls, by inserting the corresponding gap $\Delta(L)$. Moreover, they
are quite general and can be straightforwardly extended to any FOQT,
with the appropriate correspondences.
Below we discuss the case of neutral boundaries, for which the
scaling functions can be computed by resorting to the two-level
framework. Results for other BC can be found in Ref.~\cite{PRV-20}.

Within the two-level reduction of the problem, the effective
time-dependent Hamiltonian reads
\begin{equation}
  \hat{H}_{r} = \varepsilon(t) \, \hat \sigma^{(3)} + \zeta \,
  \hat \sigma^{(1)}\,,\qquad \varepsilon(t) = {m_0 \, w(t) \, L^d}
  = m_0 t L^d / t_s\,,\qquad \zeta = \Delta(L) / 2\,.
  \label{hrdef2t}
\end{equation}
The dynamics is analogous to that governing a two-level quantum
mechanical system in which the energy separation of the two levels is
a function of time, which is known as the Landau-Zener
problem~\cite{Landau-32, Zener-32}.  Therefore, the two-component
time-dependent wave function for the quantum Ising ring can be derived
from the corresponding solutions~\cite{VG-96}.  If $|\Psi_2(t)\rangle$
is the solution of the Schr\"odinger equation with the initial condition
$|\Psi_2(t_i)\rangle = |+\rangle$ (where $|+\rangle$ is the positive
eigenvalue of $\hat \sigma^{(3)}$), the dynamic FSS function
of the magnetization can be computed by taking the expectation value
\begin{equation}
  {\cal M}(\Upsilon, \Phi_{\rm KZ}) = \langle \Psi_2(t)
  | \hat \sigma^{(3)}|\Psi_2(t)\rangle \,,
  \label{fsigmasol}
\end{equation}
so that~\cite{PRV-18-loc, PRV-20}
\begin{equation}
  {\cal M}(\Upsilon, \Phi_{\rm KZ}) = 1 - \tfrac12 \Upsilon e^{-{\pi
      \Upsilon\over 8}} \left| D_{-1+i{\Upsilon\over 4}} (e^{i{3\pi\over
      4}}\sqrt{\Upsilon} \,\Phi_{\rm KZ}) \right|^2 \,, \nonumber
\end{equation}
where $D_\nu(x)$ is the parabolic cylinder function~\cite{AS-1964}.
In this case, the approach to dynamic FSS should be controlled
by the ratio between $\Delta\sim e^{-cL}$ and $\Delta_2=O(1)$,
thus it is expected to be exponentially fast~\cite{PRV-18-loc}.

The above results are expected to extend to generic FOQTs for which
the two-level reduction apply, such as quantum Ising models along
their FOQT line, in any dimension, when the BC do not favor any phase
separated by the FOQT. In these cases, the dynamic behavior arising
from a time-dependent magnetic field matches the dynamics of the
single-spin problem. It is worth mentioning that an analogous dynamic
scaling behavior is driven by time-dependent defects, such as a
magnetic field localized in one site of the Ising chain.  In a sense,
closed systems behave rigidly at FOQTs with neutral BC.  However, this
rigidity may get lost when considering other BC, such as those giving
rise to a domain wall in the ground state.  Numerical checks of the
main dynamic FSS features arising from KZ protocols at FOQTs have been
reported in Refs.~\cite{PRV-18-loc, PRV-20} for Ising chains with
global and local time-dependent perturbations, and various BC.

We point out that the dynamic scaling behaviors discussed in
the section have been observed in relatively small systems.
Therefore, given the need for high accuracy without necessarily
reaching scalability to large sizes, we believe that the available
technology for probing the coherent quantum dynamics of interacting
systems, such as with ultracold atoms in optical
lattices~\cite{Bakr-etal-10, Bloch-08, Simon-etal-11}, trapped
ions~\cite{Kim-etal-10, Edwards-etal-10, Islam-etal-11, LMD-11,
  Kim-etal-11, Richerme-etal-14, Jurcevic-etal-14, Debnath-etal-16},
as well as Rydberg atoms in arrays of optical
microtraps~\cite{Labuhn-etal-16, Keesling-etal-19}, could offer possible
playgrounds where the envisioned behaviors at FOQTs can be observed.

\section{Decoherence dynamics of qubits coupled to many-body systems}
\label{centralspin}

Decoherence generally arises when a given quantum system interacts
with an environmental many-body system. This issue is crucially
related to the emergence of classical behaviors in quantum
systems~\cite{Zurek-03, JP-09}, quantum effects such as interference
and entanglement~\cite{AFOV-08, FSDGS-18}, and is particularly
relevant for the efficiency of quantum-information
protocols~\cite{NC-book}.  There is a raising interest in monitoring
the coherent quantum dynamics of mutually coupled systems, with the
purpose of addressing aspects related to the relative decoherence
properties and the energy interchanges among the various
subsystems~\cite{Zurek-03}.  This issue is relevant both to understand
whether quantum mechanics can enhance the efficiency of energy
conversion in complex networks~\cite{CCDHP-09, Lambert-etal-13}, and
to devise novel quantum technologies which are able to optimize energy
storage in subportions of the whole system~\cite{BVMG-15,
  Campaioli-etal-17, LLMPP-18, FCAPP-18, JSRBL-20}.  Energy flows may
be influenced by the different quantum phases of the system.

\begin{figure}
  \begin{center}
    \includegraphics[width=0.4\columnwidth]{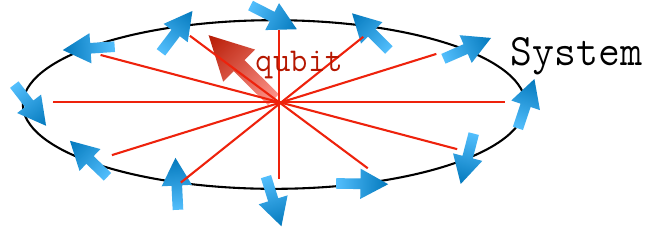}
    \caption{Sketch representing a qubit $\tt{q}$ (red arrow) coupled
      to an environment system $\tt{S}$ modeled by a quantum many-body 
      spin chain (blue arrows). In the figure, homogeneous couplings
      are considered, however more general inhomogeneous (partial or local)
      couplings can be assumed.}
    \label{Sketch_cntspin}
  \end{center}
\end{figure}

The decoherence dynamics of coupled systems has been investigated in
some paradigmatic models, such as two-level (qubit) systems
interacting with many-body systems, in particular the so-called {\em
  central spin models} (see, e.g., Refs.~\cite{Zurek-82, NP-93, SKL-02,
  CPZ-05, QSLZS-06, RCGMF-07, YZL-07a, YZL-07b, CFP-07, CP-08, LSZ-10,
  DQZ-11, NDD-12, MSD-12, HGMPM-12, SND-16, JJ-17b}), where a single
qubit ${\tt q}$ is globally, or partially, coupled to the environment
system ${\tt S}$ (for a sketch, see Fig.~\ref{Sketch_cntspin}).  In
this context, a typical problem concerns the coherence loss of the
qubit during the entangled quantum evolution of the global system,
starting from a pure state of the qubit and from the ground state of
$\tt{S}$.  The decoherence rate may significantly depend on the
quantum phase of $\tt{S}$ and, in particular, whether it develops
critical behaviors arising from QTs. Indeed the response of many-body
systems at QTs is generally amplified by {\em critical} quantum
fluctuations. At QTs, small variations of the driving parameter give
rise to significant changes of the ground-state and low-excitation
properties of many-body systems.  Therefore, environmental systems at
QTs may significantly drive the dynamics of the qubit decoherence.
QTs give rise to a substantial enhancement of the growth rate of the
decoherence dynamics with respect to noncritical
systems~\cite{QSLZS-06, RCGMF-07, YZL-07a, YZL-07b, DQZ-11, Vicari-18,
  RV-19-dec}.  In particular, as discussed below, the decoherence
growth rate at CQTs turns out to be characterized by power laws
$L^\zeta$ of the size $L$, with exponents $\zeta$ that are generally
larger than that $(L^d)$ of the volume law, expected for systems in
normal conditions, such as a gapped quantum disordered phase. The rate
enhancement of the qubit coherence loss is even more substantial at
FOQTs, characterized by an exponentially large decoherence growth
rate, related to the exponentially suppressed difference of the lowest
levels in finite-size many-body systems at FOQT.

We finally mention that the decoherence arising for the interaction
with an environmental system is considered as the fundamental
mechanism to understand the passage from quantum to classical
behaviors (see, e.g., the review articles~\cite{Zurek-03, Lykken-20}
and references therein).

\subsection{Setting of the problem}
\label{gset}

We consider a $d$-dimensional quantum many-body system ($\tt{S}$)
of size $L^d$, whose Hamiltonian is the same as in Eq.~\eqref{hlam},
\begin{equation}
  \hat{H}_{\tt S}(w) = \hat{H}_c + w \hat{H}_p \,,
  \label{hsdef}
\end{equation}
where, again, we assume that $\hat{H}_c$ is tuned at its critical point,
the term $\hat{H}_p$ is the spatial integral of local operators, such
that $[\hat{H}_c,\hat{H}_p]\neq 0$, and the parameter $w$ drives a QT
located at $w=0$.  A paradigm example is the quantum Ising
model~\eqref{hisdef} with $w \hat{H}_p$ corresponding to the
longitudinal-field term, such that $\hat{H}_p = - \sum_{\bm x}
\hat\sigma^{(1)}_{\bm x}$ and $w=h$.  The parameter $w$ drives FOQTs along the
$w=0$ line for any $g<g_c$. At the CQT point $g=g_c$, such term is one
of the relevant perturbations driving the critical behavior, the other
one being associated with the transverse-field term, identifying
$\hat{H}_p=-\sum_{\bm x} \hat \sigma^{(3)}_{\bm x}$ and $w=g-g_c$.

In addition, we consider a qubit ($\tt{q}$) whose two-level Hamiltonian
can be generally written as
\begin{equation}
  \hat{H}_q = \sum_{k=\pm} \epsilon_k | k \rangle \langle k | =
  a \hat I_2 + \tfrac12 s \, \hat\Sigma^{(3)}\, ,
      \label{Hqdef}
\end{equation}
where $\hat I_2$ is the $2 \times 2$ identity matrix, and the spin-1/2
Pauli operator $\hat\Sigma^{(3)}$ is associated with the two states
$|\pm\rangle$ of the qubit, so that $\hat\Sigma^{(3)} | \pm \rangle =
\pm | \pm \rangle$.  The Hamiltonian eigenvalues are $\epsilon_\pm = a
\pm s/2$, and $s = \epsilon_+ - \epsilon_-$.  The qubit is globally
and homogeneously coupled to the many-body system, through the
Hamiltonian term
\begin{equation}
  \hat{H}_{\tt{qS}} = \big( u \,\hat\Sigma^{(3)}
  + v \, \hat\Sigma^{(1)} \big) \, \hat{H}_p\,,
\label{hqsdef}
\end{equation}
where $\hat{H}_p$ is the operator appearing in Eq.~\eqref{hsdef}.
Putting all the terms together, we obtain the global Hamiltonian
\begin{equation}
  \hat{H} = \hat{H}_{\tt S} + \hat{H}_{\tt q} + \hat{H}_{\tt{qS}}\,.
  \label{hlamu}
\end{equation}

We consider a dynamic protocol arising from a sudden switching of the
interaction $\hat{H}_{\tt{qS}}$ between the qubit and the many-body
system, at time $t=0$, i.e., by instantaneously changing one or both
of the control parameters $u$ and $v$ in Eq.~\eqref{hqsdef}, from zero
to some finite value. More precisely:
\begin{itemize}
\item[$\bullet$] At $t=0$ the global system is prepared in the state
  \begin{equation}
    |\Psi_0 \rangle_{\tt{qS}} = | q_0 \rangle_{\tt q} \otimes | \psi_0(w)
    \rangle_{\tt S}\,,
    \label{psit0}
  \end{equation}
  where $|q_0\rangle_{\tt q}$ is a generic pure state of the qubit,
  \begin{equation}
    |q_0 \rangle_{\tt q} = c_+ |+\rangle  + c_- |-\rangle\,,\qquad 
    |c_+|^2 + |c_-|^2 = 1\,,
    \label{iqstate}
  \end{equation} 
  and $|\psi_0(w)\rangle_{\tt S}$ is the ground state of the many-body system
  with Hamiltonian $\hat{H}_{\tt S}(w)$~\footnote{In the following, we will
    indicate the subscripts ${}_{\tt q}$ and ${}_{\tt S}$ after the kets only
    when they refer to a specific subpart of the global system, omitting the
    double subscript ${}_{\tt qS}$ which refers to the global system.}.
  
\item[$\bullet$] Then, the global wave function describing the quantum
  evolution for $t>0$ must be the solution of the Schr\"odinger equation
  \begin{equation}
    {{\rm d} \,|\Psi(t)\rangle \over {\rm d} t} = -i \, \hat{H}
    |\Psi(t)\rangle\,, \qquad |\Psi(t=0)\rangle \equiv |\Psi_0\rangle\,,
    \label{sceq1}
  \end{equation}
  where the global Hamiltonian $\hat{H}$ is the one reported in
  Eq.~\eqref{hlamu} and includes the interaction~\eqref{hqsdef} between
  the qubit and the system (i.e., $\hat{H}_{\tt{qS}}\neq 0$, with the
  parameters $u$ and/or $v$ different from zero).
\end{itemize}

The above setting can be immediately extended to $N$-level systems
coupled to an environment ${\tt S}$, and also to the case the initial
qubit state is mixed, thus described by a nontrivial density matrix.

\subsection{Qubit decoherence, work, and qubit-system energy exchanges}
\label{definitions} 

An interesting issue arising from the dynamics of the problem outlined
above concerns the coherence properties of the qubit during the global
quantum evolution.  Starting from a pure state, the interaction with
the many-body system may give rise to a loss of coherence of the
qubit, depending on the properties of its density matrix
\begin{equation}
  \rho_{\tt q}(t) = {\rm Tr}_{\tt S} [\rho(t)]\,, 
  \label{rhoq}
\end{equation}
where ${\rm Tr}_{\tt S}[\:\cdot\:]$ denotes the trace over the
${\tt S}$-degrees-of-freedom of the global (pure) quantum state
\begin{equation}
  \rho(t)=|\Psi(t)\rangle \langle\Psi(t)|\,, 
  \label{rhot}
\end{equation}
with $|\Psi(t)\rangle$ given by the solution of Eq.~\eqref{sceq1}.  The
coherence of the qubit during its quantum evolution can be quantified
by its purity [see Sec.~\ref{sec:otherEnt}, Eq.~\eqref{eq:Purity}], 
\begin{equation}
  {\rm Tr}\, \big[ \rho_{\tt q}(t)^2 \big] \equiv 1 - D(t) \,,
  \label{ddef}
\end{equation}
where we have introduced the decoherence function $D$, such that $0\le
D \le 1$.  This function measures the quantum decoherence, quantifying
the departure from a pure state.  Indeed $D=0$ implies that the qubit
is in a pure state, thus $D(t=0)=0$ initially.  The other extreme value $D=1$
indicates that the qubit is totally unpolarized.

The initial quench, arising from turning on the interaction between
the qubit $\tt{q}$ and the many-body system $\tt{S}$, can be also
characterized via the quantum work $W$ done on the global
system~\cite{CHT-11, GPGS-18}. After the quench, the energy is
conserved during the global unitary evolution, thus we may only
observe some energy flow between the qubit and the environment system.
The work performed by changing the control parameters $u$ and $v$
does not generally have a definite value, while it can be defined as
the difference of two projective energy measurements~\cite{CHT-11}.
The first one at $t=0$ projects onto the eigenstates
$|\Psi_m(i)\rangle$ of the initial Hamiltonian
$\hat{H}_{i} = \hat{H}_{\tt q}+\hat{H}_{\tt S}$ with a
probability $p_{m,i}$ given by the initial density matrix.
The second energy measurement projects onto the eigenstates
$|\Psi_n(f)\rangle$ of the post-quench Hamiltonian
$\hat H_{f} = \hat{H}$~\footnote{Since the energy
  is conserved after the quench, the latter measurement can be
  performed at any time $t$ during the evolution, ruled by the unitary
  operator $e^{-i \hat{H}_{f} t}$, without changing the distribution.
  In particular, it can be performed at $t\to 0^+$.}.  The work
probability distribution $P(W)$ associated with this quantum quench
can be defined analogously to Eq.~\eqref{pwdefft}, so that the average
quantum work $\langle W \rangle$ is given by
\begin{equation}
  \langle W \rangle =
  \int dW\, W\, P(W) = 
  \langle \Psi(t)| \hat{H} |\Psi(t)\rangle -
  \langle \Psi_0| \hat{H}_{\tt q}+\hat{H}_{\tt S} |\Psi_0\rangle
  = \langle \Psi_0| \hat{H}_{\tt{qS}}| \Psi_0\rangle\,,
\label{workdef}
\end{equation}
where $|\Psi_0\rangle$ is the initial state in Eq.~\eqref{psit0}. An
analogous expression can be derived for the average of the square
work, obtaining $\langle W^2 \rangle = \langle \Psi_0|
\hat{H}_{\tt{qS}}^2 | \Psi_0\rangle$.  These relatively simple expressions
for the first two moments of the work distribution do not extend
to higher moments, whose expressions become
more complicated, requiring the computation of the whole spectrum, due
to the fact that $\hat{H}_{\tt{qS}}$ does not commute with the other
Hamiltonian terms.

In order to study the qubit-system energy exchanges, one may consider
the energy-difference distribution of the qubit, associated with two
energy measurements at $t=0$ and at a generic time $t$.  This is
defined as
\begin{equation}
  P_{\tt q}(U, t) = \! \sum_{n,a,b} \delta \big[ U -
    (E_{qb}-E_{qa})\big] \, \big| \langle b, \psi_n | e^{-i\hat{H} t}
  | a, \psi_0 \rangle \big|^2 p_a\,.
  \label{pdedefq}
\end{equation}
Here $|b,\psi_n\rangle\equiv |b\rangle_{\tt q} \otimes
|\psi_n\rangle_{\tt S}$, where $\{|a\rangle_{\tt q}, \, |b\rangle_{\tt
  q} \}$ and $\{E_{{\tt q}a}, E_{{\tt q}b} \}$ denote the eigenstates
($|\pm\rangle$) and eigenvalues of the qubit Hamiltonian $\hat{H}_{\tt
  q}$, while $\{ |\psi_n \rangle_{\tt S} \}$ indicate the eigenstates
of the system Hamiltonian $\hat{H}_{\tt S}$~\footnote{We assume a
  discrete spectrum for $\hat{H}_{\tt S}$, as generally appropriate
  for finite-size systems.}, and $p_\pm = |c_\pm|^2$ are the
probabilities of the initial qubit state at $t=0$.  According to the
definition~\eqref{pdedefq}, the average energy variation of the qubit
reads
\begin{equation}
  \langle U \rangle \equiv \int dU \,U\,P_{\tt q}(U,t) =
          {\rm Tr} \big[ \rho_{\tt q}(t) \, \hat{H}_{\tt q} \big]
          - {\rm Tr}\big[ \rho_{\tt q}(t\!=\!0) \, \hat{H}_{\tt q} \big]
          \equiv E_{\tt q}(t) - E_{\tt q}(t\!=\!0) \, .
          \label{deltastpet}
\end{equation}
More general initial distributions may be also considered, for example
associated with a bath at temperature $T$, replacing the initial
density matrix $\rho_0 = (| q\rangle_{\tt q} \otimes
|\psi_0\rangle_{\tt S}) \, ({}_{\tt q}\langle q| \otimes {}_{\tt S}
\langle \psi_0|)$ with the corresponding density matrices.  A
definition analogous to Eq.~\eqref{pdedefq} associated with the qubit
can be considered for the environment system~\cite{RV-19-dec}.

\subsection{Dynamic FSS ansatz for the qubit-system setup}
\label{dfss}

The dynamic processes arising from the instantaneous turning on of the
interaction term $\hat{H}_{\tt{qS}}$ can be described within a dynamic
FSS framework, extending the frameworks outlined in Secs.~\ref{dynqts}
and~\ref{foqtdynamics}, to take into account the interaction of the
many-body system $\tt{S}$ with the qubit $\tt{q}$.  Besides the scaling
variable associated with the Hamiltonian parameter $w$ of the system,
we also need to consider scaling variables associated with the other
parameters of the global Hamiltonian $\hat{H}$ (i.e., $u$, $v$, and
$s$).
Since both the $u$- and the $v$-term are coupled to the operator
$\hat{H}_p$, the corresponding scaling variables are expected to
behave analogously
to $w$ in the Hamiltonian $\hat{H}_S$. Therefore, at both CQTs and
FOQTs of the system $\tt{S}$, assuming that $\delta E_w(L,w)$ is the energy
variation arising from the longitudinal term $w$ coupled to $\hat{H}_p$,
we introduce the scaling variables
\begin{equation}
  \Phi_w \equiv {\delta E_w(L,w)\over \Delta(L)}\,,\qquad
  \Phi_u \equiv {\delta E_w(L,u)\over \Delta(L)}\,,\qquad
  \Phi_v \equiv {\delta E_w(L,v)\over \Delta(L)}\,,
  \label{phihuvdef}
\end{equation}
where $\Delta(L)$ is the energy difference of the two lowest states
of $\tt{S}$ at $w=0$.  A scaling variable must also be
associated with the energy difference $s$ of the eigenstates of the
qubit Hamiltonian $\hat{H}_{\tt q}$.  Since $s$ is a further energy
scale of the problem, we define
\begin{equation}
  \Lambda \equiv \frac{s}{\Delta(L)} \,.
  \label{varepsilon}
\end{equation}
Finally, the scaling variable $\Theta \equiv \Delta(L) \,t$ is associated
with the time variable.  The dynamic FSS limit that we consider is
defined as the large-$L$ limit keeping the above scaling variables
fixed.  The above scaling variables can be adopted for both CQTs and
FOQTs, with the appropriate, power-law or exponential, size dependence
of $\Delta(L)$ and $\delta E_w(L,w)$, as discussed at the end of
Sec.~\ref{fssfoqt}.

For the sake of clarity in the presentation, below we consider two
limiting situations: (i) the simpler case $v=0$ and $u \neq 0$, for
which $[\hat H_{\tt q}, \hat H_{\tt{qS}}]=0$, and (ii) the more
complicated case $u=0$ and $v \neq 0$, for which $[\hat H_{\tt q},
  \hat H_{\tt{qS}}]\neq 0$.  In both cases, the dynamic FSS behavior
of a generic observable is given by~\cite{Vicari-18, RV-19-dec}
\begin{equation}
  O(t;L,w,u/v,s) \approx L^{y_o} \, {\cal O}(\Theta, \Phi_w,
  \Phi_{u/v},\Lambda)\,,
  \label{cqteq}
\end{equation}
where, again, $y_o$ is the RG dimension of $\hat O$ at CQTs ($y_o=0$ at
FOQTs), and ${\cal O}$ is a dynamic FSS function.  Note that, in the
case (i) $v=0$, the dependence on the qubit spectrum $s$, and thus on
the scaling variable $\Lambda$, disappears.  Of course, the
equilibrium FSS behavior must be recovered in the limit $u/v \to 0$.

\subsection{Qubit decoherence functions}
\label{decoqubit}

We now discuss the decoherence properties of the qubit in some detail.
Results for the work associated with the initial quench and the energy
flow between qubit and environment are reported in Ref.~\cite{RV-19-dec}.

\subsubsection{Central spin systems with  
$[\hat H_{\tt q}, \hat H_{\tt{qS}}]=0$}
\label{v0case}

The time evolution of the global system gets significantly simplified
when $[\hat{H}_{\tt q},\hat{H}_{\tt{qS}}]=0$, i.e., when $v=0$ in
Eq.~\eqref{hqsdef}, since the qubit exhibits pure dephasing in time,
while the environment system evolves in two independent branches with
the same Hamiltonian form and different fields~\cite{QSLZS-06,
  RCGMF-07}. Indeed, the evolution of the global state can be written
in terms of dynamic evolutions of the system $\tt{S}$ only, as
\begin{equation}
  |\Psi(t)\rangle  = e^{- i \epsilon_+ t} 
  c_+ | + \rangle_{\tt q} \otimes | \phi_{w+u}(t) \rangle_{\tt S} 
  + e^{- i \epsilon_- t} c_- |- \rangle_{\tt q}
  \otimes | \phi_{w-u}(t) \rangle_{\tt S}\,,
  \label{psitcomv0}
\end{equation}
where
\begin{equation}
  | \phi_{w\pm u}(t)\rangle_{\tt S} = e^{-i\hat{H}_{\tt S}(w\pm u) \, t}
  |\psi_0(w)\rangle_{\tt S}\,
\label{phipm}
\end{equation}
are pure states of the environment system $\tt{S}$.  Notable relations
can be obtained focusing on the evolution of the qubit $\tt{q}$ only.
The elements of its $2 \times 2$ reduced density matrix,
cf. Eq.~\eqref{rhoq}, read:
\begin{equation}
  \rho_{{\tt q},11}(t) = |c_+|^2\,, \qquad 
  \rho_{{\tt q},22}(t) = |c_-|^2\,, \qquad\rho_{{\tt q},12}(t) =
  e^{-i s t} c_-^* c_+ \;{}_{\tt S}\langle \phi_{w-u}(t)|
  \phi_{w+u}(t)\rangle_{\tt S} = \rho_{{\tt q},21}(t)^*\,.
  \label{qdmco}
\end{equation}
The decoherence function $D(t)$, cf. Eq.~\eqref{ddef}, can be written as
\begin{equation}
  D(t) = 2 \, |c_+|^2 \, |c_-|^2 F_D\,,\qquad
  F_D \equiv 1 - L_D(t)^2\,, \qquad L_D \equiv \big|
  {}_{\tt S}\langle \phi_{w-u}(t) | \phi_{w+u}(t) \rangle_{\tt S} \big|\,.
  \label{purity}
\end{equation}
The function $F_D$, such that $0\le F_D(t)\le 1$ is a measure of {\em
  quantum decoherence}, quantifying the departure from a pure state:
$F_D(t)=0$ implies that the qubit is in a pure state, while $F_D(t)=1$
indicates that the qubit is maximally entangled, corresponding to a
diagonal density matrix $\rho_{\tt q} = {\rm diag}\big[
  |c_+|^2,|c_-|^2 \big]$.  Note that $D(t)$ does not depend on the
spectrum of the qubit Hamiltonian $\hat{H}_{\tt q}$, and in particular
on its parameter $s$.

The overlap function $L_D$ entering Eq.~\eqref{purity} can be
interpreted as the fidelity, or the related Loschmidt echo, of the
${\tt S}$-states associated with two different quench protocols
involving the isolated system ${\tt S}$. For both of them, the
environment system starts from the ground state of the Hamiltonian
$\hat{H}_{\tt S}(v)$ as $t=0$; then one considers, and compares using
$L_D$, the quantum evolutions at the same $t$, arising from the sudden
change of the Hamiltonian parameter $w$ to $w-u$ and to $w+u$.  The
dynamics FSS behavior of $L_D$ can be inferred from the theory
developed in Sec.~\ref{Loschecho} at quench protocols:
\begin{equation}
  L_D(t;L,w,u) \approx {\cal L}_D(\Theta, \Phi_w, \Phi_u)\,,
  \label{decqtsca}
\end{equation}
with ${\cal L}_D(\Theta, \Phi_w, \Phi_u\!=\!0)=1$, and ${\cal
  L}_D(\Theta\!=\!0,\Phi_w,\Phi_u)=1$~\footnote{Note that the dynamic
  FSS limit requires that the coupling $u$ between the qubit and the
  environment system is sufficiently small, since $\Phi_u$ of
  Eq.~\eqref{phihuvdef} must be kept constant in the large-$L$
  limit.}.

Since $L_D(t;L,w,u)$ is an even function of $u$ and $D(t;L,w,0)=0$,
we can write
\begin{equation}
  D(t;L,w,u) = \tfrac12 u^2 Q(t;L,w) + O(u^4)\,,
  \qquad
  Q(t;L,w) = \frac{\partial^2 D}{\partial u^2}\bigg|_{u=0} =
  2 \, |c_+|^2 \, |c_-|^2 \, \frac{\partial^2 F_D}{\partial u^2}\bigg|_{u=0} \,,
\label{dtmu2}
\end{equation}
which also assumes an analytic behavior around $u=0$ (at finite $L$
and $t$).  The function $Q$ quantifies the growth rate of decoherence
in the limit of small qubit-environment coupling $u$, thus measuring the
sensitivity of the coherence properties of the subsystems to $u$.
The scaling behavior~\eqref{decqtsca} allows us to derive the
corresponding dynamic FSS of $Q$ as
\begin{equation}
  Q(t;L,w) \approx \bigg(
  \frac{\partial \Phi_u}{\partial u} \bigg)^2 {\cal Q}(\Theta,\Phi_w) \,.
%  \approx L^{2y_w} {\cal Q}(\Theta, \Phi_h)\,.
\label{cfhlt}
\end{equation}

In the case of environment systems at CQTs, one has $\Phi_u = u \,
L^{y_w}$, therefore it specializes into
\begin{equation}
  Q(t;L,w) \approx L^{2y_w} {\cal Q}(\Theta, \Phi_w)\,.
  \label{qtlwcat}
  \end{equation}
This scaling equation characterizes the
amplified $O(L^{2y_w})$ rate of departure from coherence of the qubit.
Indeed, in the case of systems out of criticality, one generally
expects $Q\sim L^d$ and
\begin{equation}
  2y_w = d+3-\eta>d\,.
  \label{2yrel}
\end{equation}
For example, $\eta=1/4$ for 1$d$ quantum Ising-like chains, and
$\eta\approx 0.036$ for 2$d$ quantum Ising systems (see
Sec.~\ref{isingmodels}).  Therefore these results show a substantial
enhancement of decoherence when the environment system is at a CQT,
speeding up the decoherence by a large $O(L^{3-\eta})$ factor.

The rate enhancement of the qubit coherence loss is even more
substantial at FOQTs, characterized by an exponentially large
decoherence growth rate, related to the exponentially suppressed
difference of the lowest levels in finite-size many-body systems at
FOQTs. This occurs when the environment system ${\tt S}$ is at a FOQT
and its BC are neutral, so that $\Delta(L)\sim e^{-c L^d}$.  For
example, one may consider quantum Ising systems at their FOQT line
driven by $h$, identifying $w$ with $h$. In this case, since $\Phi_u
\sim u L^d/\Delta(L)$, one has
\begin{equation}
  \left( {\partial \Phi_u\over \partial u} \right)^2
  \approx {L^{2d}\over \Delta(L)^2}
  \sim L^{2d} e^{2c L^d}\,,\label{expbeh}
\end{equation}
to be compared again with the $O(L^d)$ behavior expected
for normal conditions, far from QTs.  The corresponding scaling function
${\cal Q}$ can be computed by exploiting the two-level framework for
the environment system at FOQTs~\cite{Vicari-18}:
\begin{equation}
  {\cal Q}(\Theta,\Phi_w) = 2 \, |c_+|^2 \, |c_-|^2 \, \frac{ 2 \big[
      1 - {\rm cos}(\Theta\sqrt{1+\Phi_w^2}) \big]}{(1+\Phi_w^2)^2}\,.
\label{cdanres}
\end{equation}

\subsubsection{Central spin systems with $[\hat H_{\tt q}, \hat H_{\tt{qS}}]\neq0$}
\label{u0case}

\begin{figure}
  \begin{center}
    \includegraphics[width=0.47\columnwidth]{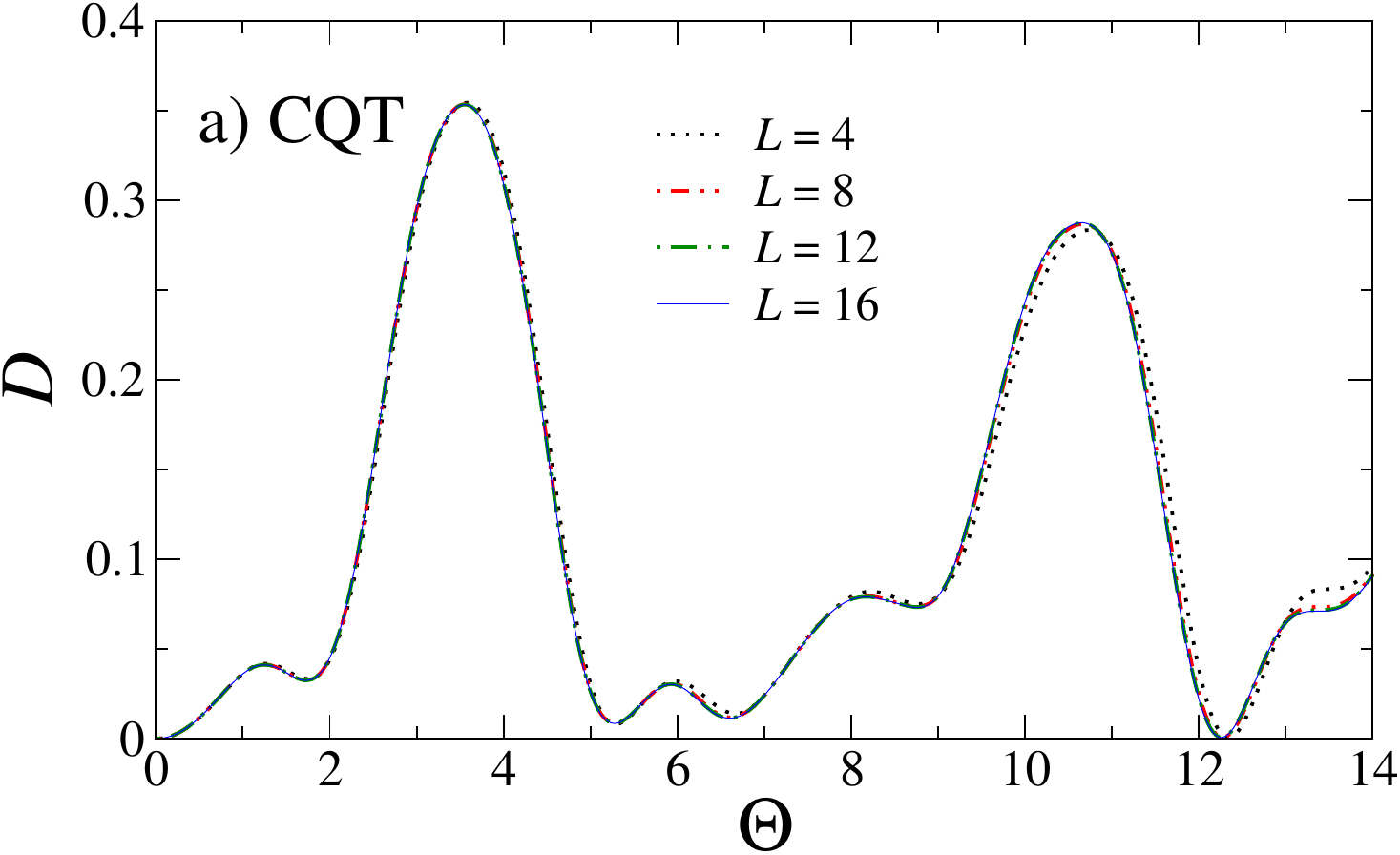}
    \hspace*{5mm}
    \includegraphics[width=0.47\columnwidth]{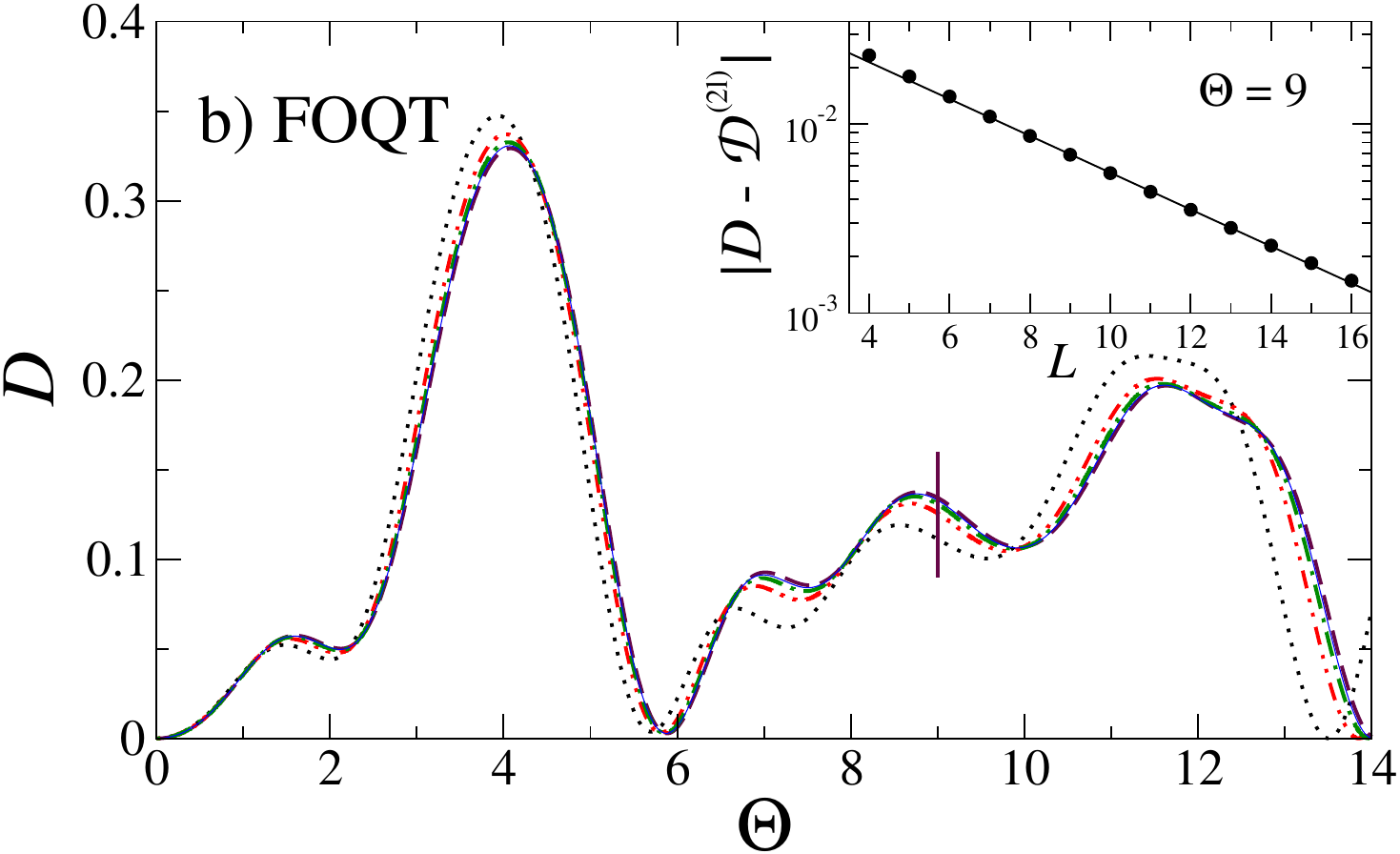}
    \caption{Decoherence function $D$ versus the rescaled time
      $\Theta$, for a qubit coupled to Ising spin-chain systems with
      PBC and of different lengths $L$ at the CQT, $g=g_c=1$ [left
        panel $(a)$], or at the FOQT, $g=0.9$ [right panel $(b)$].
      All numerical data are for a qubit-system coupling realized
      through $u=0$ and $v \neq 0$ in Eq.~\eqref{hqsdef}, such that
      $[\hat H_{\tt q} ,\hat H_{\tt{qS}}] \neq 0$.  We fix
      $\Phi_w=0.8$, $\Phi_v=1$, $\Lambda=0.5$, while the qubit is
      initialized with $c_+=\sqrt{2/3}$.  The inset in the right panel
      $(b)$ shows numerical data for $\Theta=9$, supporting an
      exponential convergence to the prediction ${\cal D}^{(2l)}$
      given by the two-level approximation for the Ising chain (thick
      brown line in the main frame).  Adapted from Ref.~\cite{RV-19-dec}.}
    \label{Purity_CQT}
  \end{center}
\end{figure}

Analogous scaling behaviors are conjectured in the more complex case
(ii), $v \neq 0$, where $[\hat H_{\tt q}, \hat H_{\tt{qS}}]\neq
0$~\footnote{For simplicity we assume $u=0$, thus only consider the
  coupling term associated with the parameter $v$ in
  $\hat{H}_{\tt{qS}}$, cf.~Eq.~\eqref{hqsdef}.}.  In that case, the
scaling functions are also expected to depend on the scaling variable
$\Lambda$ in Eq.~\eqref{varepsilon}, associated with the energy
spectrum of the qubit, in such a way that
\begin{equation}
  D(t;L,w,v,s) \approx {\cal D}(\Theta, \Phi_w,\Phi_v,\Lambda)\,.
  \label{calpscalnc}
\end{equation}
This dynamic FSS law has been numerically checked within the 1$d$
Ising chain at its CQT and along its FOQT line. The corresponding
results, reported in Fig.~\ref{Purity_CQT}, nicely support the
conjectured behavior.  In the case of environment systems ${\tt S}$ at
FOQTs with neutral BC (as in the case of the figure),
one can again exploit a two-level truncation of ${\tt S}$,
thus leading to an effective model constituted by two
coupled qubits, which is much simpler to handle~\cite{RV-19-dec}.

Now, assuming analyticity for $D$ at $v=0$ and finite volume, and since
$D\geq 0$, one expects an expansion analogous to Eq.~\eqref{dtmu2},
\begin{equation}
  D(t;L,w,v,s) = \tfrac12 v^2 Q(t;L,w,s) + O(v^3) \,,
  \label{calqdefv}
\end{equation}
which can be matched with the scaling behavior in
Eq.~\eqref{calpscalnc}, to obtain
\begin{equation}
  Q(t;L,w,s) \approx \left( {\partial\Phi_v\over\partial
    v}\right)^2 {\cal Q}(\Theta,\Phi_w,\Lambda) \,.
  \label{claQscalnc}
\end{equation}

\begin{figure}
  \begin{center}
    \includegraphics[width=0.5\columnwidth]{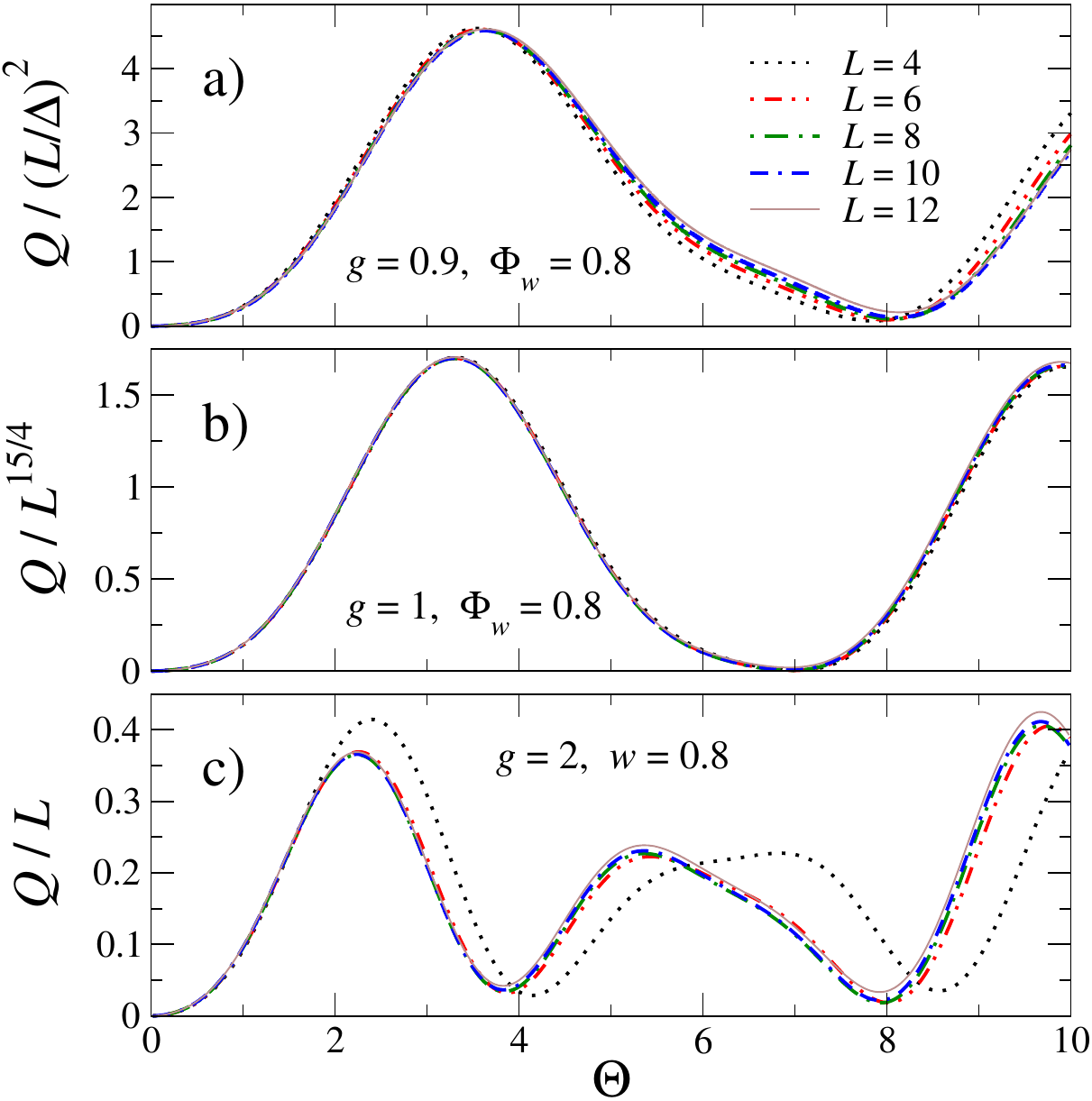}
    \caption{Scaling of the function $Q$ vs.~the rescaled
      time $\Theta$, for three distinct situations of the Ising system
      ${\tt S}$: (a) on the FOQT line,
      for $g=0.9$; (b) at the CQT, for $g=g_c=1$; (c) in the disordered
      phase, for $g=2$.  The evaluated growth-rate function quantifies the
      sensitivity of the qubit coherence to the coupling $v$.
      As in Fig.~\ref{Purity_CQT}, we fix $\Lambda = 0.5$ and
      $c_+ = \sqrt{2/3}$. The longitudinal field $h = w$ has been chosen,
      without loss of generality, in such a way that $\Phi_w = 0.8$
      for panels (a) and (b), while $w=0.8$ in panel (c).
      Adapted from Ref.~\cite{RV-19-dec}.}
    \label{GrowthRate}
  \end{center}
\end{figure}

Like the case discussed in the previous Sec.~\ref{v0case},
a power law arises when the environment system ${\tt S}$ is at
a CQT, $Q \sim L^{2 y_w}$.
Conversely, an exponential law emerges at a FOQT (when considering
neutral BC), such as $Q \sim\exp(b L^d)$.  One may
compare these results with those expected when $\tt{S}$ is not close
to a QT, for example for $g>g_c$ in the case of quantum Ising models,
for which $Q$ is expected to increase as the volume of the system, i.e
$Q\sim L^d$.  A numerical validation of the above conjectures for the
growth-rate function at CQTs, FOQTs, and far from QTs, within
the quantum Ising chain, is reported in Fig.~\ref{GrowthRate}.

\section{Master equations for open quantum systems}
\label{sec:MEq}

In Sec.~\ref{centralspin} we have addressed the role of dissipative
mechanisms on a single quantum spin ${\tt q}$ (the qubit), induced by
the Hamiltonian coupling to another quantum system ${\tt S}$ (the
environment).  In that framework, the dynamics of the whole universe
$\tt{q}+\tt{S}$ (qubit+environment) can be entirely described
as a unitary evolution
according to the Schr\"odinger equation with the global
Hamiltonian~\eqref{hlamu}.  If one is interested in the analysis of
the qubit alone, the time behavior of the corresponding reduced
density matrix $\rho_{\tt q}(t)$ can be obtained by tracing over the
environment system.

The above reasoning can be put on much more general grounds, by
microscopically deriving a {\em quantum master equation} able to
describe the continuous temporal evolution of a generic open quantum
system itself (i.e., what have been previously called the qubit, and
hereafter referred to as the system ${\tt s}$), without the necessity
to keep track of the full environment dynamics (hereafter labelled
with ${\tt E}$). In this section we give a brief overview on how to
perform such derivation, concentrating on the so-called
Gorini-Kossakowski-Lindblad-Sudarshan (GKLS) framework~\cite{GKS-76,
  Lindblad-76}.  Further details can be found, e.g., in
Refs.~\cite{BP-book, RH-book, GZ-book, Chruscinski-14}.

\subsection{Evolution of closed and open quantum systems}

We start by recalling some general properties of the temporal
evolution of closed quantum systems. Under the hypothesis in which the
system at time $t$ is in the pure state $|\psi(t)\rangle$, its time
evolution is dictated by the usual Schr\"odinger equation. If the
system Hamiltonian $\hat H$ is time-independent, its solution can be
formally written as $|\psi(t)\rangle = \hat U(t, t_0) |\psi(t_0)\rangle$,
where $\hat U(t,t_0) = \exp [-i \hat H (t-t_0)]$ is a unitary
operator.~\footnote{In contrast, if $\hat H(t)$ is time-dependent, the
shape of $\hat U$ is more complicated, although its unitary property
is maintained.  Here we limit ourselves to time-independent
Hamiltonians, however the formalism can be extended to the general
case with the use of appropriate time-ordering operations.}  In the
case the quantum system is in a statistical mixture, the state is
generally written as a density matrix $\rho(t)$ and its time evolution
is dictated by the von Neumann equation
\begin{equation}
  \frac{{\rm d} \rho(t)}{{\rm d}t} = -i \big[ \hat H, \rho(t) \big] \,,
  \qquad \rho(t)  = \hat U(t,t_0) \, \rho(t_0) \, \hat U^\dagger(t,t_0) \equiv
  \mathbb{U}_{t_0 \to t} \, \rho(t_0)\,,
  \label{eq:vN-eq}
\end{equation}
where $[\hat A, \hat B] \equiv \hat A \hat B - \hat B \hat A$ denotes the
commutator and $\mathbb{U}_{t_0 \to t}$ is an operator on the operator
space of the Hilbert space of the system, also named {\em superoperator}.

Now consider a system interacting with the environment (also known
as a bath or reservoir). The most general Hamiltonian describing this
situation is
\begin{equation}
  \hat H = \hat H_{\tt s} + \hat H_{\tt E} + \hat H_{\tt sE} \,,
\end{equation}
where $\hat H_{\tt s}$, $\hat H_{\tt E}$, and $\hat H_{\tt sE}$
respectively describe the system, the environment, and the interaction
among the two.  Usually the system-bath interaction is weak, so that
the {\em fast} motion due to $\hat H_0 \equiv \hat H_{\tt s} + \hat
H_{\tt E}$ can be reasonably separated from the {\em slow} motion due
to $\hat H_{\tt sE}$.  Following this reasoning, one can exploit the
interaction picture and define
\begin{equation}
  \hat{\tilde{A}}_{\tt sE}(t) \equiv \hat U^\dagger(t) \, \hat A_{\tt sE}
  \, \hat U(t)\,, \qquad
  \hat U(t) = e^{-i \hat H_{\tt s} t} \otimes e^{-i \hat H_{\tt E} t} \,,
\end{equation}
where $\hat A_{\tt sE}$ is a given operator acting on the whole
system~\footnote{This implies that, for operators acting
  only on the system ${\tt s}$ or on the environment ${\tt E}$,
  one has: $\hat{\tilde{S}}_{\tt s}(t) = \hat U_{\tt s}^\dagger(t) \,
  \hat S_{\tt s} \, \hat U_{\tt s}(t)$ and $\hat{\tilde{E}}_{\tt E}(t)
  =  \hat U_{\tt E}^\dagger(t) \, \hat E_{\tt E} \, \hat U_{\tt E}(t)$,
  where $\hat U_{\tt s}(t) = e^{-i \hat H_{\tt s} t}$ and
  $\hat U_{\tt E}(t) = e^{-i \hat H_{\tt E} t}$.}.
Then, the von Neumann equation~\eqref{eq:vN-eq} for the full
density matrix $\rho_{\tt sE}(t)$ reads 
\begin{equation}
  \frac{{\rm d} \tilde \rho_{\tt sE}(t)}{{\rm d}t} = 
-i \big[ \hat{\tilde{H}}_{\tt sE}(t),
    \tilde \rho_{\tt sE}(t) \big] \,,
  \label{eq:vnE-2}
\end{equation}
which is equivalent to 
\begin{equation}
  \tilde \rho_{\tt sE}(t) = \tilde \rho_{\tt sE}(0)
  - i \int_0^t {\rm d} \tau \, \big[ \hat{\tilde{H}}_{\tt sE}(\tau),
    \tilde \rho_{\tt sE}(\tau) \big]\,, \qquad
  \tilde \rho_{\tt sE}(0) = \rho_{\tt sE}(0) \,.
\end{equation}
Inserting the latter expression back into Eq.~\eqref{eq:vnE-2}, we obtain
\begin{equation}
  \frac{{\rm d} \tilde \rho_{\tt sE}(t)}{{\rm d}t}
  = -i \big[ \hat{\tilde{H}}_{\tt sE}(t),
    \rho_{\tt sE}(0) \big] - \int_0^t {\rm d} \tau \,
  \Big[ \hat{\tilde{H}}_{\tt sE}(t), \big[ \hat{\tilde{H}}_{\tt sE}(\tau),
      \tilde \rho_{\tt sE}(\tau) \big] \Big] \,.
\end{equation}
Now, tracing over the degrees of freedom of ${\tt E}$, and using
the fact that 
\begin{equation}
{\rm Tr}_{\tt E} \big[ \tilde \rho_{\tt sE} \big] = {\rm Tr}_{\tt E}
\big[ \hat U^\dagger \, \rho_{\tt sE} \, \hat U \big] = \hat U^\dagger_{\tt s} \:
{\rm Tr}_{\tt E} \big[ U^\dagger_{\tt E} \, \rho_{\tt sE} \, \hat U_{\tt E} \big] \,
\hat U_{\tt s} = \hat U^\dagger_{\tt s} \, {\rm Tr}_{\tt E} \big[\rho_{\tt sE} \big]
\, \hat U_{\tt s} = \tilde \rho_{\tt s}\,, 
\label{eqeqeq}
\end{equation}
we obtain the following
integro-differential equation for $\rho_{\tt s}$:
\begin{equation}
  \frac{{\rm d} \tilde \rho_{\tt s}(t)}{{\rm d}t} = -i \, {\rm Tr}_{\tt E} \Big\{
  \big[ \hat{\tilde{H}}_{\tt sE}(t), \rho_{\tt sE}(0) \big] \Big\}
  - \int_0^t {\rm d} \tau \, {\rm Tr}_{\tt E} \Big\{
  \Big[ \hat{\tilde{H}}_{\tt sE}(t), \big[ \hat{\tilde{H}}_{\tt sE}(\tau),
      \tilde \rho_{\tt sE}(\tau) \big] \Big] \Big\} \,.
  \label{eq:vN-integral}
\end{equation}

\subsection{Markovian quantum master equations}

Equation~\eqref{eq:vN-integral} is exact but practically useless,
being extremely difficult to handle.  In order to proceed further with
the derivation of a simpler master equation for $\rho_{\tt s}(t)$, two
important approximations are usually invoked:

\begin{itemize}
\item {\em Born approximation}. It consists in taking an environment
  ${\tt E}$ much larger than the system ${\tt s}$, so that it is
  practically unaffected by the interactions with the system.
\item {\em Markov approximation}. The environment ${\tt E}$ may
  acquire information on the system, which can also flow back (at
  least in part) into ${\tt s}$. One generally assumes that the state
  of the environment at time $t_0$ is essentially unaffected by the
  history of the system.  This implies that the information flow is
  one-way, from the system to the environment $\tt{s} \to \tt{E}$ (and
  not vice-versa).  More precisely, the knowledge of the system
  density matrix $\rho_{\tt s}(t_0)$ at a given time $t_0$ is
  sufficient to determine $\rho_{\tt s}(t)$ at any time $t > t_0$.
\end{itemize}
In practice, one starts by assuming that, at time $t=0$, the system
and the environment are uncorrelated:
\begin{equation}
  \rho_{\tt sE}(0) = \rho_{\tt S}(0) \otimes \rho_{\tt E}(0) \,,
\end{equation}
thus the interaction Hamiltonian $\hat H_{\tt sE}$ does not
significantly affect the global ground state of the whole universe
(i.e., system+environment).  We also suppose that
\begin{equation}
  {\rm Tr}_{\tt E} \Big\{ \big[ \hat{\tilde{H}}_{\tt sE}(t), 
\rho_{\tt sE}(0) \big]
  \Big\} = 0\,,
  \label{eq:condME1}
\end{equation}
thus eliminating the first term in the right-hand side of
Eq.~\eqref{eq:vN-integral}.  In fact it can be shown that, if this is
not the case, one could always redefine $\hat H_{\tt s}$ and $\hat
H_{\tt sE}$, while keeping the global Hamiltonian $\hat H$ constant,
in order to fulfill Eq.~\eqref{eq:condME1}.

Within the Born approximation, the system-bath coupling is so weak and
the reservoir so large that its state is essentially unaffected by the
interaction $\hat H_{\tt sE}$. In practice, this amounts to assuming
that correlations between the system and the bath remain negligible at
all times, therefore
\begin{equation}
  \tilde \rho_{\tt sE}(t) \approx \tilde \rho_{\tt s}(t)
  \otimes \tilde \rho_{\tt E}\,, \qquad \quad \mbox{since} \quad
  \rho_{\tt E} \equiv \rho_{\tt E}(t) \approx \rho_{\tt E}(0)\,.
  \label{eq:Born-me}
\end{equation}
Of course, this assumption relies on the fact that the bath
equilibrates on a time scale $\tau_{\tt E}$ much faster than that of
the system evolution $\tau_{\tt s}$, namely $\tau_{\tt s} \gg
\tau_{\tt E}$.
Using Eq.~\eqref{eq:condME1} and plugging~\eqref{eq:Born-me}
in Eq.~\eqref{eq:vN-integral}, we obtain
\begin{equation}
    \frac{{\rm d} \tilde \rho_{\tt s}(t)}{{\rm d}t} = 
  - \int_0^t {\rm d} \tau \, {\rm Tr}_{\tt E} \Big\{
  \Big[ \hat{\tilde{H}}_{\tt sE}(t), \big[ \hat{\tilde{H}}_{\tt sE}(\tau),
      \tilde \rho_{\tt s}(\tau) \otimes \tilde \rho_{\tt E}\big] \Big] \Big\} \,.
  \label{eq:vN-integral2}
\end{equation}
The Markov approximation guarantees that one can safely substitute
$\tilde \rho_{\tt s}(\tau) \to \tilde \rho_{\tt s}(t)$ in
Eq.~\eqref{eq:vN-integral2}, meaning that the density matrix $\tilde
\rho_{\tt s}$ remains approximately constant over all the time
interval $0 \leq \tau \leq t$.  As a consequence, we obtain
\begin{equation}
    \frac{{\rm d} \tilde \rho_{\tt s}(t)}{{\rm d}t} = 
  - \int_0^t {\rm d} \tau \, {\rm Tr}_{\tt E} \Big\{
  \Big[ \hat{\tilde{H}}_{\tt sE}(t), \big[ \hat{\tilde{H}}_{\tt sE}(\tau),
      \tilde \rho_{\tt s}(t) \otimes \tilde \rho_{\tt E}\big] \Big] \Big\} \,,
  \label{eq:vN-Markov1}
\end{equation}
thus reducing to an equation of motion for the system ${\tt s}$ that
is simply differential (and no longer integro-differential), since
$\tilde \rho_{\tt s}$ in the right-hand side has not to be integrated
in time. Equation~\eqref{eq:vN-Markov1} is known as the {\em Redfield
  master equation}.

In fact, the Redfield equation is still non-Markovian, as it retains
memory of the initial-state preparation of the system, $\rho_{\tt s}(0)$.
To obtain a Markovian equation, a coarse graining in time
has to be performed, by extending the upper limit of the integration
to infinity, with no real change in the outcome:
\begin{equation}
    \frac{{\rm d} \tilde \rho_{\tt s}(t)}{{\rm d}t} = 
  - \int_0^{+\infty} {\rm d} \tau \, {\rm Tr}_{\tt E} \Big\{
  \Big[ \hat{\tilde{H}}_{\tt sE}(t), \big[ \hat{\tilde{H}}_{\tt sE}(t-\tau),
      \tilde \rho_{\tt s}(t) \otimes \tilde \rho_{\tt E}\big] \Big] \Big\} \,,
  \label{eq:vN-Markov2}
\end{equation}
where, for the sake of convenience, we changed the integral variable
$\tau \to t-\tau$.  The two-step approximation described in
Eqs.~\eqref{eq:vN-Markov1} and~\eqref{eq:vN-Markov2} is known as the
Markov approximation.  The final result has been thus derived for weak
coupling and quickly decaying reservoir correlations (i.e., memoryless
dynamics).

\subsubsection{Lindblad master equations}

Unfortunately the master equation~\eqref{eq:vN-Markov2},
holding under the Born-Markov approximation, does not warrant
the positivity of $\rho_{\tt s}(t)$ at a generic time $t$,
sometimes giving rise to density matrices that are non-positive.
To ensure complete positivity, one further approximation is needed:
\begin{itemize}
\item {\em Secular approximation}.  As we shall see below, it involves
  discarding fast oscillating terms in the Redfield master equation,
  which can be neglected if the characteristic time of the system is
  much larger than that of the bath.  In the context of quantum
  optics, this is equivalent to the so-called rotating wave
  approximation, which consists in dropping all fast oscillating
  frequencies of a combined atom-photon system near the
  electromagnetic resonance.
\end{itemize}
To be explicit, we decompose the interaction Hamiltonian $\hat H_{\tt
  sE}$ in terms of operators of the system and of the environment,
according to
\begin{equation}
  \hat H_{\tt sE} = \sum_\alpha \hat S_\alpha \otimes \hat E_\alpha \,,
\end{equation}
where $\hat S_\alpha$ and $\hat E_\alpha$ are Hermitian operators acting
on the system ${\tt s}$ and the environment ${\tt E}$, respectively.
Assuming that $\hat H_{\tt s}$ has a discrete spectrum, one can define
the eigenoperators of the system as
\begin{equation}
  \hat S_\alpha(\omega) = \sum_{\epsilon, \epsilon', \,\epsilon'-\epsilon = \omega}
  \, \hat \Pi(\epsilon) \, \hat S_\alpha \, \hat \Pi(\epsilon') \,,
\end{equation}
where the sum is extended over the couples of eigenvalues of
$\hat H_{\tt s}$ whose difference is equal to $\omega$,
while $\hat \Pi(\epsilon)$ is the projector on the corresponding
$\epsilon$-th eigenspace.  These operators satisfy the
properties
\begin{subequations}
  \begin{align}
    & \sum_\omega \hat S_\alpha(\omega) = \hat S_\alpha\,, \qquad
    \hat S^\dagger_\alpha(\omega) = \hat S_\alpha(-\omega)\,, \qquad
    \hat{\tilde{S}}_\alpha(\omega) = e^{-i \omega t} \hat S_\alpha(\omega)\,, \\
    & \big[ \hat H_{\tt s}, \hat S_\alpha(\omega) \big] =
    -\omega \hat S_\alpha(\omega) \,, \qquad
    \big[ \hat H_{\tt s}, \hat S^\dagger_\alpha(\omega) \big] =
    \omega \hat S^\dagger_\alpha(\omega) \,.
  \end{align}
\end{subequations}
Using them, the interaction Hamiltonian can be written in
the interaction picture as
\begin{equation}
  \hat{\tilde{H}}_{\tt sE} (t) =
  \sum_{\alpha, \omega} e^{-i \omega t} \hat S_\alpha(\omega) \otimes
  \hat{\tilde{E}}_\alpha(t) = \sum_{\alpha, \omega} e^{i \omega t}
  \hat S^\dagger_\alpha(\omega) \otimes \hat{\tilde{E}}^\dagger_\alpha(t) \,.
  \label{eq:eigenop1}
\end{equation}

Now, expanding the commutators in the Redfield
equation~\eqref{eq:vN-Markov2} and using the
expression~\eqref{eq:eigenop1}, it is possible to obtain the
following:
\begin{equation}
  \frac{{\rm d} \tilde \rho_{\tt s}(t)}{{\rm d}t} =
  \sum_{\omega, \omega'} \sum_{\alpha, \beta} \left\{ e^{i (\omega'-\omega)t} \,
  \Gamma_{\alpha, \beta}(\omega) \, \big[ \hat S_\beta(\omega) \tilde \rho_{\tt S}(t),
    \hat S^\dagger_\alpha (\omega') \big] + {\rm h.c.} \right\} \,.
  \label{eq:vN-Markov3}
\end{equation}
The quantity $\Gamma_{\alpha, \beta}(\omega)$ denotes the one-sided Fourier
transform of the reservoir two-point correlation functions,
\begin{equation}
  \Gamma_{\alpha, \beta}(\omega) = \int_0^\infty {\rm d} \tau \, e^{i \omega \tau} \:
        {\rm Tr}_{\tt E} \, \Big[ \hat{\tilde{E}}_\alpha^\dagger(t) \,
          \hat{\tilde{E}}_\beta(t-\tau) \, \rho_{\tt E} \,\Big] \,.
\end{equation}
Note that $\Gamma$ does not depend on time $t$ since, under
the hypothesis of a stationary reservoir, its correlation functions
are homogeneous in time and the argument of the trace can be shifted
into: $\big[ \hat{\tilde{E}}_\alpha^\dagger(\tau) \,
  \hat{\tilde{E}}_\beta(0) \, \rho_{\tt E} \big]$.

At this point, we can invoke the secular approximation.  Given the
characteristic evolution time $\tau_{\tt s}$ of the system (that is,
the time scale over which the properties of ${\tt s}$ change
appreciably), this is generally of the order of $\tau_{\tt s} \approx
|\omega'-\omega|^{-1}$, for $\omega' \neq \omega$.  If $\tau_{\tt s}
\gg \tau_{\tt E}$ (holding for weak system-bath couplings), the terms
in~\eqref{eq:vN-Markov3} involving large energy differences ($\omega'
\neq \omega$) will be fastly oscillating, thus averaging out to zero.
According to the secular approximation, only the resonant terms
$\omega = \omega'$ significantly contribute to the dynamics, while all
the others are neglected, so that the sum over $\omega'$ and the
imaginary exponential in Eq.~\eqref{eq:vN-Markov3} drop.

We finally decompose the environment correlation matrix $\Gamma$ into
its Hermitian and non-Hermitian parts, $\Gamma_{\alpha, \beta}(\omega)
=\tfrac12 \gamma_{\alpha,\beta}(\omega) + i \pi_{\alpha, \beta}
(\omega)$, with
\begin{equation}
  \gamma_{\alpha,\beta}(\omega)= \Gamma_{\alpha, \beta}(\omega) +
  \Gamma^*_{\beta, \alpha}(\omega)\,, \qquad \quad
  \pi_{\alpha,\beta}(\omega) = \frac{1}{2i} \Big[
    \Gamma_{\alpha, \beta}(\omega) - \Gamma^*_{\beta, \alpha}(\omega) \Big] \,.
\end{equation}
With these definitions, it is possible to separate the Hermitian and
non-Hermitian parts of the dynamics ruled by Eq.~\eqref{eq:vN-Markov3}
and then transform back to the Schr{\"o}dinger picture, to obtain
\begin{equation}
  \frac{{\rm d} \rho_{\tt s}(t)}{{\rm d}t} = -i \big[ \hat H + \hat H_{Ls},
    \rho_{\tt s}(t) \big] + \mathbb{D}[\rho_{\tt s}(t)] \,,
\label{eq:Lindblad1}
\end{equation}
where the first term in the right-hand side provides the coherent
Hamiltonian dynamics of the system.  The second term, given by the
superoperator
\begin{equation}
  \mathbb{D}[\rho_{\tt s}(t)] = \sum_\omega \sum_{\alpha,\beta} \gamma_{\alpha,\beta}
  \Big[ \hat S_\beta(\omega) \, \rho_{\tt s}(t) \, \hat S^\dagger_\alpha(\omega)
    - \frac12 \big\{ \hat S^\dagger_\alpha(\omega) \, \hat S_\beta(\omega),
    \rho_{\tt s}(t) \big\} \Big] \,,
\end{equation}
with $\{ \hat A, \hat B \} \equiv \hat A \hat B + \hat B \hat A$ denoting
the anticommutator, accounts for the interaction between the system
and the environment. Note that the Hamiltonian dynamics becomes
influenced by an additional term
\begin{equation}
  \hat H_{Ls} =\sum_{\omega} \sum_{\alpha, \beta} \hat \pi_{\alpha,\beta}(\omega) \,
  \hat S^\dagger_\alpha(\omega) \, \hat S_\beta(\omega) \,,
\end{equation}
usually called {\em Lamb-shift} Hamiltonian, whose role is to
renormalize the system energy levels due to the interaction with the
environment.

Since $\gamma_{\alpha,\beta}(\omega)$ can be seen as the Fourier
transform of the positive function ${\rm Tr}\, [\hat
  E^\dagger_\alpha(\tau) \, \hat E_\beta(0) \, \rho_{\tt E}]$, the matrix
composed of its entries can be diagonalized.  Thus we finally arrive
at the GKLS master equation (hereafter shortened into the {\em
  Lindblad} master equation), given by Eq.~\eqref{eq:Lindblad1} with
the dissipator
\begin{equation}
  \mathbb{D} [\rho_{\tt s}(t)] = \sum_\omega  \sum_{\alpha} \gamma_{\alpha}
  \Big[ \hat L_\alpha(\omega) \, \rho_{\tt s}(t) \, \hat L^\dagger_\alpha(\omega)
    - \frac12 \big\{ \hat L^\dagger_\alpha(\omega) \, \hat L_\alpha(\omega),
    \rho_{\tt s}(t) \big\} \Big] \,.
  \label{eq:Lindblad2}
\end{equation}
The operators $\hat L_\alpha(\omega)$, usually referred to as the
Lindblad jump operators, are obtained through a unitary transformation
which diagonalizes $\gamma_{\alpha,\beta}(\omega)$,
while $\gamma_\alpha  \geq 0$ are its eigenvalues.~\footnote{Hereafter,
  when dealing with the Lindblad master equation, we will omit
  the subscript ${}_{\tt s}$ since the environment is no longer
  explicitly present in Eqs.~\eqref{eq:Lindblad1}-\eqref{eq:Lindblad2}.}

The Lindblad master equation~\eqref{eq:Lindblad1}-\eqref{eq:Lindblad2}
can be shown to be the most general form of continuous-time evolutor
which guarantees that its solution $\rho(t)$
is a well-defined density operator at all times,
preserving its trace and complete positivity.
More precisely, it defines the Liouvillian superoperator $\mathbb{L}$,
\begin{equation}
  \frac{{\rm d} \rho(t)}{{\rm d}t} = \mathbb{L} [\rho(t)]\,, \qquad
  \rho(t) = e^{\mathbb{L} t} \rho(0) \,,
  \label{eq:liouvillian}
\end{equation}
which represents the most general generator of a quantum dynamical
semigroup, i.e., a continuous, one-parameter family of dynamical maps
satisfying the semigroup property:
$\mathbb{L}(t_1) \, \mathbb{L}(t_2) = \mathbb{L}(t_1+t_2)$,
$t_1, t_2 \geq 0$.

\subsubsection{Local Lindblad master equations}

Finding suitable expressions for the jump operators $\hat L_\alpha (\omega)$,
the Lamb-shift correction to the Hamiltonian, and the couplings
$\gamma_\alpha$ entering Eqs.~\eqref{eq:Lindblad1}-\eqref{eq:Lindblad2}
can be a laborious problem, especially for complex systems.
In most situations, {\em local} system-bath coupling schemes are assumed
from phenomenological grounds, so that typically the $\hat H_{Ls}$ correction
is neglected, while the Lindblad operators $\hat L_j$ are taken as acting
on an appropriate spatial coordinate (e.g., a single site of a quantum
lattice model). This represents the standard framework for quantum
many-body systems, giving reliable results for quantum optical
devices~\cite{SBD-16}. In such context, one of the techniques that
have been employed is the Keldysh path-integral
approach~\cite{DDLSS-13, SBD-16}.
Several paradigmatic examples presented in the next sections
of this review are based on local system-environment couplings.
However we wish to warn the reader that, strictly speaking,
the eigenoperators $\hat S_\alpha(\omega)$ of the system
(and, consequently, the jump operators) are nonlocal.
As a result, the local approach may lead to contradictory results
and to a violation of the second principle of thermodynamics~\cite{LK-14}.

More in detail, a local shape can be recovered only after assuming
that there is no coupling between the different subsystems
(e.g., the sites) composing ${\tt s}$. On the other hand, for generic
many-body Hamiltonians, there is no reason to assume that
the local approximation is valid.
The crucial point of the derivation leading to
Eqs.~\eqref{eq:Lindblad1}-\eqref{eq:Lindblad2}
is the spectral decomposition of $\hat H_{\tt s}$, which is required
to find the eigenoperators $\hat S_\alpha(\omega)$. For interacting 
quantum many-body systems, this can be an extremely hard task,
therefore the full microscopic derivation of the
master equation is rarely used in physical situations,
apart from very specific cases~\cite{ABLZ-12}.
Among them we mention the infinite bosonic tight-binding chain~\cite{SL-16}
and the class of quadratic quantum many-body systems~\cite{DR-21},
which can be however mapped into free quasiparticle systems.

\subsection{Steady states}

The long-time dynamics described by a given master equation
deals with the possible convergence towards a time-independent
steady state: $\lim_{t \to \infty} \rho(t) = \rho_{ss}$.
This is defined as a state which is annihilated by the Liouvillian
superoperator~\eqref{eq:liouvillian},
\begin{equation}
  \mathbb{L} [\rho_{ss}] = 0 \,.
\end{equation}
It can be proven that every Lindblad master equation admits
at least one steady state.
Determining whether this steady state is unique or not is a subtle issue:
although a number of theorems have been proposed to characterize
the conditions under which a unique steady state is expected,
a conclusive statement on this subject has not been reached
so far~\cite{Davies-69, Davies-70, Evans-77, SW-10, Nigro-19}.

Nevertheless, an important result put forward by Spohn~\cite{Spohn-77}
states that, if the set of Lindblad operators $\{ \hat L_i \}$
is self-adjoint and its bicommutant $\{ \hat L_i \}''$ equals
the entire operator space, then the steady state is
unique~\footnote{We recall that the commutant $\{ \hat L_i \}'$
  is defined as the set of operators which commute with all the $\hat L_i$
  operators, while the bicommutant $\{ \hat L_i \}''$ is the commutant
  of the commutant.}.
This ensures that a relevant class of many-body Lindblad master equations
admits a unique steady state, as for quadratic fermionic or bosonic
systems linearly coupled to an environment modeled
by a set of independent thermal baths~\cite{DR-21}.
Other many-body systems can be proven to belong to this category,
such as those made by finite-dimensional subsystems, and described by
local Lindblad master equations in which the jump operators
are associated either to raising or to lowering operators~\cite{Nigro-19}.
Nonetheless, even if one is sure that the steady state is unique,
there is no reason to believe that the relaxation process must be
of the thermal kind: the open-system dynamics
may originate nontrivial nonequilibrium steady states.

\subsection{Validity of the Born-Markov approximations}

We conclude this overview on open quantum systems by mentioning
that, while justified in certain contexts, the approximations
leading to the Redfield and the Lindblad master
equations are not necessarily satisfied in physical situations.
In particular, they fail when the system-environment interaction
is not very weak and leads to long-lasting and non-negligible
correlations~\cite{BP-book}.
This is the most common scenario in several realistic cases
(e.g., in solid-state systems), where the open-system dynamics is
inevitably non-Markovian and is affected by important memory effects.

Although worth being investigated, we will not pursue further
in this direction, due to the lack of substantial results
in the many-body context (see however Sec.~\ref{oscibaths}, where we
discuss a different system-bath interaction scheme, in which
the environment is modeled by an infinite set of harmonic
oscillators, that may lead to rather different physical behaviors).
The reader who is interested in further details on the mathematical
derivation of such situations and of non-Markovianity
witnesses can have a look at Refs.~\cite{RH-book, RHP-14, BLPV-16, DA-17},
based on a quantum-information perspective.

\section{Dissipative perturbations at quantum transitions}
\label{dissQT}

\begin{figure}
  \begin{center}
    \includegraphics[width=0.95\columnwidth]{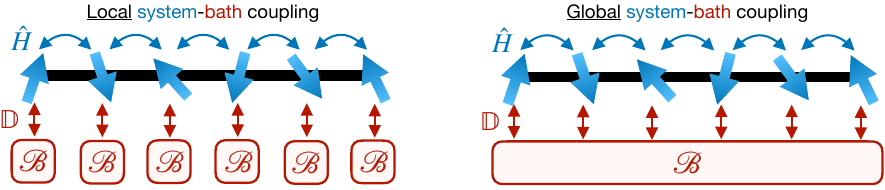}
    \caption{Sketch of a quantum spin-chain model locally and globally
      coupled to external baths.  Neighboring spins are coupled
      through a coherent Hamiltonian $\hat H$ (bidirectional blue
      arrows). Each spin is also homogeneously and weakly coupled to
      some external bath ${\cal B}$ via a set of dissipators
      $\mathbb{D}$ (vertical red arrows), whose effect is to induce
      incoherent dissipation.  The environment can be modeled either
      as a sequence of local independent baths, each for any spin of
      the chain (left drawing), or as a single common bath to which
      each spin is supposed to be uniformly coupled (right
      drawing). Adapted from Ref.~\cite{DRV-20}.}
    \label{fig:sketch}
  \end{center}
\end{figure}

In this section we discuss the dynamics of a many-body system in
proximity of a QT, in the presence of dissipative perturbations
arising from the interaction with the environment.  We consider a
class of dissipative mechanisms whose dynamics can be effectively
described through a Lindblad master equation governing the time
evolution of the system's density matrix~\cite{BP-book, RH-book} (see
Sec.~\ref{sec:MEq}).  Within this framework, the evolution of the
density matrix $\rho(t)$ of the many-body system is described by the
Lindblad master equation~\cite{BP-book}
\begin{subequations}
\label{eq:lind}
\begin{equation}
  {{\rm d} \rho\over {\rm d} t} = {\cal L}[\rho]:=
  -i \big[ \hat H,\rho \big] + u
  \,{\mathbb D}[\rho]\,, \qquad {\mathbb D}[\rho] = \sum_o {\mathbb
    D}_o[\rho]\,,
  \label{lindblaseq}
\end{equation}
where ${\cal L}[\rho]$ represents the generator of the dissipative
dynamics, its first term provides the coherent driving, while its
second term accounts for the coupling to the environment.  Its form
depends on the nature of the dissipation arising from the interaction
with the bath, which is effectively described by a set of dissipators
${\mathbb D}_o$, and a global coupling strength $u>0$.  In the case of
systems weakly coupled to Markovian baths, the trace-preserving
superoperator ${\mathbb D}_o[\rho]$ can be generally written
as~\cite{Lindblad-76, GKS-76}
\begin{equation}
  {\mathbb D}_o[\rho] = \hat L_o \rho \hat L_o^\dagger - \tfrac{1}{2}
  \big( \rho\, \hat L_o^\dagger \hat L_o + \hat L_o^\dagger \hat L_o
  \rho \big)\,,
  \label{dL}
\end{equation}
\end{subequations}
where $\hat L_o$ is the Lindblad jump operator associated with the
system-bath coupling scheme.  In the following discussion we 
restrict to homogeneous dissipation mechanisms, preserving
translational invariance, as depicted, for example, in
Fig.~\ref{fig:sketch}.  In quantum optical implementations, the
conditions leading to Eqs.~\eqref{eq:lind} are typically
satisfied~\cite{SBD-16}, therefore this formalism constitutes the
standard choice for theoretical investigations of this kind of
systems.

It is worth mentioning that an increasing interest is developing
on the study of driven-dissipative QTs in many-body systems governed
by the Lindblad master equation~\eqref{eq:lind}, i.e., on QTs
that may emerge in the steady-state density matrix of the dynamics
generated by the Liouvillian ${\cal L}[\rho]$.
A number of works recently started to tackle this problem under
different perspectives, both theoretical (see, e.g.,
Refs.~\cite{DTMFZ-10, LHC-11, Kessler-etal-12, LGL-13, Jin-etal-16,
  MG-16, RSBFC-17, Jin-etal-18, MBBC-18, Biella-etal-18, LSM-20, VRSW-20})
and experimental (see, e.g., Refs.~\cite{RSSTH-14, FSLKH-17, TNDTT-17}).
However, we will not pursue this topic further in this review.

\subsection{Quench protocols to test coherent and dissipative dynamics}
\label{settingdiss}

The interplay between the critical coherent dynamics and the dissipative
mechanisms can be addressed within a simple dynamic protocol.  We assume
again that the many-body Hamiltonian can be written as in
Eq.~\eqref{hlam}, i.e., $\hat{H}(w) = \hat{H}_{c} + w \hat{H}_{p}$, where
the term associated with the Hamiltonian parameter $w$ represents a
relevant perturbation to the critical Hamiltonian $\hat{H}_c$.  We
consider the following protocol:
\begin{itemize}
\item[$\bullet$] The system is prepared in the ground state of the
  Hamiltonian in Eq.~\eqref{hlam}, associated with a given
  value $w_i$, that is, the initial state at $t=0$ is $\rho(0) =
  |\Psi_0(w_i)\rangle \langle \Psi_0(w_i)|$. Alternatively, one may
  consider an initial state at temperature $T$, described by the Gibbs
  distribution $\propto \exp[-\hat H(w_i)/T]$.

\item[$\bullet$] Then, the dissipative interaction with the environment
  is suddenly turned on, so that the system evolves according to the
  Lindblad master equation~\eqref{eq:lind} with $u>0$.  One may
  also allow for an instantaneous quench of the Hamiltonian parameter
  $w$, from $w_i$ to $w\neq w_i$.
\end{itemize}

The approach to the asymptotic stationary states is generally
controlled by the Liouvillian gap $\Delta_{\cal L}$ associated with
the generator ${\cal L}$ of the dissipative dynamics [see
Eq.~\eqref{lindblaseq}]~\cite{BP-book, RH-book, Znidaric-15, MBBC-18, SK-20}.
In fact, the asymptotic stationary state is provided by the eigenstate of
${\cal L}$ with vanishing eigenvalue, $\Lambda_0=0$, while all the other
eigenstates have eigenvalues $\Lambda_i$ with negative real part
[i.e., ${\rm Re}\,(\Lambda_i)<0$ for any $i>0$]. The convergence to the
stationary state is ruled by the Liouvillian eigenvalue with the largest
nonzero real part, that is,
\begin{equation}
  \Delta_{\cal L} = - {\rm Max}_{i>0} \, {\rm Re}\,(\Lambda_i)\,.
  \label{deltadeb}
\end{equation}

The dissipator ${\mathbb D} = \sum_o {\mathbb D}_o$ can drive the
system to a steady state, which is generally noncritical, even when
the Hamiltonian parameters are critical.  However, one can identify a
dynamic regime where the dissipation is sufficiently small to compete
with the coherent evolution driven by the critical Hamiltonian,
leading to dynamic scaling behaviors somehow controlled by the
universality class of the QT of the isolated system.  This occurs
within a low-dissipation regime, where the decay rate of the
dissipation is comparable with the ground-state energy gap of the
Hamiltonian.  Such low-dissipation regime naturally emerges within
dynamic scaling frameworks, also involving 
the strength $u$ of the coupling with the dissipators in the Lindblad
equation~\cite{NRV-19-cd, RV-19-ss}.  Analogously to the scaling laws
of closed systems at QTs, the dynamic scaling behavior in the presence
of dissipation is expected to be universal, i.e., largely independent
of the microscopic details.

We remind the reader that hereafter we only focus on situations close to
a QT point for the corresponding closed system, so that the
dissipation acts as a perturbation. On the other hand, we are not
going to address nonequilibrium phenomena leading to novel critical
behaviors and/or QTs in the steady states, which are genuinely induced
by the mutual interplay of the unitary and the dissipative dynamics.

Other different aspects concerning localized dissipative interactions
with the environment have been discussed in the literature
(see, e.g.,
Refs.~\cite{PP-08, Prosen-08, BCPRZ-09, Prosen-11, Znidaric-15, VCG-18,
  FCKD-19, TFDM-19, BMA-19, SK-20, WSDK-20, FMKCD-20, Rossini-etal-21,
  AC-21, TV-21}),
and, in particular, for the case of quantum Ising-like and fermionic
chains with dissipative interactions at the ends of the
chain~\cite{PP-08, Prosen-08, BCPRZ-09, Prosen-11, Znidaric-15, VCG-18,
  SK-20, TV-21}. Hereafter we will not deal with such issues.

\subsection{Quantum thermodynamics of the dynamic process}
\label{qutherm}

The out-of-equilibrium quantum dynamics associated with the
above-mentioned protocol can be monitored by considering standard
observables, such as those defined in Eqs.~(\ref{magnt}-\ref{gct}).
Other useful quantities are related to the thermodynamics of the
process, such as the average work and heat characterizing the quantum
thermodynamic properties of the out-of-equilibrium dynamics associated
with the dissipative interaction with the environment.

The first law of thermodynamics describing the energy flows of the
global system, including the environment, can be written
as~\cite{DC-book, BCGAA-book, GMM-book}
\begin{equation}
  {{\rm d} E_s \over {\rm d}t} = W^\prime(t) + Q^\prime(t) \,, \label{filaw}
\end{equation}
where $E_s$ is the average energy of the open system,
\begin{equation}
  E_s \equiv {\rm Tr}\, \big[ \rho(t) \, \hat H(t) \big]\,,
  \label{esdef}
\end{equation}
and
\begin{equation}
  W^\prime(t) \equiv  {{\rm d} W\over {\rm d}t} = {\rm Tr} \bigg[ \rho(t)\, 
    {{\rm d} \hat H(t)\over {\rm d}t} \bigg]\,,\qquad
  Q^\prime(t)  \equiv  {{\rm d} Q\over {\rm d}t} = {\rm Tr} \bigg[
    {{\rm d}\rho(t)\over {\rm d}t}\, \hat H(t) \bigg]  
  \,,\label{heatwork}
\end{equation}
with $W$ and $Q$ respectively denoting the average work done on the
system and the average heat interchanged with the environment.

In the quench protocol outlined in Sec.~\ref{settingdiss}, a
nonvanishing work is only done at $t=0$, when the longitudinal-field
parameter suddenly changes from $w_i$ to $w$. After that, the
Hamiltonian is kept fixed for any $t>0$, therefore the average work is
simply given by the static expectation value
\begin{equation}
  W = \langle \Psi_0(w_i) | \hat H(w) - \hat H(w_i) |
  \Psi_0(w_i)\rangle = (w - w_i) \, \langle \Psi_0(w_i)
  | \hat{H}_p | \Psi_0(w_i)\rangle \,.
  \label{wokih}
\end{equation}
%where $|\Psi_0(w_i)\rangle$ is the starting ground state associated with
%the Hamiltonian parameter $w_i$.
Note that the average work of this
protocol is the same of that arising at sudden quenches of closed
systems, already discussed in Sec.~\ref{workflu}.  On the other hand,
the heat interchange with the environment is strictly related to the
dissipative mechanism. Indeed one can easily derive the relation
\begin{equation}
  Q^\prime(t) = u \, {\rm Tr} \big[ {\mathbb D}[\rho]\, \hat H(t)
    \big]\,,
  \label{qteq}
\end{equation}
by substituting the right-hand side of the Lindblad
equation~\eqref{lindblaseq} into the expression~\eqref{heatwork}.

\subsection{Dynamic scaling laws in the presence of homogeneous dissipation}
\label{dissdynprot}

To account for the role of the dissipators, the dynamic FSS theory
outlined in the previous sections has to be extended. 
In fact, the effect of a sufficiently low dissipation can be taken into
account by adding a further dependence on the variable associated with
$u$ in the dynamic homogeneous laws~\eqref{scadyn}.
For example, one may write~\footnote{To simplify the equations, we assume
  again translation invariance, so that only the dependence on
  ${\bm x}={\bm x}_2 - {\bm x}_1$ is required. However the equations
  can be easily generalized to systems with boundaries where
  translation invariance does not strictly hold.}
\begin{equation}
  G_{12}(t,{\bm x};L,w_i,w,u) \equiv \big\langle 
    \hat{O}_1({\bm x}_1) \, \hat{O}_2({\bm x}_2) \big\rangle_t \approx
    b^{-\varphi_{12}} \, {\cal G}_{12}(b^{-\zeta} t, b^{-1}{\bm x}, b^{-1}L,
    b^{y_w} w_i, b^{y_w} w, b^\zeta u)\,, \label{g12scadyndi}
\end{equation}
in which we added the argument $b^\zeta u$, where $\zeta$ is a
suitable exponent, to ensure the substantial balance (thus
competition) with the critical coherent driving.  As argued in
Refs.~\cite{YMZ-14, NRV-19-cd}, the exponent $\zeta$ appropriate for
homogeneous dissipation should generally coincide with the dynamic
exponent $z$.  This is a natural conjecture, due to the fact that the
parameter $u$ in Eq.~\eqref{lindblaseq} plays the role of a decay
rate, i.e., of an inverse relaxation time for the associated
dissipative process~\cite{BP-book}, and any relevant time scale at a
QT scales as $\Delta^{-1}$.

As usual, to derive dynamic FSS laws one can set $b=L$, such that
\begin{equation}
  G_{12}(t,{\bm x}; L, w_i, w, u) \approx L^{-\varphi_{12}} \, {\cal
    G}_{12}(\Theta, {\bm X}, \Phi_i, \Phi,\Gamma)\,, \label{g12scadyndi2}
\end{equation}
where the scaling variables $\Theta,{\bm X},\Phi,\delta_w$ (or $\Phi_i$)
have been defined
within the dynamic FSS of closed systems, cf.~Eq.~\eqref{scalvarque},
i.e., $\Theta = L^{-z} t$, ${\bm X} = {\bm x}/L$, $\Phi = L^{y_w} w$,
and $\delta_w = {w/w_i}-1$, while the new scaling variable associated
with the dissipation is given by
\begin{equation}
  \Gamma \equiv L^z \, u \,.
  \label{Gammadef}
  \end{equation}
This implies that, to observe a nontrivial competition between
critical coherent dynamics and dissipation, one should consider a
sufficiently small coupling $u\sim L^{-z}$, so that its size is
comparable with the energy difference $\Delta\sim L^{-z}$ of the
lowest energy levels of the Hamiltonian.  Therefore, the dissipative mechanism
dominates for larger values $u$, and in particular when keeping $u$
constant in the large-$L$ limit, while it becomes negligible when
$u\ll L^{-z}$, thus $\Gamma\ll 1$.

Scaling laws in the thermodynamic limit, holding for $|w|\neq 0$, are
obtained by fixing $b$ as in Eq.~\eqref{bla1}, i.e.,
$b=\lambda=|w|^{-1/y_w}$ with $b\to\infty$ when $w$ approaches the
critical point $w=0$, and taking the limit $L/\lambda\to \infty$.
This leads to
\begin{equation}
  G_{12}(t,{\bm x};w_i,w,u) \approx \lambda^{-\varphi_{12}} \,
  {\cal G}_\infty(\lambda^{-z} t, \lambda^{-1}{\bm x}, \lambda^{y_w} w_i ,
  \lambda^{z} u) \,.
%  \,,\qquad \lambda=|w|^{-1/y_w}\,.
  \label{g12scaqueildis}
\end{equation}
Assuming that the above dynamic scaling laws apply to the whole
dynamic evolution, including the approach to the stationary states
(i.e., no other dynamic regimes emerge at late times), we may
consider the large-time limit of the scaling
equation~\eqref{g12scaqueildis} to derive scaling laws for the
asymptotic stationary states.  If the asymptotic stationary state is
unique (i.e., independent of the initial conditions of the protocol),
as is the case for several classes of dissipators~\cite{Davies-70,
  Evans-77, SW-10, Nigro-19}, the stationary state should appear when
$t\gg\lambda^z$.  Therefore, the following large-time scaling law
follows:
\begin{equation}
  \lim_{t\to\infty} G_{12}(t,{\bm x};w_i,w,u) \equiv
  G_{12,\infty}({\bm x};w,u) \approx \lambda^{-\varphi_{12}} \, {\cal
    G}_{\infty}(\lambda^{-1} {\bm x}, \lambda^{y_w} w, \lambda^{z} u)
  \label{scagiti}\,,
\end{equation}
assuming that, under the hypothesis of uniqueness of the stationary state,
the dependence on the initial parameter $w_i$ disappears.
From the exponential large-distance decay of the correlation function
$G_{12,\infty}$, it is also possible to define a correlation length
$\xi_s$.  Indeed, since we expect that $G_{12,\infty}\sim
e^{-x/\xi_s}$ (here $x = |{\bm x}|$), we may define
\begin{equation}
  \xi_s^{-1}(w,u) \equiv -{\rm lim}_{x\to\infty} \, {{\rm ln} \,
    G_{12,\infty}(x;w,u)\over x} \,.
  \label{largedistdef}
\end{equation}
Using the scaling laws derived for $G_{12,\infty}$, we finally obtain
\begin{equation}
  \xi_s \approx \lambda \, \widetilde {\cal L}(\lambda^{z} u)
  \approx u^{-1/z} \, {\cal L}(\lambda^{z} u)\,.
  \label{xiscai1}
\end{equation}

These scaling equations are formally analogous to the equilibrium
scalings at finite temperature, after replacing $u$ with the
temperature $T$, see Sec.~\ref{freeen}.  However, we stress that the
above scaling arguments extend to cases where the asymptotic
stationary state does not coincide with a Gibbs equilibrium state.
In fact, this is the case for the Kitaev model subject to
incoherent particle decay or pumping, which we shall discuss below.
They are expected to hold for homogeneous local or global external
baths (see Fig.~\ref{fig:sketch}) described within a Lindblad
approach.  We also expect that a straightforward extension may apply
for non-Markovian baths~\cite{DA-17}.

\subsection{Numerical evidence of scaling at continuous
  quantum transitions}
\label{kitaevmod}

The behavior of many-body systems at QTs in the presence of
dissipative mechanisms has been investigated numerically, for quantum
Ising chains and related Kitaev quantum wires.  In particular,
Refs.~\cite{YMZ-14, YLC-16} report an analysis for the critical
quantum Ising chain (around $g=g_c=1$ and $h=0$) in the presence of
specific dissipators that guarantee standard thermalization. This
substantially confirms the dissipation scaling laws outlined above,
using systems up to $L=O(10)$, due to the high complexity of the
numerical computations in the presence of dissipators. A substantial
progress can be achieved by considering the Kitaev quantum
wire~\eqref{kitaev2} with translation-invariant boundaries,
such as ABC and PBC, assuming local dissipative mechanisms described
by Lindblad equations.  Indeed, this model enables to simulate very
large system sizes, up to $L=O(10^3)$~\cite{Prosen-08}.

We now summarize the latter results on Fermi lattice gases
described by 1$d$ Kitaev models in the presence of a
dissipator ${\mathbb D}[\rho]$, defined as a sum of local
(single-site) terms. These can be taken of the form
\begin{equation}
  {\mathbb D}_x[\rho] = \hat L_x \rho \hat L_x^\dagger - \tfrac{1}{2}
  \big( \rho\, \hat L_x^\dagger \hat L_x + \hat L_x^\dagger \hat L_x
  \rho \big)\,,
  \label{dLj}
\end{equation}
where $\hat L_x$ denotes the Lindblad jump operator associated with
the system-bath coupling scheme, and the index $x$ corresponds to a
lattice site.  The onsite Lindblad operators $\hat L_x$ describe the
coupling of each site with an independent bath.  Dissipation operators
associated with either particle losses (l), pumping (p), or dephasing
(d), are defined as~\cite{HCG-13, KMSFR-17, NRV-19-cd, RV-19-ss}
\begin{equation}
  \hat L_{{\rm l},x} \equiv \hat c_x \,, \qquad \hat L_{{\rm p},x} \equiv
  \hat c_x^\dagger \,, \qquad \hat L_{{\rm d},x} \equiv \hat n_x\,,
  \label{loppe}
\end{equation}
respectively.  The choice of such dissipators turns out to be
particularly convenient for the numerical analysis, allowing to scale
the difficulty of the problem linearly with $L$, and thus to obtain
results for systems with thousands of sites.  In the presence of
particle losses (l) or pumping (p) and for translationally invariant
systems, the driven-dissipative quantum dynamics ruled by
Eq.~\eqref{eq:lind} can be exactly solved by decoupling in Fourier
space the various sectors with different momenta, analogously to
fermionic Gaussian Hamiltonian models.  On the other hand, although
the quantum dynamics with a dephasing (d) mechanism cannot be simply
obtained, two-point observables are still fully captured by a set of
coupled linear differential equations, whose number increases only
linearly with the number of sites $L$ (see, e.g., the appendix in
Ref.~\cite{NRV-19-cd} for details).  Therefore the apparent
exponential complexity of the Kitaev chain with ABC gets
semi-analytically reduced to a polynomial one~\cite{Prosen-08,
  Eisler-11}.

\begin{figure}
  \begin{center}
    \includegraphics[width=0.47\columnwidth]{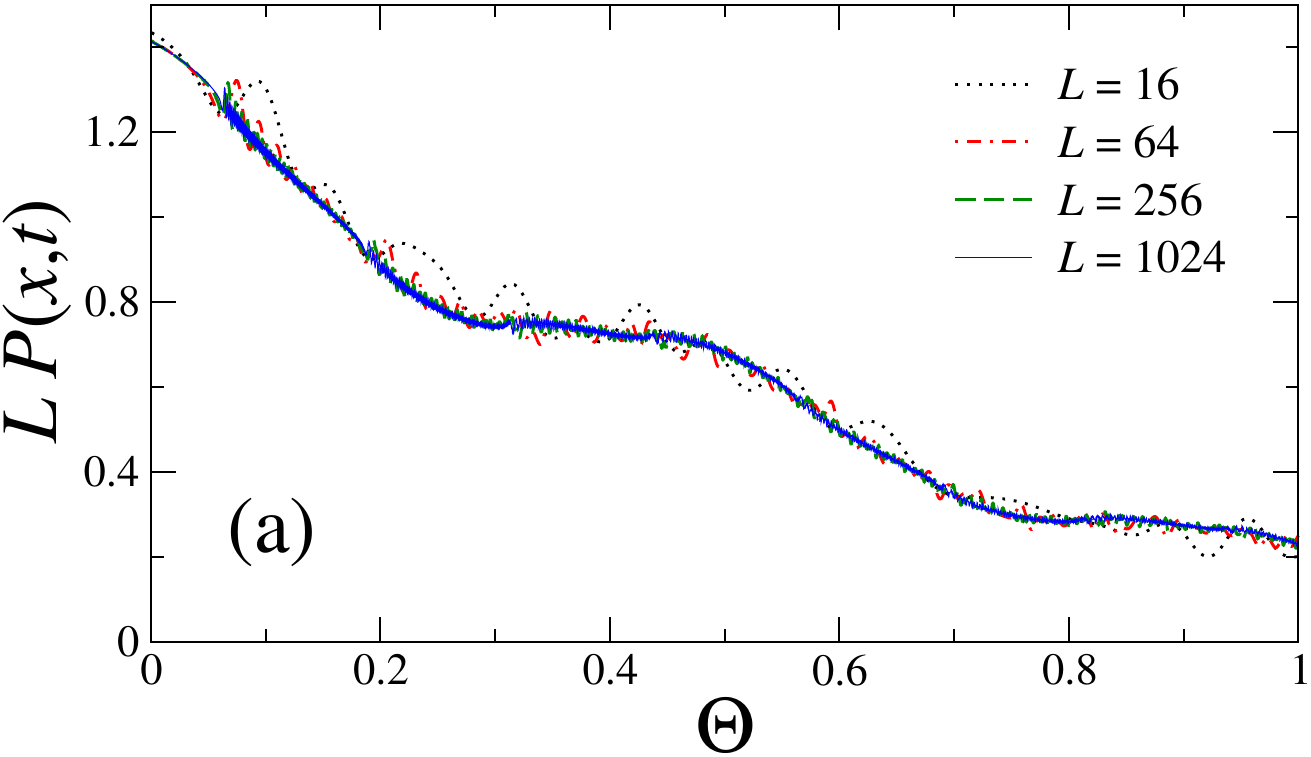}
    \hspace*{5mm}
    \includegraphics[width=0.47\columnwidth]{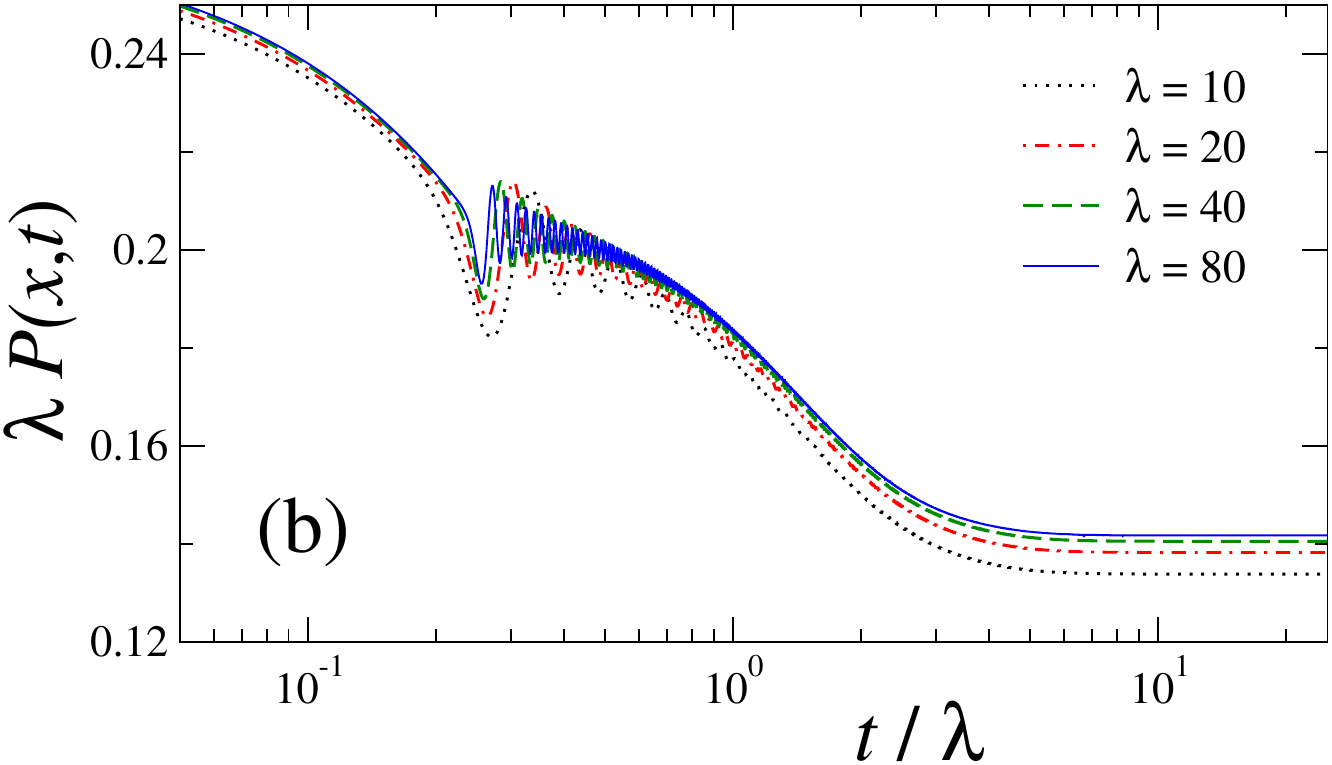}
    \caption{Dynamic scaling of the correlation function $P(x,t) =
      {\rm Tr} \big\{ \rho(t) \, \big[ \hat c^\dagger_j \hat
        c^\dagger_{j+x} + \hat c_{j+x} \hat c_j \big] \big\}$, for a
      Kitaev quantum wire subject to dissipation induced by incoherent
      particle losses, for $\Gamma =u L = 1$.  Left panel $(a)$: test
      of dynamic FSS for a system in the presence of a quench in the
      Hamiltonian parameter, from $\Phi_i =0$ to $\Phi = 2$. We fix
      $X=1/4$. The curves for different system sizes clearly approach
      a scaling function with increasing $L$, thus supporting the law
      in Eq.~\eqref{g12scadyndi2}.  Right panel $(b)$: test of dynamic
      thermodynamic-limit scaling for a static Hamiltonian, with $w_i
      = w = -1/\lambda$. We fix $x/\lambda=1$.  The curves for
      different values of $\lambda$ approach a scaling function,
      following the scaling law~\eqref{g12scaqueildis}.  Adapted from
      Refs.~\cite{NRV-19-cd, RV-19-ss}.}
    \label{Correl_L4_QD}
  \end{center}
\end{figure}

To characterize the dynamic properties of the evolution described by
the Lindblad equation, we consider again the fixed-time correlations
defined in Sec.~\ref{kitaevKZ}, cf. Eqs.~\eqref{gpcntf}, evaluated on
the mixed state $\rho(t)$. The expected dynamic scaling behavior is
that reported in Eqs.~\eqref{g12scadyndi2}, \eqref{g12scaqueildis},
and~\eqref{scagiti}, with corresponding exponents $\varphi_{12} = 1$
for the correlations $P$ and $C$, and $\varphi_{12} = 2$ for $G$, as
also mentioned in Sec.~\ref{kitaevKZ}.
The numerical results of Fig.~\ref{Correl_L4_QD} for the correlation
function $P(x,t)$ in Eq.~\eqref{eq:corrP} support: $(a)$ the dynamic
FSS predicted by Eq.~\eqref{g12scadyndi2} and thus the fact that a
nontrivial competition between critical coherence and dissipation can
be only observed for $u\sim L^{-z}$, when $L$ increases in the dynamic
FSS limit; $(b)$ the dynamic scaling in the thermodynamic limit, as
reported in Eq.~\eqref{g12scaqueildis}~\footnote{The onset of the
thermodynamic limit can be verified by comparing the numerics at fixed
$\lambda$ and $u$, with increasing $L$. The data presented in panel
$(b)$ should be considered in the infinite-size limit with great
accuracy, since finite-size effects are invisible on the scale of the
figure.}.  In particular, the curves converge to an asymptotic
nontrivial behavior when increasing $L$ or $\lambda$.  In this latter
case, the long-time limit of such a curve for $\lambda \, P$ is
different from zero, thus signaling the approach to a nontrivial
stationary state.  Therefore, in the case of homogenous dissipative
schemes, the conjectured dynamic scaling, when keeping $\gamma=L^z u$
fixed, also describes the approach to the asymptotic stationary
states. Indeed, for homogenous local dissipative mechanisms such as
pumping or decay, $\Delta_{\cal L}$ scales as $\Delta_{\cal L} \sim
1/L$ when keeping $\gamma=L u$ fixed, analogously to the critical gap
$\Delta_L \sim 1/L$ at the CQT of the Kitaev wire.  Therefore, the
dynamic scaling can follow the whole dynamic process from $t=0$ to the
asymptotic stationary states.

The dissipative scaling scenario has been numerically
challenged~\cite{DRV-20} also within the quantum Ising
chain~\eqref{hisdef} at its CQT, using the longitudinal field $h$ as
driving Hamiltonian parameter, and local Lindblad operators $\hat
L_x^{\pm} = \hat\sigma_x^{\pm} = \tfrac12 \big[ \hat\sigma_x^{(1)} \pm
  i \hat\sigma_x^{(2)} \big]$. The computations of the dissipative
dynamics are definitely harder for this problem, thus limiting them to
much smaller systems, up to size $L\approx 10$.~\footnote{An exact
numerical solution of the Lindblad master equation~\eqref{eq:lind} for
a system as the one in Eq.~\eqref{hedef} generally requires a huge
computational effort, due to the large number of states in the
many-body Hilbert space, which increases exponentially with the system
size, as $2^L$.  More precisely, the time evolution of the density
matrix $\rho(t)$, which belongs to the space of the linear operators
on that Hilbert space, can be addressed by manipulating a Liouvillian
superoperator of size $2^{2L} \times 2^{2L}$. This severely limits the
accessible system size to $L \lesssim 10$ sites, unless the model is
amenable to a direct solvability.  A notable example in this respect
is the Kitaev chain with one-body Lindblad operators, whose
corresponding Liouvillian operator is quadratic in the fermionic
creation and annihilation operators (see, e.g.,
Ref.~\cite{NRV-19-cd}).  Unfortunately, this is not the case for
dissipative Ising spin chain, in which the Jordan-Wigner mapping of
Lindblad operators constructed with the longitudinal spin operators
$\hat \sigma_x^{(1)}$ and $\hat \sigma_x^{(2)}$ produces a nonlocal
string operator forbidding an analytic treatment.} Nevertheless, the
obtained results are sufficient to support the dynamic scaling theory
outlined in this section.

At this stage, we point out that notable differences emerge when the
dissipative interactions with the environment are only localized at
the boundary of a system at a CQT~\cite{TV-21}.  In particular the
quantum evolution of fermionic Kitaev wires in the presence of
dissipative (loss and pumping) boundaries, arising from protocols
analogous to those considered in the presence of bulk dissipation,
show two different dynamic regimes.  There is an early-time regime for
times $t\sim L$, where the competition between coherent and incoherent
drivings develops a dynamic FSS analogous to that applying to bulk
dissipations, but the boundary dissipative parameter (decay rate) does
not apparently require to be tuned to zero. Then, there is a
large-time regime for $t\sim L^3$, whose dynamic scaling describes the
late quantum evolution leading to the $t\to\infty$ stationary
states. The large time scales $t\sim L^3$ are essentially related to
the fastest decay of the Lindbladian gap $\Delta_{\cal L}\sim L^{-3}$,
which characterize several quantum spin chains and fermionic wires
with boundary dissipations~\cite{PP-08, Prosen-08, Znidaric-15,
  SK-20}.

\subsection{Dissipative dynamics at first-order quantum transitions}
\label{dissfoqt}

The study of the effects of dissipative perturbations at QTs has been
extended to FOQTs, where, analogously to what happens at CQTs, it is
possible to recover a regime in which a nontrivial dynamic scaling
behavior is developed~\cite{DRV-20}. This occurs when the dissipation
parameter $u$ (globally controlling the decay rate of the dissipation
within the master Lindblad equation) scales as the energy difference
$\Delta$ of the lowest levels of the Hamiltonian of the many-body
system, i.e., $u\sim \Delta$.  However, unlike CQTs where $\Delta$ is
power-law suppressed, at FOQTs the energy difference
$\Delta$ is exponentially suppressed
with increasing the system size (provided the BC do not favor any
particular phase~\cite{CNPV-14, PRV-20}).

\subsubsection{Quantum Ising models with local dissipators}
\label{loglsiddi}

Dissipative effects at FOQTs were studied within the paradigmatic 1$d$
Ising chain with PBC, along its FOQT line~\cite{DRV-20}, thus with
an exponentially suppressed energy difference of the two lowest
levels, cf.~Eq.~\eqref{deltapbc}.  The same quench protocol described
in Sec.~\ref{settingdiss} has been considered, starting from the
ground state of the Hamiltonian $\hat{H}_{\rm Is}(g<g_c,h_i)$, which
is then subject to the time evolution ruled by Eq.~\eqref{eq:lind},
where $\hat H = \hat{H}_{\rm Is}(g<g_c,h)$ and ${\mathbb D}[\rho]$ is
the one reported in Eq.~\eqref{dLj} with 
\begin{equation}
  \hat L_x = \hat L_x^\pm = \hat{\sigma}_x^\pm = 
  \tfrac{1}{2} \big[ \hat\sigma_x^{(1)} \pm i \hat\sigma_x^{(2)} \big]\,,
  \label{isdissi}
\end{equation}
corresponding to mechanisms of incoherent raising ($+$) or lowering
($-$) for each spin of the chain.
The out-of-equilibrium evolution, for $t>0$, can be monitored by
measuring certain fixed-time observables, such as the longitudinal
magnetization $M(t)$ of Eq.~\eqref{magnt}, evaluated on the
density matrix $\rho(t)$ of the system at time $t$, or by considering
fixed-time spin correlations.

\subsubsection{Dissipative dynamic scaling}
\label{dissdynsca}

As discussed in Sec.~\ref{dissdynprot}, at CQTs the dissipator
${\mathbb D}[\rho]$ typically drives the system to a noncritical
steady state, even when the Hamiltonian parameters are close to those
leading to a QT.  However, one may identify a regime where the
dissipation is sufficiently small to compete with the coherent
evolution driven by the critical Hamiltonian, leading to potentially
novel dynamic behaviors.  At such low-dissipation regime, a dynamic
scaling framework can be observed after appropriately rescaling the
global dissipation parameter $u$, cf. Eq.~\eqref{lindblaseq},
introducing the corresponding scaling variable $\Gamma = u /
\Delta(L)$.  Within FOQTs, an analogous scaling scenario can be put
forward, by extending the dynamic FSS of closed systems
(Sec.~\ref{foqtdynamics}), adding a further dependence on the scaling
variable $\Gamma$~\cite{NRV-19-cd, RV-19-ss}. For example, the dynamic
scaling equation~\eqref{mcheckfoqt} for the magnetization is expected
to change into
\begin{equation}
  M(t;L,h_i,h,u) \approx {\cal M}(\Theta,\Phi_i,\Phi,\Gamma)\,,
  \qquad \Gamma = {u \over \Delta(L)} \,,
  \label{scamagu}
\end{equation}
where the scaling variables $\Theta$, $\Phi_i$ and $\Phi$, are defined
in Eqs.~\eqref{thetadeffo} and~\eqref{kappah}.

\begin{figure}
  \begin{center}
    \includegraphics[width=0.47\columnwidth]{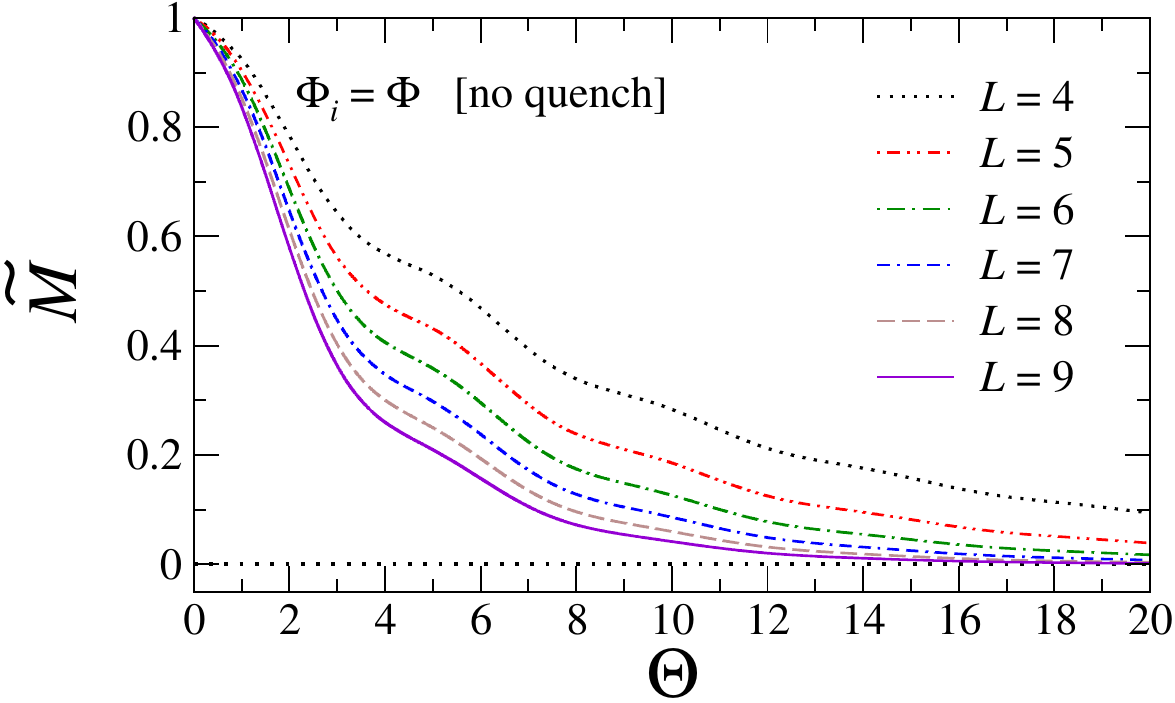}
    \hspace*{5mm}
    \includegraphics[width=0.47\columnwidth]{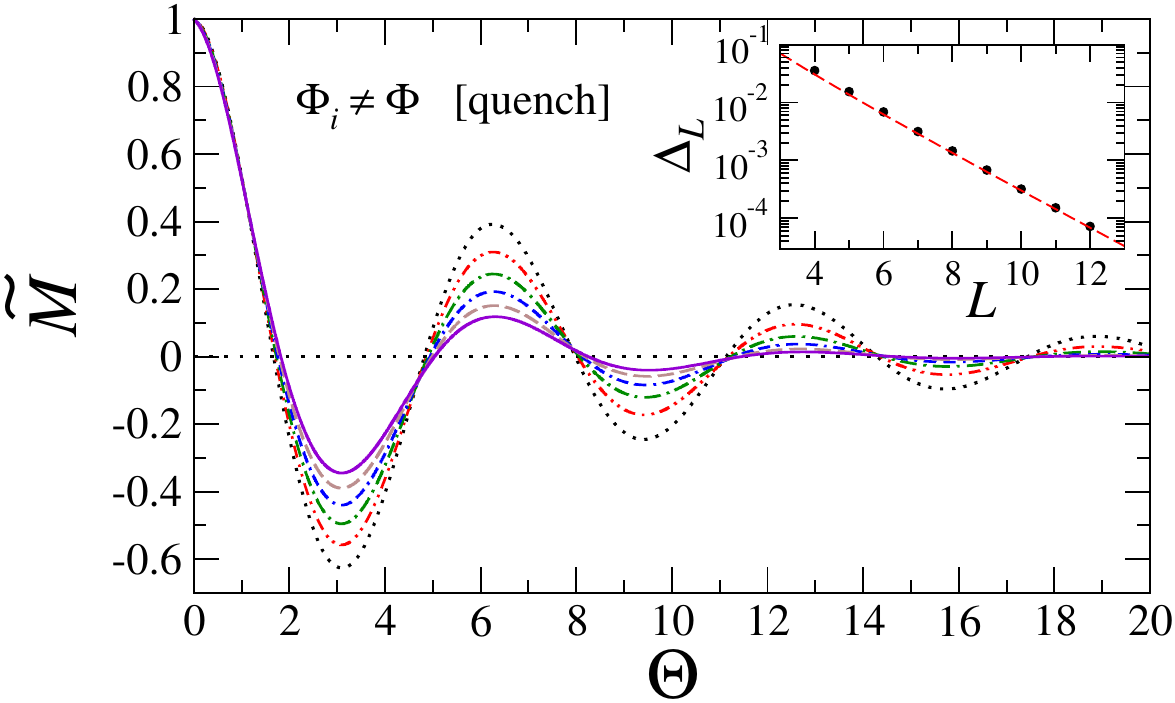}
    \caption{Time behavior of the longitudinal magnetization for a
      quantum Ising ring close to the FOQT, with $g=0.5$, in the
      presence of homogeneous dissipation described by local Lindblad
      operators $\hat L^-_x$. In particular, the figure shows the
      rescaled quantity $\widetilde{M}\equiv M(t;L,h_i,h,u)/
      M(0;L,h_i,h,u)$ versus the rescaled time $\Theta$, for different
      values of the system size $L$ (see legend).  We fix the scaling
      variables $\Phi_i=1$ and $\Gamma=0.1$.  The left panel is for
      $\Phi = \Phi_i$ (i.e., without quenching the Hamiltonian
      parameter $h$), while the right panel is for $\Phi = 0$.
      The inset of the right panel shows the energy gap $\Delta_L$ as
      a function of $L$: black circles are the results obtained from
      the numerics, while the dashed red line denotes the estimate in
      Eq.~\eqref{deltapbc}.  Adapted from Ref.~\cite{DRV-20}.}
    \label{fig:MagnLongT_FOQT_g05}
  \end{center}
\end{figure}

The numerical results reported in Fig.~\ref{fig:MagnLongT_FOQT_g05}
for the time evolution of the longitudinal magnetization $M(t)$
support the above nontrivial dynamic scaling behavior~\cite{DRV-20}.
The scaling law~\eqref{scamagu} has been checked by varying the Hamiltonian
and the dissipation parameters of the protocol with increasing size $L$,
so that the scaling variables $\Phi_i$, $\Phi$, and $\Gamma$ are kept fixed.
For any nonzero value of the dissipation variable $\Gamma$,
the longitudinal magnetization appears to asymptotically vanish
in the large-time limit, although with different qualitative trends.

We mention that the dynamic behavior under global dissipators (right
drawing of Fig.~\ref{fig:sketch}), given by
\begin{equation}
  {\mathbb D}[\rho] = 
  \hat L \rho \hat L^\dagger - \tfrac{1}{2}
  \big( \rho\, \hat L^\dagger \hat L + \hat L^\dagger \hat L \rho \big)\,,
  \qquad
  \hat L^{\pm} \equiv \hat\Sigma^{\pm}\,,\qquad
  \hat\Sigma^{\pm} \equiv {1\over L} \sum_x  \hat\sigma_x^{\pm} \,,
  \label{lglobdef}
\end{equation}
turns out to be well described by a single-spin model in the presence
of a corresponding dissipator, showing that the system behaves
rigidly, as already observed in other dynamic problems at FOQTs.

The arguments leading to the above scaling scenario at FOQTs are quite
general.  Analogous phenomena are expected to develop in any
homogeneous $d$-dimensional many-body system at a FOQT, whose Markovian
interactions with the bath can be described by local or extended
dissipators within a Lindblad equation~\eqref{eq:lind}.

The dynamic scaling behavior turns out to become apparent already for
relatively small systems, such as chains with $L\lesssim 10$.  This
makes such dynamic scaling phenomena relevant even from an
experimental point of view, where the technical difficulties in
manipulating and controlling such systems can be probably faced with
up-to-date quantum simulation platforms.

\subsection{Dissipation from the coupling with oscillator baths}
\label{oscibaths}

One may consider an alternative mechanism leading to dissipation, in
which the environment is effectively described by an infinite set of
harmonic oscillators~\cite{CL-83, Leggett-etal-87}.  For example, in
the case of a spin system, such as the quantum Ising
model~\eqref{hisdef}, one may homogeneously couple each spin with an
infinite number of quantum oscillators, through the coupling
Hamiltonian term
\begin{equation}
  \hat H_{{\rm Is-O}} = \sum_{\bm x} \hat\sigma^{(1)}_{\bm x} \sum_k \lambda_{k}\,
  (\hat{b}_{{\bm x},k}^\dagger + \hat{b}_{{\bm x},k})\,, \label{isenvh}
\end{equation}
where $\hat{b}_{{\bm x},k}^{(\dagger)}$ are annihilation (creation) operators
associated with the oscillators, whose Hamiltonian reads
\begin{equation}
  \hat H_{{\rm O}} = \sum_{\bm x} \hat H_{{\rm O}{\bm x}}\,,
  \qquad
  \hat H_{{\rm O}{\bm x}} =  \sum_k \omega_{k}\,
  \hat{b}_{{\bm x},k}^\dagger \hat{b}_{{\bm x},k}^{\phantom\dagger} \,.
  \label{heham}
\end{equation}
The dissipation arising from these couplings is essentially determined
by the spectral distribution of the oscillators coupled to
the spin located at the site ${\bm x}$, and in particular on their
power-law spectral density,
\begin{eqnarray}
  J(\omega) = \pi \sum_k \lambda_{k} \,\delta(\omega -
  \omega_{k}) = {\pi\over 2} \alpha \,\omega_c^{1-s} \, \omega^s\,, \qquad
  {\rm for}\;\; \omega<\omega_c\,,
  \label{powerlawEs}
\end{eqnarray}
and vanishing for $\omega>\omega_c$, therefore $\omega_c$ plays the role
of  high-energy cutoff, and $\alpha$ is a dimensionless measure of the
dissipation strength.  The value of the exponent $s$ determines the
qualitative features of the dissipation.  The experimentally important
Ohmic dissipation, described by the case $s=1$, limits the super-Ohmic
region $s > 1$ where the bath effects are assumed to be weak.

These dissipative mechanisms have been investigated by several studies,
in particular for the Ohmic case, showing that they can drive
QTs. In particular, by exploiting the quantum-to-classical mapping
into a statistical model with a further imaginary-time dimension,
and integrating out the
bath degrees of freedom, one obtains an effective action with
long-range interactions along the imaginary time. The simplest models
given by dissipative two-state (Ising) systems can be mapped into a
classical $1d$ Ising model with long-range interactions decaying as
$1/r^{1+s}$.  As shown by several studies (see, e.g.,
Refs.~\cite{Leggett-etal-87, AF-09, WRVB-09, GWDV-12, Vojta-12}), they
develop a phase transition between localized and delocalized phases.
New classes of phase transitions (generally different from those of
the isolated systems) are also observed for many-body
systems~\cite{WTS-05, WVTC-05, WT-05, SWT-04, PFGKS-04}, such as the
quantum Ising chain coupled to oscillator baths as in
Eqs.~\eqref{isenvh} and~\eqref{heham}, which can be mapped into a
classical $2d$ system with long-range interactions along one of the
directions.

We finally note that the dissipative mechanisms based on the coupling
with infinite sets of oscillators tend to suppress the critical
behavior at the QTs of isolated systems, analogously to the
dissipative perturbations described within the Lindblad framework. The
effects of the above oscillator baths at the QTs of isolated systems
may be treated within dynamic scaling frameworks analogous to those
outlined in this section.  However, unlike the Lindblad
frameworks, such mechanisms may themselves lead to QTs belonging to
different universality classes, corresponding to statistical systems
with long-range interactions. On the other hand, the nature of the
phase transition that dissipative Lindblad mechanisms may stabilize in
the stationary large-time limit is substantially different, more
similar to the finite-temperature transitions~\cite{DDLSS-13}.
For this reason, one does not expect phase transitions driven by
Lindblad dissipative mechanisms within $1d$ quantum models with
short-range interactions.

\section{Dissipation in the out-of-equilibrium Kibble-Zurek dynamics}
\label{KZdiss}

As already discussed in Sec.~\ref{KZdynamics}, slow passages through
QTs allow to probe some universal features of quantum fluctuations.
In this respect, we recall the KZ problem~\cite{Kibble-76, Zurek-85,
  ZDZ-05, CEGS-12}, related to the amount of final defects,
after slow (quasi-adiabatic) passages through CQTs, from the disordered
to the ordered phase.  Since dissipative mechanisms are expected
to give rise to relevant perturbations at the quantum criticality of
closed systems, such as the temperature, they do not generally
preserve the universal dynamic properties of QTs.  Therefore, the open
nature of quantum systems leads to a departure from the dynamic KZ
scaling behavior predicted for the isolated case~\cite{FFO-07,
  PSAFS-08, PASFS-09, NVC-15, DRD-16, GZYZ-17, KMSFR-17, SVPKD-17,
  HL-17, ABRS-18, GTD-19, PSHP-19, RV-20-kz, FFCQE-20, KAVY-21}.
In particular, it has been observed that slow quenches in open systems,
or subject to noisy controls, may generate an overabundance of defects,
when approaching the adiabatic limit in KZ protocols~\cite{Griffin-etal-12}.
Nonetheless, taking into account the role of dissipation is crucial
to understand the outcomes of experimental results for quantum
annealing protocols (see Sec.~\ref{sec:QAnnealing}),
as recently shown during the simulation of quasi-adiabatic passages
across the CQT of quantum Ising systems,
in the presence of a dissipative dynamics modeled by the coupling
to an Ohmic bath of harmonic
oscillators~\cite{Weinberg-etal-20, Bando-etal-20}.

Due to the general relevance of the perturbations associated with
dissipative mechanisms, slower protocols favor the dissipation
effects, in that they give them more time to act.  Therefore, unlike
closed systems, the dynamic behaviors arising from slow changes of the
Hamiltonian parameters, across their critical values, do not anymore
develop universal critical features controlled by the QT of the closed
system. One thus expects that only an appropriate tuning of the
dissipation strength may originate a nontrivial interplay with the
critical unitary dynamics.
This issue is best discussed within a dynamic scaling framework, based
on the extension of the scaling theory for KZ dynamics in closed
systems (Sec.~\ref{KZdynamics}), to allow for the dissipative
perturbations.  Open dissipative systems may still present a universal
regime controlled by the universality class of the QT of the closed
system, provided the system-environment interaction strength is
suitably tuned analogously to the quench dynamics in the presence of
dissipative perturbations, that we have already discussed in the
previous section~\ref{dissQT}.

\subsection{Dynamic Kibble-Zurek scaling for open quantum systems}
\label{dissint}

The KZ protocol in the presence of dissipative mechanisms is analogous
to that considered for closed systems, and discussed in
Sec.~\ref{KZprot}, with the only addition of dissipation during the
whole dynamics.  Namely, we assume again that the many-body
Hamiltonian depends on a parameter $w(t)$, which is linearly modulated
in time as in Eq.~\eqref{wtkz}, and can be written as
\begin{equation}
  \hat{H}(t) = \hat{H}_{c} + w(t) \hat{H}_{p}\,,\qquad w(t) = t/t_s\,.
  \label{KZprotham}
  \end{equation}
The many-body system is supposed to be initialized in the ground state
of $\hat H(t_i)$ at $t_i<0$ corresponding to $w_i = w(t_i) <0$.
Subsequently, at time $t>t_i$, it evolves according to the Lindblad
master equation~\eqref{eq:lind}, with the above time-dependent
Hamiltonian $\hat H(t)$ and with a fixed dissipation strength.

On these grounds, a dynamic KZ scaling theory allowing for the
dissipative perturbation can be obtained by extending the one for
closed systems outlined in Sec.~\ref{dynscaKZ}. For this purpose, one
should also allow for a further dependence of the various scaling
functions on a new scaling variable associated with the dissipation
parameter $u$ in the dynamic scaling relations~\eqref{scadynwt}.
In practice, this can be taken as a power law
$b^z \,u$, where the dynamic exponent $z$ ensures the substantial
balance (competition) with the critical coherent driving.

\subsubsection{Dissipative scaling in the thermodynamic limit}
\label{disskzscainfvol}

To obtain the dissipative KZ scaling in the thermodynamic limit, one
may again start from the homogeneous laws in terms of the arbitrary
length scale $b$, fix it as in Eq.~\eqref{sbla}, and then take the
limit $L/b\to\infty$. The net result are scaling equations analogous
to those reported in Sec.~\ref{dynkzinfL}, with a further dependence
on the scaling variable
\begin{equation}
  \Gamma_s = t_s^\kappa \,u \,,\qquad \kappa= {z\over y_w+z}\,,
  \label{hatgdef}
\end{equation}
associated with the dissipation strength $u$.
Thus, we obtain
\begin{equation}
  G_{12}(t,t_s,{\bm x};w_i,u) \approx \lambda^{-\varphi_{12}} \,
  {\cal G}_\infty({\bm x}/\lambda, \Omega_t,\lambda^{y_w} w_i,\Gamma_s)\,.
  \label{inflimdisG}
\end{equation}
This is expected to provide the asymptotic behavior in the
$t_s\to\infty$ limit, while keeping the scaling variables fixed,
including $\Gamma_s$. For $\Gamma_s\to 0$, we must recover the scaling
laws of the closed systems subject to the unitary evolution only.
More importantly, the above scaling law tells us that the dissipation
effects are expected to be negligible when $u \ll t_s^{-\kappa}$,
and dominant when $u \gg t_s^{-\kappa}$.

Note that, like for closed systems, the large-$t_s$ limit of KZ
protocols starting from finite and fixed $w_i<0$ corresponds to the
limit $\lambda^{y_w} w_i \to -\infty$. Indeed, the dissipation with
coupling strength $u\sim \lambda^{-z}$ is not expected to play any
relevant role at finite $w_i<0$, where the gap is $\Delta = O(1)$,
while it should compete with the unitary evolution only very close to
$w=0$ where $u\sim \Delta \sim \lambda^{-z}$.  Therefore, under such
conditions, the scaling behavior simplifies to
\begin{equation}
  G_{12}(t,t_s,{\bm x};w_i,u) \approx \lambda^{-\varphi_{12}} \, {\cal
    G}_\infty ({\bm x}/\lambda,\Omega_t,\Gamma_s)\,,
  \label{sinflimdisG}
\end{equation}
becoming independent on the initial $w_i$. The dissipative KZ scaling
limit also implies that the scaling law associated with the number of
defects, cf.~Eq.~\eqref{rhodef}, should be replaced with~\cite{RV-20-kz}
\begin{equation}
  \rho_{\rm def} \approx \lambda^{-d} \,
  \widetilde{\cal R}_{\rm def}(\Omega_t,\Gamma_s) =
  t_s^{-{d\over y_\mu+z}}\,{\cal R}_{\rm def}(\Omega_t,\Gamma_s)\,,
  \label{rhodefdis}
\end{equation}
where the dependence on the dissipative coupling $u$ enters the
scaling function ${\cal R}_{\rm def}$ through the scaling variable
$\Gamma_s$.  Of course, for $u=0$ (thus $\Gamma_s=0$) one recovers
the scaling law for closed systems [cf.~Eq.~\eqref{rhodef}].

\subsubsection{Dissipative scaling in the finite-size scaling limit}
\label{disskzfss}

Analogously to the dynamics of closed systems, it is possible to
derive scaling laws in the FSS limit, by fixing $b=L$ in the
homogeneous scaling laws, such as Eqs.~\eqref{scadynwt} with the
further dependence on $b^z \, u$.  The resulting scaling ansatz for
two-point correlation functions turns out to be analogous to that of
closed systems, i.e., Eq.~\eqref{kzfssG2}, with in addition a
dependence on the scaling variable $\Gamma=L^z \,u$ related to the
dissipative strength,
\begin{equation}
  G_{12}(t,t_s,{\bm x};L,w_i,u) = \big\langle \hat{O}_1({\bm x}_1) \,
  \hat{O}_2({\bm x}_2) \big\rangle_t \approx L^{-\varphi_{12}} \,
      {\cal G}_{12}(\Omega_t, {\bm X}, \Upsilon, \Omega_{t_i}, \Gamma)\,.
  \label{g12scadyndi2kz}
\end{equation}
Moreover, assuming again that the quantum phase for $w_i<0$ is
gapped, with $\Delta\sim\xi^{-z}$, and the ground-state length scale
$\xi$ diverges only at the critical point $w=0$, KZ protocols
associated with any finite initial $w_i<0$ develop the same dynamic
FSS independently of their actual values, corresponding to
$\Omega_{t_i}\to -\infty$, and therefore its dependence can be dropped
when considering a dynamic KZ scaling at fixed $w_i$.

\subsection{Results for the Kitaev quantum wire subject to dissipation}
\label{reskitaev}

\begin{figure}[!t]
  \begin{center}
    \includegraphics[width=0.95\textwidth]{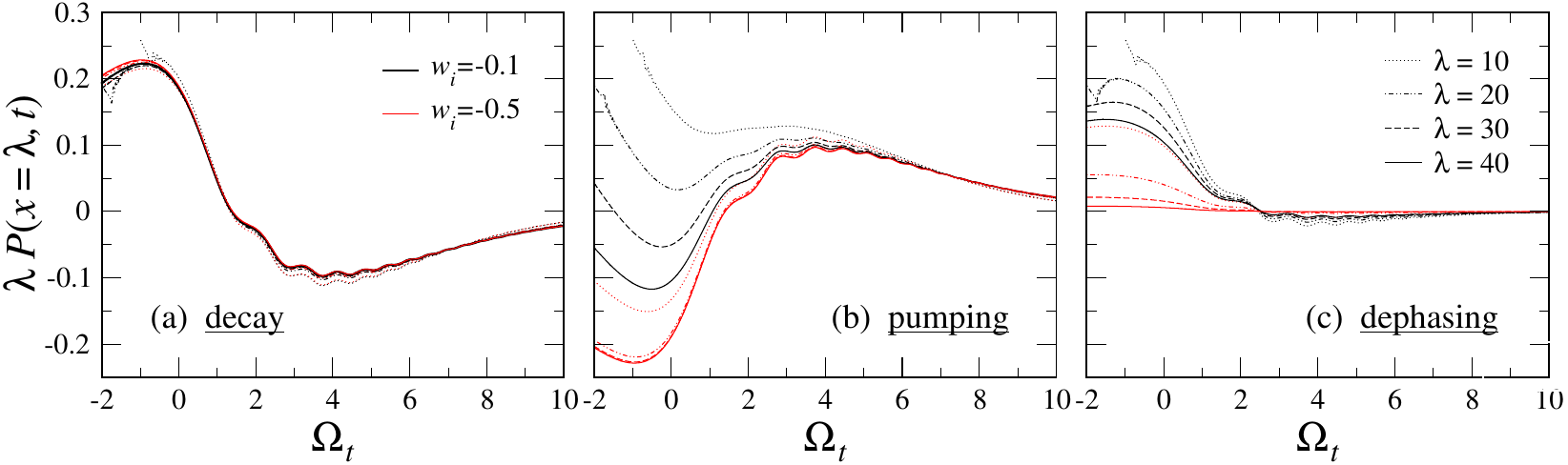}
    \caption{Rescaled correlations $\lambda \, P(x,t)$ at fixed
      initial $w_i$, for the dissipative Kitaev quantum wire in the
      thermodynamic limit, as a function of the scaling variable
      $\Omega_t$.  Here we fix $x/\lambda=1$ and $\Gamma_s =
      0.5$. Each panel refers to a specific type of dissipation
      mechanism [see Eq.~\eqref{loppe}]: decay $(a)$, pumping $(b)$,
      and dephasing $(c)$.  The color code refers to $w_i = -0.1$
      (black) and $w_i = -0.5$ (red).  Different line styles are for
      various values of $\lambda$, from $10$ to $40$ (see legend).
      Adapted from Ref.~\cite{RV-20-kz}.}
    \label{fig:PC_TL_diss_mu}
  \end{center}
\end{figure}

Accurate numerical checks of the dissipative KZ scaling behaviors have
been reported for the Kitaev quantum wire in the presence of local
dissipative mechanisms, such as those already considered in
Sec.~\ref{kitaevmod}, related to local pumping, decay, and dephasing
dissipative mechanisms [cf.~Eqs.~\eqref{loppe}].  As already mentioned,
this model is amenable to a direct solvability for systems with
$O(10^3)$ sites, thus representing the ideal playground for open
quantum lattice problems, given the remarkable difficulty to simulate
the dynamics of many-body quantum systems coupled to an external bath.
To monitor the dynamics arising from the dissipative KZ protocol, we
consider again the fixed-time correlation functions defined in
Eq.~\eqref{gpcntf}. They are expected to show the dissipative KZ
scaling reported in Eq.~\eqref{sinflimdisG} in the thermodynamic limit,
and in Eq.~\eqref{g12scadyndi2kz} in the FSS limit, similarly to
the dissipative scaling in quench protocols (see Sec.~\ref{kitaevmod}).

\begin{figure}
  \begin{center}
    \includegraphics[width=0.65\columnwidth]{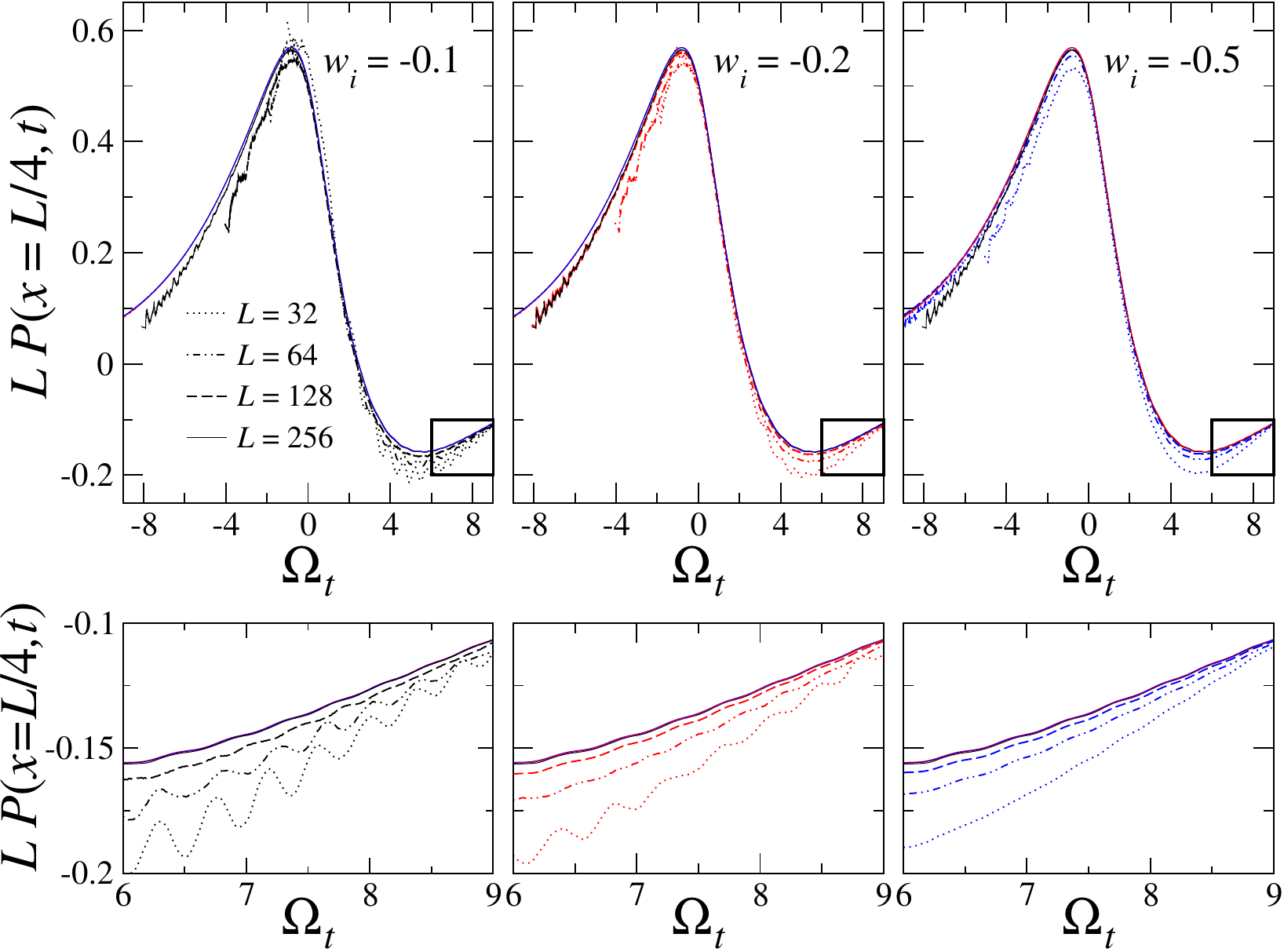} 
    \caption{The rescaled correlation $L \, P(x,t)$, with $x/L = 1/4$,
      as a function of $\Omega_t$, fixing the initial Hamiltonian
      parameter $w_i = -0.1$ (left panels, black curves), $-0.2$
      (central panels, red curves), and $-0.5$ (right panels, blue
      curves).  Different line styles stand for various $L$ (see
      legend).  The lower panels show magnifications of the upper
      ones, for $6 \leq \Omega_t \leq 9$.  We fix $\Upsilon = 0.1$ and
      $\Gamma = 1$, with dissipation given by incoherent decay.  The
      three continuous curves, corresponding to the largest size
      $L=256$ and different $w_i$, are plotted in each of the six
      panels for reference, and cannot be distinguished on the scale
      of the plots reported. Adapted from Ref.~\cite{RV-20-kz}.}
    \label{P_MuIn}
  \end{center}
\end{figure}

Fig.~\ref{fig:PC_TL_diss_mu} shows some numerical outcomes for the
correlation function $P(x,t)$ in the thermodynamic limit,
cf.~Eq.~\eqref{eq:corrP}, for a fixed value of $\Gamma_s$, two
different values of $w_i=\mu_i-\mu_c$, and various values of
$\lambda$.  The dephasing seems to be the most disruptive dissipation
mechanism.  However in all cases we observe that, with increasing
$\lambda$, they approach the same asymptotic behavior, irrespective of
the choice of $w_i$; for the incoherent decay, this appears to be much
faster than for the other kinds of dissipation.  It can be also shown
that the convergence to the asymptotic behavior with $\lambda$, for
given $\Omega_t$, is power law as expected.
Results for the same correlation function at finite size are provided
in Fig.~\ref{P_MuIn}, supporting the dynamic FSS put forward in
Eq.~\eqref{g12scadyndi2kz}. The various panels, for different initial
values of $w_i$, show convergence with $L$ to a scaling function,
which appears to be the same in all cases: the magnified bottom panels
at large $\Omega_t$ unveil how discrepancies in the temporal
behavior, starting from different $w_i$, get suppressed in the
large-$L$ limit.

We emphasize that, to obtain the correct asymptotic KZ scaling
behavior, it is crucial to allow for the dissipation to be suitably
rescaled in the large-$t_s$ limit.  That is, the parameter $u$
entering the dissipator in the Lindblad master
equation~\eqref{lindblaseq} must be scaled as $u \sim t_s^{-\kappa}$,
meaning that the scaling variable $\Gamma_s = u \, t_s^\kappa$ has to
be kept fixed.  In contrast, if the dissipation strength $u$ is not
changed with $t_s$, no dynamic scaling can be observed using KZ-like
protocols~\cite{RV-20-kz}.

\section{Measurement-induced dynamics at quantum transitions}
\label{measQT}

In general, while the unitary evolution enhances the entanglement,
measurements of observables disentangle degrees of freedom and thus
tend to decrease quantum correlations, similarly to decoherence.  A
{\em projective quantum measurement} is physically realized when the
interaction with a macroscopic object makes a quantum mechanical
system rapidly collapse into an eigenstate of a specific operator, and
the resulting time evolution appears to be a non-unitary projection.
Such process is referred to as a projective
measurement~\cite{Zurek-03, vonNeumann-book}.  When the system is
projected into an eigenstate of a local operator, the corresponding
local degree of freedom is disentangled from the rest of the system.
Moreover, if projective measurements are performed frequently, the
quantum state gets localized in the Hilbert space near a trivial
product state, leading to the quantum Zeno effect~\cite{MS-77, FP-08}.
In strongly correlated systems, QTs give rise to a peculiar dynamic
regime where long-range correlations set in; the impact of projective
quantum measurements on the decay rate of quantum correlations should
be thus substantially affected by the emerging critical dynamics.

\begin{figure}
  \begin{center}
    \includegraphics[width=0.6\columnwidth]{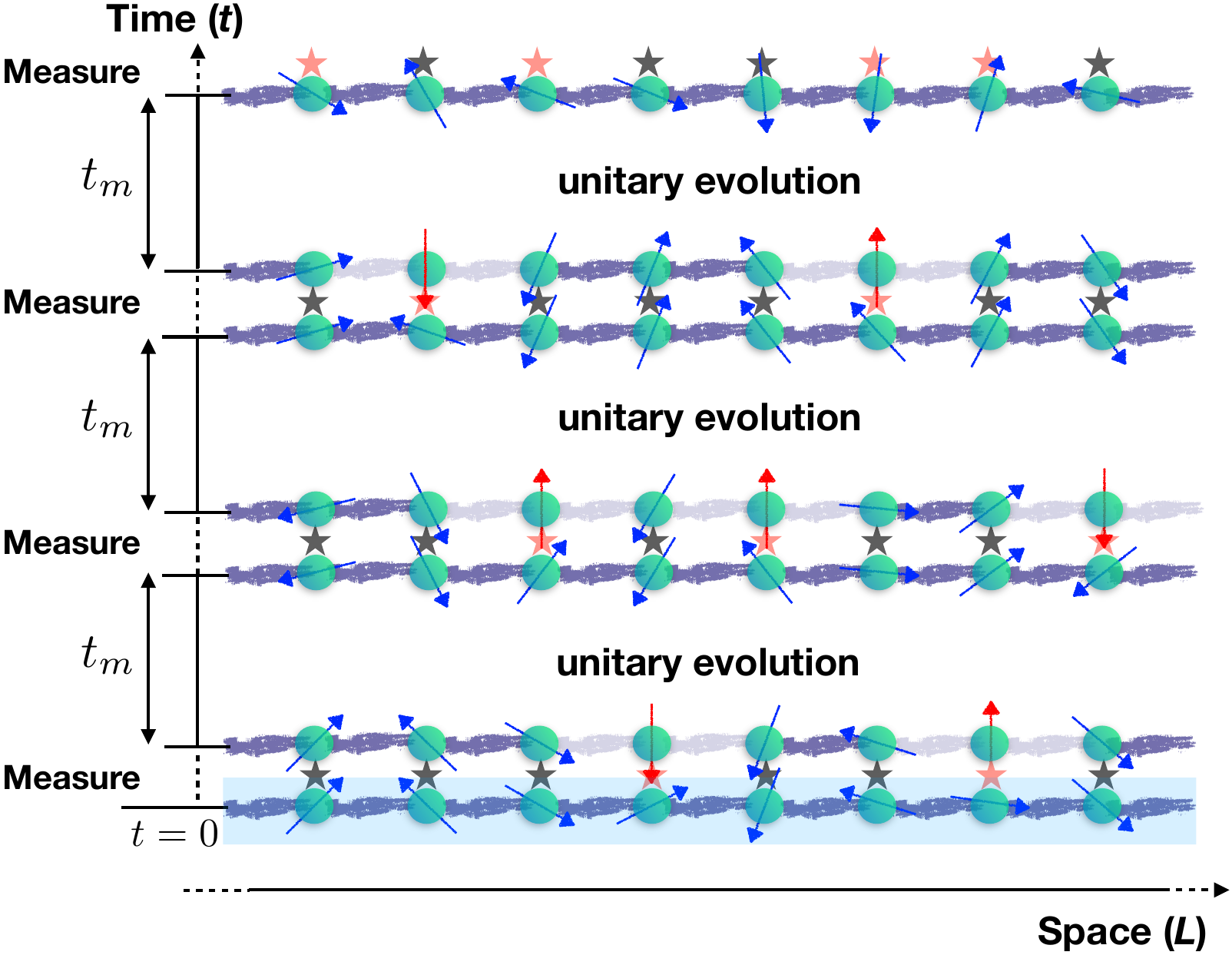}
    \caption{Sketch of a quantum-measurement protocol: A quantum spin
      system, initially frozen in its ground state at quantum
      criticality ($t=0$), is perturbed with local projective
      measurements (stars) occurring after every time interval $t_m$,
      with a homogeneous probability $p$ per site. In between two
      measurement steps, the system evolves unitarily, according to
      its Hamiltonian.  Red stars denote the occurrence of a
      measurement on a given site (for the sake of clarity, in the
      figure we consider $\hat \sigma^{(3)}$-type measures: spins
      colored in red are projected along the $z$-axis).
      From Ref.~\cite{RV-20-meas}.}
    \label{sketch}
  \end{center}
\end{figure}

To address the interplay of unitary and projective dynamics in
experimentally viable many-body systems at QTs, such as quantum spin
networks, one may consider dynamic problems arising from protocols
combining the unitary Hamiltonian and local measurement drivings (for
a cartoon, see Fig.~\ref{sketch}).  In such conditions, different
regimes emerge, depending on the measurement schemes and their
parameters.  If every site were measured during each projective step,
the system would be continually reset to a tensor product state. More
intriguing scenarios should hold when local measurements are spatially
dilute.

Most of the work done in this context focuses on the investigation of
entanglement transitions genuinely driven by local measurements,
either in random circuits~\cite{LCF-18, CNPS-19, SRN-19, LCF-19, SRS-19,
  GH-20-prx, BCA-20, JYVL-20, GH-20-prl, Zabalo-etal-20, BBCAY-21},
in the BH model~\cite{TZ-20, GD-20}, in quantum spin
systems~\cite{DD-16, LB-20, BS-20, TBFDS-21, BDM-21, Sierant-etal-21},
or in free-fermionic chains~\cite{CTD-19, MDB-21, ZLJC-21, ABD-21, MSKS-21},
and on measurement-induced state preparation~\cite{RCGG-20}.
In noninteracting models, continuous local measurements were shown
to largely suppress entanglement. Here we review a substantially different
dynamic problem, related to the effects of local random measurements
on the quantum critical dynamics of many-body systems, i.e., when a QT
is driven by the Hamiltonian parameters~\cite{RV-20-meas}.

Different regimes arise from the interplay between unitary and
projective dynamics in critical systems.  One of them is dominated by
local random measurements, for example for any finite probability $p$
of making the local measurement.  In contrast, for sufficiently small
$p$ values (decreasing as a sufficiently large power of the inverse
diverging length scale $\xi$), the measurements are irrelevant.  These
two regimes turn out to be separated by dynamic conditions leading to
peculiar dynamic scaling behaviors, controlled by the universality
class of the QT.  For a $d$-dimensional critical system, scaling
arguments lead to the exact power law 
\begin{equation}
  p\sim \xi^{-(z+d)}\,.
  \label{plawsca}
\end{equation}

Therefore, local measurements generally suppress quantum correlations,
even in the dynamic scaling limit, with scaling laws that are
qualitatively different when being far from criticality.  The
corresponding time scale at a QT is indeed expected to behave as
\begin{equation}
  \tau_m\sim \xi^z\sim p^{-\kappa}\,,\qquad \kappa={z\over z+d}<1\,,
  \label{taumsca}
\end{equation}
to be compared with the noncritical case $\tau_m \sim p^{-1}$.

\subsection{Measurement protocols for quantum Ising models}
\label{sec:measure}

We discuss quantum lattice spin systems, assuming that only one
relevant parameter $w$ of the Hamiltonian $\hat H(w)$ (with RG
dimension $y_w>0$) can deviate from the QCP, located at $w_c=0$. For
example, we may consider the paradigmatic $d$-dimensional quantum
Ising Hamiltonian~\eqref{hisdef} of size $L^d$ with PBC, at its CQT
point driven by the relevant parameters $r=g-g_c$ and $h$, which we
may identify with the driving parameter $w$.  The following
measurement protocol can be adopted:
\begin{itemize}
\item[$\bullet$] The system is initialized, at $t=0$, in the ground
  state $|\Psi_0(w)\rangle$ close to criticality, thus $|w| \ll 1$.

\item[$\bullet$] Random local projective measurements are performed at
  every time interval $t_m$, so that each site has a (homogeneous)
  probability $p$ to be measured. For the quantum Ising model, one may
  consider local measurements of the spin components, along transverse
  [$\hat \sigma_{\bm x}^{(3)}$] or longitudinal [$\hat \sigma_{\bm
      x}^{(1)}$] directions.

\item[$\bullet$] The updated many-body wave function is obtained by
  projecting onto the spin state at site ${\bm x}$ corresponding to
  the measured value of the spin component, and normalizing it.
  
\item[$\bullet$] In between two measurement steps, the system evolves
  according to the unitary operator $e^{-i \hat H(w) t}$.
\end{itemize}
If $p\to 1$, each spin gets measured every $t_m$, and the effects of
projections are expected to dominate over those of the unitary
evolution.  In contrast, for $p$ sufficiently small, the time
evolution may result unaffected by measurements. In between these two
regimes, there is a competing unitary vs.~projective dynamics,
characterized by controllable dynamic scaling behaviors associated
with the universality class of the QT.

The evolution along the above protocol can be monitored by fixed-time
averages of observables, as the longitudinal magnetization $M(t)$ and
the susceptibility $\chi(t)$,
\begin{equation}
  m(t) = L^{-d} \sum_{\bm x} \big\langle \hat \sigma_{\bm x}^{(1)}
  \big\rangle_t \,, \qquad \chi(t) = L^{-d} \sum_{{\bm x},{\bm y}}
  \big\langle \hat \sigma_{\bm x}^{(1)} \, \hat \sigma_{\bm y}^{(1)} 
\big\rangle_t \,,
\end{equation}
averaging over the trajectories. The time scale $\tau_m$ of the
suppression of quantum correlations can be estimated from the halving
time of the ratio
\begin{equation}
  R_\chi(t)\equiv \frac{\chi(t)-1}{\chi(t=0)-1} \,, \qquad  
R_\chi(t) \in [0,1] \,,
\end{equation}
which is expected to go from one ($t=0$) to zero ($t\to\infty$).
Indeed note that, since projective measurements generally suppress
quantum correlations, $\lim_{t\to \infty}\chi(t) = 1$, corresponding
to an uncorrelated state.

\subsection{Phenomenological dynamic scaling theory}
\label{phedynscathe}

We now present a phenomenological scaling theory for the
out-of-equilibrium dynamics arising from random local projective
measurements during the evolution of a many-body system at a QT.  The
starting point is a homogeneous scaling law allowing for the
measurement process, and in particular for the parameters associated
with measurement protocol. We write
\begin{equation}
  O(t;L,w;t_m,p) \approx 
  b^{-y_o} \, {\cal O}(b^{-z} t, b^{-1} L, b^{y_w} w,
  b^{\zeta} t_m ,b^{\varepsilon} p)\,,
  \label{sdynscab}
\end{equation}
where $\zeta$ and $\varepsilon$ are appropriate exponents associated
with the measurement process.  The above scaling equation is supposed
to provide the power-law asymptotic behavior in the large-$b$ limit,
neglecting further dependences on other parameters, which are supposed
to be suppressed (and thus irrelevant) in such limit.

The arbitrariness of the scale parameter $b$ in Eq.~\eqref{sdynscab}
can be fixed by setting $b = \lambda \equiv |w|^{-1/y_w}$, where
$\lambda \sim \xi$ is the length scale of critical modes.  The scaling
variable associated with the time interval $t_m$ should be given by
the product $\Delta \, t_m$, where $\Delta^{-1}\sim \lambda^z$ is the time
scale of the critical models (this implies $\zeta=-z$).  Keeping $t_m$
fixed in the large-$\lambda$ limit, the dependence on $t_m$ disappears
asymptotically, originating only $O(\lambda^{-z})$ scaling corrections.
Moreover, noticing that $p$ is effectively a probability per unit of time
and space, a reasonable guess would be its correct scaling to compete
with the critical modes, that is $p\sim \lambda^{-z-d}$, thus
\begin{equation}
  \varepsilon = z+d\,.
  \label{kappaguess}
\end{equation}
This leads to the dynamic scaling equation
\begin{equation}
  O(t;w;t_m,p) \approx
  \lambda^{-y_o} \, {\cal O}(\lambda^{-z} t, \lambda^\varepsilon p) \,.
  \label{oscala}
\end{equation}
The value of $\varepsilon$ in Eq.~\eqref{kappaguess} is crucial, since
it provides the condition that separates the measurement-irrelevant regime
$p = o(\lambda^{-\varepsilon})$ from the measurement-dominant regime
$\lambda^{\varepsilon} \,p\to\infty$.  Since
$p \sim \lambda^{-\varepsilon}$ and $t \sim \lambda^z$, the dynamic scaling
ansatz predicts that the time scale $\tau_m$ associated with the
suppression of quantum correlations behaves as
$\tau_m\sim p^{-\kappa}$ with $\kappa = z/\varepsilon<1$.

The above scaling theory holds in the thermodynamic limit
$L/\lambda\to\infty$, that is expected to be well defined for any $w
\neq 0$, for which $\lambda$ is finite.  Nonetheless, for most
practical purposes, both experimental and numerical, one typically has
to face with systems of finite length.  Such situations can be framed
in the FSS framework, where the scale parameter in
Eq.~\eqref{sdynscab} is set to $b=L$.  Fixing again $t_m$, one obtains
the scaling law
\begin{equation}
  O(t;L,w;t_m,p) \approx  
  L^{-y_o} \, {\cal O}( \Theta, \Phi, L^{\varepsilon} p)\,,
  \label{dynfssmeas}
\end{equation}
so that the proper dynamic FSS behavior is obtained in the
thermodynamic limit, taking the scaling variables $\Theta=L^{-z} \,t$,
$\Phi=L^{y_w} \,w$, and $L^{\varepsilon} \,p$ fixed.

Analogous scaling ansatze for more general observables, as fixed-time
correlation functions, are obtainable with the same arguments and
assumptions.  They can be extended to include an initial quench of the
Hamiltonian parameter $w_i \to w$ [by adding a further dependence on
  $b^{y_w} \, w_i$ in Eq.~\eqref{sdynscab}, corresponding to the
  further scaling variable $\Phi_i = L^{y_w} \, w_i$ in
  Eq.~\eqref{dynfssmeas}], to consider finite-temperature initial
Gibbs states [by adding a dependence on $b^z \,T$ in
  Eq.~\eqref{sdynscab}, corresponding to the further scaling
  variable $\Xi = T/\Delta(L)$ in Eq.~\eqref{dynfssmeas}], and also
allowing for weak dissipation.  Note also that the scaling arguments
do not depend on the type of local measurement, therefore they are
expected to be generally independent of them, to some extent.

\subsection{Numerical evidence in favor of the dynamic scaling theory}
\label{numevi}

\begin{figure}
  \begin{center}
    \includegraphics[width=0.47\columnwidth]{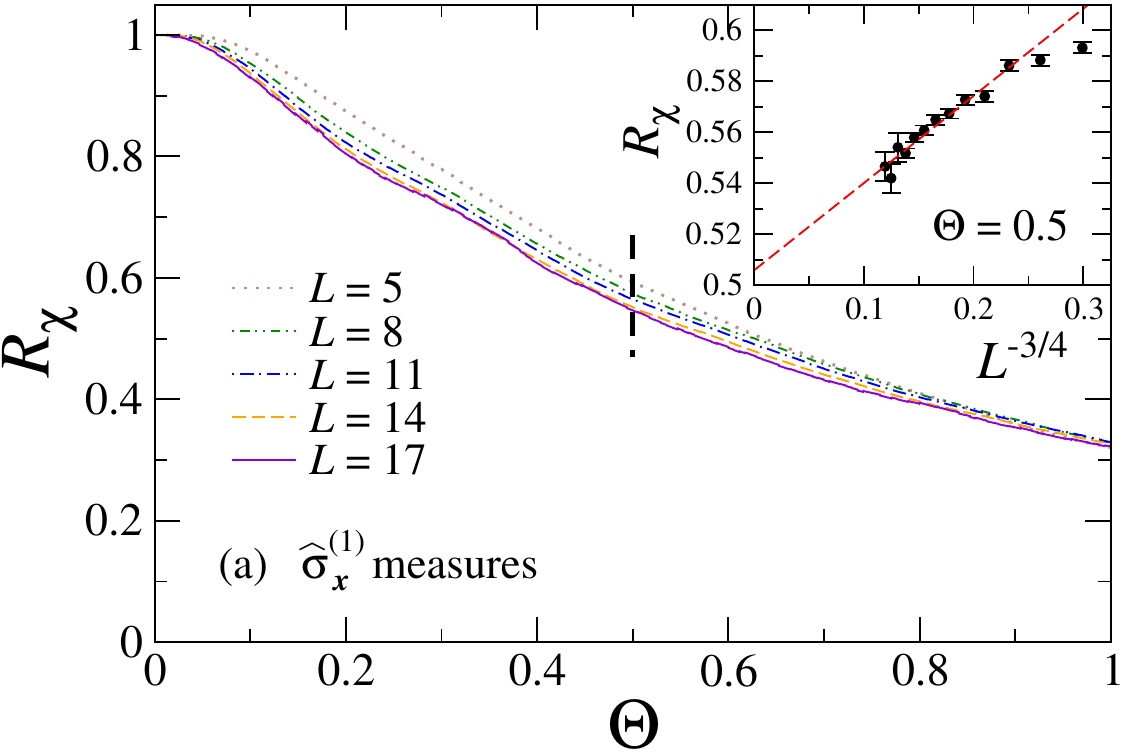}
    \hspace*{5mm}
    \includegraphics[width=0.47\columnwidth]{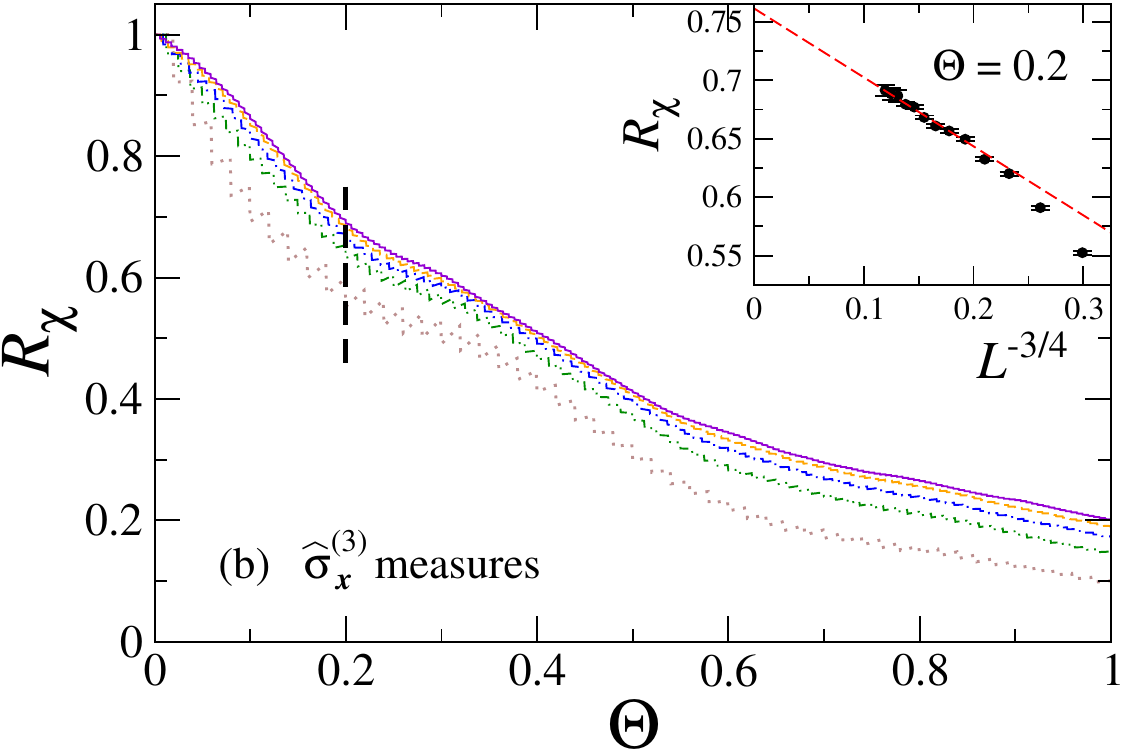}
    \caption{Time behavior of $R_\chi$ in the quantum Ising chain at
      criticality, for various sizes.  Random measurements are either
      along the longitudinal [panel $(a)$] or the transverse [panel
        $(b)$] direction.  Here $t_m = 0.1$, while $p=1/L^2$ has been
      fixed according to the guess~\eqref{kappaguess}.  The two insets
      display data for specific values of $\Theta$ (dashed lines in the
      main frames), showing that the convergence to the asymptotic
      behavior is compatible with a $O(L^{-3/4})$ approach (dashed red
      lines).  Adapted from Ref.~\cite{RV-20-meas}.}
    \label{fig:susc}
  \end{center}
\end{figure}

The above phenomenological scaling theory has been checked on the
quantum Ising chain.  The dynamic FSS laws for the magnetization and
its susceptibility follow Eq.~\eqref{dynfssmeas}, in which the
parameter $w$ corresponds to either $r = g-g_c$ or $h$ in
Eq.~\eqref{hedef}.  In particular, for $r = h = 0$, one obtains
$m(t)=0$ by symmetry, and
\begin{equation}
  R_\chi(t;L;t_m,p) \approx {\cal R}_\chi(L^{-z}t,L^\varepsilon p)\,.
  \label{rchiising}
\end{equation}
Results for a system at its QCP, with random local longitudinal and
transverse spin measurements, are shown in Fig.~\ref{fig:susc}.  Data
for $R_\chi$ versus $\Theta = L^{-z} \,t$ nicely agree with
Eq.~\eqref{rchiising}, and corrections to the scaling are consistent
with a power-law suppression as $L^{-3/4}$ approach, as
expected~\cite{RV-20-kz}.  Analogous scaling results are obtained in
the small-$t_m$ limit, and also for the magnetization at $h \neq 0$,
keeping $\Phi = L^{y_h} \,h$ constant.

The above scaling behaviors can be compared with situations far from
QTs, for example in the quantum disordered phase $g > g_c$ of quantum
Ising models, where the length scale $\xi$ of quantum correlations and
the gap $\Delta$ remain finite with increasing $L$.  Away from
criticality, the characteristic time $\tau_m$ of the measurements
scales as $\tau_m\sim p^{-1}$, unlike the critical behavior, where
$\tau_m \sim p^{-\kappa}$ with $\kappa = z/\varepsilon<1$.  The
smaller power $\kappa$ at criticality (i.e., faster decay rate) can be
explained by the fact that the relevant probability ($p_r$) driving
the measurement process is the one to perform a local measurement
within the critical volume $\xi^d$, therefore $p_r=\xi^d \,p$.  The
time rate thus behaves as $\tau_m \sim p_r^{-1}$, similarly to the
noncritical case, where $\xi=O(1)$. Of course, this implies that the
critical conditions significantly speed up the decoherence process
arising from the local quantum measurements~\cite{RV-20-kz}.

\section{Outlook and Applications}
\label{applications}

We have presented a series of issues concerning the equilibrium and
out-of-equilibrium behavior of quantum many-body systems in proximity
of quantum phase transitions.  The competition between quantum
fluctuations, either at equilibrium or along the out-of-equilibrium
unitary dynamics, and decohering effects induced by the temperature,
the environment, or quantum measurements, probably represents the most
intricate dynamic regime where complex many-body phenomena may appear.
Notwithstanding the apparent difficulty in treating this kind of
situations, a universal behavior forcefully emerges if the system
stays close to a zero-temperature quantum transition, of both
continuous and first-order types.

We wrote this review with the purpose to collect and elucidate the
scaling behaviors that may naturally emerge in the thermodynamic and
the FSS limits, providing a unified framework where both continuous
and first-order quantum transitions can be quantitatively investigated
with a satisfactory accuracy.
The emerging dynamic scaling scenario is quite general, in the sense
that it is expected to hold for generic quantum transitions in any
spatial dimension.  Although, at the time of writing this review, most
of the out-of-equilibrium dynamic scaling arguments, and related
predictions, have been explicitly verified by computations within
quantum $1d$ Ising-like systems or Kitaev fermionic wires, the general
ideas should apply to other quantum many-body systems, as well. We
mention, for example, fermionic or bosonic Hubbard-like models, with
any geometry, and also in the presence of different dissipative
schemes, such as in a non-Markovian setting.  Moreover, one could
consider the presence of defects, with various possible geometries.
It would be thus tempting to investigate and carefully verify the
dynamic scaling behavior in other situations, where novel features may
arise.

\subsection{Experimental feasibility}

In view of the impressive experimental advances that have been
achieved in this new millennium era in quantum simulation and quantum
computing~\cite{CZ-12, GAN-14}, we hope this review will stimulate
further progress and foster the investigations in a field that,
according to us, can be legitimately considered as one of the great
challenges of contemporary physics.  In fact, we hope that the FSS
predictions described are amenable to a direct verification in the
laboratory, by means of small-size quantum simulators operating on a
limited amount of quantum objects (of the order of ten, or few tens).

A number of remarkable experiments already demonstrated their
capability to faithfully reproduce and control the dynamics of
many-body quantum systems, such as Ising- or Hubbard-like models.
Some platforms are best suited for specific conditions, however the
cross-fertilization between the various available possibilities is
likely to provide us further precious insight in this fascinating
field of research in physics.  Among the others, we mention the
following platforms:
\begin{itemize}
\item
  Ultracold atoms in artificial optical lattices~\cite{BDZ-08,
    Bloch-08, BDN-12}.  Optical lattices are formed by the
  interference pattern of counter-propagating laser beams in different
  directions, which create a regular pattern mimicking the crystalline
  structure of a solid. Atoms feel the periodic potential landscape
  induced by lasers via the Stark shift.  This kind of setup is
  extremely clean and controllable, such that the dynamics of isolated
  quantum many-body systems can be studied, keeping coherence for a
  significant amount of time.  In practice, the Hamiltonian parameters
  can be tuned close to quantum transition points and also easily
  varied in time.  Single-site addressability is not a severe
  concern~\cite{BGPFG-09, Bakr-etal-10, Sherson-etal-10}.  For all
  these reasons, cold atoms are one of the best options to study
  quantum many-body systems at equilibrium and in nondissipative
  out-of-equilibrium conditions (see, e.g., Refs.~\cite{GMEHB-02,
    KWW-04, KWW-06, Endres-etal-11, Cheneau-etal-12, Gring-etal-12,
    GUJTE-13, Mancini-etal-15, Schreiber-etal-15}).  Dissipation can
  be included by suitable bath engineering and/or using Rydberg
  atoms~\cite{MDPZ-12}, which also enable to tune the range of
  interactions~\cite{Bernien-etal-17}.
 
\item
  Trapped ions~\cite{BR-12, Monroe-etal-21}.  It is possible to form
  ion crystals by balancing the Coulomb repulsion between ions and the
  trap confinement force, allowing them to be accurately controlled
  and manipulated, as well as to produce long-range and tunable
  interactions.  They already proved their great potential in the
  simulation of Ising-like spin systems, where each spin is
  represented by suitable internal energy levels of a single
  ion~\cite{Edwards-etal-10, Kim-etal-10, Kim-etal-11, Islam-etal-11,
    Zhang-etal-17, LMD-11, Richerme-etal-14}.  Ions can be addressed
  and measured locally, at the level of a fully-programmable quantum
  simulator~\cite{Debnath-etal-16}.  The range of interactions and the
  addition of disorder can be adjusted in a controllable
  way~\cite{Smith-etal-16, Zhang-etal-17a}.
  
\item
  Nuclear and electronic spins in solid-state structures~\cite{BW-08,
    HA-08, Yao-etal-12, PGLMV-13}.  They demonstrated their
  versatility in reproducing the physics of non-equilibrium many-body
  quantum systems and of quantum criticality, using different
  platforms. We mention:
  \begin{itemize}
  \item
    Nuclear magnetic resonance, which permits to manipulate a dozen of
    interacting spins, showing long coherence times and a high degree
    of controllability~\cite{YZL-07b, ZPRS-08, ZYLAB-12}.
  \item
    Nitrogen vacancy centers in diamond~\cite{Yao-etal-12, Choi-etal-17}.
  \item
    Arrays of semiconductor quantum
    dots~\cite{Singha-etal-11, Hensgens-etal-17}.
  \end{itemize}
  
\item
  Coupled quantum electrodynamics (QED).  Using hybrid platforms, one
  can induce strong effective interaction between photons~\cite{TF-10,
    CC-13, CVL-14, Hartmann-16, NA-16}.  Collective light-matter
  excitations in these systems are called polaritons. Dissipation is
  naturally present in the photonic components of the polaritons,
  therefore it is possible to simulate archetypal lattice models under
  driven nonequilibrium conditions. Different experimental platforms
  have been taken into account:
  \begin{itemize}
    \item
      Photonic crystals or fibers.  The confinement of photons in
      optical systems, as fibers or arrays of microcavities filled
      with atoms, enables the generation of large, tunable
      nonlinearities.  It is thus possible to simulate strongly
      interacting quantum many-body systems as Hubbard-like models or
      spin chains~\cite{HBP-07, Chang-etal-08, Douglas-etal-15}.
    \item
      Semiconductor microcavities. The confined photons are strongly
      coupled to electronic excitations, leading to the creation of
      exciton polaritons.  So far, most of the experimental results
      have been obtained in the mean-field regime, originating from
      the collective behavior of a large number of polaritons,
      described via a single wave function.  Macroscopic quantum
      coherence has been observed in nonequilibrium Bose-Einstein
      condensation~\cite{Kasprzak-etal-06} or polariton
      superfluidity~\cite{Amo-etal-09}.
    \item
      Circuit QED~\cite{HTK-12}.  Superconducting circuits are among
      the most promising platforms to engineer quantum simulators and
      quantum computers, being easily scalable and highly flexible in
      the nanofabrication.  Photonic modes are realized from a
      co-planar transmission line or an LC circuit acting as a
      microwave cavity; nonlinearities can be induced through
      Josephson junctions.  Several aspects of the many-body quantum
      physics, including phase transitions, quantum criticality,
      Kibble-Zurek physics, many-body localization, have been
      addressed (see, e.g., Refs.~\cite{Barends-etal-15, Roushan-etal-17,
        FSLKH-17, Harris-etal-18, Hamerly-etal-19}).
    \item
      Coupled optomechanical cavities~\cite{LM-13, AKM-14}.  Inside
      each cavity, a localized mechanical mode interacts with a
      laser-driven cavity mode via radiation pressure; neighboring
      cavities are effectively coupled through photon and phonon
      hopping.
  \end{itemize}
\end{itemize}

\appendix

\section*{Acknowledgments}
We thank Luigi Amico, Claudio Bonati, Pasquale Calabrese, Massimo
Campostrini, Giacomo Ceccarelli, Francesco Delfino, Giovanni Di
Meglio, Rosario Fazio, Mihail Mintchev, Jacopo Nespolo, Davide Nigro,
Haralambos Panagopoulos, Andrea Pelissetto, Subir Sachdev, and
Christian Torrero for collaborating with us on some issues considered
in this review.
We also thank Adolfo del Campo and Subir Sachdev for useful suggestions.

\bibliography{bibfile}

\end{document}